# Small Price Changes, Sales Volume, and Menu Cost*

Doron Sayag
Department of Economics, Bar-Ilan University
Ramat-Gan 5290002, Israel
Doronsayag2@gmail.com

Avichai Snir
Department of Economics, Bar-Ilan University
Ramat-Gan 5290002, Israel
Snirav@biu.ac.il

Daniel Levy**
Department of Economics, Bar-Ilan University
Ramat-Gan 5290002, Israel,
Department of Economics, Emory University
Atlanta, GA 30322, USA,
ICEA, ISET at TSU, and RCEA
Daniel.Levy@biu.ac.il

Revised: February 29, 2024

\* This is a substantially revised version of the manuscript that was presented at the 2021 ECB-Federal Reserve Bank of Cleveland Conference on Inflation: Drivers and Dynamics, at the 2021 ICEA's 3rd Warsaw Money-Macro-Finance Conference, at the 2022 annual conference of the Royal Economic Society, at the 2022 annual conference of the Armenian Economic Association, at the 2023 annual conference of the Israeli Economic Association, at the 2023 International Conference on Empirical Economics at Pennsylvania State University, and at the Bank of Israel Research Seminar. We thank the participants of these conferences for their helpful and constructive comments and suggestions. All errors are ours.

Declarations of interest: None

\*\* Corresponding author: Daniel Levy, Daniel.Levy@biu.ac.il

# Small Price Changes, Sales Volume, and Menu Cost

### *Abstract*

The finding of small price changes in many retail price datasets is often viewed as a puzzle. We show that a possible explanation for the presence of small price changes is related to sales volume, an observation that has been overlooked in the existing literature. Analyzing a large retail scanner price dataset that contains information on both prices and sales volume, we find that small price changes are more frequent when products' sales volume is high. This finding holds across product categories, within product categories, and for individual products. It is also robust to various sensitivity analyses such as measurement errors, the definition of "small" price changes, the inclusion of measures of price synchronization, the size of producers, the time horizon used to compute the average sales volume, the revenues, the competition, shoppers' characteristics, etc.

## 1. Introduction

Extensive empirical analyses of price-setting behavior using various micro-level price datasets show that individual prices tend to change at a significantly lower frequency than the corresponding market conditions.[1] A leading model offered to explain the sluggish response of prices to underlying shocks is the menu cost model, which posits that each time a firm changes a price, it incurs a lump sum cost that is independent of the size or the direction of the price change.[2]

A key prediction of the simple menu cost model is that firms make infrequent but relatively large price changes because making frequent small price changes is less economical (Caplin and Spulber 1987). However, empirical studies find that 20%–44% of the observed price changes are small, which many authors see as evidence against the simple menu cost model.[3]

To reconcile the existence of small price changes with menu costs, Dotsey et al. (1999) model stochastic menu costs, which lead to small price changes when the realized menu cost is small. Lach and Tsiddon (2007), Klenow and Malin (2011), Midrigan (2011), Alvarez and Lippi (2014) and Alvarez et al. (2016) suggest economies of scope in price adjustments, allowing both small and large price changes, as long as the average price change is larger than the menu cost.[4] Chakraborty et al. (2015), Rotemberg (1982), and Chen et al. (2008) suggest that consumer inattention can explain small price changes.

In this paper, which is primarily empirical, we use a large scanner retail price dataset with over 98 million weekly observations to show that sales volumes can be another explanation for small price changes. The empirical evidence we present suggests that small price changes are significantly more likely for products with high sales volumes than for products with low sales volumes. This result is robust. It holds across product categories, within product categories, or at the level of individual products across stores.

---

[1] Examples include Carlton (1986), Cecchetti (1986), Lach and Tsiddon (1992, 1996, 2007), Kashyap (1995), Blinder et al. (1998), Slade (1998), Eden (2001, 2018), Dutta et al. (1999, 2002), Fisher and Konieczny (2000, 2006), Owen and Trzepacz (2002), Chevalier et al. (2003), Baharad and Eden (2004), Bils and Klenow (2004), Levy and Young (2004), Zbaracki et al. (2004), Álvarez et al. (2006), Dhyne et al. (2006), Knotek (2008, and forthcoming), Nakamura and Steinsson (2008), Campbell and Eden (2014), Kehoe and Midrigan (2015), Konieczny and Skrzypacz (2005, 2015), Gorodnichenko and Talavera (2017), Anderson et al. (2015, 2017), and studies cited therein. For older surveys, see Romer (1993), Weiss (1993), Taylor (1999), Willis (2003), and Wolman (2007). More recent surveys include Klenow and Malin (2011), Leahy (2011), and Nakamura and Steinsson (2013).

[2] See, for example, Barro (1972), Sheshinski and Weiss (1977 and 1992), Akerlof and Yellen (1985), Mankiw (1985), and Konieczny and Rumler (2006).

[3] See, for example, Bils and Klenow (2004), Nakamura and Steinsson (2008), Chen et al. (2008), Klenow and Kryvtsov (2008), Msidrigan (2011), Bhattarai and Schoenle (2011), Klenow and Malin (2011), and Gautier et al. (forthcoming).

[4] Alvarez et al. (2014), Eichenbaum et al. (2014), Cavallo and Rigobon (2016), and Cavallo (2018) suggest that many of the reported small price changes are due to measurement errors. Even these studies, however, find a non-negligible share of small price changes that cannot be explained by measurement errors.

It is also robust to various sensitivity analyses such as the definition of "small" price changes in relative vs. absolute terms, measurement errors, the time horizon used to compute the average sales volume, and the inclusion of controls for competition, markups, pricing zones, producers' size, etc.

Sales volumes as an explanation for small price changes seem to have been overlooked by the existing literature, although it has a straightforward intuition. Under a menu cost (i.e., non-convex lump-sum price adjustment cost), the firm incurs the same price adjustment cost regardless of the number of units sold because it pays the menu cost once to change the price of all units sold. If it sells one unit, it will change the price only if the benefit from the change exceeds the menu cost. If it sells many units, the benefit from changing the price is accumulated across all units sold, while the menu cost is the same, which will likely make a small price change more profitable. Thus, comparing products that differ only in sales volume, we expect that products with higher sales volumes would have more price changes and that their price changes would be smaller, on average, than products with lower sales volumes. In the appendix, we show that the data is consistent with both predictions.

Particularly relevant to our work are Bhattarai and Schoenle (2014) and Kang and Usher (2023). Bhattarai and Schoenle (2014) use the BLS micro-level price data underlying the PPI to show that the average size of price changes is negatively correlated with the number of products offered by a producer. They also present evidence suggesting that producers that offer many products tend to have high sales volume.

We check if our results may be driven by large firms having higher sales volume or more price changes. We find that controlling for (a) the number of products offered by each producer, (b) the percentage of prices that change, and (c) the average size of all price changes (excluding the current price change), has little effect on the estimated coefficients of the sales volume.

Kang and Usher (2023) construct a model based on the assumption that the size of price changes is negatively correlated with the revenue and therefore, small price changes are possible if the revenue is sufficiently large. Because there is a strong correlation between sales volume and revenues (in our data, the average correlation is 0.85), our model and findings are consistent with their model and findings. Indeed, when we test the correlation between the likelihood of small changes and revenue, we find a positive and significant correlation. However, our empirical results further suggest that the positive correlation between revenue and small price changes is driven by the sales volume component of the revenue, and not by the price component.

Our results are likely to hold in other datasets as well. For example, the strong correlation between sales volume and revenue suggests that our results are likely to hold in Kang and Usher's (2023) data. In addition, although the Bhattarai and Schoenle (2014) data does not allow direct analysis of the correlation between small price changes and sales volume, the observation that our results hold along with their results, suggests that such a correlation exists in their data.

Further, our results suggest that a lack of correlation between the price gap and the likelihood of a price change, as reported by Karadi et al. (forthcoming), for example, is not necessarily evidence against selection. If the likelihood of a small price change depends on sales volumes, then retailers might select to adjust prices of products with high sales volumes even if the deviation of their price from the optimal price is small. Thus, even when selection is present, in the wake of a monetary shock we are likely to observe both large and small price changes.

We proceed as follows. To motivate our empirical analyses, in section 2, we extend Barro's (1972) model to derive a relationship between the width of the $(S, s)$ band and the sales volume. In section 3, we discuss the data. In section 4, we present the empirical findings. In section 5, we discuss robustness. We conclude in section 6.

## 2. Sales volume and the width of the optimal $(S, s)$ band

To motivate the empirical analyses of the relationship between sales volumes and the prevalence of small price changes, we extend Barro's (1972) model. Although the model is highly stylistic, it is useful because one criticism of the canonical menu cost model is that it fails to predict small price changes. We show that conditional on sales volume, even this highly stylistic model can predict small price changes.

Following Barro (1972), consider a profit-maximizing monopolist producing a homogenous good. The linear demand and the quadratic cost functions are given by $Y = \alpha - \beta P + u$ and $C(Y) = a + bY + cY^2$, respectively, where $u$ is a symmetric demand disturbance/shifter, $C'(Y) > 0$, and $a, b, c, \alpha, \beta > 0$. The producer's maximization problem is thus given by:

$$\begin{cases} \max \left[ PY - \left( a + bY + cY^2 \right) \right] \\ \text{s.t. } Y = \alpha - \beta P + u \end{cases} \tag{1}$$

Setting $MR = MC$, and solving for $P$ and $Y$, we obtain

$$P^* = \left[\frac{\alpha + \beta\left(2c\alpha + b\right)}{2\beta\left(1 + c\beta\right)}\right] + \left[\frac{1 + 2c\beta}{2\beta\left(1 + c\beta\right)}\right]u \tag{2}$$

and

$$Y^* = \left[\frac{\alpha - \beta b}{2\left(1 + c\beta\right)}\right] + \left[\frac{1}{2\left(1 + c\beta\right)}\right]u \tag{3}$$

The second-order condition for a maximum is given by $1 + c\beta > 0$.

In the absence of a disturbance, i.e. if $u = 0$, the profit-maximizing output is given by

$$Y^*\big|_{u=0} = \frac{\alpha - \beta b}{2\left(1 + c\beta\right)} \tag{4}$$

where $\alpha - \beta b > 0$, which is required for the output to be positive in the disturbance-free equilibrium. We can think of $Y^*\big|_{u=0}$ as the expected output.

Following Barro (1972, p. 19), suppose that the value of the disturbance changes from 0 to $u$. Assuming that the firm continuously adjusts its price and output to the change in $u$, the resulting change in the firm's profit, as Barro shows, is given by

$$\begin{aligned}
\Delta\pi_{(0,u)} &= \int_0^u \left(\frac{d\pi}{du}\right) du \\
&= \int_0^u \left[P - C'(Y)\right] du \\
&= \left[\frac{\alpha - \beta b}{2\beta\left(1 + c\beta\right)}\right]u + \left[\frac{1}{4\beta\left(1 + c\beta\right)}\right]u^2
\end{aligned} \tag{5}$$

Next, assume that the firm's price is sticky, stuck at $\hat{P}$, which denotes the optimal price in the disturbance-free equilibrium, such that $d\hat{P}/du = 0$. Then, (2) implies that

$$\hat{P} = \frac{\alpha + \beta\left(2c\alpha + b\right)}{2\beta\left(1 + c\beta\right)} \tag{6}$$

We follow Barro (1972, p. 20) to assume that the disturbance is not "too small" or "too large", i.e., $u_{min} \le u \le u_{max}$. This is necessary to avoid the situations of no production, which will be the case if $u < u_{min}$, or a shortage, which will be the case if $u > u_{max}$. Then,

$$\begin{aligned}
\Delta\hat{\pi}_{(0,u)} &= \int_0^u \left(\frac{d\hat{\pi}}{du}\right) du \\
&= \int_0^u \left[\hat{P} - C'(\hat{Y})\right] du \\
&= \left[\frac{\alpha - \beta b}{2\beta\left(1 + c\beta\right)}\right]u - cu^2
\end{aligned} \tag{7}$$

The expression in (7) is the change in the profit when the disturbance value changes from 0 to $u$, but the firm does not adjust its price, i.e. when the price is stuck at $\hat{P}$.

The firm's profit gain, if it adjusts its price to the demand shock, is therefore given by

$$\Delta \pi_{(0,u)} - \Delta \hat{\pi}_{(0,u)} = \theta u^2 \qquad (8)$$

where

$$\theta = \frac{(1+2c\beta)^2}{4\beta(1+c\beta)} > 0 \qquad (9)$$

The expression in (8) can be interpreted as the loss the firm incurs for not adjusting its price in response to the demand shock. As Barro (1972, p. 20) notes, the symmetry of this loss means that what matters is the size of the demand shock, not its sign. It follows that the optimal price adjustment rule $(S, s)$, is symmetric. Also, for a given disturbance $u$, the loss from not adjusting the price decreases with the price sensitivity of demand $\beta$, and increases with the slope of the marginal cost curve $C''(Y) = 2c$.

Barro's (1972) main conclusion is that if $u$ follows a symmetric random walk, then the optimal $(S, s)$ band is symmetric, given by $\left(\hat{h}, -\hat{h}\right)$, where

$$\hat{h} = \sqrt{\sigma} \left(\frac{6\gamma}{\theta}\right)^{0.25} \qquad (10)$$

where $\gamma$ is a fixed, lump-sum menu cost, $\sigma^2$ is the variance of the Bernoulli process driving the symmetric random walk, and $\theta$ is given by (9).[5]

According to (10), the higher the menu cost, the wider the band of inaction. On the other hand, a high $\theta$ implies a narrow band of inaction. That is because according to (8)–(9), a high $\theta$ means a greater profit loss from not adjusting the price.

In models with CES demand, the optimal value of the barrier $\hat{h}$ is independent of the output produced, because of the constant price elasticity assumption. Here, however, the demand is assumed linear and thus its price elasticity is not constant. We can therefore take advantage of this property by extending the model to derive the relationship between the optimal barrier, i.e., the optimal $(S, s)$ band, and the output. Rewrite (9) as

$$\theta = \frac{(1+2c\beta)^2}{4\beta(1+c\beta)}$$
$$= \left[\frac{\alpha-\beta b}{2(1+c\beta)}\right]\left[\frac{(1+2c\beta)^2}{2\beta(\alpha-\beta b)}\right] \qquad (11)$$

---

[5] The expression for the barrier $\hat{h}$ as given above, is identical to the expression for the barrier that Dixit (1991, p. 144) derives and reports in his equation (11).

By (4), the term in the first brackets is the optimal level of output in the disturbance-free equilibrium $Y*\big|_{u=0}$. Therefore, (11) can be written as a function of $Y*\big|_{u=0}$,

$$\theta\left(Y*\big|_{u=0}\right) = \left(Y*\big|_{u=0}\right)\left[\frac{(1+2c\beta)^2}{2\beta(\alpha-\beta b)}\right] \tag{12}$$

which shows that $\theta = \theta\left(Y*\big|_{u=0}\right)$, i.e., $\theta$ is a function of $Y*\big|_{u=0}$.

To show the effect of changes in output $Y*\big|_{u=0}$ on the frequency of small price changes, consider a change in $\beta$ which by (4) affects the output $Y*\big|_{u=0}$ because of its effect on the demand, while in parallel it also affects $\theta$ by (11). Note that from the first part of (11), it follows that

$$\begin{aligned}
\frac{\partial\theta}{\partial\beta} &= \frac{\partial}{\partial\beta}\left[\frac{(1+2c\beta)^2}{4\beta(1+c\beta)}\right] \\
&= -\frac{1+2c\beta}{4\beta^2(1+c\beta)^2} \\
&< 0
\end{aligned} \tag{13}$$

while from (10) it follows that

$$\begin{aligned}
\frac{\partial\hat{h}}{\partial\theta} &= \frac{\partial}{\partial\theta}\left[\sqrt{\sigma}\left(\frac{6\gamma}{\theta}\right)^{\frac{1}{4}}\right] \\
&= -\frac{\sqrt{\sigma}(6\gamma)^{\frac{1}{4}}}{4\theta^{\frac{5}{4}}} \\
&< 0
\end{aligned} \tag{14}$$

Now consider a situation where there is an increase in demand because of a decrease in $\beta$, which leads to higher $Y*\big|_{u=0}$ because, using (4), we have

$$\begin{aligned}
\frac{\partial\left(Y*\big|_{u=0}\right)}{\partial\beta} &= \frac{\partial}{\partial\beta}\left[\frac{\alpha-\beta b}{2(1+c\beta)}\right] \\
&= -\frac{(\alpha c+b)}{2(c\beta+1)^2} \\
&< 0
\end{aligned} \tag{15}$$

Then, the decrease in $\beta$ which by (15) increases $Y*\big|_{u=0}$, increases $\theta$ by (13), which by (14) decreases $\hat{h}$, making the (s, S) band narrower.

Another way to see this, is to find directly the sign of the partial derivative of $\hat{h}$ with respect to $\beta$, by first substituting (9) in (10). Then we obtain the partial derivative

$$\frac{\partial \hat{h}}{\partial \beta} = \frac{\partial}{\partial \beta} \left[ \sqrt{\sigma} \left( \frac{6\gamma}{\theta} \right)^{\frac{1}{4}} \right]$$

$$= \frac{\partial}{\partial \beta} \left[ \sqrt{\sigma} \left\{ \frac{6\gamma}{\left[ \dfrac{(1+2c\beta)^2}{4\beta(1+c\beta)} \right]} \right\}^{\frac{1}{4}} \right]$$

$$= \frac{\sqrt[4]{\dfrac{3}{2}} \left( \dfrac{c\gamma\beta}{(1+2c\beta)^2} - \dfrac{4c\gamma\beta(1+c\beta)}{(1+2c\beta)^3} + \dfrac{\gamma(1+c\beta)}{(1+2c\beta)^2} \right) \sqrt{\sigma}}{2 \left[ \dfrac{\gamma\beta(1+c\beta)}{(1+2c\beta)^2} \right]^{\frac{3}{4}}} \qquad (16)$$

$$= \frac{\sqrt[4]{\dfrac{3}{2}} \, \gamma \sqrt{\sigma}}{2(1+2c\beta)^3 \left[ \dfrac{\gamma\beta(1+c\beta)}{(1+2c\beta)^2} \right]^{\frac{3}{4}}}$$

$$> 0$$

The expressions in (13)–(15), or alternatively (15)–(16), constitute our main analytical result. Consider, for example, a situation where there is an increase in demand because of a decrease in $\beta$. Then, according to (15), output will increase. But according to (16), the decrease in $\beta$ will reduce $\hat{h}$, leading to a narrower optimal $(S, s)$ band. In other words, there is an inverse relationship between the level of output (as determined by the demand) and the width of the $(S, s)$ band. If the output of the monopolist is high (low), then the $(S, s)$ band will be narrower (wider), which means that we will see more (less) frequent smaller price changes. Trivially, this will be true for small menu costs. However, our result is independent of the size of the menu cost. That is, the model predicts that we will likely see small price changes even if the menu cost is large.

Another implication of the reduction in the width of the $(S, s)$ band is that as the sales volume increases, the frequency of price changes, irrespective of their size, should also increase. We test this second prediction in Appendix O, in the Online Supplementary

Appendix, and find that it is supported by the data.

## 3. Data

We use data from Dominick's, a large US retail food chain with 93 stores in the greater Chicago area with a market share of 25%. The data contain more than 98 million weekly observations over 8 years, from September 14, 1989, to May 8, 1997, for 13,504 products in 29 categories, including food, cleaning products, hygienic products, and pharmaceutical products.[6] Each weekly observation includes the retail price, the number of units sold, the revenue, the retailer's markup, and some product attributes. These features make Dominick's dataset especially well-suited for our analysis.

An important attribute of Dominick's data is that its prices were set on a weekly basis. Thus, each week there was one price. If manufacturer coupons were used, we cannot account for these. During the sample period, however, the use of such coupons was limited (Barsky et al. 2003, Chen et al. 2008, Levy et al. 2010 and 2011).

## 4. Empirical findings

### *4a. Results of pooled analysis*

To study the correlation between sales volumes and small price changes, we follow Chen et al. (2008) to define a small price change as a price change of 10¢ or less. We choose to focus on the absolute rather than the relative size of price changes because our hypothesis implies that a price change of a given size is profitable if the change in cents multiplied by the sales volume is greater than the menu cost. For example, a change of 1 cent in a price adds 1 cent to the revenue if the firm sells 1 unit and $10 if the firm sells 1,000 units, irrespective of the price prior to the change. For our purpose, therefore, it makes sense to define small price changes in cents. In addition, a "unit" in our data may be composed of multiple units, e.g., a six-pack of beer. In such cases, we count each pack as a single unit, because the consumer pays once for the entire pack.

In the appendix, we show that our results hold (and are even stronger), if we exclude multi-units packs. In the appendix, we also show that our findings are robust to alternative definitions of small price changes, including 5¢, 15¢, 2%, and 5%, as well as relative to the average product-level price change (Midrigan 2011, and Bhattarai and Schoenle 2014). Thus, although we focus here on absolute price changes, the results also apply to relative price changes, as in Alvarez et al. (2016).

---

[6] Dominick's data contains observations on 18,035 UPCs. However, some of the UPCs are re-launches of the same product. See Mehrhoff (2018), and Dominick's Data Manual (p. 9), available at https://www.chicagobooth.edu/-/media/enterprise/centers/kilts/datasets/dominicks-dataset/dominicks-manual-and-codebook_kiltscenter.aspx.

Another possible source of noise in our data is measurement errors (Alvarez et al., 2016). As Eichenbaum et al. (2014) note, prices reported in scanner datasets are weekly average prices. This can result in spurious small price changes when shoppers pay different prices, for example when some shoppers use coupons.[7] To control for this, we use observations on price changes only if the post-change price lasted for at least two consecutive weeks. As noted by Strulov-Shlain (2023), the likelihood that a spurious price change would persist for more than one week is very low. In the appendix, we report the results of two other robustness tests. In the first, we exclude observations where price changes are $\leq 2\cent$ (Alvarez et al., 2016). In the second, we exclude observations where Dominick's dataset indicates that coupons were used.

As a first test of the correlation between small price changes and sales volumes, we merge the observations in all 29 categories. Across all categories, 26.6% of all price changes are small (i.e., smaller, or equal to $10\cent$), and the average sales volume is 10.0 units per week. We then divide product-stores into deciles according to their sales volume. Figure 1 depicts the results. An increase in the sales volume is associated with a significant increase in the percentage of small price changes. The percentage of small price changes in the 10th decile, 33.36%, is 2.4 times higher than the percentage of small price changes in the 1st decile, 13.89%.

As a formal test, we estimate the following fixed effect regression model using the pooled data:

$$small\ price\ change_{i,s,t} = \alpha + \beta ln\big(average\ sales\ volume_{i,s}\big) + \gamma X_{i,s,t} +$$

$$month_t + year_t + \kappa_i + \delta_s + \mu_i + u_{i,s,t} \tag{17}$$

where small price change is a dummy that equals 1 if a price change of product $i$ in store $s$ in week $t$ is less or equal to $10\cent$, and 0 otherwise. The average sales volume is for product $i$ in store $s$ over the sample period.[8] By taking the average over a long period, we obtain an estimate of the expected sales volume that does not depend on transitory shocks or sales. **X** is a matrix of other control variables. Month and year are fixed effects for the month (to control for seasonality) and the year of the price change. To control for the differences across stores and products, we add $\kappa$, $\delta$ and $\mu$ which are fixed effects for categories, stores and products, respectively. $u$ is an i.i.d error term.

---

[7] Consider the following example. In week $t$, the price of a good was $1.99 and all units were sold at the posted price. In week $t+1$, the posted price remained $1.99. 9 consumers bought at the posted price, and one used a coupon and paid 1.79. The price that would be recorded in the scanner dataset is $1.97, i.e, a 2 cents change relative to week $t$.

[8] In calculating the average sales volume, we need to account for missing observations, because a missing observation in week $t$ implies that the product was either out of stock or had 0 sales on that week. Thus, averaging over the available observations can lead to an upward bias for products that are sold in small numbers. Therefore, for each product in each store, we calculate the average by first determining the total number of units sold over all available observations. We then identify the first and last week for which we have observations, and calculate the average for each product-store as $\frac{total\ no.\ of\ units\ sold}{last\ week\ -\ first\ week}$. The resulting figure is smaller than we would obtain if we averaged over all available observations (which would not include obsservations on weeks with 0 sales).

Table 1 reports the estimates of the coefficients of the key variable, average sales volume. Column 1 reports the results of a baseline regression that includes only the average sales volume and fixed effects for months, years, stores, and products. The coefficient of the sales volume is 0.026, and it is statistically significant. This result suggests that a 1% increase in the sales volume is associated with a 2.6 percentage points increase in the likelihood of a small price change. In column 2, the matrix **X** includes the following control variables: the log of the average price, to control for the price level effect on the size of price changes, the percentage change in the wholesale price, and control for sale- and bounce-back prices. The latter is important as price changes associated with sales tend to be large (Nakamura and Steinsson 2008).[9]

The coefficient of the sales volume remains positive and statistically significant. Its value is 0.017, suggesting that a 1% increase in the sales volume is associated with a 1.7 percentage points increase in the likelihood of a small price change.

In column 3, we add a dummy for 9-ending prices as an additional control because when the pre-change price is 9-ending, price changes tend to be larger than when the pre-change price ends in other digits (Levy et al. 2020). Thus, if products with high sales volume tend to have non-9-ending prices, then it might lead to small changes in their prices. This has only a marginal effect on the coefficient; its value remains 0.017 and statistically significant.

In column 4, we keep the same control variables as in column 3, but we focus on regular prices by excluding the sale- and bounce-back prices. We do this for two reasons. First, sale- and bounce-back prices tend to be large, and therefore, we need to account for them properly. Second, it is often argued that changes in sale prices have smaller effect on inflation than changes in regular prices (Nakamura and Steinsson 2008, Midrigan 2011, Anderson et al. 2017, Ray et al. 2023).

We find that the estimate of the coefficient of the sales volume is 0.033, and is statistically significant. The pooled results, therefore, suggest that there is a positive and statistically significant correlation between small price changes and sales volumes.

---

[9] To identify sale prices, we do not use the sales' flag included in the Dominick's data because it was not set on a consistent basis (Peltzman 2000). Instead, we use the sales filter algorithm of Fox and Syed (2016) to identify sales. This algorithm has the advantage that it was calibrated using Dominick's data and, consequently, it is particularly useful for identifying sales in the Dominick's data.

### *4b. Results of category-level analyses*

Estimation using pooled data can hide large differences across categories. Therefore, we study the category-level correlation between small price changes and sales volumes.

As a first test, for each category, we group the products into high, medium, and low sales volume products, according to the average sales volumes over the sample period. Low sales volume products are products with average sales volume in the bottom third of the distribution, high sales volume products have sales volume in the top third of the distribution, and medium sales volume products have sales volume in between.

Figure 2 shows, for every category, the frequency of price changes for each size of price change from 1¢ to 50¢. The red dashed line depicts the frequency of price changes for high sales volume products, the black dotted line depicts the frequency of price changes for medium sales volume products, and the blue solid line depicts the frequency of price changes for low sales volume products. The shaded area marks the range of small price changes, $\Delta P \leq 10¢$.

The figure shows that the most common price changes are multiples of 10¢ (as reported also by Chen et al. 2008 and Levy et al. 2011). It can also be observed that in all categories except cigarettes (which are highly regulated), price changes are far more common among high sales volume products than among low sales volume products.

Focusing on the shaded area, we see that the frequency of small price changes is in general far greater among the high sales volume products than among low sales volume products. Indeed, for high sales volume products, in most categories, the frequency of small price changes exceeds the frequency of large price changes. This is less common, and less dramatic, among low sales volume products. For the medium sales volume products, the frequency of price changes, and the frequency of *small* price changes in particular, fall in between the frequencies of the low and high sales-volume products.

As a formal test, we estimate a series of category-level fixed-effect regressions, similar to (17). The only difference is that we now exclude the category fixed effects. Table 2 reports the coefficients of the key variable, average sales volume, for each product category. Column 1 reports the results of baseline regressions that exclude the **X** matrix. I.e, the regressions include only the average sales volume and fixed effects for months, years, stores, and products.

We find that in all 29 product categories, the coefficients are positive, and 27 are statistically significant. One more is marginally significant. In other words, in 28 of 29 product categories, there is a positive and statistically significant correlation between the likelihood that a price change is small and the average sales volume. The effect is economically significant. The average coefficient is 0.026, suggesting that an increase of 1% in the sales volume is associated with an increase of 2.6 percentage points in the likelihood that a price change will be small.

In column 2, we add the **X** matrix which includes the following control variables: the log of the average price to control for the price level effect, the percentage change in the wholesale price, and control for sale- and bounce-back prices, all as defined above. The results are similar to column 1. The coefficients of the average sales volume are positive and statistically significant in 27 categories, and marginally significant in 2 more. The average coefficient is 0.019. Thus, even after including the controls, we still find that increasing the average sales volume by 1% is associated with an increase of 1.9 percentage points in the likelihood of a small price change.

In column 3, we add a dummy for 9-ending prices as an additional control. Adding this dummy does not change the main result appreciably. All 29 coefficients remain positive. 27 are statistically significant, and 2 more marginally significant. Controlling for 9-ending prices, increasing the average sales volume by 1% is associated with a 2.0 percentage points increase in the likelihood of a small price change, on average.

In column 4, we focus on regular prices by excluding the sale- and bounce-back prices. We find that all the coefficients remain positive. 27 are statistically significant, and 1 more is marginally significant. The average coefficient is 0.038, implying that for regular prices, an increase of 1% in the average sales volume is associated with an increase of 3.8 percentage points in the likelihood of a small price change.

### 4c. Results of product-level analyses

A possible explanation for the correlation between sales volume and small price changes is that products with high sales volume have some unobserved attributes that make them prone to small price changes. We explore this possibility by estimating for each product a separate regression. If the correlation between sales volume and small price changes is found at the level of individual products, then it cannot be explained by unobserved attributes, since in each regression we have data on only one product.

Before presenting the full regression results, consider as an example the bathroom tissue category. In Figure 3, we show a scatter plot for each one of the 13 bathroom-tissue products that have data for all 93 stores at Dominick's. In each of the 13 panels, there are 93 dots, one for each store. In each figure, the *x*-axis in the figures gives the average weekly sales volume of the product in a store, and the *y*-axis gives the share of small price changes of the product in a store. The straight lines are regression lines.

According to the plots, the correlation between sales volume and the share of small price changes is positive for 11 of the 13 individual products. None of the negative correlations is statistically significant, while 8 of the 11 positive correlations (marked with solid black regression lines) are statistically significant. The regression lines that are not statistically significant are marked with red dotted lines.

For a more formal analysis, we calculate for each product in each of the 29 product categories the average weekly sales volume and the share of small price changes in each of the stores it was offered. Many products in the sample were offered for only short periods or only in a small number of stores. To avoid biases, we drop products for which we do not have information for at least 30 stores.

Using these data, we estimate for each product in each category an OLS regression with robust standard errors. The dependent variable is the share of small price changes for the product in each store. The independent variable is the average sales volume of the product in each store. The estimation results are summarized in Table 3.[10]

Column 1 gives, for each product category, the average of the estimated coefficients. Columns 2–5 give information on the sign of the estimated coefficients: the total number of coefficients, the % of positive coefficients, the total number of coefficients that are statistically significant at the 5% level, and the % of coefficients that are both positive and statistically significant at the 5% level.

According to the figures in the table, the average coefficients are positive in 28 of the 29 product categories. The only exception is the highly regulated cigarettes category, which is often excluded from the analyses (Chen et al. 2008, p. 729, footnote 2). In addition, the number of positive coefficients far exceeds the number of negative coefficients. On average, the former is 3.2 times larger than the latter. Ignoring the cigarettes category, more than 74.5% of the coefficients are positive.

Focusing on statistically significant coefficients, we find a far greater number of positive coefficients than negative coefficients that are significant. Except for the cigarettes category, in all categories, 81.40%–100% of the statistically significant coefficients are positive. In other words, for the overwhelming majority of the individual products in our sample, we find a positive relationship between sales volume and the share of small price changes.

---

[10] For robustness, we have also conducted an analysis using LPM regressions, as in the previous section. See the discussion in Online Supplementary Web Appendix F.

To summarize, we find that the correlation between sales volume and the share of small price changes is positive whether we look across categories, within categories, and for individual products across stores. It seems unlikely, therefore, that the correlation is due to unobserved characteristics of the products or the product categories.

### 4d. Sales volume versus revenue

Kang and Usher (2023) find that the size of price changes is negatively correlated with the revenue. In column 1 of Table 4, we report the Pearson correlation coefficient between the average sales volume and the average revenue for product-stores for each product category. The average correlation is 0.85, suggesting that at the category level, the correlation is very strong.

In column 2, we replicate one of Kang and Usher's (2023) key findings by reporting for each product category, the coefficient estimates of regression (16) where we replace the log of the average sales volume with the log of the average revenue. The controls include fixed effects for months, years, stores, and products. The estimated coefficients are positive for all 29 product categories, and 28 of these are statistically significant. We thus confirm that Kang and Usher's (2023) findings hold in our data: there is a strong positive correlation between revenue and the likelihood of a small price change.

Revenue, however, is a product of the sales volume and the price. Our hypothesis implies that the revenue is correlated with the likelihood of a small price change via the sales volume, rather than via the price. To test this, in columns 3 and 4, we show the results of regressions that include both the sales volume and the revenue as independent variables. Once we add the sales volume to the regression, the coefficients of the revenue turn negative in 22 of the 29 categories. The coefficients of the sales volume, on the other hand, are positive in 23 of the 29 categories.

These results suggest that holding the sales volume constant, the higher the price level, the less frequent small price changes are. In other words, the positive correlation between revenues and the frequency of small price changes seems to materialize through the sales volume and not through the price.

### 4e. Sales volume, producer size, and price synchronization

According to Lach and Tsiddon (2007), Midrigan (2011), Alvarez and Lippi (2014), Letterie and Nilsen (2014), and Alvarez et al. (2016), small price changes can be explained by economies of scale in price adjustment, which makes small price changes profitable if the average price change exceeds the menu cost. Lach and Tsiddon (2007) and Bhattarai and Schoenle (2014) show that: (a) producers offering a large number of products are more likely to make small price changes than producers offering a small number of products, (b) small price changes are more likely when price changes are more synchronized, and (c) small price changes are more likely when the average size of all other contemporaneous price changes is high. In addition, Bhattarai and Schoenle (2014) provide evidence suggesting that producers offering more products are more likely to have a large sales volume. It is therefore of interest to examine whether our results hold when accounting for economies of scope by adding controls for price changes synchronization and the number of products per producer.

In column 1 of Table 5, we report the coefficients of the log of sales volume in regression (17), where we also include the average number of products per category offered by the same producer, as a proxy to the producer's size.[11] The regression also includes fixed effects for months, years, stores, and products. We find that all coefficients of the sales volume are positive and 27 are statistically significant. In addition, the coefficients of the sales volume are almost unaffected in comparison to the figures we report in column 1 of Table 3. Thus, it appears that the effects of sales volume on the likelihood of small price changes are unrelated to the effects of the size of the producers.

In column 2, we add a control for the percentage of the products that changed the price in the same week, excluding the current observation. Again, the coefficients remain almost unchanged in comparison to the figures we report in column 1 of Table 3.

In column 3, we further add the average size of contemporaneous price changes, excluding the current observation. Lach and Tsiddon (2007) show that when price changes are synchronized, a small price change is correlated with a large average contemporaneous price change. However, we find that adding the average size of price changes has little effect on the size of the coefficients of the sales volume. Also in this specification, 27 of the 29 coefficients are statistically significant.

Finally, in column 4, we add the percentage of the products that are produced by the same producer and changed price in the same week, excluding the current observation. This forces us to drop some observations because we can only use an observation if the producer offers at least two products on the relevant week. The upshot is that most of the remaining observations are of products produced by relatively large producers which are most likely to make small price changes (Bhattarai and Schoenle 2014).

---

[11] To calculate the average number of products offered by a producer, we follow Bhattarai and Schoenle (2014), by first determining the nubmer of products offered by a producer each week, and then averaging over all weeks.

Therefore, if our results are driven by the size of the producers rather than by the sales volume, then focusing on large producers should lead to a substantial drop in the coefficients of the sales volume. We find, however, that the results do not change significantly. All the coefficients are positive, 27 of the 29 of them are statistically significant, and 2 more are marginally significant.

As another test, we follow Bonomo et al. (2022). They show that a large share of price changes takes place on "peak days." We therefore follow their methodology and create a dummy for peak days and then add it as a control to the regressions. The estimation results, which we present and discuss in Online Supplementary Web Appendix P, show that the correlation between sales volumes and small price changes holds also when we control for peak days.

## 5. Robustness

To assess the robustness of our findings, we conducted 19 sets of robustness tests:

1) ***Measurement errors***: Eichenbaum et al. (2014) conclude that prices based on scanner data might include spurious small price changes. We mitigate this concern by focusing only on price changes in which the post change price remained unchanged for at least 2 weeks. As further tests, we exclude all 1¢ and 2¢ price changes (Eichenbaum et al. 2014), and remove observations on price changes if Dominick's sale flag indicates that there was a coupon use because if some consumers used coupons, this could result in measurement errors (Eichenbaum et al. 2014). We also estimate the regressions using observations on all price changes, conditional on observing the prices in weeks *t* and *t*+1.

2) ***Definition of small price changes***: We define a small price change as a price change of 5¢ or less, and 15¢ or less. Because the size of price changes may be larger for more expensive products, we re-run the analyses by defining a small price change in percent as a price change of 2% or less, and 5% or less. Following Midrigan (2011) and Bhattarai and Schoenle (2014), we also define small price changes relative to the average price change at the store-product level. I.e., a price change is small if it is smaller or equal to $\kappa|\overline{\Delta p_{i,s}}|$, where $\overline{\Delta p_{i,s}}$ is the average price change for product $i$ in store $s$, and $\kappa$ attains the values (8) 0.50, (9) 0.33, (10) 0.25 and (11) 0.10.

3) ***Time horizon for computing the average sales volume***: Above, we use the average sales volume over the entire period, which can be thought of as a proxy for the expected sales volume that is based on 8 years of data. However, this implicitly assumes that in the first years, the retailer can predict the future sales volume. An alternative is that the retailer makes decisions based on recent sales data. We, therefore repeat the analyses by calculating the average sales volume based on a rolling 52-week window of past observations.

4) ***Competition effect***: As another control for the effect of competition, we add control for Dominick's pricing zones. We re-estimate the product-level regressions by augmenting the data with demographic information of the consumers that live in the proximity of each store, including their median income, the share of minorities, and the share of unemployed.

5) ***Controlling for revenue***: In section 4d, we estimate regressions with both the sales volume and the revenue at the category level. For robustness, we explore the effect of adding further controls. In addition, at the category level, the high correlation between the sales volume and the revenue may render the results suspect. Therefore, we pool data from all categories together and re-estimate the regressions. We also test an alternative definition of the average revenue, defining it as the product of the average sales volume and the average price for estimating category-level regressions.

6) ***Controlling for the producers' size***: In section 4e, we calculate the number of products offered by each producer at the category level. To test that the results do not change for firms of different sizes, we follow Bhattarai and Schoenle (2014) in dividing producers into bins according to the number of products they offer in each category and estimate the correlation between sales volume and small price changes for each bin separately. In addition, it is possible that producers that offer products in more than one category synchronize price changes across categories, We, therefore, pool the data together and repeat the analysis.

7) ***Controlling for profit margins***: Some products have high sales volumes, yet they have few, if any, small price changes. One example is iPhones, which have high sales volumes, yet most of their price changes are large. A possible explanation is that small price changes are less likely for products with large markups because large markups imply that small price changes have, in percentage terms, only a small effect on total profits. We, therefore, add Dominick's measure of markups to the category-level regressions as a further control variable.

8) ***Graphic illustration of the category-level correlation between small price changes and sales volume***: We include a figure similar to Figure 1, which shows that small price changes are correlated with high sales volume at the category level as well as at the aggregate level.

9) ***Reproduce Figure 2 using % price changes***: In Figure 2, the most common price changes are multiples of 10 cents. This is consistent with a large literature on price points (Levy et al., 2011, 2020). However, because price changes that are multiples of 10 cents might be large for some products, we reproduce Figure 2 using the size of price changes in % terms rather than in cents.

10) ***National brands vs. private labels***: A possible explanation for our results is that the correlation between sales volume and small price changes is an artifact of differences in demand. We, therefore, separate the data into two groups: national brands and private labels, since the demand for private labels is likely to exhibit different patterns than for national brands.

11) ***Holiday price rigidity***: Levy et al. (2010) provide evidence suggesting that menu costs are higher than normal during the Thanksgiving-Christmas holiday period. It is, therefore, possible that there are fewer small price changes during the Thanksgiving-Christmas period, leading to a decline in the correlation between sales volumes and small price changes. We, therefore, re-estimate equation 1, using only the observations from the Thanksgiving-Christmas period.

12) ***Another prediction of Barro's (1972) model***: In the paper, we study the correlation between small price changes and sales volumes. However, Barro's (1972) model, as shown in section 2, also predicts that sales volume should also be correlated with the number of price changes. In other words, prices of products with high sales volumes should change more frequently than the prices of products with low sales volumes. We test this hypothesis in the appendix.

13) ***Controlling for peak days***: Bonomo et al. (2022) show that in their data, the majority of price changes occur during "peak days." Following their definition, for each category in each store, we identify peak days as the subset of the most active days that jointly account for one-half of all price changes in a store over the entire sample period. We then control for peak days in the regressions.

14) ***Cross-category comparisons***: We show that there are large variations in both the average sales volume and the likelihood of small price changes across categories. We then show that as we hypothesize, some of the variation in the likelihood of small price changes can be explained by the average sales volumes.

15) ***Excluding observations on the multi-unit package***: Some of the products in our

dataset are composed of several units. For example, we have products such as 6-packs of beer, or a 4-pack of canned tuna. Above, we treat such packages as a single good, because the consumer pays once for the entire package. However, these may add noise to the regression, because it is unclear how consumers perceive such packages. We, therefore, exclude such multi-unit packaged goods and focus on products that are sold in single units.

16) ***Product level regressions***: We show that the results we obtained for the product level correlation between small price changes and sales volumes hold when we add further controls and when we estimate the correlations using linear probability models.

17) ***Storable vs. non-storable products***: Retailers might employ different strategies for storable vs. non-storable products. We therefore test for the robustness of our results by assessing them separately for storable and non-storable products.

18) ***Asymmetry in the correlations***: Our model predicts a symmetric correlation between sales volumes and the likelihood of small price changes. However, it is possible that the correlation is asymmetric, for example, if shoppers are not attentive to small price changes (Chen et al., 2008, Chakraborty et al., 2015). We therefore estimate separate regressions for price increases and price decreases.

19) ***Cross-category analysis***: We conduct further analysis comparing the likelihood of small price changes and the sales volumes across categories.

The Online Supplementary Appendix contains the details of these analyses. Overall, our main results are broadly unchanged, and are robust across the different specifications.

## 6. Conclusion and policy implications

The finding of frequent small price changes in many retail price datasets has been interpreted by authors as prima facie evidence against simple menu cost models. We find, however, that sales volume can explain some of the small price changes found in many datasets. When a retailer expects to sell many units, then small price changes can be profitable even in the presence of lump sum menu costs.

We use Dominick's scanner price dataset to show a strong positive correlation between the frequency of small price changes and products' sales volume. This finding is robust. It holds across product categories, within product categories, and for individual products. It is also robust to a variety of sensitivity analyses such as the definition of "small" price changes, measurement errors, the inclusion of control variables, firms' size, price synchronization, and the time horizon used to compute the average sales volume.

Our findings hold irrespective of how we measure price changes—in absolute or relative terms. The latter is useful if one wants to assess the "Calvo-ness" of the relevant model.

Our findings are consistent with the findings reported by Bhattarai and Schoenle (2014) and Kang and Usher (2023), who employ more recent datasets. However, the advantage of Dominick's dataset, in comparison to more recent datasets is its richness. In addition to prices and sales volume—two critical variables for our analyses, it also contains data on wholesale prices, promotions and sales, pricing zones, shoppers' socio-demographics, markups, etc. We employ these additional variables to check robustness.

Above, we use Barro's (1972) model to illustrate the theoretical correlation between sales volumes and the size of price changes. However, Barro's (1972) model is too stylistic for deriving predictions about the distribution of the size of price changes. In addition, it cannot predict the number of small price changes observed in many datasets unless we assume unrealistically high sales volumes. Therefore, we believe that there is a need for models that will account for sales volumes, perhaps along the lines of Golosov and Lucas (2007). It would then be possible to evaluate the implications of heterogeneity in sales volumes for the macro-level price rigidity.

Such a model might yield important insights. First, it might yield a non-trivial distribution of the size of price changes even if all producers sell one product and the menu costs are fixed. Second, if the size of price changes depends on the sales volume, then the existence of small price changes does not rule out selection. Thus, unlike many of the existing models, in such a model, the kurtosis of the distribution of the size of price changes might not necessarily indicate selection (Alvarez et al., 2016). Therefore, for a given frequency of price changes, we might obtain a result where a monetary shock has only a small real effect even in the presence of many small price changes (Kang and Usher, 2023). Related to this, Karadi et al. (forthcoming) find that the likelihood of a price change does not depend on the gap between the price and the optimal price. They interpret their finding as evidence against selection. However, in a model where price changes depend on sales volumes, we may have selection together with both small and large price changes in response to a monetary shock.

Table 1. Pooled regressions of small price changes and sales volume

|  | (1) | (2) | (3) | (4) |
|---|---|---|---|---|
| Average sales volume | 0.026*** | 0.017*** | 0.017*** | 0.033*** |
|  | (0.001) | (0.001) | (0.001) | (0.001) |
| Observations | 9,553,542 | 9,553,542 | 9,553,542 | 2,328,405 |

Notes: The dependent variable is "small price change," which equals 1 if a price change of product $i$ in store $s$ at time $t$ is less or equal to 10¢, and 0 otherwise. The main independent variable is the log of the average sales volume of product $i$ in store $s$ over the sample period. Column 1 reports the results of a baseline regression that includes only the average sales volume and the fixed effects for months, years, stores, and products. In column 2, we add the following controls: the log of the average price, the log of the absolute change in the wholesale price, and a control for the sale- and bounce-back prices, which we identify using a sales filter algorithm. In column 3, we add a dummy for 9-ending prices as an additional control. In column 4, we focus on regular prices by excluding the sale- and bounce-back prices. All regressions also include fixed effects for categories, stores, products, years, and months. We estimate separate regressions for each product category, clustering the errors by product. * $p < 10\%$, ** $p < 5\%$, *** $p < 1\%$

Table 2. Category-level regressions of small price changes and sales volume

| Category | | (1) | (2) | (3) | (4) |
|---|---|---|---|---|---|
| Analgesics | Coefficient (Std.) | 0.0262*** (0.0034) | 0.019*** (0.0032) | 0.0188*** (0.0031) | 0.029*** (0.0068) |
| | Observations | 144,461 | 144,461 | 144,461 | 44,950 |
| Bath Soap | Coefficient (Std.) | 0.0293*** (0.008) | 0.0277*** (0.0082) | 0.0285*** (0.0081) | 0.0972*** (0.0192) |
| | Observations | 15,295 | 15,295 | 15,295 | 3,208 |
| Bathroom Tissues | Coefficient (Std.) | 0.0408*** (0.007) | 0.0179** (0.0058) | 0.0184*** (0.0057) | 0.0386*** (0.0084) |
| | Observations | 149,441 | 149,441 | 149,441 | 47,041 |
| Beer | Coefficient (Std.) | 0.013*** (0.0012) | 0.0147*** (0.0012) | 0.0147*** (0.0012) | 0.0699*** (0.0063) |
| | Observations | 290,620 | 290,620 | 290,620 | 27,348 |
| Bottled Juice | Coefficient (Std.) | 0.0329*** (0.0053) | 0.0239*** (0.0044) | 0.0238*** (0.0045) | 0.0304*** (0.0063) |
| | Observations | 496,557 | 496,557 | 496,557 | 133,714 |
| Canned Soup | Coefficient (Std.) | 0.0158*** (0.0056) | 0.0108* (0.005) | 0.0134** (0.0049) | 0.0144** (0.0049) |
| | Observations | 495,543 | 495,543 | 495,543 | 176,235 |
| Canned Tuna | Coefficient (Std.) | 0.0237*** (0.0054) | 0.0134** (0.0046) | 0.0131** (0.0045) | 0.0225*** (0.0058) |
| | Observations | 213,043 | 213,043 | 213,043 | 64,161 |
| Cereals | Coefficient (Std.) | 0.0204*** (0.0037) | 0.0149*** (0.0034) | 0.0148*** (0.0034) | 0.0188*** (0.0046) |
| | Observations | 357,120 | 357,120 | 357,120 | 155,367 |
| Cheese | Coefficient (Std.) | 0.0201*** (0.0028) | 0.0113*** (0.0025) | 0.0112*** (0.0025) | 0.0113*** (0.0033) |
| | Observations | 796,150 | 796,150 | 796,150 | 224,889 |
| Cigarettes | Coefficient (Std.) | 0.0084** (0.0046) | 0.0073 (0.0045) | 0.0074 (0.0044) | 0.0069 (0.0054) |
| | Observations | 36,157 | 36,157 | 36,157 | 30,262 |
| Cookies | Coefficient (Std.) | 0.0267*** (0.0018) | 0.022*** (0.0017) | 0.0223*** (0.0017) | 0.046*** (0.0035) |
| | Observations | 688,761 | 688,761 | 688,761 | 132,488 |
| Crackers | Coefficient (Std.) | 0.0379*** (0.0031) | 0.0302*** (0.0026) | 0.0306*** (0.0026) | 0.0467*** (0.0072) |
| | Observations | 245,185 | 245,185 | 245,185 | 50,029 |
| Dish Detergent | Coefficient (Std.) | 0.0393*** (0.0044) | 0.028*** (0.0036) | 0.0279*** (0.0035) | 0.0361*** (0.0041) |
| | Observations | 189,633 | 189,633 | 189,633 | 53,289 |
| Fabric Softener | Coefficient (Std.) | 0.0238*** (0.0048) | 0.0123*** (0.0043) | 0.0126*** (0.0043) | 0.0325*** (0.0057) |
| | Observations | 181,056 | 181,056 | 181,056 | 56,234 |
| Front-End-Candies | Coefficient (Std.) | 0.0047 (0.0041) | 0.0057 (0.0033) | 0.0059 (0.0032) | 0.0053 (0.0031) |
| | Observations | 278,853 | 278,853 | 278,853 | 111,635 |
| Frozen Dinners | Coefficient (Std.) | 0.0475*** (0.0034) | 0.036*** (0.0028) | 0.0389*** (0.0027) | 0.0795*** (0.0066) |
| | Observations | 203,191 | 203,191 | 203,191 | 37,527 |

Table 2. (Cont.)

| Category | | (1) | (2) | (3) | (4) |
|---|---|---|---|---|---|
| Frozen Entrees | Coefficient (Std.) | 0.028*** (0.0021) | 0.0259*** (0.0019) | 0.0266*** (0.0019) | 0.0443*** (0.0034) |
| | Observations | 864,832 | 864,832 | 864,832 | 213,545 |
| Frozen Juices | Coefficient (Std.) | 0.0273*** (0.0048) | 0.0198*** (0.0042) | 0.0206*** (0.0041) | 0.0293*** (0.0062) |
| | Observations | 308,817 | 308,817 | 308,817 | 87,919 |
| Grooming Products | Coefficient (Std.) | 0.0187*** (0.0023) | 0.0209*** (0.0024) | 0.021*** (0.0024) | 0.0417*** (0.0069) |
| | Observations | 269,873 | 269,873 | 269,873 | 51,819 |
| Laundry Detergents | Coefficient (Std.) | 0.0196*** (0.0032) | 0.0093*** (0.0028) | 0.0097*** (0.0028) | 0.02*** (0.0045) |
| | Observations | 272,765 | 272,765 | 272,765 | 85,184 |
| Oatmeal | Coefficient (Std.) | 0.028*** (0.008) | 0.0152* (0.0067) | 0.0154* (0.0067) | 0.0338*** (0.0098) |
| | Observations | 79,983 | 79,983 | 79,983 | 36,043 |
| Paper Towels | Coefficient (Std.) | 0.0454*** (0.0105) | 0.0316*** (0.0118) | 0.0321*** (0.0119) | 0.0378*** (0.0115) |
| | Observations | 116,204 | 116,204 | 116,204 | 29,280 |
| Refrigerated Juices | Coefficient (Std.) | 0.0357*** (0.0047) | 0.0209*** (0.0039) | 0.0209*** (0.0039) | 0.033*** (0.0056) |
| | Observations | 306,865 | 306,865 | 306,865 | 72,031 |
| Shampoos | Coefficient (Std.) | 0.0162*** (0.0015) | 0.02*** (0.0016) | 0.0201*** (0.0015) | 0.046*** (0.0052) |
| | Observations | 261,778 | 261,778 | 261,778 | 40,996 |
| Snack Crackers | Coefficient (Std.) | 0.0319*** (0.0032) | 0.0282*** (0.003) | 0.0284*** (0.003) | 0.0518*** (0.0052) |
| | Observations | 398,665 | 398,665 | 398,665 | 78,581 |
| Soaps | Coefficient (Std.) | 0.0374*** (0.0055) | 0.0226*** (0.005) | 0.0237*** (0.005) | 0.049*** (0.0076) |
| | Observations | 152,379 | 152,379 | 152,379 | 46,829 |
| Soft Drinks | Coefficient (Std.) | 0.0211*** (0.0017) | 0.024*** (0.0013) | 0.023*** (0.0012) | 0.0517*** (0.0037) |
| | Observations | 1,350,618 | 1,350,618 | 1,350,618 | 156,004 |
| Toothbrushes | Coefficient (Std.) | 0.0204*** (0.0028) | 0.02*** (0.0028) | 0.0195*** (0.0028) | 0.0498*** (0.0076) |
| | Observations | 125,380 | 125,380 | 125,380 | 24,955 |
| Toothpastes | Coefficient (Std.) | 0.0123*** (0.0026) | 0.0111*** (0.0022) | 0.0111*** (0.0022) | 0.0393*** (0.0063) |
| | Observations | 264,317 | 264,317 | 264,317 | 56,842 |
| **Average coefficients** | | **0.0260** | **0.0195** | **0.0198** | **0.0384** |

Notes: The table reports the results of category-level fixed effect regressions of the probability of a small price change. The dependent variable is "small price change," which equals 1 if a price change of product $i$ in store $s$ at time $t$ is less or equal to 10¢, and 0 otherwise. The main independent variable is the log of the average sales volume of product $i$ in store $s$ over the sample period. Column 1 reports the results of baseline regression that includes only the average sales volume and the fixed effects for months, years, stores, and products. In column 2, we add the following controls: the log of the average price, the log of the absolute change in the wholesale price, and control for sale- and bounce-back prices, which we identify using a sales filter algorithm. In column 3, we add a dummy for 9-ending prices as an additional control. In column 4, we focus on regular prices by excluding the sale- and bounce-back prices. We estimate separate regressions for each product category, clustering the errors by product.  * $p < 10\%$, ** $p < 5\%$, *** $p < 1\%$

Table 3. Product-level regressions of the % of small price changes and sales volume by categories

| Product Category | Average coefficient estimate (1) | Number of coefficients (2) | Percentage of positive coefficients (3) | Number of significant coefficients (4) | % of positive and significant coefficients (5) |
|---|---|---|---|---|---|
| Analgesics | 0.031 | 213 | 83.20% | 87 | 98.25% |
| Bath Soaps | 0.034 | 33 | 69.77% | 14 | 100.00% |
| Bathroom tissues | 0.057 | 100 | 79.31% | 34 | 90.24% |
| Beers | 0.019 | 202 | 91.57% | 142 | 97.80% |
| Bottled juices | 0.051 | 370 | 83.67% | 192 | 84.56% |
| Canned soups | 0.041 | 348 | 77.41% | 145 | 90.37% |
| Canned tuna | 0.037 | 181 | 76.16% | 60 | 80.00% |
| Cereals | 0.047 | 345 | 80.13% | 99 | 89.32% |
| Cheese | 0.036 | 474 | 81.76% | 224 | 90.26% |
| Cigarettes | -0.020 | 107 | 71.43% | 4 | 40.00% |
| Cookies | 0.034 | 667 | 82.42% | 304 | 94.71% |
| Crackers | 0.044 | 212 | 88.41% | 118 | 98.84% |
| Dish detergents | 0.041 | 199 | 85.71% | 84 | 91.89% |
| Fabric softeners | 0.043 | 226 | 85.35% | 69 | 91.67% |
| Front end candies | 0.053 | 275 | 75.28% | 92 | 91.03% |
| Frozen dinners | 0.053 | 215 | 92.06% | 131 | 97.59% |
| Frozen entrees | 0.052 | 671 | 89.32% | 363 | 97.46% |
| Frozen juices | 0.032 | 142 | 75.00% | 62 | 97.67% |
| Grooming products | 0.011 | 531 | 80.94% | 229 | 90.65% |
| Laundry detergents | 0.018 | 406 | 75.00% | 101 | 81.40% |
| Oatmeal | 0.047 | 69 | 76.81% | 20 | 84.62% |
| Paper towels | 0.052 | 90 | 73.61% | 33 | 90.91% |
| Refrigerated juices | 0.036 | 176 | 73.08% | 77 | 85.92% |
| Shampoos | 0.022 | 614 | 82.78% | 282 | 96.72% |
| Snack crackers | 0.041 | 282 | 87.36% | 172 | 92.55% |
| Soaps | 0.039 | 217 | 80.55% | 421 | 84.31% |
| Soft drinks | 0.032 | 902 | 82.49% | 72 | 94.87% |
| Toothbrushes | 0.026 | 204 | 82.84% | 89 | 95.83% |
| Toothpastes | 0.015 | 337 | 79.67% | 99 | 92.75% |
| **Average** | **0.035** | **303** | **73.29%** | **96** | **90.08%** |

Notes: For each product category, column 1 presents the average estimated coefficients. Column 2 presents the total number of coefficients. Column 3 presents the % of positive coefficients out of all coefficients. Column 4 presents the total number of coefficients that are statistically significant at the 5% level. Column 5 presents the % of coefficients that are positive and statistically significant, at the 5% level. Products are identified by their UPCs.

Table 4. Category-level regressions of small price changes, sales volume, and revenue

| Category | | Correlation | Revenue | Sales volume and revenue | |
|---|---|---|---|---|---|
| | | (1) | (2) | (3) | (4) |
| Analgesics | Coefficient (Std.) | 0.8088 | 0.0244*** (0.0035) | 0.2246*** (0.0463) | -0.201*** (0.0467) |
| | Observations | | 144,461 | | |
| Bath Soap | Coefficient (Std.) | 0.7757 | 0.0319*** (0.0082) | -0.3214** (0.1271) | 0.3527* (0.1279) |
| | Observations | | 15,295 | | |
| Bathroom Tissues | Coefficient (Std.) | 0.7838 | 0.036*** (0.0062) | 0.8456*** (0.1168) | -0.8195*** (0.1169) |
| | Observations | | 149,441 | | |
| Beer | Coefficient (Std.) | 0.8642 | 0.0131*** (0.0012) | -0.0562* (0.0288) | 0.0692** (0.0287) |
| | Observations | | 290,620 | | |
| Bottled Juice | Coefficient (Std.) | 0.9138 | 0.0305*** (0.0052) | 0.8371*** (0.1025) | -0.8065*** (0.1027) |
| | Observations | | 496,557 | | |
| Canned Soup | Coefficient (Std.) | 0.9458 | 0.0148*** (0.0056) | 0.5075*** (0.0785) | -0.4943*** (0.0791) |
| | Observations | | 495,543 | | |
| Canned Tuna | Coefficient (Std.) | 0.863 | 0.0201*** (0.0054) | 0.5276*** (0.1175) | -0.5093*** (0.1183) |
| | Observations | | 213,043 | | |
| Cereals | Coefficient (Std.) | 0.9311 | 0.0203*** (0.0038) | 0.0634 (0.0429) | -0.0433 (0.0437) |
| | Observations | | 357,120 | | |
| Cheese | Coefficient (Std.) | 0.86 | 0.018*** (0.0029) | 0.4048*** (0.1085) | -0.3846*** (0.1091) |
| | Observations | | 796,150 | | |
| Cigarettes | Coefficient (Std.) | 0.9938 | 0.0102** (0.0046) | -0.3209*** (0.0622) | 0.3303*** (0.062) |
| | Observations | | 36,157 | | |
| Cookies | Coefficient (Std.) | 0.9485 | 0.0273*** (0.0018) | -0.0045 (0.022) | 0.0318 (0.0221) |
| | Observations | | 688,761 | | |
| Crackers | Coefficient (Std.) | 0.9536 | 0.038*** (0.0031) | 0.1257** (0.0554) | -0.0881 (0.056) |
| | Observations | | 245,185 | | |
| Dish Detergent | Coefficient (Std.) | 0.8123 | 0.038*** (0.0043) | 0.4862*** (0.1382) | -0.4511*** (0.1376) |
| | Observations | | 189,633 | | |
| Fabric Softener | Coefficient (Std.) | 0.7951 | 0.0181*** (0.0048) | 0.9508*** (0.2067) | -0.9307*** (0.208) |
| | Observations | | 181,056 | | |
| Front−End−Candies | Coefficient (Std.) | 0.8611 | 0.0045 (0.0042) | 0.8471*** (0.075) | -0.8406*** (0.0755) |
| | Observations | | 278,853 | | |
| Frozen Dinners | Coefficient (Std.) | 0.726 | 0.0471*** (0.0033) | 0.2111*** (0.0446) | -0.162*** (0.044) |
| | Observations | | 203,191 | | |

Table 4 (Cont.)

| Category | | Correlation | Revenue | Sales volume and Revenue | |
|---|---|---|---|---|---|
| | | (1) | (2) | (3) | (4) |
| Frozen Entrees | Coefficient (Std.) | 0.8584 | 0.0271*** (0.002) | 0.1611*** (0.0155) | -0.1321*** (0.0153) |
| | Observations | | | 864,832 | |
| Frozen Juices | Coefficient (Std.) | 0.9651 | 0.0255*** (0.005) | 0.2732*** (0.068) | -0.2456*** (0.0687) |
| | Observations | | | 308,817 | |
| Grooming Products | Coefficient (Std.) | 0.811 | 0.0188*** (0.0024) | 0.0026 (0.0378) | 0.0162 (0.0382) |
| | Observations | | | 269,873 | |
| Laundry Detergents | Coefficient (Std.) | 0.9153 | 0.0174*** (0.0031) | 0.3272*** (0.061) | -0.311*** (0.0604) |
| | Observations | | | 272,765 | |
| Oatmeal | Coefficient (Std.) | 0.8775 | 0.0293*** (0.0082) | -0.0121 (0.0248) | 0.0416 (0.0255) |
| | Observations | | | 79,983 | |
| Paper Towels | Coefficient (Std.) | 0.7593 | 0.0395*** (0.0107) | 0.8975*** (0.204) | -0.8574*** (0.2028) |
| | Observations | | | 116,204 | |
| Refrigerated Juices | Coefficient (Std.) | 0.9201 | 0.0351*** (0.0048) | 0.1247** (0.0586) | -0.09 (0.0592) |
| | Observations | | | 306,865 | |
| Shampoos | Coefficient (Std.) | 0.7226 | 0.0162*** (0.0015) | 0.0171 (0.0162) | -0.0008 (0.0162) |
| | Observations | | | 261,778 | |
| Snack Crackers | Coefficient (Std.) | 0.9124 | 0.0324*** (0.0033) | 0.0736 (0.083) | -0.0414 (0.0839) |
| | Observations | | | 398,665 | |
| Soaps | Coefficient (Std.) | 0.6725 | 0.0359*** (0.0055) | 0.4935*** (0.0435) | -0.4684*** (0.0429) |
| | Observations | | | 152,379 | |
| Soft Drinks | Coefficient (Std.) | 0.756 | 0.0207*** (0.0017) | 0.5154*** (0.1761) | -0.4784*** (0.1763) |
| | Observations | | | 1,350,618 | |
| Toothbrushes | Coefficient (Std.) | 0.82 | 0.0202*** (0.0029) | 0.045 (0.0418) | -0.0247 (0.0419) |
| | Observations | | | 125,380 | |
| Toothpastes | Coefficient (Std.) | 0.8902 | 0.0137*** (0.0026) | -0.1035*** (0.0373) | 0.1181*** (0.0373) |
| | Observations | | | 264,317 | |
| **Average coefficients** | | **0.8523** | **0.03397** | **0.1902** | **-0.1705** |

Notes: Column 1 reports the Pearson correlation coefficient between the sales volume and revenues for each category. Columns 2–4 report the results of category-level fixed effect regressions of the probability of a small price change. The dependent variable in all columns is "small price change," which equals 1 if a price change of product $i$ in store $s$ at time $t$ is less or equal to 10¢, and 0 otherwise. In column 2, the main independent variable is the log of average revenue for product $i$ in store $s$ over the sample period. In columns 3 and 4, we report the results of a regression that includes both the log of the sales volume and the log of the revenue as independent variables. Column 3 reports the coefficients of the sales volume. Column 4 reports the coefficients of the revenue. All regressions also include fixed effects for months, years, stores, and products. We estimate separate regressions for each product category, clustering the errors by product. * $p < 10\%$, ** $p < 5\%$, *** $p < 1\%$

Table 5. Category-level regressions of small price changes and synchronization

| Category | | **(1)** | **(2)** | **(3)** | **(4)** |
|---|---|---|---|---|---|
| Analgesics | Coefficient (Std.) | 0.026*** (0.0034) | 0.0258*** (0.0034) | 0.024*** (0.0034) | 0.0245*** (0.0033) |
| | Observations | 144,461 | 144,461 | 143,780 | 139,228 |
| Bath Soap | Coefficient (Std.) | 0.0313*** (0.0077) | 0.0307*** (0.0077) | 0.0266*** (0.0086) | 0.0243*** (0.0088) |
| | Observations | 15,295 | 15,288 | 13,228 | 10,538 |
| Bathroom Tissues | Coefficient (Std.) | 0.0406*** (0.007) | 0.0402*** (0.007) | 0.0398*** (0.007) | 0.0436*** (0.0073) |
| | Observations | 149,441 | 149,441 | 148,926 | 139,207 |
| Beer | Coefficient (Std.) | 0.0123*** (0.0013) | 0.0114*** (0.0012) | 0.0114*** (0.0012) | 0.0119*** (0.0013) |
| | Observations | 290,620 | 290,617 | 290,524 | 267,233 |
| Bottled Juice | Coefficient (Std.) | 0.0357*** (0.0052) | 0.0356*** (0.0052) | 0.0348*** (0.0052) | 0.0357*** (0.0054) |
| | Observations | 496,557 | 496,555 | 496,461 | 485,670 |
| Canned Soup | Coefficient (Std.) | 0.0174*** (0.0056) | 0.0174*** (0.0056) | 0.0172*** (0.0055) | 0.0169*** (0.0056) |
| | Observations | 495,543 | 495,543 | 495,276 | 490,981 |
| Canned Tuna | Coefficient (Std.) | 0.0229*** (0.0053) | 0.0226*** (0.0053) | 0.023*** (0.0053) | 0.0234*** (0.0056) |
| | Observations | 213,043 | 213,043 | 212,567 | 202,922 |
| Cereals | Coefficient (Std.) | 0.0205*** (0.0037) | 0.0205*** (0.0037) | 0.0195*** (0.0035) | 0.0205*** (0.0034) |
| | Observations | 357,120 | 357,120 | 357,077 | 352,500 |
| Cheese | Coefficient (Std.) | 0.02*** (0.0028) | 0.0198*** (0.0028) | 0.0198*** (0.0028) | 0.0198*** (0.003) |
| | Observations | 796,150 | 796,148 | 796,142 | 758,753 |
| Cigarettes | Coefficient (Std.) | 0.007 (0.0051) | 0.0076 (0.005) | 0.0081 (0.0051) | 0.0091* (0.0047) |
| | Observations | 36,157 | 36,152 | 35,824 | 35,408 |
| Cookies | Coefficient (Std.) | 0.0272*** (0.0019) | 0.0278*** (0.0019) | 0.0275*** (0.0019) | 0.0277*** (0.0019) |
| | Observations | 688,761 | 688,759 | 688,726 | 681,886 |
| Crackers | Coefficient (Std.) | 0.0378*** (0.0031) | 0.0379*** (0.0031) | 0.0369*** (0.003) | 0.0359*** (0.003) |
| | Observations | 245,185 | 245,183 | 244,898 | 236,163 |
| Dish Detergent | Coefficient (Std.) | 0.0396*** (0.0045) | 0.0387*** (0.0045) | 0.0372*** (0.0043) | 0.0394*** (0.0041) |
| | Observations | 189,633 | 189,633 | 189,182 | 185,996 |
| Fabric Softener | Coefficient (Std.) | 0.0255*** (0.0048) | 0.0252*** (0.0048) | 0.0227*** (0.0047) | 0.0239*** (0.0048) |
| | Observations | 181,056 | 181,056 | 180,721 | 168,434 |
| Front-End-Candies | Coefficient (Std.) | 0.0042 (0.0041) | 0.0032 (0.0041) | 0.0054 (0.0034) | 0.0064* (0.0035) |
| | Observations | 278,853 | 278,853 | 278,019 | 267,951 |
| Frozen Dinners | Coefficient (Std.) | 0.0439*** (0.0037) | 0.0439*** (0.0037) | 0.0426*** (0.0036) | 0.0426*** (0.0036) |
| | Observations | 203,191 | 203,191 | 203,064 | 202,953 |

Table 5. (Cont.) Category-level regressions of small price changes, synchronization

| Category | | **(1)** | **(2)** | **(3)** | **(4)** |
|---|---|---|---|---|---|
| Frozen Entrees | Coefficient (Std.) | 0.0317*** (0.0023) | 0.0318*** (0.0023) | 0.0309*** (0.0022) | 0.0305*** (0.0022) |
| | Observations | 864,832 | 864,832 | 864,819 | 862,193 |
| Frozen Juices | Coefficient (Std.) | 0.0277*** (0.0048) | 0.0279*** (0.0048) | 0.0273*** (0.0046) | 0.0273*** (0.0045) |
| | Observations | 308,817 | 308,817 | 308,802 | 298,899 |
| Grooming Products | Coefficient (Std.) | 0.0188*** (0.0024) | 0.019*** (0.0024) | 0.0186*** (0.0024) | 0.0186*** (0.0025) |
| | Observations | 269,873 | 269,872 | 269,780 | 268,124 |
| Laundry Detergents | Coefficient (Std.) | 0.02*** (0.0033) | 0.0198*** (0.0033) | 0.0178*** (0.0032) | 0.0176*** (0.0033) |
| | Observations | 272,765 | 272,765 | 272,695 | 269,543 |
| Oatmeal | Coefficient (Std.) | 0.0281*** (0.0079) | 0.0281*** (0.008) | 0.0282*** (0.0079) | 0.0255*** (0.0081) |
| | Observations | 79,983 | 79,983 | 78,341 | 71,261 |
| Paper Towels | Coefficient (Std.) | 0.0447*** (0.0104) | 0.0447*** (0.0104) | 0.0436*** (0.0103) | 0.048*** (0.0109) |
| | Observations | 116,204 | 116,204 | 115,754 | 108,011 |
| Refrigerated Juices | Coefficient (Std.) | 0.035*** (0.0046) | 0.0352*** (0.0046) | 0.0335*** (0.0044) | 0.0344*** (0.0043) |
| | Observations | 306,865 | 306,865 | 306,841 | 293,807 |
| Shampoos | Coefficient (Std.) | 0.0146*** (0.0016) | 0.0138*** (0.0016) | 0.0128*** (0.0016) | 0.0127*** (0.0016) |
| | Observations | 261,778 | 261,778 | 261,740 | 257,886 |
| Snack Crackers | Coefficient (Std.) | 0.0338*** (0.0032) | 0.0335*** (0.0032) | 0.0328*** (0.0031) | 0.0328*** (0.003) |
| | Observations | 398,665 | 398,665 | 398,573 | 389,240 |
| Soaps | Coefficient (Std.) | 0.0375*** (0.0056) | 0.0372*** (0.0055) | 0.0348*** (0.0055) | 0.0345*** (0.0056) |
| | Observations | 152,379 | 152,379 | 152,104 | 149,407 |
| Soft Drinks | Coefficient (Std.) | 0.02*** (0.0019) | 0.0197*** (0.0019) | 0.0199*** (0.0019) | 0.02*** (0.0019) |
| | Observations | 1,350,618 | 1,350,617 | 1,350,613 | 1,337,747 |
| Toothbrushes | Coefficient (Std.) | 0.023*** (0.0028) | 0.0219*** (0.0028) | 0.0204*** (0.003) | 0.0205*** (0.0029) |
| | Observations | 125,380 | 125,380 | 124,743 | 122,787 |
| Toothpastes | Coefficient (Std.) | 0.0112*** (0.0026) | 0.0093*** (0.0026) | 0.0062** (0.0026) | 0.0067*** (0.0026) |
| | Observations | 264,317 | 264,317 | 264,156 | 260,282 |
| **Average coefficients** | | **0.0261** | **0.0259** | **0.0249** | **0.0253** |

<u>Notes</u>: The table reports the results of category-level fixed effect regressions of the probability of a small price change. The dependent variable is "small price change," which equals 1 if a price change of product $i$ in store $s$ at time $t$ is less or equal to 10¢, and 0 otherwise. The main independent variable is the log of the average sales volume of product $i$ in store $s$ over the sample period. Column 1 reports the results of a regression that includes the log of average sales volume and the average number of products offered in the category by the same producer. In column 2, we add the percentage of the products whose prices changed in the same week, excluding the current observation. In column 3, we add the average size of contemporaneous price changes, excluding the current observation. In column 4, we add the percentage of the products that are produced by the same producer and that changed price in the same week, excluding the current observation. All regressions include fixed effects for months, years, stores, and products. We estimate separate regressions for each product category, clustering the errors by product. * $p < 10\%$, ** $p < 5\%$, *** $p < 1\%$

Figure 1. Frequency of small price changes by sales volume deciles

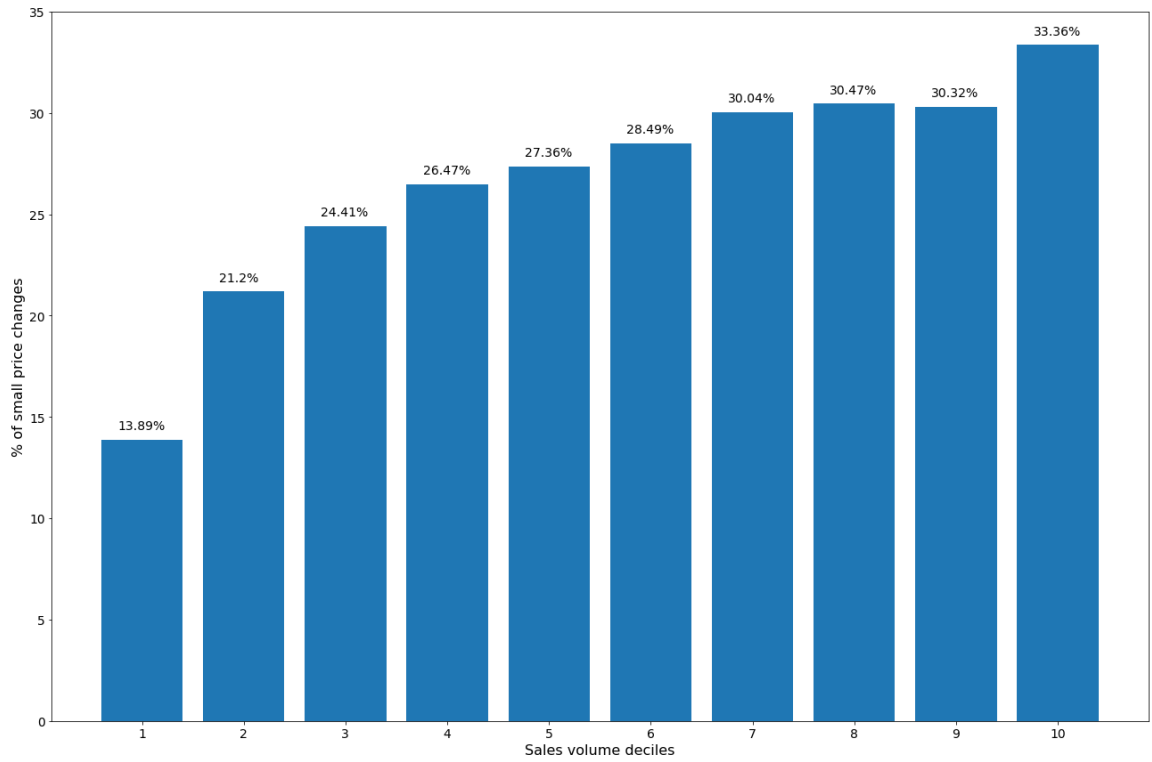

**Notes**: The chart was obtained by merging all 29 product categories and dividing the resulting data into deciles according to the products' sales volume. The % of small price changes $(\Delta P \leq 10¢)$ was calculated for each decile as a ratio of the number of small price changes to the number of total price changes in each decile.

Figure 2. Frequency of price changes by size for high, medium, and low sales volume products

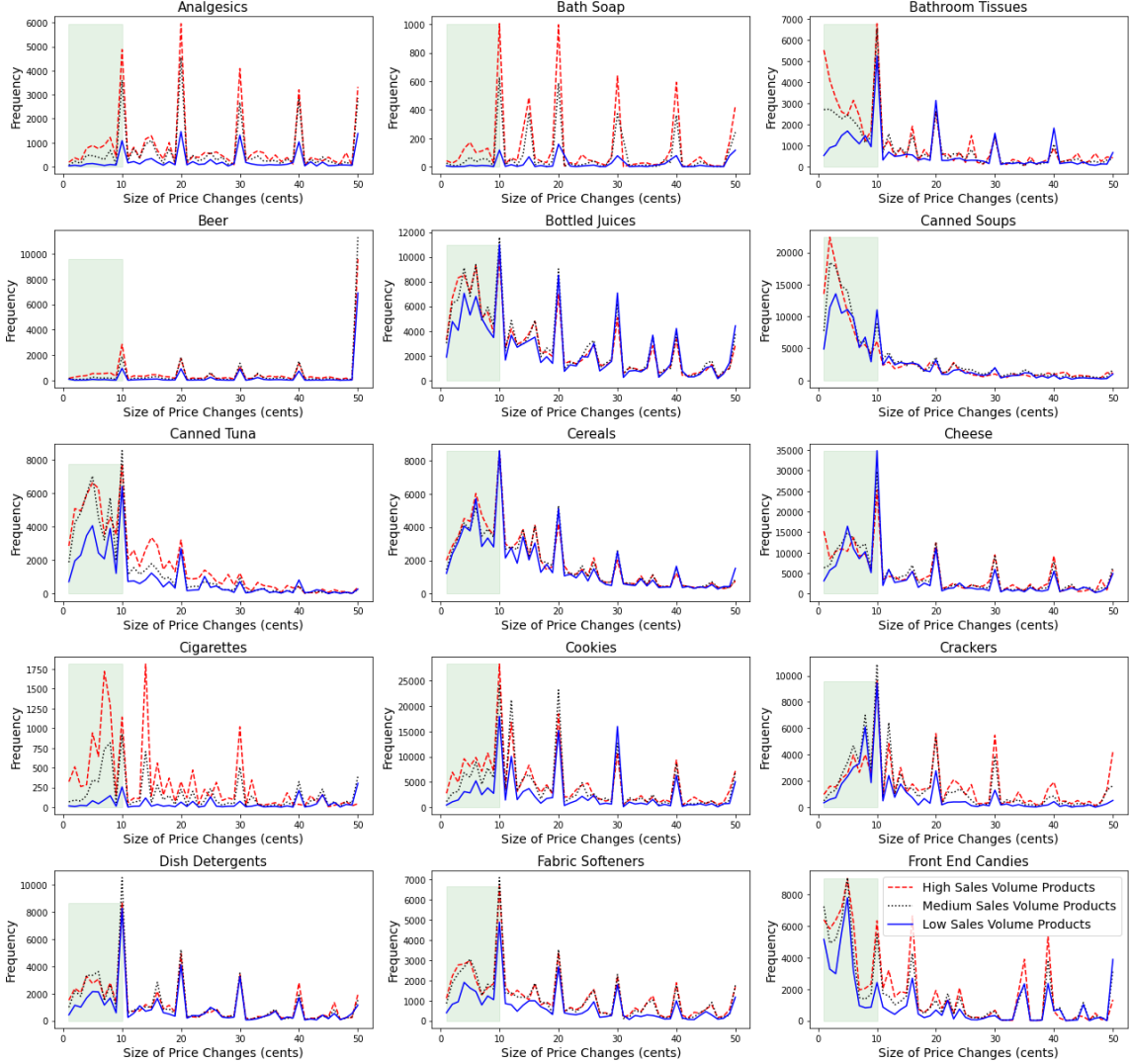

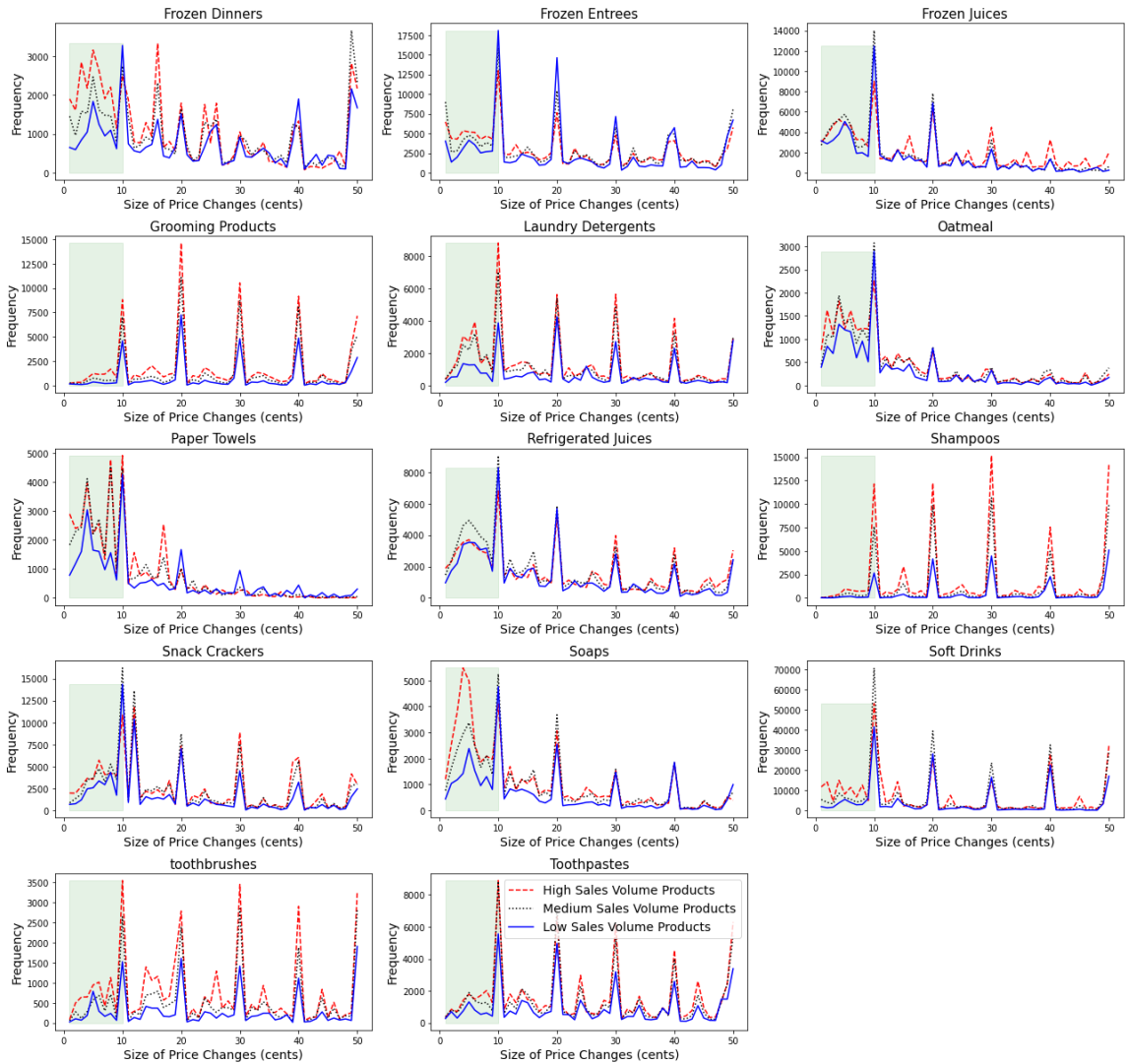

Notes: For each category, the figure shows the frequency of price changes for each size of price change from 1¢ to 50¢, comparing high sales volume products to medium sales volume products, and low sales volume products. To obtain the figures, we compute the average sales volume over the entire sample period for each product, in each store. We then group the products into high, medium, and low sales volume products. High (low) sales volume products are products in the high (low) third of the distribution. Medium sales volume products are in the middle third of the distribution. The y-axis shows the frequency of price changes. The red dashed line depicts the frequency of price changes for the high sales-volume products, the purple dotted line depicts the frequency of price changes for the medium sales-volume products, and the blue solid line depicts the frequency of price changes for the low sales volume products. The green shaded area marks the range of small price changes, $\Delta P \leq 10$¢ .

Figure 3. Product-level correlations between sales volume and small price changes in the Bathroom Tissues Category

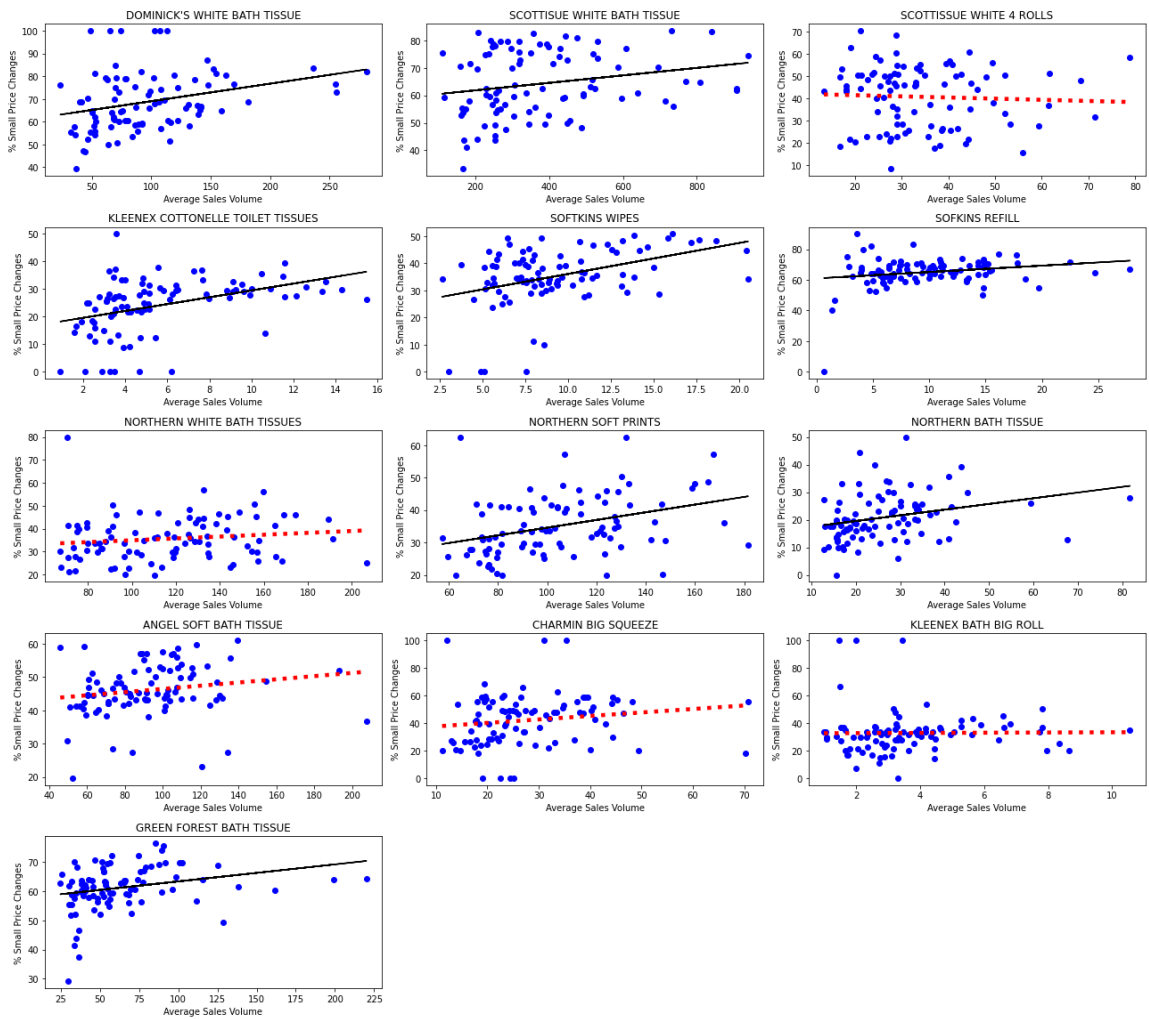

Note: The figure depicts the correlation between average sales volume (*x*-axis) and the percentage of small price changes for various products in the bathroom tissues category. Each dot in the figures represents the data for the product in a specific store. There are 93 dots in each figure, one for each store. The straight lines in the figures are the regression lines. Black solid regression lines indicate that the regression coefficient is significant at the 5% significance level, which is the case for 8 of the 13 products. The regression lines that are not statistically significant are marked with red dotted lines.

# Small Price Changes, Sales Volume, and Menu Cost

Doron Sayag

Department of Economics, Bar-Ilan University

Ramat-Gan 5290002, Israel

DoronSayag2@gmail.com

Avichai Snir

Department of Economics, Bar-Ilan University

Ramat-Gan 5290002, Israel

snirav@biu.ac.il

Daniel Levy

Department of Economics, Bar-Ilan University

Ramat-Gan 5290002, Israel,

Department of Economics, Emory University

Atlanta, GA 30322, USA,

ICEA, ISET at TSU (Georgia), and RCEA

Daniel.Levy@biu.ac.il

**Table of Contents**

## Appendix A. Controlling for measurement errors

We use a scanner dataset. As Eichenbaum et al. (2014) note, the distribution of the size of price changes in scanner datasets is prone to measurement errors. The errors may arise because, in scanner datasets, the price of a product in a given week is calculated as the ratio of the sales revenue to the quantity sold. Thus, if the price has changed during the week, or if some consumers used coupons, the price in the dataset might differ from the actual transaction price in that week.

This type of error is less of a concern in Dominick's dataset, because prices at Dominick's are set on a weekly basis, and the use of coupons in the period we study was limited. See Barsky et al. (2003), Chen et al. (2008), and Levy et al. (2010, 2011). Nevertheless, to mitigate possible concerns, we use in the paper only those price change observations after which the price has remained unchanged for at least 2 weeks. If a price remains unchanged for more than one week, then it is unlikely to be a mistake, since it is unlikely that the same error occurred two weeks in a row.

In this appendix, we conduct two more robustness checks. First, Table A1 presents the results of regressions equivalent to the regressions we present in Table 3 in the paper. This time, however, we include all price changes but exclude price changes smaller or equal to 2¢. Eichenbaum et al. (2014) suggest that such small price changes could be the result of measurement errors. Alvarez et al. (2016) also use Dominick's data, and they remove observations on price changes of 1¢. We, therefore, are more conservative by using a stricter rule than Alvarez et al. (2016). The regressions take the following form:

$$small\ price\ change_{i,s,t} = \alpha + \beta \ln(average\ sales\ volume_{i,s}) + \gamma \mathbf{X}_{i,s,t}$$
$$+ month_t + year_t + \delta_s + \mu_i + u_{i,s,t} \tag{A1}$$

where *small price change* is a dummy that equals 1 if a price change of product *i* in store *s* at time *t* is less or equal to 10¢ and larger than 2¢, and 0 otherwise. The *average sales volume* is the average sales volume of product *i* in store *s* over the sample period. $\mathbf{X}$ is a matrix of other control variables. *Month* and *year* are fixed effects for the month and the year of the price change. $\delta$ and $\mu$ are fixed effects for stores and products. *u* is an i.i.d error term. We estimate separate regressions for each product category, clustering the errors by product.

The values in Table A1 are the coefficients of the log of the average sales volume. In column 1, the only control variables are the log of the average sales volume, and the dummies for months, years, stores, and products. Consistent with the results we report in the paper, we find that all the coefficients of the log of the average sales volume are positive. 28 of the 29 coefficients are statistically significant. The average coefficient is 0.030, suggesting that a 1% increase in the sales volume is associated with a 3% increase in the likelihood of a small price change.

In column 2, we add controls for the log of the average price, the log of the absolute change in the wholesale price, and for sale- and bounce-back prices, which we identify using the sales filter algorithm of Fox and Syed (2016). All the coefficients are positive and statistically significant: 26 at the 1% level, two at the 5% level, and one at the 10% level. The average coefficient is 0.025, suggesting that a 1% increase in the sales volume is associated with a 2.5% increase in the likelihood of a small price change.

In column 3, we also add control for 9-ending prices. All coefficients remain positive and statistically significant: 26 at the 1% level, one at the 5%, and two at the 10% level. The average coefficient is 0.023, suggesting that a 1% increase in the sales volume is associated with a 2.3% increase in the likelihood of a small price change.

As a further control for the possible effects of sales on the results, in column 4 we focus on regular prices by excluding all sale- and bounce-back prices. When we focus on regular prices, the results are even stronger. All the coefficients are positive and statistically significant at the 1% level. The average coefficient is 0.045, suggesting that a 1% increase in the sales volume is associated with a 4.5% increase in the likelihood of a small price change.

As a second robustness test, we re-run the above regressions, after dropping observations if Dominick's sales flag indicated that there was a coupon use in either the week the price changed or in the preceding week because according to Eichenbaum et al. (2014), that might lead to spurious small price changes.

Table A2 reports the estimation results. The coefficient estimates are similar in sign, magnitude, and statistical significance, to the corresponding figures in Table A1. In all columns, the coefficient estimates are positive and significant, ranging between 0.025 and 0.046, on average.

   To summarize, the exclusion of (a) very small price changes, or (b) the exclusion of observations with coupon use, do not change our main result. The likelihood of small price changes remains strongly correlated with the average sales volume. We, therefore, conclude that our results are not likely to be driven by measurement errors.

Table A1. Category-level regressions of small price changes $(\Delta P \leq 10\cent)$ and sales volume

| Category | | **(1)** | **(2)** | **(3)** | **(4)** |
|---|---|---|---|---|---|
| Analgesics | Coefficient (Std.) | 0.0343*** (0.0031) | 0.0271*** (0.0026) | 0.0224*** (0.0025) | 0.044*** (0.0057) |
| | Observations | 275,225 | 275,225 | 275,225 | 73,576 |
| Bath Soap | Coefficient (Std.) | 0.0396*** (0.0085) | 0.044*** (0.0088) | 0.0417*** (0.0086) | 0.0897*** (0.0162) |
| | Observations | 35,377 | 35,377 | 35,377 | 6,362 |
| Bathroom Tissues | Coefficient (Std.) | 0.0311*** (0.0051) | 0.019*** (0.0055) | 0.0165*** (0.0052) | 0.04*** (0.0071) |
| | Observations | 288,963 | 288,963 | 288,963 | 58,189 |
| Beer | Coefficient (Std.) | 0.0212*** (0.0013) | 0.023*** (0.0011) | 0.0198*** (0.001) | 0.0673*** (0.005) |
| | Observations | 456,740 | 456,740 | 456,740 | 54,870 |
| Bottled Juice | Coefficient (Std.) | 0.0457*** (0.0042) | 0.0343*** (0.0031) | 0.0304*** (0.0032) | 0.0395*** (0.0049) |
| | Observations | 881,264 | 881,264 | 881,264 | 188,079 |
| Canned Soup | Coefficient (Std.) | 0.0202*** (0.0041) | 0.0109*** (0.0037) | 0.0121*** (0.0036) | 0.0268*** (0.0051) |
| | Observations | 814,575 | 814,575 | 814,575 | 191,616 |
| Canned Tuna | Coefficient (Std.) | 0.0322*** (0.0051) | 0.0246*** (0.0045) | 0.0216*** (0.0043) | 0.0361*** (0.0057) |
| | Observations | 330,897 | 330,897 | 330,897 | 89,246 |
| Cereals | Coefficient (Std.) | 0.0177*** (0.0025) | 0.0144*** (0.0023) | 0.0135*** (0.0024) | 0.0251*** (0.0038) |
| | Observations | 685,899 | 685,899 | 685,899 | 227,390 |
| Cheese | Coefficient (Std.) | 0.0301*** (0.0029) | 0.0189*** (0.0024) | 0.0159*** (0.0024) | 0.0123*** (0.0044) |
| | Observations | 1,615,593 | 1,615,593 | 1,615,593 | 371,129 |
| Cigarettes | Coefficient (Std.) | 0.0131 (0.0081) | 0.014* (0.0071) | 0.014* (0.007) | 0.0148*** (0.005) |
| | Observations | 15,395 | 15,395 | 15,395 | 9,130 |
| Cookies | Coefficient (Std.) | 0.0348*** (0.0016) | 0.0317*** (0.0016) | 0.0273*** (0.0015) | 0.05*** (0.0031) |
| | Observations | 1,305,448 | 1,305,448 | 1,305,448 | 205,310 |
| Crackers | Coefficient (Std.) | 0.044*** (0.0029) | 0.0365*** (0.0027) | 0.0334*** (0.0026) | 0.0516*** (0.0052) |
| | Observations | 453,298 | 453,298 | 453,298 | 78,127 |
| Dish Detergent | Coefficient (Std.) | 0.0377*** (0.0037) | 0.03*** (0.003) | 0.0273*** (0.0029) | 0.04*** (0.0047) |
| | Observations | 374,089 | 374,089 | 374,089 | 76,206 |
| Fabric Softener | Coefficient (Std.) | 0.0246*** (0.0036) | 0.0176*** (0.0033) | 0.0153*** (0.0034) | 0.0378*** (0.0053) |
| | Observations | 357,746 | 357,746 | 357,746 | 86,846 |
| Front-End-Candies | Coefficient (Std.) | 0.0238*** (0.0036) | 0.0146*** (0.0028) | 0.0139*** (0.0028) | 0.0121*** (0.0033) |
| | Observations | 415,331 | 415,331 | 415,331 | 121,111 |
| Frozen Dinners | Coefficient (Std.) | 0.0457*** (0.0026) | 0.0381*** (0.0025) | 0.0373*** (0.0025) | 0.101*** (0.0067) |
| | Observations | 477,997 | 477,997 | 477,997 | 57,704 |

Table A1. (Cont.)

| Category | | (1) | (2) | (3) | (4) |
|---|---|---|---|---|---|
| Frozen Entrees | Coefficient (Std.) | 0.0321*** (0.0016) | 0.0284*** (0.0015) | 0.0279*** (0.0015) | 0.0572*** (0.0034) |
| | Observations | 1,768,979 | 1,768,979 | 1,768,979 | 295,796 |
| Frozen Juices | Coefficient (Std.) | 0.0262*** (0.0036) | 0.0216*** (0.0033) | 0.0196*** (0.0032) | 0.0308*** (0.0059) |
| | Observations | 602,210 | 602,210 | 602,210 | 112,532 |
| Grooming Products | Coefficient (Std.) | 0.0394*** (0.0022) | 0.0426*** (0.002) | 0.0379*** (0.002) | 0.0637*** (0.0061) |
| | Observations | 658,707 | 658,707 | 658,707 | 95,757 |
| Laundry Detergents | Coefficient (Std.) | 0.0156*** (0.0029) | 0.0133*** (0.0025) | 0.0112*** (0.0023) | 0.025*** (0.0047) |
| | Observations | 580,679 | 580,679 | 580,679 | 135,575 |
| Oatmeal | Coefficient (Std.) | 0.0241*** (0.007) | 0.0153** (0.0059) | 0.0139*** (0.0059) | 0.0416*** (0.0073) |
| | Observations | 154,817 | 154,817 | 154,817 | 51,510 |
| Paper Towels | Coefficient (Std.) | 0.0306*** (0.0119) | 0.0255** (0.0125) | 0.0245* (0.0127) | 0.0359*** (0.0119) |
| | Observations | 215,951 | 215,951 | 215,951 | 36,645 |
| Refrigerated Juices | Coefficient (Std.) | 0.0239*** (0.0032) | 0.0179*** (0.0028) | 0.0158*** (0.0027) | 0.029*** (0.0047) |
| | Observations | 749,239 | 749,239 | 749,239 | 127,091 |
| Shampoos | Coefficient (Std.) | 0.0293*** (0.0013) | 0.0337*** (0.0013) | 0.0295*** (0.0012) | 0.0644*** (0.0042) |
| | Observations | 708,002 | 708,002 | 708,002 | 83,652 |
| Snack Crackers | Coefficient (Std.) | 0.0365*** (0.0029) | 0.0338*** (0.0028) | 0.0305*** (0.0026) | 0.0621*** (0.0041) |
| | Observations | 770,442 | 770,442 | 770,442 | 127,881 |
| Soaps | Coefficient (Std.) | 0.0231*** (0.0011) | 0.0205*** (0.0009) | 0.0171*** (0.0008) | 0.0442*** (0.0023) |
| | Observations | 4,243,492 | 4,243,492 | 4,243,492 | 305,545 |
| Soft Drinks | Coefficient (Std.) | 0.0415*** (0.0058) | 0.0342*** (0.0042) | 0.0286*** (0.004) | 0.0562*** (0.0069) |
| | Observations | 300,763 | 300,763 | 300,763 | 71,459 |
| Toothbrushes | Coefficient (Std.) | 0.0248*** (0.0029) | 0.0276*** (0.003) | 0.0237*** (0.0029) | 0.0562*** (0.0059) |
| | Observations | 291,093 | 291,093 | 291,093 | 42,658 |
| Toothpastes | Coefficient (Std.) | 0.0242*** (0.0029) | 0.0247*** (0.0025) | 0.0218*** (0.0024) | 0.0507*** (0.0059) |
| | Observations | 584,401 | 584,401 | 584,401 | 84,802 |
| **Average coefficients** | | **0.0299** | **0.0255** | **0.0229** | **0.0450** |

<u>Notes</u>: The table reports the results of category-level fixed effect regressions of the probability of a small price change. The dependent variable is "small price change," which equals 1 if a price change of product $i$ in store $s$ at time $t$ is less or equal to 10¢ and larger than 2¢, and 0 otherwise. The main independent variable is the log of the average sales volume of product $i$ in store $s$ over the sample period. Column 1 reports the results of the baseline regression that includes only the log of the average sales volume and the fixed effects for months, years, stores, and products. In column 2, we add the following controls: the natural log of the average price, the natural log of the absolute change in the wholesale price, and control for sale- and bounce-back prices, which we identify using a sales filter algorithm. In column 3, we add a dummy for 9-ending prices as an additional control. In column 4, we focus on regular prices by excluding the sale- and bounce-back prices. We estimate separate regressions for each product category, clustering the errors by product. * $p < 10\%$, ** $p < 5\%$, *** $p < 1\%$

Table A2. Category-level regressions of small price changes and sales volume, excluding coupon sales

| Category | | **(1)** | **(2)** | **(3)** | **(4)** |
|---|---|---|---|---|---|
| Analgesics | Coefficient (Std.) | 0.0388*** (0.0033) | 0.0305*** (0.0027) | 0.0248*** (0.0025) | 0.0475*** (0.0057) |
| | Observations | 278,043 | 278,043 | 278,043 | 75,945 |
| Bath Soap | Coefficient (Std.) | 0.0409*** (0.0093) | 0.0452*** (0.0095) | 0.0422*** (0.0091) | 0.0871*** (0.016) |
| | Observations | 35,795 | 35,795 | 35,795 | 6,555 |
| Bathroom Tissues | Coefficient (Std.) | 0.0372*** (0.0056) | 0.0203*** (0.0053) | 0.0177*** (0.0049) | 0.0351*** (0.0069) |
| | Observations | 326,382 | 326,382 | 326,382 | 81,914 |
| Beer | Coefficient (Std.) | 0.023*** (0.0015) | 0.0249*** (0.0012) | 0.0208*** (0.0012) | 0.0691*** (0.005) |
| | Observations | 459,669 | 459,669 | 459,669 | 56,427 |
| Bottled Juice | Coefficient (Std.) | 0.0554*** (0.0043) | 0.0393*** (0.003) | 0.0343*** (0.0031) | 0.0368*** (0.0045) |
| | Observations | 959,958 | 959,958 | 959,958 | 244,198 |
| Canned Soup | Coefficient (Std.) | 0.0272*** (0.004) | 0.0151*** (0.0034) | 0.0158*** (0.0033) | 0.0217*** (0.0038) |
| | Observations | 947,633 | 947,633 | 947,633 | 278,451 |
| Canned Tuna | Coefficient (Std.) | 0.037*** (0.0052) | 0.0266*** (0.0044) | 0.0225*** (0.0041) | 0.0334*** (0.0047) |
| | Observations | 375,343 | 375,343 | 375,343 | 116,170 |
| Cereals | Coefficient (Std.) | 0.0216*** (0.0026) | 0.0168*** (0.0023) | 0.0156*** (0.0024) | 0.0263*** (0.0035) |
| | Observations | 724,232 | 724,232 | 724,232 | 260,035 |
| Cheese | Coefficient (Std.) | 0.0374*** (0.0029) | 0.0208*** (0.0022) | 0.0168*** (0.0022) | 0.0116*** (0.0031) |
| | Observations | 1,811,792 | 1,811,792 | 1,811,792 | 519,225 |
| Cigarettes | Coefficient (Std.) | 0.019*** (0.0082) | 0.0203*** (0.0068) | 0.0197*** (0.0067) | 0.0215*** (0.0045) |
| | Observations | 15,862 | 15,862 | 15,862 | 9,593 |
| Cookies | Coefficient (Std.) | 0.0429*** (0.0017) | 0.0372*** (0.0017) | 0.0315*** (0.0015) | 0.0542*** (0.0031) |
| | Observations | 1,356,845 | 1,356,845 | 1,356,845 | 229,139 |
| Crackers | Coefficient (Std.) | 0.0544*** (0.0033) | 0.0431*** (0.0031) | 0.0389*** (0.0029) | 0.0563*** (0.0061) |
| | Observations | 475,368 | 475,368 | 475,368 | 89,210 |
| Dish Detergent | Coefficient (Std.) | 0.0481*** (0.0038) | 0.0357*** (0.003) | 0.0315*** (0.0029) | 0.0417*** (0.0043) |
| | Observations | 401,001 | 401,001 | 401,001 | 95,477 |
| Fabric Softener | Coefficient (Std.) | 0.0342*** (0.0038) | 0.0245*** (0.0034) | 0.0209*** (0.0035) | 0.0428*** (0.0049) |
| | Observations | 378,773 | 378,773 | 378,773 | 101,926 |
| Front-End-Candies | Coefficient (Std.) | 0.0165*** (0.0039) | 0.0091*** (0.0028) | 0.0082*** (0.0028) | 0.0113*** (0.0031) |
| | Observations | 490,220 | 490,220 | 490,220 | 155,203 |
| Frozen Dinners | Coefficient (Std.) | 0.0536*** (0.0027) | 0.0408*** (0.0025) | 0.0394*** (0.0025) | 0.0907*** (0.006) |
| | Observations | 502,792 | 502,792 | 502,792 | 72,693 |

Table A2. (Cont.)

| Category | | **(1)** | **(2)** | **(3)** | **(4)** |
|---|---|---|---|---|---|
| Frozen Entrees | Coefficient (Std.) | 0.0354*** (0.0019) | 0.0301*** (0.0017) | 0.0292*** (0.0017) | 0.0602*** (0.0032) |
| | Observations | 1,848,166 | 1,848,166 | 1,848,166 | 353,120 |
| Frozen Juices | Coefficient (Std.) | 0.0342*** (0.0037) | 0.0253*** (0.0031) | 0.0227*** (0.003) | 0.03*** (0.0048) |
| | Observations | 659,295 | 659,295 | 659,295 | 150,129 |
| Grooming Products | Coefficient (Std.) | 0.0426*** (0.0024) | 0.0455*** (0.0022) | 0.039*** (0.0021) | 0.0673*** (0.0061) |
| | Observations | 668,809 | 668,809 | 668,809 | 99,252 |
| Laundry Detergents | Coefficient (Std.) | 0.0185*** (0.0031) | 0.0155*** (0.0027) | 0.0126*** (0.0025) | 0.0264*** (0.0047) |
| | Observations | 594,247 | 594,247 | 594,247 | 145,167 |
| Oatmeal | Coefficient (Std.) | 0.0288*** (0.0071) | 0.0172*** (0.0052) | 0.0151*** (0.0052) | 0.0319*** (0.0094) |
| | Observations | 168,988 | 168,988 | 168,988 | 63,575 |
| Paper Towels | Coefficient (Std.) | 0.0376*** (0.0114) | 0.0296*** (0.0116) | 0.0284*** (0.0117) | 0.0376*** (0.0096) |
| | Observations | 244,037 | 244,037 | 244,037 | 52,321 |
| Refrigerated Juices | Coefficient (Std.) | 0.031*** (0.0032) | 0.0209*** (0.0027) | 0.0182*** (0.0026) | 0.0305*** (0.0041) |
| | Observations | 800,176 | 800,176 | 800,176 | 161,074 |
| Shampoos | Coefficient (Std.) | 0.0323*** (0.0014) | 0.0368*** (0.0014) | 0.032*** (0.0013) | 0.0674*** (0.0043) |
| | Observations | 713,652 | 713,652 | 713,652 | 86,458 |
| Snack Crackers | Coefficient (Std.) | 0.0435*** (0.0032) | 0.0382*** (0.003) | 0.0338*** (0.0027) | 0.066*** (0.004) |
| | Observations | 801,599 | 801,599 | 801,599 | 143,154 |
| Soaps | Coefficient (Std.) | 0.0306*** (0.0014) | 0.0265*** (0.001) | 0.0223*** (0.0009) | 0.0586*** (0.0027) |
| | Observations | 4,372,346 | 4,372,346 | 4,372,346 | 346,106 |
| Soft Drinks | Coefficient (Std.) | 0.0545*** (0.006) | 0.0413*** (0.0044) | 0.0336*** (0.0042) | 0.0555*** (0.0057) |
| | Observations | 333,170 | 333,170 | 333,170 | 94,295 |
| Toothbrushes | Coefficient (Std.) | 0.0291*** (0.0032) | 0.0317*** (0.0034) | 0.0265*** (0.0032) | 0.0618*** (0.006) |
| | Observations | 295,275 | 295,275 | 295,275 | 44,690 |
| Toothpastes | Coefficient (Std.) | 0.0289*** (0.0032) | 0.028*** (0.0027) | 0.0241*** (0.0026) | 0.0561*** (0.0063) |
| | Observations | 596,900 | 596,900 | 596,900 | 91,759 |
| **Average coefficients** | | **0.0356** | **0.0288** | **0.0254** | **0.0461** |

<u>Notes:</u> The table reports the results of category-level fixed effect regressions of the probability of a small price change. The dependent variable is "small price change," which equals 1 if a price change of product $i$ in store $s$ at time $t$ is less or equal to 10¢, and 0 otherwise. We exclude observations on price changes if Dominick's sales flag indicates a coupon sale in either week $t$ or $t-1$. The main independent variable is the log of the average sales volume of product $i$ in store $s$ over the sample period. Column 1 reports the results of the baseline regression that includes only the log of the average sales volume and fixed effects for months, years, stores, and products. In column 2, we add the following controls: the natural log of the average price, the natural log of the absolute change in the wholesale price, and control for sale- and bounce-back prices, which we identify using a sales filter algorithm. In column 3, we add a dummy for 9-ending prices as an additional control. In column 4, we focus on regular prices by excluding the sale- and bounce-back prices. We estimate separate regressions for each product category, clustering the errors by product. * $p < 10\%$, ** $p < 5\%$, *** $p < 1\%$

## Appendix B. Alternative definitions of small price changes

In the paper, we define a small price change as a price change smaller than or equal to 10¢. In this appendix, we repeat our main analyses using 8 alternative definitions of small price changes. First, we define small price changes as (1) price changes up to, and including, 5¢, (2) price changes up to, and including, 15¢, (3) price changes up to, and including 2%, and (4) price changes up to, and including, 5%.

Second, we follow Midrigan (2011) and Bhattarai and Schoenle (2014) to define small price changes relative to the average price change in the corresponding category. I.e., a price change is small if it is smaller than or equal to $\kappa |\overline{\Delta p_{i,s}}|$, where $\overline{\Delta p_{i,s}}$ is the average price change of product $i$ in store $s$, and $\kappa$ attains the values 0.5, 0.33, 0.25 and 0.10.

As we do in the paper, we use observations on price changes only if we observe the price in both weeks $t$ and $t + 1$ and the post-change price remained unchanged for at least 2 weeks.

Table B1 presents the results of regressions equivalent to the regressions in Table 3 in the paper. The regressions take the following form:

$$small\ price\ change_{i,s,t} = \alpha + \beta \ln(average\ sales\ volume_{i,s}) + \gamma \mathbf{X}_{i,s,t}$$
$$+ month_t + year_t + \delta_s + \mu_i + u_{i,s,t} \tag{B1}$$

where *small price change* is a dummy that equals 1 if a price change of product $i$ in store $s$ at time $t$ is less or equal to 5¢, and 0 otherwise. The *average sales volume* is the average sales volume of product $i$ in store $s$ over the sample period. $\mathbf{X}$ is a matrix of other control variables. *Month* and *year* are fixed effects for the month and the year of the price change. $\delta$ and $\mu$ are fixed effects for stores and products, respectively, and $u$ is an i.i.d error term. We estimate separate regressions for each product category, clustering the errors by product.

The values in the table are the coefficients of the log of the average sales volume. In column 1, the only control variables are the log of the average sales volume, and the dummies for months, years, stores, and products. We find that 28 of the coefficients of the log of the average sales volume are positive, 16 of them are statistically significant. The average coefficient is 0.010, suggesting that a 1% increase in the sales volume is associated with a 1.0% increase in the likelihood of a small price change.

In column 2, we add controls for the log of the average price, the log of the absolute change in the wholesale price, and a control for sale- and bounce-back prices, which we identify using the sales filter algorithm of Fox and Syed (2016). 27 of the coefficients are positive, 13 of the 27 are statistically significant, and 2 more are marginally statistically significant. The average coefficient is 0.007, suggesting that a 1% increase in the sales volume is associated with a 0.7% increase in the likelihood of a small price change.

In column 3, we also add control for 9-ending prices. 27 of the coefficients are positive, 14 of them statistically significant, and 3 more are marginally statistically significant. The average coefficient is 0.008, suggesting that a 1% increase in the sales volume is associated with a 0.8% increase in the likelihood of a small price change.

As a further control for the effects of sales on the results, in column 4 we focus on regular prices by excluding all sale- and bounce-back prices. We find that 28 coefficients are positive. 20 of the positive coefficients are statistically significant, and 2 more are statistically significant at the 10% level. The average coefficient is 0.019, suggesting that a 1% increase in the sales volume is associated with a 1.9% increase in the likelihood of a small price change.

Table B2 presents the results of similar regressions, where we define small price changes as price changes of up to, and including, 15¢. In column 1, the only control variables are the log of the average sales volume, and the dummies for months, years, stores, and products. We find that 26 of the coefficients of the log of the average sales volume are positive. 17 of the 26 are statistically significant, and 2 more are marginally statistically significant. The average coefficient is 0.016, suggesting that a 1% increase in the sales volume is associated with a 1.6% increase in the likelihood of a small price change.

In column 2, we add controls for the log of the average price, the log of the absolute change in the wholesale price, and a control for sale- and bounce-back prices, which we identify using the sales filter algorithm of Fox and Syed (2016). 25 of the coefficients are positive. 13 of the positive coefficients are statistically significant, and 2 more are marginally statistically significant. The average coefficient is 0.012, suggesting that a 1% increase in the sales volume is associated with a 1.2% increase in the likelihood of a small price change.

In column 3, we also add control for 9-ending prices. 25 of the coefficients are positive. 13 of the positive coefficients are statistically significant, and 5 more are marginally statistically significant. The average coefficient is 0.012, suggesting that a 1% increase in the sales volume is associated with a 1.2% increase in the likelihood of a small price change.

As a further control for the effects of sales on the results, in column 4 we focus on regular prices by excluding all sale- and bounce-back prices. 26 of the coefficients are positive. 17 of the 26 are statistically significant, and 2 more are statistically significant at the 10% level. The average coefficient is 0.019, suggesting that a 1% increase in the sales volume is associated with a 1.9% increase in the likelihood of a small price change.

Table B3 presents the results where we define small price changes as price changes of up to 2%. In column 1, the only control variables are the log of the average sales volume, and the dummies for months, years, stores, and products. We find that 27 of the 29 coefficients of the log of the average sales volume are positive. Out of the 27, 20 are statistically significant, and 3 more are statistically significant at the 10% level. The average coefficient is 0.009, suggesting that a 1% increase in the sales volume is associated with a 0.9% increase in the likelihood of a small price change.

In column 2, we add controls for the log of the average price, the log of the absolute change in the wholesale price, and a control for sale- and bounce-back prices, which we identify using the sales filter algorithm of Fox and Syed (2016). We find that 27 of the 29 coefficients are positive. 17 of the 27 are statistically significant, and 6 more are statistically significant at the 10% level. The average coefficient is 0.007, suggesting that a 1% increase in the sales volume is associated with a 0.7% increase in the likelihood of a small price change.

In column 3, we also add a control for 9-ending prices. 27 of the 29 coefficients are still positive. 17 of the 27 are statistically significant, and 5 more are statistically significant at the 10% level. The average coefficient is 0.007, suggesting that a 1% increase in the sales volume is associated with a 0.7% increase in the likelihood of a small price change.

As a further control for the effects of sales on the results, in column 4 we focus on regular prices by excluding all sale- and bounce-back prices. When we focus on regular prices, 28 of the 29 coefficients are positive. 17 of the positive coefficients are statistically significant, and 3 more are marginally statistically significant. The average coefficient is 0.015, suggesting that a 1% increase in the sales volume is associated with a 1.5% increase in the likelihood of a small price change.

Table B4 presents the results where we define small price changes as price changes up to 5%. In column 1, the only control variables are the log of the average sales volume, and the dummies for months, years, stores, and products. We find that 23 coefficients of the log of the average sales volume are positive. 17 of them are statistically significant, and 2 more are marginally statistically significant. The average coefficient is 0.015, suggesting that a 1% increase in the sales volume is associated with a 1.5% increase in the likelihood of a small price change.

In column 2, we add controls for the log of the average price, the log of the absolute change in the wholesale price, and a control for sale- and bounce-back prices, which we identify using the sales filter algorithm of Fox and Syed (2016). 23 of the coefficients are positive. 15 of the positive coefficients are statistically significant, and 2 more are marginally statistically significant. The average coefficient is 0.010, suggesting that a 1% increase in the sales volume is associated with a 1.0% increase in the likelihood of a small price change.

In column 3, we also add a control for 9-ending prices. 23 of the coefficients are positive. 15 of the 23 are statistically significant, and 2 more are marginally statistically significant. The average coefficient is 0.010, suggesting that a 1% increase in the sales volume is associated with a 1.0% increase in the likelihood of a small price change.

As a further control for the effects of sales on the results, in column 4 we focus on regular prices by excluding all sale- and bounce-back prices. When we focus on regular prices, 28 of the coefficients are positive. 16 of the 28 are statistically significant, and 2 more are marginally statistically significant.  The average coefficient is 0.021, suggesting that a 1% increase in the sales volume is associated with a 2.1% increase in the likelihood of a small price change.

Table B5 presents the results where we define small price changes as price changes of up to 50% of the average price change of the product-store. In other words, a price change, $\Delta p_{i,s,t}$, of product $i$ in store $s$ in week $t$ is small if $\left|\Delta p_{i,s,t}\right| \leq 0.50\left|\overline{\Delta p_{i,s}}\right|$, where $\overline{\Delta p_{i,s}}$ is the average size of a price change of product $i$ in store $s$. In column 1, the only control variables are the log of the average sales volume, and dummies for months, years, stores, and products. We find that 26 of the coefficients are positive. 22 of the positive coefficients are statistically significant, and 1 more is marginally statistically significant. The average coefficient is 0.021, suggesting that a 1% increase in the sales volume is associated with a 2.1% increase in the likelihood of a small price change.

In column 2, we add controls for the log of the average price, the log of the absolute change in the wholesale price, and a control for sale- and bounce-back prices, which we identify using the sales filter algorithm of Fox and Syed (2016). We find that 26 of the coefficients are positive. 18 of the 26 are statistically significant, and 2 more are marginally statistically significant. The average coefficient is 0.020, suggesting that a 1% increase in the sales volume is associated with a 2.0% increase in the likelihood of a small price change.

In column 3, we also add a control for 9-ending prices. We find that 27 of the coefficients are positive. 18 of the 27 are statistically significant, and 2 more are marginally statistically significant. The average coefficient is 0.020, suggesting that a 1% increase in the sales volume is associated with a 2.0% increase in the likelihood of a small price change.

As a further control for the effects of sales on the results, in column 4 we focus on regular prices by excluding all sale- and bounce-back prices. When we focus on regular prices, 28 of the 29 are statistically significant. 23 of the positive coefficients are statistically significant, and 3 more are marginally significant. The average coefficient is 0.041, suggesting that a 1% increase in the sales volume is associated with a 4.1% increase in the likelihood of a small price change.

Table B6 presents the results where we define small price changes as price changes up to 33% of the average price in the category. In other words, a price change, $\Delta p_{i,s,t}$, of product $i$ in store $s$ in week $t$ is small if $\left|\Delta p_{i,s,t}\right| \leq 0.33\left|\overline{\Delta p_{i,s}}\right|$, where $\overline{\Delta p_{i,s}}$ is the average size of a price change of product $i$ in store $s$. In column 1, the only control variables are the log of the average sales volume, and dummies for months, years, stores, and products. We find that 27 of the 29 coefficients of the log of the average sales volume are positive.

22 of them are statistically significant, and 2 more is marginally significant. The average coefficient is 0.016, suggesting that a 1% increase in the sales volume is associated with a 1.6% increase in the likelihood of a small price change.

In column 2, we add controls for the log of the average price, the log of the absolute change in the wholesale price, and a control for sale- and bounce-back prices, which we identify using the sales filter algorithm of Fox and Syed (2016). We find that 27 of the 29 coefficients are positive. 18 of them are statistically significant, and 3 more are marginally significant. The average coefficient is 0.015, suggesting that a 1% increase in the sales volume is associated with a 1.5% increase in the likelihood of a small price change.

In column 3, we also add a control for 9-ending prices. We find that 27 of the 29 coefficients are positive. 18 of them are statistically significant, and 3 more are marginally significant. The average coefficient is 0.015, suggesting that a 1% increase in the sales volume is associated with a 1.5% increase in the likelihood of a small price change.

As a further control for the effects of sales on the results, in column 4 we focus on regular prices by excluding all sale- and bounce-back prices. We find that all the coefficients are positive and 23 of them are statistically significant. 3 more coefficients are marginally significant. The average coefficient is 0.031, suggesting that a 1% increase in the sales volume is associated with a 3.1% increase in the likelihood of a small price change.

Table B7 presents the results where we define small price changes as price changes up to 25% of the average price in the category. In other words, a price change, $\Delta p_{i,s,t}$, of product $i$ in store $s$ in week $t$ is small if $\left|\Delta p_{i,s,t}\right| \leq 0.25\left|\overline{\Delta p_{i,s}}\right|$, where $\overline{\Delta p_{i,s}}$ is the average size of a price change of product $i$ in store $s$. In column 1, the only control variables are the log of the average sales volume, and the dummies for months, years, stores, and products. We find that 28 of the 29 coefficients of the log of the average sales volume are positive. 22 of them are statistically significant, and 1 more is marginally significant. The average coefficient is 0.013, suggesting that a 1% increase in the sales volume is associated with a 1.3% increase in the likelihood of a small price change.

In column 2, we add controls for the log of the average price, the log of the absolute change in the wholesale price, and a control for sale- and bounce-back prices, which we identify using the sales filter algorithm of Fox and Syed (2016). We find that 27 of the 29 coefficients are positive. 22 of them are statistically significant. The average coefficient is 0.012, suggesting that a 1% increase in the sales volume is associated with a 1.2% increase in the likelihood of a small price change.

In column 3, we also add a control for 9-ending prices. We find that 27 of the 29 coefficients are positive. 22 of them are statistically significant. The average coefficient is 0.012, suggesting that a 1% increase in the sales volume is associated with a 1.2% increase in the likelihood of a small price change.

As a further control for the effects of sales on the results, in column 4 we focus on regular prices by excluding all sale- and bounce-back prices. We find that all the coefficients are positive and 25 of them are statistically significant at the 1% level. One more coefficient is marginally significant. The average coefficient is 0.025, suggesting that a 1% increase in the sales volume is associated with a 2.5% increase in the likelihood of a small price change.

Table B8 presents the results where we define small price changes as price changes up to 10% of the average price in the category. In other words, a price change, $\Delta p_{i,s,t}$, of product $i$ in store $s$ in week $t$ is small if $|\Delta p_{i,s,t}| \leq 0.10|\overline{\Delta p_{i,s}}|$, where $\overline{\Delta p_{i,s}}$ is the average size of a price change of product $i$ in store $s$. In column 1, the only control variables are the log of the average sales volume, and the dummies for months, years, stores, and products. We find that 24 of the 29 coefficients of the log of the average sales volume are positive. 19 of them are statistically significant, and 2 more are marginally significant. The average coefficient is 0.004, suggesting that a 1% increase in the sales volume is associated with a 0.4% increase in the likelihood of a small price change.

In column 2, we add controls for the log of the average price, the log of the absolute change in the wholesale price, and a control for sale- and bounce-back prices, which we identify using the sales filter algorithm of Fox and Syed (2016). We find that 25 of the 29 coefficients are positive. 20 of them are statistically significant, and 2 more are marginally significant. The average coefficient is 0.005, suggesting that a 1% increase in the sales volume is associated with a 0.5% increase in the likelihood of a small price change.

In column 3, we also add a control for 9-ending prices. We find that 25 of the 29 coefficients are positive. 20 of them are statistically significant, and 2 more are marginally significant. The average coefficient is 0.005, suggesting that a 1% increase in the sales volume is associated with a 0.5% increase in the likelihood of a small price change.

As a further control for the effects of sales on the results, in column 4 we focus on regular prices by excluding all sale- and bounce-back prices. We find that 25 of the coefficients are positive. 20 of them are statistically significant, and 3 more are marginally significant. The average coefficient is 0.010, suggesting that a 1% increase in the sales volume is associated with a 1.0% increase in the likelihood of a small price change.

Table B1. Category-level regressions of small price changes ($\Delta P \leq 5¢$) and sales volume

| Category | | (1) | (2) | (3) | (4) |
|---|---|---|---|---|---|
| Analgesics | Coefficient (Std.) | 0.0033 (0.0023) | 0.0022 (0.0022) | 0.0019 (0.0022) | 0.0068 (0.0056) |
| | Observations | 74,451 | 74,451 | 74,451 | 24,729 |
| Bath Soap | Coefficient (Std.) | 0.0128** (0.0053) | 0.0114** (0.0051) | 0.0113** (0.0044) | 0.0198 (0.0129) |
| | Observations | 6,650 | 6,650 | 6,650 | 1,466 |
| Bathroom Tissues | Coefficient (Std.) | 0.0481*** (0.0079) | 0.0235*** (0.0066) | 0.0233*** (0.0063) | 0.0466*** (0.0074) |
| | Observations | 56,458 | 56,458 | 56,458 | 19,285 |
| Beer | Coefficient (Std.) | 0.0003 (0.0002) | 0.0005** (0.0002) | 0.0005** (0.0002) | 0.0051** (0.0021) |
| | Observations | 187,691 | 187,691 | 187,691 | 12,080 |
| Bottled Juice | Coefficient (Std.) | 0.0067 (0.0068) | 0.0017 (0.0059) | 0.0017 (0.006) | 0.0398*** (0.0079) |
| | Observations | 224,857 | 224,857 | 224,857 | 60,015 |
| Canned Soup | Coefficient (Std.) | 0.0055 (0.0086) | 0.0028 (0.0084) | 0.0052 (0.0082) | 0.0148* (0.0078) |
| | Observations | 233,779 | 233,779 | 233,779 | 95,310 |
| Canned Tuna | Coefficient (Std.) | 0.0097** (0.0047) | 0.004 (0.0045) | 0.0039 (0.0045) | 0.0201*** (0.0064) |
| | Observations | 112,629 | 112,629 | 112,629 | 31,922 |
| Cereals | Coefficient (Std.) | 0.0042 (0.0036) | 0.0045 (0.0034) | 0.0045 (0.0033) | 0.019*** (0.0059) |
| | Observations | 141,087 | 141,087 | 141,087 | 72,789 |
| Cheese | Coefficient (Std.) | 0.0004 (0.0029) | -0.0001 (0.0023) | -0.0003 (0.0023) | 0.018*** (0.0051) |
| | Observations | 357,679 | 357,679 | 357,679 | 92,758 |
| Cigarettes | Coefficient (Std.) | 0.0114*** (0.0021) | 0.0107*** (0.0021) | 0.0108*** (0.0021) | 0.0114*** (0.0025) |
| | Observations | 24,553 | 24,553 | 24,553 | 20,692 |
| Cookies | Coefficient (Std.) | 0.0043*** (0.001) | 0.0037*** (0.001) | 0.0036*** (0.001) | 0.0149*** (0.0026) |
| | Observations | 317,932 | 317,932 | 317,932 | 66,087 |
| Crackers | Coefficient (Std.) | 0.0006 (0.0016) | 0.0001 (0.0016) | 0.0002 (0.0016) | 0.0141*** (0.0044) |
| | Observations | 115,658 | 115,658 | 115,658 | 24,771 |
| Dish Detergent | Coefficient (Std.) | 0.0175*** (0.003) | 0.0123*** (0.0026) | 0.0126*** (0.0024) | 0.0289*** (0.0052) |
| | Observations | 85,222 | 85,222 | 85,222 | 26,735 |
| Fabric Softener | Coefficient (Std.) | 0.01** (0.0044) | 0.0027 (0.0039) | 0.0032 (0.0039) | 0.0157** (0.0079) |
| | Observations | 85,337 | 85,337 | 85,337 | 27,488 |
| Front-End-Candies | Coefficient (Std.) | -0.0038 (0.0034) | -0.0046 (0.0031) | -0.0045 (0.0031) | 0.0011 (0.0028) |
| | Observations | 148,200 | 148,200 | 148,200 | 77,323 |
| Frozen Dinners | Coefficient (Std.) | 0.0252*** (0.0061) | 0.0186*** (0.0052) | 0.0198*** (0.0051) | 0.0524*** (0.012) |
| | Observations | 52,893 | 52,893 | 52,893 | 12,287 |

Table B1. (Cont.)

| Category | | (1) | (2) | (3) | (4) |
|---|---|---|---|---|---|
| Frozen Entrees | Coefficient (Std.) | 0.0132*** (0.0022) | 0.0123*** (0.0019) | 0.0125*** (0.0019) | 0.0294*** (0.0028) |
| | Observations | 345,223 | 345,223 | 345,223 | 117,044 |
| Frozen Juices | Coefficient (Std.) | 0.0185*** (0.0051) | 0.0162*** (0.0044) | 0.0168*** (0.0044) | 0.0249*** (0.0073) |
| | Observations | 118,582 | 118,582 | 118,582 | 40,517 |
| Grooming Products | Coefficient (Std.) | 0.0028** (0.0011) | 0.0029*** (0.0011) | 0.0031*** (0.0011) | 0.0058 (0.0039) |
| | Observations | 101,944 | 101,944 | 101,944 | 22,102 |
| Laundry Detergents | Coefficient (Std.) | 0.0053** (0.0024) | 0.0024 (0.0022) | 0.0027 (0.0022) | 0.0073* (0.0042) |
| | Observations | 121,566 | 121,566 | 121,566 | 42,121 |
| Oatmeal | Coefficient (Std.) | 0.0421*** (0.0139) | 0.0327*** (0.011) | 0.0308*** (0.0104) | 0.0564*** (0.0153) |
| | Observations | 25,523 | 25,523 | 25,523 | 13,605 |
| Paper Towels | Coefficient (Std.) | 0.0113 (0.0101) | 0.0095 (0.0097) | 0.0107 (0.0095) | -0.006 (0.0108) |
| | Observations | 48,199 | 48,199 | 48,199 | 9,243 |
| Refrigerated Juices | Coefficient (Std.) | 0.018*** (0.0054) | 0.0104* (0.0057) | 0.0104* (0.0056) | 0.0267** (0.0108) |
| | Observations | 108,965 | 108,965 | 108,965 | 23,705 |
| Shampoos | Coefficient (Std.) | 0.0007 (0.0006) | 0.0008 (0.0006) | 0.0008 (0.0006) | 0.0038 (0.0023) |
| | Observations | 88,193 | 88,193 | 88,193 | 16,099 |
| Snack Crackers | Coefficient (Std.) | 0.0018 (0.0013) | 0.0024* (0.0013) | 0.0025** (0.0013) | 0.0131*** (0.0034) |
| | Observations | 176,527 | 176,527 | 176,527 | 38,123 |
| Soaps | Coefficient (Std.) | 0.0216*** (0.0084) | 0.0125 (0.0082) | 0.0157*** (0.0079) | 0.0473*** (0.0109) |
| | Observations | 56,725 | 56,725 | 56,725 | 16,882 |
| Soft Drinks | Coefficient (Std.) | 0.0026 (0.0018) | 0.006*** (0.0015) | 0.0048*** (0.0015) | 0.0005 (0.0027) |
| | Observations | 243,837 | 243,837 | 243,837 | 49,989 |
| Toothbrushes | Coefficient (Std.) | 0.0066** (0.0026) | 0.0059** (0.0026) | 0.0057** (0.0025) | 0.0121** (0.0059) |
| | Observations | 52,185 | 52,185 | 52,185 | 13,695 |
| Toothpastes | Coefficient (Std.) | 0.0036 (0.0026) | 0.0036 (0.0023) | 0.0038* (0.0022) | 0.0145*** (0.0055) |
| | Observations | 100,845 | 100,845 | 100,845 | 28,039 |
| **Average coefficients** | | **0.0105** | **0.0073** | **0.0075** | **0.0195** |

<u>Notes</u>: The table reports the results of category-level fixed effect regressions of the probability of a small price change. The dependent variable is "small price change," which equals 1 if a price change of product $i$ in store $s$ at time $t$ is less or equal to 5¢, and 0 otherwise. The main independent variable is the log of the average sales volume of product $i$ in store $s$ over the sample period. Column 1 reports the results of the baseline regression that includes only the log of the average sales volume and the fixed effects for months, years, stores, and products. In column 2, we add the following controls: the log of the average price, the log of the absolute change in the wholesale price, and a control for sale- and bounce-back prices, which we identify using a sales filter algorithm. In column 3, we add a dummy for 9-ending prices as an additional control. In column 4, we focus on regular prices by excluding the sale- and bounce-back prices. We estimate separate regressions for each product category, clustering the errors by product. * $p < 10\%$, ** $p < 5\%$, *** $p < 1\%$

Table B2. Category-level regressions of small price changes ($\Delta P \le 15\rlap{/}c$) and sales volume

| Category | | **(1)** | **(2)** | **(3)** | **(4)** |
|---|---|---|---|---|---|
| Analgesics | Coefficient (Std.) | 0.0177*** (0.0052) | 0.0127*** (0.0049) | 0.0118*** (0.0049) | 0.0078* (0.0091) |
| | Observations | 74,451 | 74,451 | 74,451 | 24,729 |
| Bath Soap | Coefficient (Std.) | 0.0349** (0.0189) | 0.0296** (0.0179) | 0.0295** (0.0177) | 0.018 (0.0348) |
| | Observations | 6,650 | 6,650 | 6,650 | 1,466 |
| Bathroom Tissues | Coefficient (Std.) | 0.0592*** (0.01) | 0.0333*** (0.0092) | 0.0332*** (0.0094) | 0.0323*** (0.0084) |
| | Observations | 56,458 | 56,458 | 56,458 | 19,285 |
| Beer | Coefficient (Std.) | 0.0018*** (0.0006) | 0.0042*** (0.0007) | 0.0042*** (0.0007) | 0.0219*** (0.0052) |
| | Observations | 187,691 | 187,691 | 187,691 | 12,080 |
| Bottled Juice | Coefficient (Std.) | 0.015*** (0.0069) | 0.0108** (0.006) | 0.0109** (0.0058) | 0.0161** (0.0088) |
| | Observations | 224,857 | 224,857 | 224,857 | 60,015 |
| Canned Soup | Coefficient (Std.) | -0.0049* (0.0056) | -0.0057* (0.0056) | -0.0044 (0.0054) | 0.0009 (0.0035) |
| | Observations | 233,779 | 233,779 | 233,779 | 95,310 |
| Canned Tuna | Coefficient (Std.) | 0.0084* (0.0077) | -0.0021 (0.0066) | -0.0022 (0.0066) | 0.0098* (0.0077) |
| | Observations | 112,629 | 112,629 | 112,629 | 31,922 |
| Cereals | Coefficient (Std.) | 0.0054* (0.0058) | 0.0052* (0.0053) | 0.0052* (0.0053) | 0.0071* (0.0061) |
| | Observations | 141,087 | 141,087 | 141,087 | 72,789 |
| Cheese | Coefficient (Std.) | 0.0055** (0.0042) | 0.0052** (0.0037) | 0.0052** (0.0037) | 0.0104*** (0.0039) |
| | Observations | 357,679 | 357,679 | 357,679 | 92,758 |
| Cigarettes | Coefficient (Std.) | 0.0008 (0.0046) | 0 (0.0047) | 0.0001 (0.0047) | -0.0025 (0.005) |
| | Observations | 24,553 | 24,553 | 24,553 | 20,692 |
| Cookies | Coefficient (Std.) | 0.0117*** (0.0025) | 0.0103*** (0.0024) | 0.01*** (0.0023) | 0.016*** (0.0043) |
| | Observations | 317,932 | 317,932 | 317,932 | 66,087 |
| Crackers | Coefficient (Std.) | 0.0081*** (0.0032) | 0.0081*** (0.003) | 0.0083*** (0.0029) | 0.0131*** (0.0053) |
| | Observations | 115,658 | 115,658 | 115,658 | 24,771 |
| Dish Detergent | Coefficient (Std.) | 0.0257*** (0.0054) | 0.0194*** (0.0045) | 0.0194*** (0.0045) | 0.0219*** (0.0057) |
| | Observations | 85,222 | 85,222 | 85,222 | 26,735 |
| Fabric Softener | Coefficient (Std.) | 0.0138** (0.0086) | 0.0032 (0.0077) | 0.0035 (0.0077) | 0.0176*** (0.0086) |
| | Observations | 85,337 | 85,337 | 85,337 | 27,488 |
| Front-End-Candies | Coefficient (Std.) | -0.0027 (0.0047) | -0.0025 (0.0034) | -0.0024 (0.0034) | -0.0012 (0.003) |
| | Observations | 148,200 | 148,200 | 148,200 | 77,323 |
| Frozen Dinners | Coefficient (Std.) | 0.0434*** (0.0076) | 0.0362*** (0.0069) | 0.0354*** (0.0068) | 0.0729*** (0.0108) |
| | Observations | 52,893 | 52,893 | 52,893 | 12,287 |

Table B2. (Cont.)

| Category | | **(1)** | **(2)** | **(3)** | **(4)** |
|---|---|---|---|---|---|
| Frozen Entrees | Coefficient (Std.) | 0.0209*** (0.0032) | 0.0202*** (0.0029) | 0.0202*** (0.0029) | 0.0141*** (0.0037) |
| | Observations | 345,223 | 345,223 | 345,223 | 117,044 |
| Frozen Juices | Coefficient (Std.) | 0.0177** (0.0093) | 0.0163** (0.009) | 0.0166** (0.009) | 0.0004 (0.0091) |
| | Observations | 118,582 | 118,582 | 118,582 | 40,517 |
| Grooming Products | Coefficient (Std.) | 0.0127*** (0.0039) | 0.015*** (0.0038) | 0.015*** (0.0038) | 0.0203** (0.0118) |
| | Observations | 101,944 | 101,944 | 101,944 | 22,102 |
| Laundry Detergents | Coefficient (Std.) | 0.0302*** (0.0055) | 0.0192*** (0.0048) | 0.0195*** (0.0048) | 0.013*** (0.0058) |
| | Observations | 121,566 | 121,566 | 121,566 | 42,121 |
| Oatmeal | Coefficient (Std.) | 0.0419*** (0.0152) | 0.0366*** (0.0146) | 0.037*** (0.0147) | 0.0433*** (0.0171) |
| | Observations | 25,523 | 25,523 | 25,523 | 13,605 |
| Paper Towels | Coefficient (Std.) | 0.014 (0.0168) | 0.0084 (0.0176) | 0.0084 (0.0175) | 0.0279*** (0.0131) |
| | Observations | 48,199 | 48,199 | 48,199 | 9,243 |
| Refrigerated Juices | Coefficient (Std.) | 0.0236*** (0.0069) | 0.013** (0.007) | 0.0131** (0.0071) | 0.0283*** (0.009) |
| | Observations | 108,965 | 108,965 | 108,965 | 23,705 |
| Shampoos | Coefficient (Std.) | 0.0279*** (0.004) | 0.0278*** (0.0038) | 0.0278*** (0.0038) | 0.0424*** (0.0114) |
| | Observations | 88,193 | 88,193 | 88,193 | 16,099 |
| Snack Crackers | Coefficient (Std.) | -0.0041** (0.0034) | -0.0033** (0.0034) | -0.0031* (0.0034) | 0.0044 (0.0057) |
| | Observations | 176,527 | 176,527 | 176,527 | 38,123 |
| Soaps | Coefficient (Std.) | 0.0169*** (0.0077) | 0.007** (0.007) | 0.0094** (0.0071) | 0.0335*** (0.0113) |
| | Observations | 56,725 | 56,725 | 56,725 | 16,882 |
| Soft Drinks | Coefficient (Std.) | 0.0079*** (0.0026) | 0.0099*** (0.0025) | 0.0096*** (0.0025) | -0.0003 (0.004) |
| | Observations | 243,837 | 243,837 | 243,837 | 49,989 |
| Toothbrushes | Coefficient (Std.) | 0.0182*** (0.0061) | 0.0119** (0.0062) | 0.0113** (0.0061) | 0.0385*** (0.013) |
| | Observations | 52,185 | 52,185 | 52,185 | 13,695 |
| Toothpastes | Coefficient (Std.) | 0.0024 (0.0062) | 0.0005 (0.0059) | 0.0009 (0.0058) | 0.0172** (0.0088) |
| | Observations | 100,845 | 100,845 | 100,845 | 28,039 |
| **Average coefficients** | | **0.0163** | **0.0121** | **0.0122** | **0.0188** |

<u>Notes:</u> The table reports the results of category-level fixed effect regressions of the probability of a small price change. The dependent variable is "small price change," which equals 1 if a price change of product $i$ in store $s$ at time $t$ is less or equal to 15¢, and 0 otherwise. The main independent variable is the log of the average sales volume of product $i$ in store $s$ over the sample period. Column 1 reports the results of the baseline regression that includes only the log of the average sales volume and the fixed effects for months, years, stores, and products. In column 2, we add the following controls: the log of the average price, the log of the absolute change in the wholesale price, and a control for sale- and bounce-back prices, which we identify using a sales filter algorithm. In column 3, we add a dummy for 9-ending prices as an additional control. In column 4, we focus on regular prices by excluding the sale- and bounce-back prices. We estimate separate regressions for each product category, clustering the errors by product. * $p < 10\%$, ** $p < 5\%$, *** $p < 1\%$

Table B3. Category-level regressions of small price changes ($\Delta P \leq 2\%$) and sales volume

| Category | | (1) | (2) | (3) | (4) |
|---|---|---|---|---|---|
| Analgesics | Coefficient (Std.) | 0.0097*** (0.0034) | 0.0058** (0.003) | 0.0054** (0.003) | 0.0132** (0.0072) |
| | Observations | 74,451 | 74,451 | 74,451 | 24,729 |
| Bath Soap | Coefficient (Std.) | 0.0093** (0.0054) | 0.0085** (0.0049) | 0.0084** (0.0046) | 0.0004 (0.0127) |
| | Observations | 6,650 | 6,650 | 6,650 | 1,466 |
| Bathroom Tissues | Coefficient (Std.) | 0.022*** (0.0054) | 0.0095*** (0.0044) | 0.0095*** (0.0043) | 0.015** (0.0082) |
| | Observations | 56,458 | 56,458 | 56,458 | 19,285 |
| Beer | Coefficient (Std.) | 0.0017*** (0.0004) | 0.0034*** (0.0006) | 0.0034*** (0.0006) | 0.0196*** (0.004) |
| | Observations | 187,691 | 187,691 | 187,691 | 12,080 |
| Bottled Juice | Coefficient (Std.) | 0.0115*** (0.0039) | 0.0068*** (0.0032) | 0.0068*** (0.0031) | 0.0237*** (0.0095) |
| | Observations | 224,857 | 224,857 | 224,857 | 60,015 |
| Canned Soup | Coefficient (Std.) | -0.0032* (0.0031) | -0.0038** (0.0028) | -0.0034** (0.0028) | -0.0061* (0.006) |
| | Observations | 233,779 | 233,779 | 233,779 | 95,310 |
| Canned Tuna | Coefficient (Std.) | 0.0069*** (0.0026) | 0.0038** (0.0023) | 0.0038** (0.0023) | 0.0072* (0.0063) |
| | Observations | 112,629 | 112,629 | 112,629 | 31,922 |
| Cereals | Coefficient (Std.) | 0.01*** (0.0039) | 0.011*** (0.0033) | 0.011*** (0.0033) | 0.0237*** (0.0063) |
| | Observations | 141,087 | 141,087 | 141,087 | 72,789 |
| Cheese | Coefficient (Std.) | 0.0051** (0.0027) | 0.0043*** (0.002) | 0.0042*** (0.002) | 0.0149*** (0.0048) |
| | Observations | 357,679 | 357,679 | 357,679 | 92,758 |
| Cigarettes | Coefficient (Std.) | 0.0041** (0.0026) | 0.0049** (0.0026) | 0.0049** (0.0026) | 0.0073*** (0.0029) |
| | Observations | 24,553 | 24,553 | 24,553 | 20,692 |
| Cookies | Coefficient (Std.) | 0.0032*** (0.0008) | 0.0025*** (0.0007) | 0.0025*** (0.0007) | 0.0073*** (0.0022) |
| | Observations | 317,932 | 317,932 | 317,932 | 66,087 |
| Crackers | Coefficient (Std.) | 0.0015** (0.0008) | 0.0011** (0.0008) | 0.0012** (0.0008) | 0.0028* (0.0031) |
| | Observations | 115,658 | 115,658 | 115,658 | 24,771 |
| Dish Detergent | Coefficient (Std.) | 0.0139*** (0.0022) | 0.009*** (0.0018) | 0.0091*** (0.0018) | 0.0233*** (0.0047) |
| | Observations | 85,222 | 85,222 | 85,222 | 26,735 |
| Fabric Softener | Coefficient (Std.) | 0.0104*** (0.004) | 0.0035* (0.0033) | 0.0037* (0.0033) | 0.0082* (0.0078) |
| | Observations | 85,337 | 85,337 | 85,337 | 27,488 |
| Front-End-Candies | Coefficient (Std.) | -0.002** (0.0012) | -0.002** (0.0011) | -0.0021** (0.0011) | 0.0009 (0.0015) |
| | Observations | 148,200 | 148,200 | 148,200 | 77,323 |
| Frozen Dinners | Coefficient (Std.) | 0.0227*** (0.005) | 0.0156*** (0.0038) | 0.0162*** (0.0037) | 0.0246*** (0.0091) |
| | Observations | 52,893 | 52,893 | 52,893 | 12,287 |

Table B3. (Cont.)

| Category | | (1) | (2) | (3) | (4) |
|---|---|---|---|---|---|
| Frozen Entrees | Coefficient (Std.) | 0.0097*** (0.0018) | 0.0092*** (0.0016) | 0.0093*** (0.0015) | 0.0266*** (0.0027) |
| | Observations | 345,223 | 345,223 | 345,223 | 117,044 |
| Frozen Juices | Coefficient (Std.) | 0.0152*** (0.0035) | 0.0131*** (0.003) | 0.0131*** (0.003) | 0.0206*** (0.0068) |
| | Observations | 118,582 | 118,582 | 118,582 | 40,517 |
| Grooming Products | Coefficient (Std.) | 0.0026* (0.0023) | 0.0031** (0.0023) | 0.0033** (0.0023) | 0.0048 (0.0086) |
| | Observations | 101,944 | 101,944 | 101,944 | 22,102 |
| Laundry Detergents | Coefficient (Std.) | 0.019*** (0.0044) | 0.0101*** (0.003) | 0.0105*** (0.003) | 0.0116** (0.0059) |
| | Observations | 121,566 | 121,566 | 121,566 | 42,121 |
| Oatmeal | Coefficient (Std.) | 0.0428*** (0.0153) | 0.0314*** (0.0105) | 0.0303*** (0.0102) | 0.0511*** (0.0141) |
| | Observations | 25,523 | 25,523 | 25,523 | 13,605 |
| Paper Towels | Coefficient (Std.) | 0.0053** (0.0037) | 0.0066** (0.0035) | 0.0068** (0.0035) | 0.0255** (0.0147) |
| | Observations | 48,199 | 48,199 | 48,199 | 9,243 |
| Refrigerated Juices | Coefficient (Std.) | 0.0156*** (0.0038) | 0.0082*** (0.0029) | 0.0082*** (0.0029) | 0.0212*** (0.0081) |
| | Observations | 108,965 | 108,965 | 108,965 | 23,705 |
| Shampoos | Coefficient (Std.) | 0.0022*** (0.0011) | 0.0025*** (0.0011) | 0.0025*** (0.0011) | 0.0127*** (0.0052) |
| | Observations | 88,193 | 88,193 | 88,193 | 16,099 |
| Snack Crackers | Coefficient (Std.) | 0.0005 (0.0009) | 0.0008* (0.0009) | 0.0008* (0.0009) | 0.0035* (0.0028) |
| | Observations | 176,527 | 176,527 | 176,527 | 38,123 |
| Soaps | Coefficient (Std.) | 0.019*** (0.0059) | 0.0088** (0.0048) | 0.0101** (0.0048) | 0.0273*** (0.0111) |
| | Observations | 56,725 | 56,725 | 56,725 | 16,882 |
| Soft Drinks | Coefficient (Std.) | 0.0017*** (0.0008) | 0.0015** (0.0008) | 0.0012** (0.0008) | 0.0003 (0.0027) |
| | Observations | 243,837 | 243,837 | 243,837 | 49,989 |
| Toothbrushes | Coefficient (Std.) | 0.0056*** (0.0022) | 0.0044*** (0.0021) | 0.0044*** (0.0021) | 0.0166*** (0.0066) |
| | Observations | 52,185 | 52,185 | 52,185 | 13,695 |
| Toothpastes | Coefficient (Std.) | 0.0066*** (0.0021) | 0.0054*** (0.0019) | 0.0056*** (0.0019) | 0.0157*** (0.0055) |
| | Observations | 100,845 | 100,845 | 100,845 | 28,039 |
| **Average coefficients** | | **0.0094** | **0.0065** | **0.0066** | **0.0146** |

Notes: The table reports the results of category-level fixed effect regressions of the probability of a small price change. The dependent variable is "small price change," which equals 1 if a price change of product $i$ in store $s$ at time $t$ is less or equal to 2%, and 0 otherwise. The main independent variable is the log of the average sales volume of product $i$ in store $s$ over the sample period. Column 1 reports the results of the baseline regression that includes only the log of the average sales volume and the fixed effects for months, years, stores, and products. In column 2, we add the following controls: the log of the average price, the log of the absolute change in the wholesale price, and a control for sale- and bounce-back prices, which we identify using a sales filter algorithm. In column 3, we add a dummy for 9-ending prices as an additional control. In column 4, we focus on regular prices by excluding the sale- and bounce-back prices. We estimate separate regressions for each product category, clustering the errors by product. * $p < 10\%$, ** $p < 5\%$, *** $p < 1\%$

Table B4. Category-level regressions of small price changes ($\Delta P \leq 5\%$) and sales volume

| Category | | **(1)** | **(2)** | **(3)** | **(4)** |
|---|---|---|---|---|---|
| Analgesics | Coefficient (Std.) | 0.0266*** (0.0065) | 0.0214*** (0.0061) | 0.0204*** (0.0061) | 0.0135** (0.0103) |
| | Observations | 74,451 | 74,451 | 74,451 | 24,729 |
| Bath Soap | Coefficient (Std.) | 0.0207** (0.0148) | 0.0144* (0.0135) | 0.0144* (0.0134) | 0.0176 (0.0329) |
| | Observations | 6,650 | 6,650 | 6,650 | 1,466 |
| Bathroom Tissues | Coefficient (Std.) | 0.0236*** (0.009) | 0.0005 (0.0081) | 0.0003 (0.0082) | 0.0048 (0.0091) |
| | Observations | 56,458 | 56,458 | 56,458 | 19,285 |
| Beer | Coefficient (Std.) | 0.0066*** (0.0015) | 0.0115*** (0.0016) | 0.0115*** (0.0016) | 0.0444*** (0.0069) |
| | Observations | 187,691 | 187,691 | 187,691 | 12,080 |
| Bottled Juice | Coefficient (Std.) | -0.0025 (0.0071) | -0.0068* (0.0064) | -0.0067* (0.0066) | 0.0195*** (0.0075) |
| | Observations | 224,857 | 224,857 | 224,857 | 60,015 |
| Canned Soup | Coefficient (Std.) | -0.0201** (0.0114) | -0.0226** (0.0111) | -0.0202** (0.011) | 0.0133** (0.0086) |
| | Observations | 233,779 | 233,779 | 233,779 | 95,310 |
| Canned Tuna | Coefficient (Std.) | -0.0021 (0.0058) | -0.0064* (0.0057) | -0.0065* (0.0057) | 0.0129** (0.0088) |
| | Observations | 112,629 | 112,629 | 112,629 | 31,922 |
| Cereals | Coefficient (Std.) | -0.0003 (0.0076) | -0.0001 (0.0073) | 0 (0.0073) | 0.014** (0.0073) |
| | Observations | 141,087 | 141,087 | 141,087 | 72,789 |
| Cheese | Coefficient (Std.) | 0.0047* (0.0037) | 0.0043* (0.0034) | 0.004* (0.0034) | 0.0126*** (0.0044) |
| | Observations | 357,679 | 357,679 | 357,679 | 92,758 |
| Cigarettes | Coefficient (Std.) | -0.0007 (0.004) | -0.0037* (0.0039) | -0.0038* (0.004) | -0.0062** (0.0046) |
| | Observations | 24,553 | 24,553 | 24,553 | 20,692 |
| Cookies | Coefficient (Std.) | 0.0109*** (0.002) | 0.0097*** (0.0021) | 0.0096*** (0.002) | 0.0102*** (0.0036) |
| | Observations | 317,932 | 317,932 | 317,932 | 66,087 |
| Crackers | Coefficient (Std.) | 0.0024 (0.003) | 0.0015 (0.003) | 0.0019 (0.003) | 0.0147*** (0.0061) |
| | Observations | 115,658 | 115,658 | 115,658 | 24,771 |
| Dish Detergent | Coefficient (Std.) | 0.0308*** (0.0049) | 0.0249*** (0.0042) | 0.025*** (0.0041) | 0.0256*** (0.0055) |
| | Observations | 85,222 | 85,222 | 85,222 | 26,735 |
| Fabric Softener | Coefficient (Std.) | 0.0207*** (0.0079) | 0.0104** (0.0069) | 0.011** (0.0069) | 0.0232*** (0.0082) |
| | Observations | 85,337 | 85,337 | 85,337 | 27,488 |
| Front-End-Candies | Coefficient (Std.) | -0.0033** (0.0026) | -0.0033** (0.0025) | -0.0033** (0.0025) | 0.0015 (0.0023) |
| | Observations | 148,200 | 148,200 | 148,200 | 77,323 |
| Frozen Dinners | Coefficient (Std.) | 0.0385*** (0.0071) | 0.0296*** (0.0065) | 0.0295*** (0.0065) | 0.0684*** (0.0099) |
| | Observations | 52,893 | 52,893 | 52,893 | 12,287 |

Table B4. (Cont.)

| Category | | **(1)** | **(2)** | **(3)** | **(4)** |
|---|---|---|---|---|---|
| Frozen Entrees | Coefficient (Std.) | 0.0186*** (0.0029) | 0.0196*** (0.0026) | 0.0197*** (0.0026) | 0.0249*** (0.0035) |
| | Observations | 345,223 | 345,223 | 345,223 | 117,044 |
| Frozen Juices | Coefficient (Std.) | 0.0244*** (0.0057) | 0.0224*** (0.005) | 0.0233*** (0.0048) | 0.0323*** (0.0071) |
| | Observations | 118,582 | 118,582 | 118,582 | 40,517 |
| Grooming Products | Coefficient (Std.) | 0.0091*** (0.0038) | 0.0104*** (0.0039) | 0.0106*** (0.004) | 0.0154** (0.0097) |
| | Observations | 101,944 | 101,944 | 101,944 | 22,102 |
| Laundry Detergents | Coefficient (Std.) | 0.0273*** (0.006) | 0.0153*** (0.0052) | 0.0159*** (0.0052) | 0.0119** (0.0063) |
| | Observations | 121,566 | 121,566 | 121,566 | 42,121 |
| Oatmeal | Coefficient (Std.) | 0.0424*** (0.0138) | 0.0367*** (0.0135) | 0.0362*** (0.0133) | 0.0536*** (0.0115) |
| | Observations | 25,523 | 25,523 | 25,523 | 13,605 |
| Paper Towels | Coefficient (Std.) | 0.0232** (0.0126) | 0.0226** (0.0117) | 0.0232** (0.0115) | 0.01 (0.0167) |
| | Observations | 48,199 | 48,199 | 48,199 | 9,243 |
| Refrigerated Juices | Coefficient (Std.) | 0.0264*** (0.0065) | 0.0187*** (0.0064) | 0.0187*** (0.0064) | 0.0223** (0.0116) |
| | Observations | 108,965 | 108,965 | 108,965 | 23,705 |
| Shampoos | Coefficient (Std.) | 0.0176*** (0.0029) | 0.0185*** (0.0029) | 0.0185*** (0.0029) | 0.0318*** (0.0089) |
| | Observations | 88,193 | 88,193 | 88,193 | 16,099 |
| Snack Crackers | Coefficient (Std.) | 0.0028* (0.0026) | 0.0035** (0.0026) | 0.0035** (0.0026) | 0.0173*** (0.0047) |
| | Observations | 176,527 | 176,527 | 176,527 | 38,123 |
| Soaps | Coefficient (Std.) | 0.0295*** (0.009) | 0.0182*** (0.0085) | 0.0212*** (0.0084) | 0.0509*** (0.0102) |
| | Observations | 56,725 | 56,725 | 56,725 | 16,882 |
| Soft Drinks | Coefficient (Std.) | 0.0029*** (0.0013) | 0.0028*** (0.0012) | 0.0023** (0.0012) | 0.0012** (0.0033) |
| | Observations | 243,837 | 243,837 | 243,837 | 49,989 |
| Toothbrushes | Coefficient (Std.) | 0.0141*** (0.0059) | 0.0108** (0.006) | 0.0103** (0.006) | 0.0314*** (0.0119) |
| | Observations | 52,185 | 52,185 | 52,185 | 13,695 |
| Toothpastes | Coefficient (Std.) | 0.0083** (0.0043) | 0.0082** (0.004) | 0.0084*** (0.004) | 0.0225*** (0.0083) |
| | Observations | 100,845 | 100,845 | 100,845 | 28,039 |
| **Average coefficients** | | **0.0139** | **0.0101** | **0.0103** | **0.0212** |

Notes: The table reports the results of category-level fixed effect regressions of the probability of a small price change. The dependent variable is "small price change," which equals 1 if a price change of product $i$ in store $s$ at time $t$ is less or equal to 5%, and 0 otherwise. The main independent variable is the log of the average sales volume of product $i$ in store $s$ over the sample period. Column 1 reports the results of the baseline regression that includes only the log of the average sales volume and the fixed effects for months, years, stores, and products. In column 2, we add the following controls: the log of the average price, the log of the absolute change in the wholesale price, and a control for sale- and bounce-back prices, which we identify using a sales filter algorithm. In column 3, we add a dummy for 9-ending prices as an additional control. In column 4, we focus on regular prices by excluding the sale- and bounce-back prices. We estimate separate regressions for each product category, clustering the errors by product. * $p < 10\%$, ** $p < 5\%$, *** $p < 1\%$

Table B5. Category-level regressions of small price changes ($\Delta P \le 50\%$ of the average price change) and sales volume

| Category | | **(1)** | **(2)** | **(3)** | **(4)** |
|---|---|---|---|---|---|
| Analgesics | Coefficient (Std.) | 0.0374*** (0.0074) | 0.032*** (0.007) | 0.0314*** (0.007) | 0.0563*** (0.0108) |
| | Observations | 74,451 | 74,451 | 74,451 | 24,729 |
| Bath Soap | Coefficient (Std.) | 0.0503*** (0.0138) | 0.0433*** (0.0137) | 0.0432*** (0.0136) | 0.1405*** (0.0357) |
| | Observations | 6,650 | 6,650 | 6,650 | 1,466 |
| Bathroom Tissues | Coefficient (Std.) | 0.0184*** (0.0058) | 0.0175*** (0.0068) | 0.0174*** (0.0068) | 0.0405*** (0.0097) |
| | Observations | 56,458 | 56,458 | 56,458 | 19,285 |
| Beer | Coefficient (Std.) | 0.0057*** (0.0015) | 0.012*** (0.0014) | 0.012*** (0.0014) | 0.0648*** (0.0084) |
| | Observations | 187,691 | 187,691 | 187,691 | 12,080 |
| Bottled Juice | Coefficient (Std.) | 0.0132*** (0.0053) | 0.01** (0.0054) | 0.0101** (0.0054) | 0.0193*** (0.0087) |
| | Observations | 224,857 | 224,857 | 224,857 | 60,015 |
| Canned Soup | Coefficient (Std.) | -0.0043 (0.0066) | -0.0003 (0.0069) | 0.0017 (0.0068) | 0.0129** (0.0073) |
| | Observations | 233,779 | 233,779 | 233,779 | 95,310 |
| Canned Tuna | Coefficient (Std.) | 0.0125*** (0.0048) | 0.0118*** (0.0048) | 0.0116*** (0.0049) | 0.0222*** (0.0074) |
| | Observations | 112,629 | 112,629 | 112,629 | 31,922 |
| Cereals | Coefficient (Std.) | -0.0018 (0.0046) | -0.0012 (0.0048) | -0.0011 (0.0048) | 0.0082* (0.0074) |
| | Observations | 141,087 | 141,087 | 141,087 | 72,789 |
| Cheese | Coefficient (Std.) | 0.0078*** (0.0036) | 0.0066** (0.004) | 0.0065** (0.004) | 0.0158*** (0.0059) |
| | Observations | 357,679 | 357,679 | 357,679 | 92,758 |
| Cigarettes | Coefficient (Std.) | 0.107*** (0.0084) | 0.1075*** (0.0081) | 0.1075*** (0.0081) | 0.1225*** (0.009) |
| | Observations | 24,553 | 24,553 | 24,553 | 20,692 |
| Cookies | Coefficient (Std.) | 0.0153*** (0.0031) | 0.0127*** (0.0031) | 0.0126*** (0.0031) | 0.0367*** (0.006) |
| | Observations | 317,932 | 317,932 | 317,932 | 66,087 |
| Crackers | Coefficient (Std.) | 0.0152*** (0.0049) | 0.0138*** (0.005) | 0.0139*** (0.005) | 0.0253*** (0.0078) |
| | Observations | 115,658 | 115,658 | 115,658 | 24,771 |
| Dish Detergent | Coefficient (Std.) | 0.0092*** (0.0042) | 0.004* (0.0041) | 0.0042* (0.0042) | 0.0137** (0.0072) |
| | Observations | 85,222 | 85,222 | 85,222 | 26,735 |
| Fabric Softener | Coefficient (Std.) | 0.0125** (0.0065) | 0.0088** (0.0065) | 0.0093** (0.0066) | 0.0317*** (0.0102) |
| | Observations | 85,337 | 85,337 | 85,337 | 27,488 |
| Front-End-Candies | Coefficient (Std.) | 0.0199*** (0.0052) | 0.0239*** (0.0052) | 0.0241*** (0.0052) | 0.0128*** (0.0043) |
| | Observations | 148,200 | 148,200 | 148,200 | 77,323 |
| Frozen Dinners | Coefficient (Std.) | 0.0285*** (0.0079) | 0.0312*** (0.0085) | 0.0305*** (0.0084) | 0.0678*** (0.0128) |
| | Observations | 52,893 | 52,893 | 52,893 | 12,287 |

Table B5. (Cont.)

| Category | | (1) | (2) | (3) | (4) |
|---|---|---|---|---|---|
| Frozen Entrees | Coefficient (Std.) | 0.0286*** (0.0032) | 0.0377*** (0.0034) | 0.0375*** (0.0034) | 0.0308*** (0.0036) |
| | Observations | 345,223 | 345,223 | 345,223 | 117,044 |
| Frozen Juices | Coefficient (Std.) | 0.0143*** (0.0066) | 0.0125** (0.0076) | 0.013** (0.0077) | 0.0229*** (0.0103) |
| | Observations | 118,582 | 118,582 | 118,582 | 40,517 |
| Grooming Products | Coefficient (Std.) | 0.0417*** (0.0037) | 0.0435*** (0.0038) | 0.0434*** (0.0038) | 0.0797*** (0.009) |
| | Observations | 101,944 | 101,944 | 101,944 | 22,102 |
| Laundry Detergents | Coefficient (Std.) | 0.0159*** (0.0044) | 0.0113*** (0.004) | 0.0117*** (0.004) | 0.0198*** (0.0066) |
| | Observations | 121,566 | 121,566 | 121,566 | 42,121 |
| Oatmeal | Coefficient (Std.) | -0.0063 (0.0101) | -0.0135* (0.012) | -0.0133* (0.012) | 0.0112 (0.0167) |
| | Observations | 25,523 | 25,523 | 25,523 | 13,605 |
| Paper Towels | Coefficient (Std.) | 0.0111** (0.0071) | 0.0161*** (0.0071) | 0.0167*** (0.0071) | -0.004 (0.0188) |
| | Observations | 48,199 | 48,199 | 48,199 | 9,243 |
| Refrigerated Juices | Coefficient (Std.) | 0.0068** (0.0049) | 0.0034 (0.0052) | 0.0034 (0.0052) | 0.0211** (0.0113) |
| | Observations | 108,965 | 108,965 | 108,965 | 23,705 |
| Shampoos | Coefficient (Std.) | 0.0522*** (0.0035) | 0.0521*** (0.0035) | 0.0521*** (0.0035) | 0.1111*** (0.0112) |
| | Observations | 88,193 | 88,193 | 88,193 | 16,099 |
| Snack Crackers | Coefficient (Std.) | 0.0048* (0.0048) | 0.0007 (0.0052) | 0.0008 (0.0052) | 0.0215*** (0.0063) |
| | Observations | 176,527 | 176,527 | 176,527 | 38,123 |
| Soaps | Coefficient (Std.) | 0.0137*** (0.0067) | 0.0049 (0.0069) | 0.0077* (0.007) | 0.0332*** (0.0118) |
| | Observations | 56,725 | 56,725 | 56,725 | 16,882 |
| Soft Drinks | Coefficient (Std.) | 0.0137*** (0.0022) | 0.0119*** (0.0024) | 0.0122*** (0.0024) | 0.0152*** (0.0043) |
| | Observations | 243,837 | 243,837 | 243,837 | 49,989 |
| Toothbrushes | Coefficient (Std.) | 0.0445*** (0.0058) | 0.0394*** (0.0055) | 0.0387*** (0.0056) | 0.0881*** (0.0125) |
| | Observations | 52,185 | 52,185 | 52,185 | 13,695 |
| Toothpastes | Coefficient (Std.) | 0.0287*** (0.0046) | 0.0287*** (0.0047) | 0.0288*** (0.0047) | 0.0613*** (0.0092) |
| | Observations | 100,845 | 100,845 | 100,845 | 28,039 |
| **Average coefficients** | | **0.0213** | **0.0201** | **0.0203** | **0.0415** |

Notes: The table reports the results of category-level fixed effect regressions of the probability of a small price change. The dependent variable is "small price change," which equals 1 if a price change of product $i$ in store $s$ at time $t$ is less or equal to $0.5|\overline{\Delta p_{i,s}}|$ where $\overline{\Delta p_{i,s}}$ is the average size of a price change of product $i$ in store $s$, and 0 otherwise. The main independent variable is the log of the average sales volume of product $i$ in store $s$ over the sample period. Column 1 reports the results of the baseline regression that includes only the log of the average sales volume and the fixed effects for months, years, stores, and products. In column 2, we add the following controls: the log of the average price, the log of the absolute change in the wholesale price, and a control for sale- and bounce-back prices, which we identify using a sales filter algorithm. In column 3, we add a dummy for 9-ending prices as an additional control. In column 4, we focus on regular prices by excluding the sale- and bounce-back prices. We estimate separate regressions for each product category, clustering the errors by product. * $p < 10\%$, ** $p < 5\%$, *** $p < 1\%$

Table B6. Category-level regressions of small price changes ($\Delta P \leq 33\%$ of the average price change) and sales volume

| Category | | **(1)** | **(2)** | **(3)** | **(4)** |
|---|---|---|---|---|---|
| Analgesics | Coefficient (Std.) | 0.0293*** (0.0045) | 0.0248*** (0.0043) | 0.0242*** (0.0042) | 0.0414*** (0.0088) |
| | Observations | 74,451 | 74,451 | 74,451 | 24,729 |
| Bath Soap | Coefficient (Std.) | 0.0304*** (0.0113) | 0.026*** (0.0107) | 0.026*** (0.0105) | 0.0736*** (0.0225) |
| | Observations | 6,650 | 6,650 | 6,650 | 1,466 |
| Bathroom Tissues | Coefficient (Std.) | 0.0057* (0.005) | 0.0107** (0.0064) | 0.0106** (0.0064) | 0.0297*** (0.0123) |
| | Observations | 56,458 | 56,458 | 56,458 | 19,285 |
| Beer | Coefficient (Std.) | 0.0023*** (0.0006) | 0.0065*** (0.0008) | 0.0064*** (0.0008) | 0.0526*** (0.0069) |
| | Observations | 187,691 | 187,691 | 187,691 | 12,080 |
| Bottled Juice | Coefficient (Std.) | 0.0184*** (0.0041) | 0.0157*** (0.0044) | 0.0157*** (0.0044) | 0.0183*** (0.0081) |
| | Observations | 224,857 | 224,857 | 224,857 | 60,015 |
| Canned Soup | Coefficient (Std.) | -0.0072* (0.0057) | -0.0028 (0.006) | -0.0014 (0.0058) | 0.0121** (0.0073) |
| | Observations | 233,779 | 233,779 | 233,779 | 95,310 |
| Canned Tuna | Coefficient (Std.) | 0.0102*** (0.0036) | 0.0082*** (0.0032) | 0.0081*** (0.0032) | 0.0161*** (0.0063) |
| | Observations | 112,629 | 112,629 | 112,629 | 31,922 |
| Cereals | Coefficient (Std.) | 0.0081*** (0.0039) | 0.0088*** (0.004) | 0.0088*** (0.004) | 0.0163*** (0.0069) |
| | Observations | 141,087 | 141,087 | 141,087 | 72,789 |
| Cheese | Coefficient (Std.) | 0.0078*** (0.0033) | 0.0068** (0.0035) | 0.0068** (0.0035) | 0.0142*** (0.0061) |
| | Observations | 357,679 | 357,679 | 357,679 | 92,758 |
| Cigarettes | Coefficient (Std.) | 0.0476*** (0.0061) | 0.0492*** (0.0059) | 0.0493*** (0.0058) | 0.0588*** (0.0065) |
| | Observations | 24,553 | 24,553 | 24,553 | 20,692 |
| Cookies | Coefficient (Std.) | 0.0121*** (0.0023) | 0.0103*** (0.0025) | 0.0102*** (0.0024) | 0.0148*** (0.0043) |
| | Observations | 317,932 | 317,932 | 317,932 | 66,087 |
| Crackers | Coefficient (Std.) | 0.0104*** (0.0041) | 0.0092*** (0.0041) | 0.0093*** (0.0041) | 0.0195*** (0.006) |
| | Observations | 115,658 | 115,658 | 115,658 | 24,771 |
| Dish Detergent | Coefficient (Std.) | 0.0073*** (0.0028) | 0.0034** (0.0029) | 0.0036* (0.0029) | 0.0181*** (0.0048) |
| | Observations | 85,222 | 85,222 | 85,222 | 26,735 |
| Fabric Softener | Coefficient (Std.) | 0.0098** (0.0061) | 0.0066** (0.0059) | 0.0069* (0.006) | 0.033*** (0.0097) |
| | Observations | 85,337 | 85,337 | 85,337 | 27,488 |
| Front-End-Candies | Coefficient (Std.) | 0.0107*** (0.0046) | 0.0132*** (0.0045) | 0.0134*** (0.0045) | 0.0115*** (0.0039) |
| | Observations | 148,200 | 148,200 | 148,200 | 77,323 |
| Frozen Dinners | Coefficient (Std.) | 0.0312*** (0.0078) | 0.0339*** (0.0087) | 0.0339*** (0.0086) | 0.0534*** (0.0144) |
| | Observations | 52,893 | 52,893 | 52,893 | 12,287 |

| Category | | **(1)** | **(2)** | **(3)** | **(4)** |
|---|---|---|---|---|---|
| Frozen Entrees | Coefficient (Std.) | 0.0332*** (0.003) | 0.0405*** (0.0031) | 0.0404*** (0.0031) | 0.0311*** (0.0034) |
| | Observations | 345,223 | 345,223 | 345,223 | 117,044 |
| Frozen Juices | Coefficient (Std.) | 0.0251*** (0.0067) | 0.0228*** (0.0076) | 0.0232*** (0.0077) | 0.0289*** (0.009) |
| | Observations | 118,582 | 118,582 | 118,582 | 40,517 |
| Grooming Products | Coefficient (Std.) | 0.0279*** (0.0031) | 0.029*** (0.003) | 0.0291*** (0.003) | 0.0645*** (0.008) |
| | Observations | 101,944 | 101,944 | 101,944 | 22,102 |
| Laundry Detergents | Coefficient (Std.) | 0.0092*** (0.0033) | 0.005** (0.0032) | 0.0053** (0.0032) | 0.0104** (0.0057) |
| | Observations | 121,566 | 121,566 | 121,566 | 42,121 |
| Oatmeal | Coefficient (Std.) | -0.0023 (0.0102) | -0.0087 (0.0121) | -0.0094 (0.012) | 0.0159* (0.0172) |
| | Observations | 25,523 | 25,523 | 25,523 | 13,605 |
| Paper Towels | Coefficient (Std.) | 0.0122** (0.009) | 0.0174** (0.0092) | 0.0179** (0.0092) | 0.0161** (0.0151) |
| | Observations | 48,199 | 48,199 | 48,199 | 9,243 |
| Refrigerated Juices | Coefficient (Std.) | 0.0095** (0.005) | 0.0061* (0.0049) | 0.0061* (0.0048) | 0.0122** (0.0126) |
| | Observations | 108,965 | 108,965 | 108,965 | 23,705 |
| Shampoos | Coefficient (Std.) | 0.0328*** (0.0027) | 0.0325*** (0.0026) | 0.0325*** (0.0026) | 0.0829*** (0.0094) |
| | Observations | 88,193 | 88,193 | 88,193 | 16,099 |
| Snack Crackers | Coefficient (Std.) | 0.0037* (0.0039) | 0.0001 (0.0042) | 0.0002 (0.0042) | 0.0177*** (0.0047) |
| | Observations | 176,527 | 176,527 | 176,527 | 38,123 |
| Soaps | Coefficient (Std.) | 0.0143*** (0.0055) | 0.0083** (0.0056) | 0.01** (0.0057) | 0.0216** (0.0121) |
| | Observations | 56,725 | 56,725 | 56,725 | 16,882 |
| Soft Drinks | Coefficient (Std.) | 0.0143*** (0.0021) | 0.0121*** (0.0024) | 0.0119*** (0.0024) | 0.0137*** (0.0031) |
| | Observations | 243,837 | 243,837 | 243,837 | 49,989 |
| Toothbrushes | Coefficient (Std.) | 0.0277*** (0.0043) | 0.0238*** (0.0041) | 0.0234*** (0.0041) | 0.0611*** (0.0108) |
| | Observations | 52,185 | 52,185 | 52,185 | 13,695 |
| Toothpastes | Coefficient (Std.) | 0.0211*** (0.0034) | 0.0204*** (0.0035) | 0.0205*** (0.0035) | 0.0425*** (0.0078) |
| | Observations | 100,845 | 100,845 | 100,845 | 28,039 |
| **Average coefficients** | | **0.0160** | **0.0152** | **0.0153** | **0.0311** |

<u>Notes</u>: The table reports the results of category-level fixed effect regressions of the probability of a small price change. The dependent variable is "small price change," which equals 1 if a price change of product $i$ in store $s$ at time $t$ is less or to $0.33|\overline{\Delta p_{i,s}}|$ where $\overline{\Delta p_{i,s}}$ is the average size of a price change of product $i$ in store $s$, and 0 otherwise. The main independent variable is the log of the average sales volume of product $i$ in store $s$ over the sample period. Column 1 reports the results of the baseline regression that includes only the log of the average sales volume and the fixed effects for months, years, stores, and products. In column 2, we add the following controls: the log of the average price, the log of the absolute change in the wholesale price, and a control for sale- and bounce-back prices, which we identify using a sales filter algorithm. In column 3, we add a dummy for 9-ending prices as an additional control. In column 4, we focus on regular prices by excluding the sale- and bounce-back prices. We estimate separate regressions for each product category, clustering the errors by product. * $p < 10\%$, ** $p < 5\%$, *** $p < 1\%$

Table B7. Category-level regressions of small price changes ($\Delta P \leq 25\%$ of the average price change) and sales volume

| Category | | **(1)** | **(2)** | **(3)** | **(4)** |
|---|---|---|---|---|---|
| Analgesics | Coefficient (Std.) | 0.0211*** (0.0035) | 0.0177*** (0.0032) | 0.0172*** (0.0032) | 0.0337*** (0.0074) |
| | Observations | 74,451 | 74,451 | 74,451 | 24,729 |
| Bath Soap | Coefficient (Std.) | 0.0209*** (0.0083) | 0.0177*** (0.0077) | 0.0176*** (0.0074) | 0.0616*** (0.0241) |
| | Observations | 6,650 | 6,650 | 6,650 | 1,466 |
| Bathroom Tissues | Coefficient (Std.) | 0.0036 (0.005) | 0.012*** (0.0057) | 0.0119*** (0.0057) | 0.028*** (0.0127) |
| | Observations | 56,458 | 56,458 | 56,458 | 19,285 |
| Beer | Coefficient (Std.) | 0.0025*** (0.0006) | 0.0054*** (0.0007) | 0.0054*** (0.0007) | 0.0377*** (0.0059) |
| | Observations | 187,691 | 187,691 | 187,691 | 12,080 |
| Bottled Juice | Coefficient (Std.) | 0.016*** (0.004) | 0.0134*** (0.0041) | 0.0135*** (0.0041) | 0.0221*** (0.0071) |
| | Observations | 224,857 | 224,857 | 224,857 | 60,015 |
| Canned Soup | Coefficient (Std.) | -0.0121** (0.0062) | -0.0077* (0.0065) | -0.0064* (0.0064) | 0.0152*** (0.0065) |
| | Observations | 233,779 | 233,779 | 233,779 | 95,310 |
| Canned Tuna | Coefficient (Std.) | 0.0089*** (0.0037) | 0.0066*** (0.0031) | 0.0065*** (0.0031) | 0.014*** (0.0063) |
| | Observations | 112,629 | 112,629 | 112,629 | 31,922 |
| Cereals | Coefficient (Std.) | 0.0105*** (0.0035) | 0.0111*** (0.0035) | 0.0112*** (0.0034) | 0.0195*** (0.0059) |
| | Observations | 141,087 | 141,087 | 141,087 | 72,789 |
| Cheese | Coefficient (Std.) | 0.0085*** (0.0029) | 0.0077*** (0.0029) | 0.0077*** (0.0029) | 0.0152*** (0.0053) |
| | Observations | 357,679 | 357,679 | 357,679 | 92,758 |
| Cigarettes | Coefficient (Std.) | 0.0193*** (0.004) | 0.0209*** (0.0039) | 0.0209*** (0.0039) | 0.025*** (0.0045) |
| | Observations | 24,553 | 24,553 | 24,553 | 20,692 |
| Cookies | Coefficient (Std.) | 0.0086*** (0.0018) | 0.0073*** (0.0019) | 0.0072*** (0.0019) | 0.0143*** (0.0035) |
| | Observations | 317,932 | 317,932 | 317,932 | 66,087 |
| Crackers | Coefficient (Std.) | 0.0045** (0.0025) | 0.0038** (0.0026) | 0.0038** (0.0025) | 0.0068** (0.0041) |
| | Observations | 115,658 | 115,658 | 115,658 | 24,771 |
| Dish Detergent | Coefficient (Std.) | 0.0093*** (0.0021) | 0.006*** (0.0021) | 0.0061*** (0.0021) | 0.0183*** (0.0046) |
| | Observations | 85,222 | 85,222 | 85,222 | 26,735 |
| Fabric Softener | Coefficient (Std.) | 0.0128*** (0.0044) | 0.0098*** (0.0043) | 0.0101*** (0.0043) | 0.0278*** (0.0087) |
| | Observations | 85,337 | 85,337 | 85,337 | 27,488 |
| Front-End-Candies | Coefficient (Std.) | 0.0125*** (0.0042) | 0.0147*** (0.0042) | 0.0148*** (0.0042) | 0.0132*** (0.0041) |
| | Observations | 148,200 | 148,200 | 148,200 | 77,323 |
| Frozen Dinners | Coefficient (Std.) | 0.0373*** (0.0064) | 0.0398*** (0.0076) | 0.0401*** (0.0075) | 0.0586*** (0.0123) |
| | Observations | 52,893 | 52,893 | 52,893 | 12,287 |

Table B7. (Cont.)

| Category | | **(1)** | **(2)** | **(3)** | **(4)** |
|---|---|---|---|---|---|
| Frozen Entrees | Coefficient (Std.) | 0.0324*** (0.0032) | 0.0389*** (0.0035) | 0.0389*** (0.0034) | 0.0321*** (0.0039) |
| | Observations | 345,223 | 345,223 | 345,223 | 117,044 |
| Frozen Juices | Coefficient (Std.) | 0.0279*** (0.0062) | 0.0255*** (0.0068) | 0.0259*** (0.0068) | 0.0312*** (0.0089) |
| | Observations | 118,582 | 118,582 | 118,582 | 40,517 |
| Grooming Products | Coefficient (Std.) | 0.0149*** (0.0021) | 0.0156*** (0.0022) | 0.0157*** (0.0022) | 0.0368*** (0.0062) |
| | Observations | 101,944 | 101,944 | 101,944 | 22,102 |
| Laundry Detergents | Coefficient (Std.) | 0.0013 (0.0025) | -0.0018 (0.0024) | -0.0017 (0.0024) | 0.0011 (0.0054) |
| | Observations | 121,566 | 121,566 | 121,566 | 42,121 |
| Oatmeal | Coefficient (Std.) | 0.0065 (0.0082) | 0.0011 (0.0095) | -0.0002 (0.0094) | 0.0022 (0.0145) |
| | Observations | 25,523 | 25,523 | 25,523 | 13,605 |
| Paper Towels | Coefficient (Std.) | 0.0074 (0.0102) | 0.0116* (0.0106) | 0.0123* (0.0105) | 0.0265** (0.0132) |
| | Observations | 48,199 | 48,199 | 48,199 | 9,243 |
| Refrigerated Juices | Coefficient (Std.) | 0.0094*** (0.0041) | 0.006** (0.0041) | 0.0061** (0.0041) | 0.0151** (0.0108) |
| | Observations | 108,965 | 108,965 | 108,965 | 23,705 |
| Shampoos | Coefficient (Std.) | 0.018*** (0.0021) | 0.018*** (0.002) | 0.018*** (0.002) | 0.054*** (0.0077) |
| | Observations | 88,193 | 88,193 | 88,193 | 16,099 |
| Snack Crackers | Coefficient (Std.) | 0.0044** (0.0027) | 0.0015 (0.0027) | 0.0015 (0.0027) | 0.0126*** (0.0046) |
| | Observations | 176,527 | 176,527 | 176,527 | 38,123 |
| Soaps | Coefficient (Std.) | 0.0184*** (0.005) | 0.0138*** (0.005) | 0.015*** (0.0051) | 0.0251*** (0.0109) |
| | Observations | 56,725 | 56,725 | 56,725 | 16,882 |
| Soft Drinks | Coefficient (Std.) | 0.016*** (0.0026) | 0.0136*** (0.0027) | 0.0131*** (0.0028) | 0.0132*** (0.0032) |
| | Observations | 243,837 | 243,837 | 243,837 | 49,989 |
| Toothbrushes | Coefficient (Std.) | 0.0156*** (0.0043) | 0.013*** (0.004) | 0.0127*** (0.004) | 0.0427*** (0.0095) |
| | Observations | 52,185 | 52,185 | 52,185 | 13,695 |
| Toothpastes | Coefficient (Std.) | 0.0178*** (0.0027) | 0.0168*** (0.0026) | 0.0169*** (0.0026) | 0.0315*** (0.007) |
| | Observations | 100,845 | 100,845 | 100,845 | 28,039 |
| **Average coefficients** | | **0.0129** | **0.0124** | **0.0125** | **0.0253** |

Notes: The table reports the results of category-level fixed effect regressions of the probability of a small price change. The dependent variable is "small price change," which equals 1 if a price change of product $i$ in store $s$ at time $t$ is less or to $0.25 |\overline{\Delta p}_{i,s}|$ where $\overline{\Delta p}_{i,s}$ is the average size of a price change of product $i$ in store $s$, and 0 otherwise. The main independent variable is the log of the average sales volume of product $i$ in store $s$ over the sample period. Column 1 reports the results of the baseline regression that includes only the log of the average sales volume and the fixed effects for months, years, stores, and products. In column 2, we add the following controls: the log of the average price, the log of the absolute change in the wholesale price, and a control for sale- and bounce-back prices, which we identify using a sales filter algorithm. In column 3, we add a dummy for 9-ending prices as an additional control. In column 4, we focus on regular prices by excluding the sale- and bounce-back prices. We estimate separate regressions for each product category, clustering the errors by product. * $p < 10\%$, ** $p < 5\%$, *** $p < 1\%$

Table B8. Category-level regressions of small price changes ($\Delta P \le 10\%$ of the average price change) and sales volume

| Category | | **(1)** | **(2)** | **(3)** | **(4)** |
|---|---|---|---|---|---|
| Analgesics | Coefficient (Std.) | 0.0067*** (0.0018) | 0.0057*** (0.0017) | 0.0055*** (0.0017) | 0.0122*** (0.0047) |
| | Observations | 74,451 | 74,451 | 74,451 | 24,729 |
| Bath Soap | Coefficient (Std.) | 0.005*** (0.0021) | 0.0045*** (0.0019) | 0.0044*** (0.0019) | -0.0015 (0.0107) |
| | Observations | 6,650 | 6,650 | 6,650 | 1,466 |
| Bathroom Tissues | Coefficient (Std.) | -0.0058*** (0.0029) | 0.0069** (0.0039) | 0.0069** (0.0039) | 0.0207*** (0.0078) |
| | Observations | 56,458 | 56,458 | 56,458 | 19,285 |
| Beer | Coefficient (Std.) | 0.0015*** (0.0004) | 0.0031*** (0.0006) | 0.0031*** (0.0006) | 0.0191*** (0.004) |
| | Observations | 187,691 | 187,691 | 187,691 | 12,080 |
| Bottled Juice | Coefficient (Std.) | 0.0029** (0.0015) | 0.0019* (0.0015) | 0.0019* (0.0015) | 0.0039* (0.0045) |
| | Observations | 224,857 | 224,857 | 224,857 | 60,015 |
| Canned Soup | Coefficient (Std.) | 0.0001 (0.002) | 0.0031** (0.0023) | 0.0033** (0.0023) | 0.0128*** (0.004) |
| | Observations | 233,779 | 233,779 | 233,779 | 95,310 |
| Canned Tuna | Coefficient (Std.) | 0.0032*** (0.0013) | 0.0021*** (0.0011) | 0.0021** (0.0011) | 0.0089*** (0.0029) |
| | Observations | 112,629 | 112,629 | 112,629 | 31,922 |
| Cereals | Coefficient (Std.) | 0.0038*** (0.0015) | 0.004*** (0.0015) | 0.004*** (0.0015) | 0.0094*** (0.003) |
| | Observations | 141,087 | 141,087 | 141,087 | 72,789 |
| Cheese | Coefficient (Std.) | 0.0018*** (0.0008) | 0.0015*** (0.0008) | 0.0015** (0.0008) | 0.0075*** (0.0026) |
| | Observations | 357,679 | 357,679 | 357,679 | 92,758 |
| Cigarettes | Coefficient (Std.) | -0.001 (0.0013) | -0.0004 (0.0013) | -0.0004 (0.0013) | -0.0004 (0.0015) |
| | Observations | 24,553 | 24,553 | 24,553 | 20,692 |
| Cookies | Coefficient (Std.) | 0.0013*** (0.0005) | 0.0012*** (0.0005) | 0.0012*** (0.0004) | 0.0028*** (0.0011) |
| | Observations | 317,932 | 317,932 | 317,932 | 66,087 |
| Crackers | Coefficient (Std.) | -0.0002 (0.0005) | -0.0003 (0.0005) | -0.0002 (0.0005) | 0.0014* (0.0015) |
| | Observations | 115,658 | 115,658 | 115,658 | 24,771 |
| Dish Detergent | Coefficient (Std.) | 0.0047*** (0.0013) | 0.0036*** (0.0012) | 0.0037*** (0.0012) | 0.0087*** (0.0033) |
| | Observations | 85,222 | 85,222 | 85,222 | 26,735 |
| Fabric Softener | Coefficient (Std.) | 0.0014 (0.0018) | 0.0009 (0.0018) | 0.001 (0.0018) | 0.0034 (0.0055) |
| | Observations | 85,337 | 85,337 | 85,337 | 27,488 |
| Front-End-Candies | Coefficient (Std.) | 0.0077*** (0.0025) | 0.0093*** (0.0028) | 0.0093*** (0.0028) | 0.0079*** (0.0032) |
| | Observations | 148,200 | 148,200 | 148,200 | 77,323 |
| Frozen Dinners | Coefficient (Std.) | 0.028*** (0.0056) | 0.0279*** (0.0061) | 0.0288*** (0.006) | 0.0425*** (0.0124) |
| | Observations | 52,893 | 52,893 | 52,893 | 12,287 |

Table B8. (Cont.)

| Category | | **(1)** | **(2)** | **(3)** | **(4)** |
|---|---|---|---|---|---|
| Frozen Entrees | Coefficient (Std.) | 0.0157*** (0.0023) | 0.02*** (0.0024) | 0.0201*** (0.0024) | 0.0322*** (0.0036) |
| | Observations | 345,223 | 345,223 | 345,223 | 117,044 |
| Frozen Juices | Coefficient (Std.) | 0.0165*** (0.0041) | 0.0149*** (0.0043) | 0.0151*** (0.0042) | 0.0251*** (0.006) |
| | Observations | 118,582 | 118,582 | 118,582 | 40,517 |
| Grooming Products | Coefficient (Std.) | 0.0013*** (0.0006) | 0.0014*** (0.0006) | 0.0015*** (0.0006) | 0.0031** (0.0022) |
| | Observations | 101,944 | 101,944 | 101,944 | 22,102 |
| Laundry Detergents | Coefficient (Std.) | -0.0009 (0.0015) | -0.0016* (0.0014) | -0.0016* (0.0014) | -0.0034* (0.0036) |
| | Observations | 121,566 | 121,566 | 121,566 | 42,121 |
| Oatmeal | Coefficient (Std.) | 0.0032 (0.0042) | 0.0014 (0.0042) | 0.001 (0.0041) | 0.0056 (0.0081) |
| | Observations | 25,523 | 25,523 | 25,523 | 13,605 |
| Paper Towels | Coefficient (Std.) | 0.0058** (0.0034) | 0.0062** (0.0034) | 0.0065** (0.0034) | 0.0143*** (0.0067) |
| | Observations | 48,199 | 48,199 | 48,199 | 9,243 |
| Refrigerated Juices | Coefficient (Std.) | 0.0052*** (0.0014) | 0.0033*** (0.0013) | 0.0033*** (0.0013) | 0.0165*** (0.0054) |
| | Observations | 108,965 | 108,965 | 108,965 | 23,705 |
| Shampoos | Coefficient (Std.) | 0.0009*** (0.0003) | 0.0009*** (0.0003) | 0.0009*** (0.0003) | 0.0041*** (0.0016) |
| | Observations | 88,193 | 88,193 | 88,193 | 16,099 |
| Snack Crackers | Coefficient (Std.) | -0.0003 (0.0009) | -0.0007 (0.0009) | -0.0007 (0.0009) | 0.0024* (0.0019) |
| | Observations | 176,527 | 176,527 | 176,527 | 38,123 |
| Soaps | Coefficient (Std.) | 0.0067*** (0.0022) | 0.0057*** (0.0021) | 0.0059*** (0.0021) | 0.0107*** (0.0054) |
| | Observations | 56,725 | 56,725 | 56,725 | 16,882 |
| Soft Drinks | Coefficient (Std.) | 0.0022*** (0.0007) | 0.0014*** (0.0007) | 0.0008* (0.0007) | -0.002* (0.0021) |
| | Observations | 243,837 | 243,837 | 243,837 | 49,989 |
| Toothbrushes | Coefficient (Std.) | 0.0053*** (0.0013) | 0.0047*** (0.0013) | 0.0046*** (0.0013) | 0.0119*** (0.004) |
| | Observations | 52,185 | 52,185 | 52,185 | 13,695 |
| Toothpastes | Coefficient (Std.) | 0.0048*** (0.001) | 0.0046*** (0.001) | 0.0047*** (0.001) | 0.0123*** (0.0032) |
| | Observations | 100,845 | 100,845 | 100,845 | 28,039 |
| **Average coefficients** | | **0.0044** | **0.0047** | **0.0048** | **0.0101** |

Notes: The table reports the results of category-level fixed effect regressions of the probability of a small price change. The dependent variable is "small price change," which equals 1 if a price change of product $i$ in store $s$ at time $t$ is less or to $0.10|\overline{\Delta p}_{i,s}|$ where $\overline{\Delta p}_{i,s}$ is the average size of a price change of product $i$ in store $s$, and 0 otherwise. The main independent variable is the log of the average sales volume of product $i$ in store $s$ over the sample period. Column 1 reports the results of the baseline regression that includes only the log of the average sales volume and the fixed effects for months, years, stores, and products. In column 2, we add the following controls: the log of the average price, the log of the absolute change in the wholesale price, and a control for sale- and bounce-back prices, which we identify using a sales filter algorithm. In column 3, we add a dummy for 9-ending prices as an additional control. In column 4, we focus on regular prices by excluding the sale- and bounce-back prices. We estimate separate regressions for each product category, clustering the errors by product. * $p < 10\%$, ** $p < 5\%$, *** $p < 1\%$

**Appendix C. Using all price changes**

In the paper, we use observations on price changes only if we observe the price in both week $t$ and $t+1$ and the post change price remained unchanged for at least 2 weeks. In this appendix, we re-run the regressions we report in Table 3 in the paper, but this time: (1) using observations if we observe the price in both week $t$ and $t+1$ and (2) using observations on all price changes. As in the paper, we define small price changes as price changes smaller than, or equal to, 10¢.

Table C1 presents the results of regressions equivalent to the regressions in Table 3 in the paper. The regressions take the following form:

$$small\ price\ change_{i,s,t} = \alpha + \beta \ln(average\ sales\ volume_{i,s}) + \gamma \mathbf{X}_{i,s,t}$$
$$+month_t + year_t + \delta_s + \mu_i + u_{i,s,t}$$

$$(C1)$$

where *small price change* is a dummy that equals 1 if a price change of product $i$ in store $s$ at time $t$ is less or equal to 10¢, and 0 otherwise. The *average sales volume* is the average sales volume of product $i$ in store $s$ over the sample period. $\mathbf{X}$ is a matrix of other control variables. *Month* and *year* are fixed effects for the month and the year of the price change. $\delta$ and $\mu$ are fixed effects for stores and products, respectively, and $u$ is an i.i.d error term. We estimate separate regressions for each product category, clustering the errors by product. We use all observations on price changes if we observe the price in both week $t$ and week $t+1$.

The values in the table are the coefficients of the log of the average sales volume. In column 1, the only control variables are the log of the average sales volume, and the dummies for months, years, stores, and products. We find that all the coefficients of the log of the average sales volume are positive and statistically significant. The average coefficient is 0.030, suggesting that a 1% increase in the sales volume is associated with a 3.0% increase in the likelihood of a small price change.

In column 2, we add controls for the log of the average price, the log of the absolute change in the wholesale price, and a control for sale- and bounce-back prices, which we identify using the sales filter algorithm of Fox and Syed (2016). All the coefficients are positive and statistically significant. The average coefficient is 0.025, suggesting that a 1% increase in the sales volume is associated with a 2.5% increase in the likelihood of a small price change.

In column 3, we also add a control for 9-ending prices. All the coefficients are still positive and statistically significant. The average coefficient is 0.023, suggesting that a 1% increase in the sales volume is associated with a 2.3% increase in the likelihood of a small price change.

As a further control for the effects of sales on the results, in column 4 we focus on regular prices by excluding all sale- and bounce-back prices. When we focus on regular prices, the results are even stronger. All the coefficients are positive and statistically significant. The average coefficient is 0.045, suggesting that a 1% increase in the sales volume is associated with a 4.5% increase in the likelihood of a small price change.

Table C2 presents the results when we use observations on all price changes. In column 1, the control variables are the log of the average sales volume, and the dummies for months, years, stores, and products. We find that all the coefficients of the log of the average sales volume are positive and 28 of the 29 are statistically significant at the 1% level. The remaining coefficient is statistically significant at the 10% level. The average coefficient is 0.035, suggesting that a 1% increase in the sales volume is associated with a 3.5% increase in the likelihood of a small price change.

In column 2, we add controls for the log of the average price, the log of the absolute change in the wholesale price, and a control for sale- and bounce-back prices, which we identify using the sales filter algorithm of Fox and Syed (2016). All the coefficients are positive and statistically significant at the 1% level. The average coefficient is 0.035, suggesting that a 1% increase in the sales volume is associated with a 3.5% increase in the likelihood of a small price change.

In column 3, we add a control for 9-ending prices. All the coefficients are still positive and statistically significant. The average coefficient is 0.032, suggesting that a 1% increase in the sales volume is associated with a 3.2% increase in the likelihood of a small price change.

As a further control for the effects of sales on the estimation results, in column 4 we focus on regular prices by excluding all sale- and bounce-back prices. When we focus on regular prices, the results are even stronger. All the coefficients are positive and statistically significant. The average coefficient is 0.055, suggesting that a 1% increase in the sales volume is associated with a 5.5% increase in the likelihood of a small price change.

Table C1. Category-level regressions of small price changes and sales volume

| Category | | **(1)** | **(2)** | **(3)** | **(4)** |
|---|---|---|---|---|---|
| Analgesics | Coefficient (Std.) | 0.0388*** (0.0033) | 0.0305*** (0.0027) | 0.0248*** (0.0025) | 0.0475*** (0.0057) |
| | Observations | 278,052 | 278,052 | 278,052 | 75,945 |
| Bath Soap | Coefficient (Std.) | 0.0409*** (0.0093) | 0.0452*** (0.0095) | 0.0422*** (0.0091) | 0.0871*** (0.016) |
| | Observations | 35,795 | 35,795 | 35,795 | 6,555 |
| Bathroom Tissues | Coefficient (Std.) | 0.0372*** (0.0056) | 0.0203*** (0.0053) | 0.0177*** (0.0049) | 0.0351*** (0.0069) |
| | Observations | 326,383 | 326,383 | 326,383 | 81,914 |
| Beer | Coefficient (Std.) | 0.023*** (0.0015) | 0.0249*** (0.0012) | 0.0208*** (0.0012) | 0.0691*** (0.005) |
| | Observations | 459,669 | 459,669 | 459,669 | 56,427 |
| Bottled Juice | Coefficient (Std.) | 0.0554*** (0.0043) | 0.0393*** (0.003) | 0.0343*** (0.0031) | 0.0368*** (0.0045) |
| | Observations | 960,033 | 960,033 | 960,033 | 244,199 |
| Canned Soup | Coefficient (Std.) | 0.0272*** (0.004) | 0.0151*** (0.0034) | 0.0158*** (0.0033) | 0.0217*** (0.0038) |
| | Observations | 947,633 | 947,633 | 947,633 | 278,451 |
| Canned Tuna | Coefficient (Std.) | 0.037*** (0.0052) | 0.0266*** (0.0044) | 0.0225*** (0.0041) | 0.0334*** (0.0047) |
| | Observations | 375,343 | 375,343 | 375,343 | 116,170 |
| Cereals | Coefficient (Std.) | 0.0215*** (0.0026) | 0.0168*** (0.0023) | 0.0156*** (0.0024) | 0.0263*** (0.0035) |
| | Observations | 724,902 | 724,902 | 724,902 | 260,110 |
| Cheese | Coefficient (Std.) | 0.0374*** (0.0029) | 0.0208*** (0.0022) | 0.0169*** (0.0022) | 0.0116*** (0.0031) |
| | Observations | 1,812,016 | 1,812,016 | 1,812,016 | 519,361 |
| Cigarettes | Coefficient (Std.) | 0.019*** (0.0082) | 0.0203*** (0.0068) | 0.0197** (0.0067) | 0.0215*** (0.0045) |
| | Observations | 15,862 | 15,862 | 15,862 | 9,593 |
| Cookies | Coefficient (Std.) | 0.0429*** (0.0017) | 0.0372*** (0.0017) | 0.0315*** (0.0015) | 0.0543*** (0.0031) |
| | Observations | 1,357,300 | 1,357,300 | 1,357,300 | 229,189 |
| Crackers | Coefficient (Std.) | 0.0544*** (0.0033) | 0.0431*** (0.0031) | 0.0389*** (0.0029) | 0.0563*** (0.0061) |
| | Observations | 475,497 | 475,497 | 475,497 | 89,212 |
| Dish Detergent | Coefficient (Std.) | 0.0481*** (0.0038) | 0.0357*** (0.003) | 0.0315*** (0.0029) | 0.0417*** (0.0043) |
| | Observations | 401,332 | 401,332 | 401,332 | 95,495 |
| Fabric Softener | Coefficient (Std.) | 0.0342*** (0.0038) | 0.0245*** (0.0034) | 0.0209*** (0.0035) | 0.0428*** (0.0049) |
| | Observations | 378,836 | 378,836 | 378,836 | 101,979 |
| Front-End-Candies | Coefficient (Std.) | 0.0165*** (0.0039) | 0.0091** (0.0028) | 0.0082** (0.0028) | 0.0113*** (0.0032) |
| | Observations | 490,627 | 490,627 | 490,627 | 155,230 |
| Frozen Dinners | Coefficient (Std.) | 0.0536*** (0.0027) | 0.0408*** (0.0025) | 0.0394*** (0.0025) | 0.0907*** (0.006) |
| | Observations | 502,792 | 502,792 | 502,792 | 72,693 |

Table C1. (Cont.)

| Category | | **(1)** | **(2)** | **(3)** | **(4)** |
|---|---|---|---|---|---|
| Frozen Entrees | Coefficient (Std.) | 0.0354*** (0.0019) | 0.0301*** (0.0017) | 0.0292*** (0.0017) | 0.0602*** (0.0032) |
| | Observations | 1,848,187 | 1,848,187 | 1,848,187 | 353,136 |
| Frozen Juices | Coefficient (Std.) | 0.0342*** (0.0037) | 0.0253*** (0.0031) | 0.0227*** (0.003) | 0.0299*** (0.0048) |
| | Observations | 659,305 | 659,305 | 659,305 | 150,138 |
| Grooming Products | Coefficient (Std.) | 0.0426*** (0.0024) | 0.0455*** (0.0022) | 0.039*** (0.0021) | 0.0673*** (0.0061) |
| | Observations | 668,821 | 668,821 | 668,821 | 99,253 |
| Laundry Detergents | Coefficient (Std.) | 0.0185*** (0.0031) | 0.0155*** (0.0027) | 0.0126*** (0.0025) | 0.0264*** (0.0047) |
| | Observations | 594,258 | 594,258 | 594,258 | 145,176 |
| Oatmeal | Coefficient (Std.) | 0.0288*** (0.0071) | 0.0172*** (0.0052) | 0.0151** (0.0052) | 0.0319*** (0.0094) |
| | Observations | 168,988 | 168,988 | 168,988 | 63,575 |
| Paper Towels | Coefficient (Std.) | 0.0378*** (0.0114) | 0.0298*** (0.0116) | 0.0285** (0.0117) | 0.0376*** (0.0096) |
| | Observations | 244,068 | 244,068 | 244,068 | 52,327 |
| Refrigerated Juices | Coefficient (Std.) | 0.031*** (0.0032) | 0.0209*** (0.0027) | 0.0182*** (0.0026) | 0.0305*** (0.0041) |
| | Observations | 800,280 | 800,280 | 800,280 | 161,098 |
| Shampoos | Coefficient (Std.) | 0.0323*** (0.0014) | 0.0368*** (0.0014) | 0.032*** (0.0013) | 0.0674*** (0.0043) |
| | Observations | 713,730 | 713,730 | 713,730 | 86,458 |
| Snack Crackers | Coefficient (Std.) | 0.0434*** (0.0032) | 0.0381*** (0.003) | 0.0337*** (0.0027) | 0.0661*** (0.004) |
| | Observations | 802,462 | 802,462 | 802,462 | 143,164 |
| Soaps | Coefficient (Std.) | 0.0305*** (0.0014) | 0.0265*** (0.001) | 0.0222*** (0.0009) | 0.0585*** (0.0027) |
| | Observations | 4,378,334 | 4,378,334 | 4,378,334 | 346,632 |
| Soft Drinks | Coefficient (Std.) | 0.0545*** (0.006) | 0.0413*** (0.0044) | 0.0336*** (0.0042) | 0.0555*** (0.0057) |
| | Observations | 333,170 | 333,170 | 333,170 | 94,295 |
| Toothbrushes | Coefficient (Std.) | 0.0291*** (0.0032) | 0.0317*** (0.0034) | 0.0265*** (0.0032) | 0.0619*** (0.006) |
| | Observations | 295,403 | 295,403 | 295,403 | 44,776 |
| Toothpastes | Coefficient (Std.) | 0.0289*** (0.0032) | 0.028*** (0.0027) | 0.0241*** (0.0026) | 0.0561*** (0.0063) |
| | Observations | 596,903 | 596,903 | 596,903 | 91,760 |
| **Average coefficients** | | **0.030** | **0.025** | **0.023** | **0.045** |

<u>Notes</u>: The table reports the results of category-level fixed effect regressions of the probability of a small price change. The dependent variable is "small price change," which equals 1 if a price change of product $i$ in store $s$ at time $t$ is less or equal to 10¢, and 0 otherwise. The main independent variable is the log of the average sales volume of product $i$ in store $s$ over the sample period. Column 1 reports the results of baseline regression that includes only the average sales volume and the fixed effects for months, years, stores, and products. In column 2, we add the following controls: the log of the average price, the log of the absolute change in the wholesale price, and a control for sale- and bounce back prices, which we identify using a sales filter algorithm. In column 3, we add a dummy for 9-ending prices as an additional control. In column 4, we focus on regular prices by excluding the sale- and bounce-back prices. We estimate separate regressions for each product category, clustering the errors by product. * $p < 10\%$, ** $p < 5\%$, *** $p < 1\%$

Table C2. Category-level regressions of small price changes $(\Delta P \le 10\text{¢})$ and sales volume, using all observations

| Category | | **(1)** | **(2)** | **(3)** | **(4)** |
|---|---|---|---|---|---|
| Analgesics | Coefficient (Std.) | 0.0288*** (0.0028) | 0.0406*** (0.0028) | 0.0362*** (0.0025) | 0.0571*** (0.0045) |
| | Observations | 467,137 | 467,137 | 467,137 | 158,600 |
| Bath Soap | Coefficient (Std.) | 0.0197*** (0.0071) | 0.0479*** (0.0063) | 0.0458*** (0.0061) | 0.0701*** (0.0094) |
| | Observations | 76,548 | 76,548 | 76,548 | 22,545 |
| Bathroom Tissues | Coefficient (Std.) | 0.0377*** (0.0055) | 0.0246*** (0.0054) | 0.0218*** (0.005) | 0.0392*** (0.0074) |
| | Observations | 347,559 | 347,559 | 347,559 | 88,388 |
| Beer | Coefficient (Std.) | 0.0184*** (0.0014) | 0.0287*** (0.0014) | 0.0244*** (0.0013) | 0.0561*** (0.0038) |
| | Observations | 617,181 | 617,181 | 617,181 | 95,440 |
| Bottled Juice | Coefficient (Std.) | 0.0591*** (0.0041) | 0.0467*** (0.0031) | 0.041*** (0.0032) | 0.0553*** (0.0048) |
| | Observations | 1,044,176 | 1,044,176 | 1,044,176 | 269,990 |
| Canned Soup | Coefficient (Std.) | 0.029*** (0.0038) | 0.0194*** (0.0033) | 0.0196*** (0.0031) | 0.0339*** (0.004) |
| | Observations | 1,041,402 | 1,041,402 | 1,041,402 | 309,450 |
| Canned Tuna | Coefficient (Std.) | 0.0418*** (0.0049) | 0.0342*** (0.0044) | 0.0294*** (0.004) | 0.0461*** (0.0053) |
| | Observations | 447,946 | 447,946 | 447,946 | 142,596 |
| Cereals | Coefficient (Std.) | 0.0249*** (0.0026) | 0.0242*** (0.0024) | 0.0228*** (0.0025) | 0.0407*** (0.0035) |
| | Observations | 771,993 | 771,993 | 771,993 | 281,908 |
| Cheese | Coefficient (Std.) | 0.0407*** (0.0027) | 0.0278*** (0.0023) | 0.0232*** (0.0023) | 0.0282*** (0.0037) |
| | Observations | 1,955,416 | 1,955,416 | 1,955,416 | 557,994 |
| Cigarettes | Coefficient (Std.) | 0.024* (0.0133) | 0.0297*** (0.01) | 0.0295*** (0.01) | 0.0245** (0.0096) |
| | Observations | 71,155 | 71,155 | 71,155 | 35,156 |
| Cookies | Coefficient (Std.) | 0.0466*** (0.0017) | 0.0464*** (0.0017) | 0.0404*** (0.0015) | 0.0681*** (0.0029) |
| | Observations | 1,581,102 | 1,581,102 | 1,581,102 | 297,881 |
| Crackers | Coefficient (Std.) | 0.0601*** (0.0035) | 0.0543*** (0.0033) | 0.0494*** (0.0031) | 0.0709*** (0.0057) |
| | Observations | 567,809 | 567,809 | 567,809 | 114,425 |
| Dish Detergent | Coefficient (Std.) | 0.0459*** (0.0041) | 0.0411*** (0.0035) | 0.0372*** (0.0034) | 0.0618*** (0.0047) |
| | Observations | 497,210 | 497,210 | 497,210 | 115,037 |
| Fabric Softener | Coefficient (Std.) | 0.0333*** (0.0041) | 0.0292*** (0.0036) | 0.0247*** (0.0037) | 0.0529*** (0.0049) |
| | Observations | 478,611 | 478,611 | 478,611 | 123,818 |
| Front-End-Candies | Coefficient (Std.) | 0.0184*** (0.0036) | 0.0131*** (0.003) | 0.0122*** (0.003) | 0.0157*** (0.0034) |
| | Observations | 537,812 | 537,812 | 537,812 | 173,538 |
| Frozen Dinners | Coefficient (Std.) | 0.0584*** (0.0041) | 0.0528*** (0.0034) | 0.051*** (0.0035) | 0.1183*** (0.0065) |
| | Observations | 567,884 | 567,884 | 567,884 | 86,750 |

Table C2. (Cont.)

| Category | | (1) | (2) | (3) | (4) |
|---|---|---|---|---|---|
| Frozen Entrees | Coefficient (Std.) | 0.0359*** (0.0019) | 0.0392*** (0.0017) | 0.0382*** (0.0017) | 0.087*** (0.0038) |
| | Observations | 2,084,913 | 2,084,913 | 2,084,913 | 419,173 |
| Frozen Juices | Coefficient (Std.) | 0.0389*** (0.0038) | 0.0331*** (0.0035) | 0.0297*** (0.0033) | 0.0504*** (0.0054) |
| | Observations | 703,893 | 703,893 | 703,893 | 162,718 |
| Grooming Products | Coefficient (Std.) | 0.0307*** (0.0021) | 0.0473*** (0.002) | 0.0423*** (0.0019) | 0.0601*** (0.0042) |
| | Observations | 1,092,785 | 1,092,785 | 1,092,785 | 210,384 |
| Laundry Detergents | Coefficient (Std.) | 0.0176*** (0.0032) | 0.0191*** (0.0028) | 0.0162*** (0.0024) | 0.0367*** (0.0046) |
| | Observations | 766,390 | 766,390 | 766,390 | 183,661 |
| Oatmeal | Coefficient (Std.) | 0.0333*** (0.0069) | 0.023*** (0.0057) | 0.0205*** (0.0056) | 0.045*** (0.0115) |
| | Observations | 181,193 | 181,193 | 181,193 | 69,150 |
| Paper Towels | Coefficient (Std.) | 0.0394*** (0.0111) | 0.0327*** (0.0114) | 0.0304*** (0.0114) | 0.0482*** (0.0089) |
| | Observations | 274,918 | 274,918 | 274,918 | 58,771 |
| Refrigerated Juices | Coefficient (Std.) | 0.0348*** (0.0032) | 0.0283*** (0.0029) | 0.025*** (0.0027) | 0.047*** (0.0048) |
| | Observations | 827,359 | 827,359 | 827,359 | 169,826 |
| Shampoos | Coefficient (Std.) | 0.0185*** (0.0013) | 0.0361*** (0.0012) | 0.0328*** (0.0012) | 0.0503*** (0.0026) |
| | Observations | 1,315,278 | 1,315,278 | 1,315,278 | 272,979 |
| Snack Crackers | Coefficient (Std.) | 0.0492*** (0.0035) | 0.0483*** (0.0033) | 0.0432*** (0.003) | 0.084*** (0.0045) |
| | Observations | 903,254 | 903,254 | 903,254 | 172,655 |
| Soaps | Coefficient (Std.) | 0.0321*** (0.0013) | 0.0313*** (0.0011) | 0.0266*** (0.0009) | 0.0634*** (0.0027) |
| | Observations | 4,985,172 | 4,985,172 | 4,985,172 | 451,007 |
| Soft Drinks | Coefficient (Std.) | 0.0519*** (0.0079) | 0.0488*** (0.0044) | 0.0398*** (0.0042) | 0.0704*** (0.0059) |
| | Observations | 395,114 | 395,114 | 395,114 | 110,144 |
| Toothbrushes | Coefficient (Std.) | 0.0197*** (0.0035) | 0.0358*** (0.0034) | 0.0305*** (0.0032) | 0.0504*** (0.0056) |
| | Observations | 481,842 | 481,842 | 481,842 | 93,164 |
| Toothpastes | Coefficient (Std.) | 0.0251*** (0.0029) | 0.0349*** (0.0025) | 0.0311*** (0.0025) | 0.0684*** (0.0058) |
| | Observations | 771,084 | 771,084 | 771,084 | 136,134 |
| **Average coefficients** | | **0.0350** | **0.0351** | **0.0316** | **0.0552** |

Notes: The table reports the results of category-level fixed effect regressions of the probability of a small price change. The dependent variable is "small price change," which equals 1 if a price change of product $i$ in store $s$ at time $t$ is less or equal to 10¢, and 0 otherwise. The main independent variable is the log of the average sales volume of product $i$ in store $s$ over the sample period. Column 1 reports the results of the baseline regression that includes only the log of the average sales volume and the fixed effects for months, years, stores, and products. In column 2, we add the following controls: the log of the average price, the log of the absolute change in the wholesale price, and a control for sale- and bounce-back prices, which we identify using a sales filter algorithm. In column 3, we add a dummy for 9-ending prices as an additional control. In column 4, we focus on regular prices by excluding the sale- and bounce-back prices. We estimate separate regressions for each product category, clustering the errors by product. * $p < 10\%$, ** $p < 5\%$, *** $p < 1\%$

## Appendix D. Using a rolling 52-week window to calculate the average sales volume

In the paper, we calculate the average sales volume for each product in each store over the entire period. This has the advantage of using a long-term "expected" sales volume for each product in each store. However, it implicitly assumes that the retailer can forecast future sales.

An alternative is to assume that the retailer makes decisions based on a recent past. To control for this possibility, we calculate the average sales volume for each product in each store based on data from the previous 52 weeks.

We then use the results to re-estimate regressions similar to the ones that we report in Table 3 in the paper. The regressions take the following form:

$$small\ price\ change_{i,s,t} = \alpha + \beta \ln(average\ sales\ volume_{i,s}) + \gamma \mathbf{X}_{i,s,t}$$
$$+month_t + year_t + \delta_s + \mu_i + u_{i,s,t}$$
(D1)

where *small price change* is a dummy that equals 1 if a price change of product $i$ in store $s$ at time $t$ is less or equal to 10¢, and 0 otherwise. The *average sales volume* is the average sales volume of product $i$ in store $s$ over the 52 weeks preceding week $t$. $\mathbf{X}$ is a matrix of other control variables. *Month* and *year* are fixed effects for the month and the year of the price change. $\delta$ and $\mu$ are fixed effects for stores and products, respectively, and $u$ is an i.i.d error term. We estimate separate regressions for each product category, clustering the errors by product.

The results are summarized in Table D1. The values in the table are the coefficients of the log of the average sales volume. In column 1, the only control variables are the log of the average sales volume, and the dummies for months, years, stores, and products. We find that 24 of the 29 coefficients of the log of the average sales volume are positive and that 16 of them are statistically significant. One more coefficient is marginally significant. None of the five negative coefficients are statistically significant. The average coefficient is 0.018, suggesting that a 1% increase in the sales volume is associated with a 1.8% increase in the likelihood of a small price change.

In column 2, we add controls for the log of the average price, the log of the absolute change in the wholesale price, and a control for sale- and bounce-back prices, which we identify using the sales filter algorithm of Fox and Syed (2016). We find that 25 of the 29 coefficients are positive and that 17 of them are statistically significant. None of the four negative coefficients are statistically significant. The average coefficient is 0.017, suggesting that a 1% increase in the sales volume is associated with a 1.7% increase in the likelihood of a small price change.

In column 3, we also add a control for 9-ending prices. Again, we find that 25 of the 29 coefficients are positive and that 16 of them are statistically significant. One more coefficient is marginally statistically significant. None of the negative coefficients is statistically significant. The average coefficient is 0.015, suggesting that a 1% increase in the sales volume is associated with a 1.5% increase in the likelihood of a small price change.

As a further control for the effects of sales on the results, in column 4 we focus on regular prices by excluding all sale- and bounce-back prices. When we focus on regular prices, the results are stronger. We find that 27 of the 29 coefficients are positive and 21 of them are statistically significant. One more coefficient is marginally significant. Out of the two negative coefficients, one (cereals) is marginally significant. The average coefficient is 0.31, suggesting that a 1% increase in the sales volume is associated with a 3.1% increase in the likelihood of a small price change.

Thus, basing the estimation on the sales volume of the more recent period does not change our main results. The correlation between small price changes and sales volume holds in a large majority of the product categories.

Table D1. Category-level regressions of small price changes $(\Delta P \leq 10\cent)$, using a rolling 52-week window for sales volume

| Category | | (1) | (2) | (3) | (4) |
|---|---|---|---|---|---|
| Analgesics | Coefficient (Std.) | 0.0339*** (0.0084) | 0.0262*** (0.0074) | 0.0215*** (0.0071) | 0.0358** (0.0139) |
| | Observations | 258,282 | 258,282 | 258,282 | 71,945 |
| Bath Soap | Coefficient (Std.) | 0.0263*** (0.0076) | 0.0246*** (0.0075) | 0.0256*** (0.0072) | 0.0421*** (0.015) |
| | Observations | 31,704 | 31,704 | 31,704 | 5,186 |
| Bathroom Tissues | Coefficient (Std.) | 0.0155 (0.0134) | 0.0084 (0.0111) | 0.0051 (0.0101) | 0.0225** (0.0104) |
| | Observations | 311,206 | 311,206 | 311,206 | 76,063 |
| Beer | Coefficient (Std.) | 0.0172*** (0.0019) | 0.0146*** (0.0016) | 0.0111*** (0.0016) | 0.0318*** (0.0062) |
| | Observations | 410,854 | 410,854 | 410,854 | 50,131 |
| Bottled Juice | Coefficient (Std.) | 0.0225*** (0.007) | 0.0122** (0.0057) | 0.0102* (0.0059) | 0.001 (0.0075) |
| | Observations | 917,557 | 917,557 | 917,557 | 228,910 |
| Canned Soup | Coefficient (Std.) | -0.0107 (0.0078) | -0.0038 (0.007) | -0.0015 (0.007) | 0.0116 (0.0076) |
| | Observations | 890,145 | 890,145 | 890,145 | 256,793 |
| Canned Tuna | Coefficient (Std.) | 0.001 (0.0112) | 0.004 (0.0091) | 0.0014 (0.0088) | 0.0257*** (0.0075) |
| | Observations | 354,012 | 354,012 | 354,012 | 110,295 |
| Cereals | Coefficient (Std.) | 0.0059 (0.006) | 0.0062 (0.0049) | 0.0046 (0.0048) | -0.0006 (0.0092) |
| | Observations | 692,679 | 692,679 | 692,679 | 242,184 |
| Cheese | Coefficient (Std.) | 0.0109* (0.0065) | 0.0034 (0.0052) | 0.0014 (0.0052) | -0.0058 (0.0075) |
| | Observations | 1,725,208 | 1,725,208 | 1,725,208 | 489,617 |
| Cigarettes | Coefficient (Std.) | 0.0081 (0.0158) | 0.0078 (0.014) | 0.0067 (0.0136) | 0.0083 (0.0133) |
| | Observations | 13,712 | 13,712 | 13,712 | 8,591 |
| Cookies | Coefficient (Std.) | 0.0335*** (0.0044) | 0.0253*** (0.004) | 0.0229*** (0.0039) | 0.0392*** (0.0076) |
| | Observations | 1,286,069 | 1,286,069 | 1,286,069 | 209,459 |
| Crackers | Coefficient (Std.) | 0.0493*** (0.0076) | 0.0364*** (0.0072) | 0.0339*** (0.0072) | 0.0446*** (0.0087) |
| | Observations | 448,590 | 448,590 | 448,590 | 81,601 |
| Dish Detergent | Coefficient (Std.) | 0.0292*** (0.0071) | 0.0276*** (0.0059) | 0.0232*** (0.0058) | 0.0329*** (0.0087) |
| | Observations | 374,776 | 374,776 | 374,776 | 89,534 |
| Fabric Softener | Coefficient (Std.) | 0.0126 (0.0091) | 0.0197*** (0.007) | 0.0152** (0.0067) | 0.0312** (0.0134) |
| | Observations | 357,352 | 357,352 | 357,352 | 96,359 |
| Front-End-Candies | Coefficient (Std.) | -0.0161 (0.0098) | -0.0004 (0.0085) | -0.0006 (0.0084) | 0.0032 (0.006) |
| | Observations | 471,213 | 471,213 | 471,213 | 140,179 |
| Frozen Dinners | Coefficient (Std.) | 0.0587*** (0.0068) | 0.0516*** (0.0054) | 0.0519*** (0.0055) | 0.0751*** (0.0101) |
| | Observations | 443,557 | 443,557 | 443,557 | 58,225 |

Table D1. (Cont.)

| Category | | **(1)** | **(2)** | **(3)** | **(4)** |
|---|---|---|---|---|---|
| Frozen Entrees | Coefficient (Std.) | 0.0186*** (0.0025) | 0.0272*** (0.0026) | 0.0288*** (0.0025) | 0.0518*** (0.0058) |
| | Observations | 1,771,958 | 1,771,958 | 1,771,958 | 318,555 |
| Frozen Juices | Coefficient (Std.) | 0.0203*** (0.0074) | 0.0174*** (0.0065) | 0.013** (0.0065) | 0.0597*** (0.0086) |
| | Observations | 627,846 | 627,846 | 627,846 | 140,924 |
| Grooming Products | Coefficient (Std.) | 0.0392*** (0.0035) | 0.0364*** (0.0033) | 0.0349*** (0.0033) | 0.0493*** (0.0076) |
| | Observations | 607,229 | 607,229 | 607,229 | 84,669 |
| Laundry Detergents | Coefficient (Std.) | -0.0072 (0.0058) | 0.0049 (0.0048) | 0.0019 (0.0048) | 0.0216*** (0.0094) |
| | Observations | 557,386 | 557,386 | 557,386 | 133,678 |
| Oatmeal | Coefficient (Std.) | -0.0021 (0.0111) | -0.0071 (0.0093) | -0.0075 (0.0089) | 0.0081 (0.009) |
| | Observations | 153,883 | 153,883 | 153,883 | 56,264 |
| Paper Towels | Coefficient (Std.) | 0.0293 (0.0236) | 0.0285 (0.0193) | 0.0264 (0.0187) | 0.0382* (0.0196) |
| | Observations | 229,649 | 229,649 | 229,649 | 48,847 |
| Refrigerated Juices | Coefficient (Std.) | -0.0034 (0.0082) | -0.0012 (0.0067) | -0.0048 (0.0065) | 0.0223*** (0.008) |
| | Observations | 763,905 | 763,905 | 763,905 | 150,388 |
| Shampoos | Coefficient (Std.) | 0.029*** (0.0023) | 0.0268*** (0.0022) | 0.0231*** (0.0021) | 0.0306*** (0.0059) |
| | Observations | 605,146 | 605,146 | 605,146 | 60,380 |
| Snack Crackers | Coefficient (Std.) | 0.0367*** (0.006) | 0.0306*** (0.0055) | 0.0273*** (0.0053) | 0.0563*** (0.0092) |
| | Observations | 758,707 | 758,707 | 758,707 | 128,389 |
| Soaps | Coefficient (Std.) | 0.0139*** (0.0031) | 0.0134*** (0.0029) | 0.0117*** (0.0029) | 0.046*** (0.0047) |
| | Observations | 4,147,187 | 4,147,187 | 4,147,187 | 304,352 |
| Soft Drinks | Coefficient (Std.) | 0.0075 (0.0144) | 0.0101 (0.0109) | 0.0071 (0.0106) | 0.035*** (0.0104) |
| | Observations | 297,007 | 297,007 | 297,007 | 83,454 |
| Toothbrushes | Coefficient (Std.) | 0.0291*** (0.0065) | 0.0275*** (0.0066) | 0.023*** (0.0062) | 0.0303** (0.0123) |
| | Observations | 274,744 | 274,744 | 274,744 | 38,861 |
| Toothpastes | Coefficient (Std.) | 0.0175*** (0.0063) | 0.0177*** (0.0054) | 0.0147*** (0.0055) | 0.0456*** (0.01) |
| | Observations | 567,725 | 567,725 | 567,725 | 84,935 |
| **Average coefficients** | | **0.0181** | **0.0171** | **0.0149** | **0.0308** |

Notes: The table reports the results of category-level fixed effect regressions of the probability of a small price change. The dependent variable is "small price change," which equals 1 if a price change of product $i$ in store $s$ at time $t$ is less or equal to 10¢, and 0 otherwise. The main independent variable is the log of the average sales volume of product $i$ in store $s$ over the 52 weeks preceding time $t$. Column 1 reports the results of the baseline regression that includes only the log of the average sales volume and the fixed effects for months, years, stores, and products. In column 2, we add the following controls: the log of the average price, the log of the absolute change in the wholesale price, and a control for sale- and bounce-back prices, which we identify using a sales filter algorithm. In column 3, we add a dummy for 9-ending prices as an additional control. In column 4, we focus on regular prices by excluding the sale- and bounce-back prices. We estimate separate regressions for each product category, clustering the errors by product. * $p < 10\%$, ** $p < 5\%$, *** $p < 1\%$

## Appendix E. Adding Dominick's pricing zones

According to Dominick's data manual, Dominick's employed 16 price zones. Thus, we can use the zones as a proxy for the competition level.

We, therefore, incorporate the data on pricing zones and re-estimate regressions similar to the ones that we report in Table 3 in the paper. The regressions take the form,

$$small\ price\ change_{i,s,t} = \alpha + \beta \ln(average\ sales\ volume_{i,s}) + \gamma \mathbf{X}_{i,s,t}$$
$$+ month_t + year_t + \delta_s + \mu_i + u_{i,s,t}$$

(E1)

where *small price change* is a dummy that equals 1 if a price change of product $i$ in store $s$ at time $t$ is less or equal to 10¢, and 0 otherwise. The *average sales volume* is the average sales volume of product $i$ in store $s$ over the 52 weeks preceding week $t$. $\mathbf{X}$ is a matrix of other control variables. *Month* and *year* are fixed effects for the month and the year of the price change. $\delta$ and $\mu$ are fixed effects for stores and products, respectively, and $u$ is an i.i.d error term. We estimate separate regressions for each product category, clustering the errors by product.

The figures in Table E1 are the coefficients of the log of the average sales volume. In column 1, the only control variables are the log of the average sales volume, and the dummies for months, years, stores, and products. We find that 27 of the 29 coefficients of the log of the average sales volume are positive. 15 of the 27 are statistically significant, and 4 more are marginally significant. The average coefficient is 0.014, suggesting that a 1% increase in the sales volume is associated with a 1.4% increase in the likelihood of a small price change.

In column 2, we add controls for the log of the average price, the log of the absolute change in the wholesale price, a control for sale- and bounce-back prices (which we identify using the sales filter algorithm of Fox and Syed 2016), and Dominick's pricing zone. We find that 27 of the 29 coefficients are positive. 14 of the positive coefficients are statistically significant, and one more is marginally significant. The average coefficient is 0.010, suggesting that a 1% increase in the sales volume is associated with a 1.0% increase in the likelihood of a small price change.

In column 3, we add a control for 9-ending prices. We find that 27 of the 29 coefficients are positive. 14 of the 29 are statistically significant, and two more are marginally significant. The average coefficient is 0.010, suggesting that a 1% increase in the sales volume is associated with a 1.0% increase in the likelihood of a small price change.

As a further control for the effects of sales on the results, in column 4 we focus on regular prices by excluding all sale- and bounce-back prices. When we focus on regular prices, we find that 27 of the 29 coefficients are positive. 18 are statistically significant, and 5 more are marginally significant. The average coefficient is 0.020, suggesting that a 1% increase in the sales volume is associated with a 2.0% increase in the likelihood of a small price change.

Thus, adding a control for pricing zones does not change our main results. The correlation between small price changes and sales volume holds in all 29 product categories.

Table E1. Category-level regressions of small price changes and sales volume, controlling for Dominick's pricing zones

| Category | | (1) | (2) | (3) | (4) |
|---|---|---|---|---|---|
| Analgesics | Coefficient (Std.) | 0.0169*** (0.004) | 0.0128*** (0.0038) | 0.0126*** (0.0038) | 0.0133*** (0.0079) |
| | Observations | 74,451 | 74,451 | 74,451 | 24,729 |
| Bath Soap | Coefficient (Std.) | 0.0099 (0.0128) | 0.0059 (0.013) | 0.0058 (0.0126) | -0.0267 (0.0256) |
| | Observations | 6,650 | 6,650 | 6,650 | 1,466 |
| Bathroom Tissues | Coefficient (Std.) | 0.0479*** (0.0087) | 0.0203** (0.0084) | 0.02** (0.0083) | 0.0349*** (0.0098) |
| | Observations | 56,458 | 56,458 | 56,458 | 19,285 |
| Beer | Coefficient (Std.) | 0.002*** (0.0006) | 0.0043*** (0.0007) | 0.0043*** (0.0007) | 0.018*** (0.0055) |
| | Observations | 187,691 | 187,691 | 187,691 | 12,080 |
| Bottled Juice | Coefficient (Std.) | 0.0235*** (0.0079) | 0.0187*** (0.007) | 0.0188*** (0.007) | 0.0333*** (0.0091) |
| | Observations | 224,857 | 224,857 | 224,857 | 60,015 |
| Canned Soup | Coefficient (Std.) | -0.0023 (0.0091) | -0.004 (0.0088) | -0.0012 (0.0086) | 0.0129* (0.0072) |
| | Observations | 233,779 | 233,779 | 233,779 | 95,310 |
| Canned Tuna | Coefficient (Std.) | 0.0092 (0.0065) | -0.0006 (0.0057) | -0.0008 (0.0057) | 0.0128 (0.0083) |
| | Observations | 112,629 | 112,629 | 112,629 | 31,922 |
| Cereals | Coefficient (Std.) | 0.0051 (0.0065) | 0.0049 (0.0063) | 0.005 (0.0063) | 0.0221*** (0.0071) |
| | Observations | 141,087 | 141,087 | 141,087 | 72,789 |
| Cheese | Coefficient (Std.) | 0.0069* (0.0038) | 0.0066** (0.0033) | 0.0063* (0.0033) | 0.0124*** (0.0046) |
| | Observations | 357,679 | 357,679 | 357,679 | 92,758 |
| Cigarettes | Coefficient (Std.) | 0.0044 (0.0051) | 0.0021 (0.0049) | 0.0022 (0.0048) | 0 (0.0055) |
| | Observations | 24,553 | 24,553 | 24,553 | 20,692 |
| Cookies | Coefficient (Std.) | 0.0084*** (0.0019) | 0.0074*** (0.002) | 0.0073*** (0.0019) | 0.0063* (0.0036) |
| | Observations | 317,932 | 317,932 | 317,932 | 66,087 |
| Crackers | Coefficient (Std.) | 0.0009 (0.0033) | 0.0007 (0.0032) | 0.001 (0.0032) | 0.0114* (0.0066) |
| | Observations | 115,658 | 115,658 | 115,658 | 24,771 |
| Dish Detergent | Coefficient (Std.) | 0.0295*** (0.0068) | 0.0241*** (0.006) | 0.0244*** (0.0058) | 0.0261*** (0.0058) |
| | Observations | 85,222 | 85,222 | 85,222 | 26,735 |
| Fabric Softener | Coefficient (Std.) | 0.0147** (0.0069) | 0.0028 (0.0057) | 0.0033 (0.0057) | 0.0233*** (0.0078) |
| | Observations | 85,337 | 85,337 | 85,337 | 27,488 |
| Front-End-Candies | Coefficient (Std.) | -0.004 (0.0043) | -0.0044 (0.0034) | -0.0043 (0.0034) | -0.0005 (0.0032) |
| | Observations | 148,200 | 148,200 | 148,200 | 77,323 |
| Frozen Dinners | Coefficient (Std.) | 0.049*** (0.007) | 0.0414*** (0.0062) | 0.0431*** (0.0062) | 0.0751*** (0.0104) |
| | Observations | 52,893 | 52,893 | 52,893 | 12,287 |

Table E1. (Cont.)

| Category | | **(1)** | **(2)** | **(3)** | **(4)** |
|---|---|---|---|---|---|
| Frozen Entrees | Coefficient (Std.) | 0.0192*** (0.0027) | 0.0186*** (0.0025) | 0.0188*** (0.0025) | 0.0247*** (0.0039) |
| | Observations | 345,223 | 345,223 | 345,223 | 117,044 |
| Frozen Juices | Coefficient (Std.) | 0.0134* (0.0073) | 0.0113* (0.0068) | 0.0124* (0.0066) | 0.0213** (0.0087) |
| | Observations | 118,582 | 118,582 | 118,582 | 40,517 |
| Grooming Products | Coefficient (Std.) | 0.0097*** (0.0033) | 0.0111*** (0.0033) | 0.0113*** (0.0033) | 0.0166 (0.011) |
| | Observations | 101,944 | 101,944 | 101,944 | 22,102 |
| Laundry Detergents | Coefficient (Std.) | 0.0213*** (0.0047) | 0.0128*** (0.0039) | 0.0131*** (0.0039) | 0.0175*** (0.0057) |
| | Observations | 121,566 | 121,566 | 121,566 | 42,121 |
| Oatmeal | Coefficient (Std.) | 0.0154 (0.0124) | 0.0086 (0.0121) | 0.0067 (0.0115) | 0.059*** (0.0115) |
| | Observations | 25,523 | 25,523 | 25,523 | 13,605 |
| Paper Towels | Coefficient (Std.) | 0.0275*** (0.0156) | 0.021 (0.0178) | 0.0225 (0.0177) | 0.0325** (0.016) |
| | Observations | 48,199 | 48,199 | 48,199 | 9,243 |
| Refrigerated Juices | Coefficient (Std.) | 0.0136* (0.0077) | 0.0062 (0.0079) | 0.0063 (0.0077) | 0.0253** (0.0121) |
| | Observations | 108,965 | 108,965 | 108,965 | 23,705 |
| Shampoos | Coefficient (Std.) | 0.0098*** (0.0025) | 0.0104*** (0.0025) | 0.0104*** (0.0025) | 0.0232*** (0.008) |
| | Observations | 88,193 | 88,193 | 88,193 | 16,099 |
| Snack Crackers | Coefficient (Std.) | 0.0013 (0.0029) | 0.0033 (0.0028) | 0.0033 (0.0028) | 0.0153*** (0.0052) |
| | Observations | 176,527 | 176,527 | 176,527 | 38,123 |
| Soaps | Coefficient (Std.) | 0.0237*** (0.0088) | 0.013 (0.0082) | 0.0165** (0.0081) | 0.0509*** (0.0117) |
| | Observations | 56,725 | 56,725 | 56,725 | 16,882 |
| Soft Drinks | Coefficient (Std.) | 0.0087*** (0.0022) | 0.013*** (0.0018) | 0.0121*** (0.0018) | 0.0078** (0.0034) |
| | Observations | 243,837 | 243,837 | 243,837 | 49,989 |
| Toothbrushes | Coefficient (Std.) | 0.013*** (0.0046) | 0.0098** (0.0047) | 0.0091** (0.0046) | 0.019* (0.0102) |
| | Observations | 52,185 | 52,185 | 52,185 | 13,695 |
| Toothpastes | Coefficient (Std.) | 0.0007 (0.0039) | 0.0001 (0.0038) | 0.0003 (0.0038) | 0.0052 (0.0082) |
| | Observations | 100,845 | 100,845 | 100,845 | 28,039 |
| **Average coefficients** | | **0.0138** | **0.0097** | **0.0100** | **0.0204** |

Notes: The table reports the results of category-level fixed effect regressions of the probability of a small price change. The dependent variable is "small price change," which equals 1 if a price change of product $i$ in store $s$ at time $t$ is less or equal to 10¢, and 0 otherwise. The main independent variable is the log of the average sales volume of product $i$ in store $s$ over the 52 weeks preceding time $t$. Column 1 reports the results of the baseline regression that includes only the log of the average sales volume and the fixed effects for months, years, stores, and products. In column 2, we add the following controls: the log of the average price, the log of the absolute change in the wholesale price, a control for sale- and bounce-back prices, which we identify using a sales filter algorithm, and the pricing zone of the store. In column 3, we add a dummy for 9-ending prices as an additional control. In column 4, we focus on regular prices by excluding the sale- and bounce-back prices. We estimate separate regressions for each product category, clustering the errors by product. * $p < 10\%$, ** $p < 5\%$, *** $p < 1\%$.

**Appendix F. Robustness of the Product-level regressions of the % of small price changes and sales volume**

In the paper, we study the correlation between small price changes and sales volume at the product level using regressions that have only the sales volume as the independent variable. This can raise concerns that the results may be driven by differences between the stores rather than by differences in the sales volume.

To mitigate this concern, we augment the data with demographic information about consumers living in the neighborhood of each store, including their median income, the share of minorities, and the share of unemployed. To control for local competition, we also add a control for the pricing zone of each store, using pricing zone indicators included in Dominick's data.

We estimate for each product in each category an OLS regression with robust standard errors. The dependent variable is the share of small price changes for the product in each store. The independent variable is the average sales volume of the product in each store, the median income, the share of minorities, the share of unemployed, and the stores' pricing zone.

As we do in the paper, we use observations on price changes only if we observe the price in both weeks $t$ and $t + 1$ and the post change price remained unchanged for at least 2 weeks. The estimation results are summarized in Table F1. Column 1 presents for each product category, the average of the estimated coefficients. Column 2 presents the total number of coefficients. Column 3 presents the percentage of the positive coefficients. Column 4 presents the number of statistically significant coefficients. Column 5 presents the percentage of positive and statistically significant coefficients out of the total number of statistically significant coefficients.

According to the figures in the table, the average coefficients are positive in 28 of the 29 product categories. The only exception is in the highly regulated cigarettes category. Further, in all categories, the number of positive coefficients far exceeds the number of negative coefficients. On average, 72.19% of all the coefficients are positive.

Focusing on statistically significant coefficients, we find a far greater number of positive coefficients that are significant than negative coefficients that are significant. On average, 87.71% of all the statistically significant coefficients are positive. In other words, for the overwhelming majority of the individual products in our sample, we find a positive relationship between sales volume and the share of small price changes.

As another test, we estimate linear probability model (LPM) regressions with robust standard errors, instead of regressions at the store level. In other words, we estimate:

$$small\ price\ change_{i,s,t} = \alpha + \beta ln\big(average\ sales\ volume_{i,s}\big) + \gamma X_{i,s,t} + u_{i,s,t} \qquad \text{(F1)}$$

where small price change is a dummy that equals 1 if a price change of product $i$ in store $s$ in week $t$ is less or equal to 10¢, and 0 otherwise. The average sales volume is the average sales volume of product $i$ in store $s$ over the sample period.[1] $\mathbf{X}$ is a matrix of other control variables.

We estimate a separate regression for each product in each category, conditional on it having at least 30 price changes and at least 1 small price change over the sample period.[2] Table F2 reports the estimation results of regressions in which the $\mathbf{X}$ matrix is empty. We find that in all but the toothpaste category, the average coefficient is positive. Furthermore, on average, 72.19% of the coefficients are positive. When we focus on the positive and statistically significant coefficients, we find that, on average, 91.96% of the coefficients are positive.

Table F3 reports the estimation results of regressions in which the $\mathbf{X}$ matrix includes the following independent variables: the log of the average price, the log of the absolute change in the wholesale price, a dummy for sale- and bounce-back prices, which we identify using the sales filter algorithm of Fox and Syed (2016), and a dummy for 9-ending prices.

We find that 22 of the 28 average coefficients are positive. When we focus on all the coefficients, we find that on average, 67.61% of the coefficients in a category are positive. When we focus on positive and statistically significant coefficients, we find that,

---

[1] In calculating the average sales volume, we need to account for missing observations, because a missing observation in week $t$ implies that the product was either out of stock or had 0 sales on that week. Thus, averaging over the available observations can lead to an upward bias for products that are sold in small numbers. Therefore, for each product in each store, we calculate the average by first determining the total number of units sold over all available observations. We then identify the first and last week for which we have observations, and calculate the average for each product-store as $\frac{total\ no.\ of\ units\ sold}{last\ week - first\ week}$. The resulting figure is smaller than we would obtain if we averaged over all available observations (which would not include observations on weeks with 0 sales).

[2] Prices and price changes in the cigarettes category were heavily regulated during the sample period. Consequently, we have no product-store combination for which we have 30 or more price changes over the sample period in the cigarettes' category.

on average, 89.57% of the coefficients are positive.

We therefore conclude that changing the estimation method does not change the conclusions we report in the paper. There is a positive correlation at the product level between the likelihood of a small price change and the sales volume.

Table F1. Product-level regressions of the % of small price changes and sales volume by categories, including controls

| Product Category | Average coefficient (1) | No. of coefficients (2) | % positive coefficients (3) | No. of significant coefficients (4) | % positive and significant coefficients (5) |
|---|---|---|---|---|---|
| Analgesics | 0.025 | 212 | 74.53% | 48 | 97.92% |
| Bath Soaps | 0.036 | 33 | 72.73% | 8 | 100.00% |
| Bathroom tissues | 0.045 | 100 | 75.00% | 37 | 89.19% |
| Beers | 0.018 | 202 | 89.11% | 71 | 98.59% |
| Bottled juices | 0.042 | 370 | 73.24% | 115 | 86.96% |
| Canned soups | 0.032 | 348 | 70.98% | 112 | 86.61% |
| Canned tuna | 0.025 | 181 | 59.67% | 61 | 70.49% |
| Cereals | 0.039 | 345 | 68.41% | 110 | 81.82% |
| Cheese | 0.033 | 474 | 70.68% | 151 | 92.72% |
| Cigarettes | -0.016 | 106 | 50.94% | 9 | 0.00% |
| Cookies | 0.034 | 666 | 74.77% | 213 | 94.37% |
| Crackers | 0.040 | 212 | 79.72% | 72 | 98.61% |
| Dish detergents | 0.032 | 199 | 69.85% | 46 | 91.30% |
| Fabric softeners | 0.029 | 226 | 70.80% | 52 | 94.23% |
| Front end candies | 0.033 | 274 | 63.50% | 56 | 91.07% |
| Frozen dinners | 0.056 | 215 | 84.19% | 77 | 96.10% |
| Frozen entrees | 0.044 | 671 | 82.41% | 270 | 97.78% |
| Frozen juices | 0.045 | 142 | 79.58% | 57 | 96.49% |
| Grooming products | 0.008 | 528 | 68.75% | 89 | 94.38% |
| Laundry detergents | 0.011 | 406 | 63.30% | 74 | 77.03% |
| Oatmeal | 0.042 | 69 | 71.01% | 13 | 92.31% |
| Paper towels | 0.045 | 90 | 73.33% | 37 | 81.08% |
| Refrigerated juices | 0.031 | 176 | 68.75% | 63 | 82.54% |
| Shampoos | 0.019 | 608 | 70.39% | 97 | 97.94% |
| Snack crackers | 0.043 | 282 | 77.66% | 89 | 95.51% |
| Soaps | 0.032 | 216 | 68.52% | 43 | 81.40% |
| Soft drinks | 0.030 | 897 | 76.25% | 285 | 96.84% |
| Toothbrushes | 0.024 | 202 | 80.20% | 45 | 93.33% |
| Toothpastes | 0.013 | 336 | 65.18% | 61 | 86.89% |
| **Average** | 0.031 | 303 | 72.19% | 85 | 87.71% |

<u>Notes</u>: Results of product-level regression. The dependent variable in all regressions is the % of small price changes at each store. For each product category, column 1 presents the average estimated coefficients of the average sales volumes. The regressions also include controls for the median income, the share of ethnic minorities, the unemployment rate, and the pricing zone of the store. Column 2 presents the total number of coefficients. Column 3 presents the % of positive coefficients out of all coefficients. Column 4 presents the total number of coefficients that are statistically significant at the 5% level. Column 5 presents the % of coefficients that are positive and statistically significant, at the 5% level.

Table F2. Product-level regressions of the % of small price changes and sales volume by categories, using LPM

| Product Category | Average coefficient (1) | No. of coefficients (2) | % positive coefficients (3) | No. of significant coefficients (4) | % positive and significant coefficients (5) |
|---|---|---|---|---|---|
| Analgesics | 0.055 | 24 | 70.83% | 2 | 100.00% |
| Bath Soaps | 0.290 | 1 | 100.00% | 0 | – |
| Bathroom tissues | 0.032 | 23 | 86.96% | 15 | 86.67% |
| Beers | 0.007 | 68 | 72.06% | 17 | 100.00% |
| Bottled juices | 0.044 | 98 | 76.53% | 43 | 90.70% |
| Canned soups | 0.049 | 100 | 81.00% | 40 | 87.50% |
| Canned tuna | 0.012 | 37 | 56.76% | 14 | 42.86% |
| Cereals | 0.060 | 59 | 76.27% | 28 | 92.86% |
| Cheese | 0.027 | 161 | 83.23% | 64 | 95.31% |
| Cigarettes | – | 0 | | 0 | – |
| Cookies | 0.047 | 109 | 79.82% | 34 | 100.00% |
| Crackers | 0.236 | 51 | 88.24% | 24 | 100.00% |
| Dish detergents | 0.024 | 30 | 70.00% | 15 | 93.33% |
| Fabric softeners | 0.037 | 21 | 80.95% | 9 | 100.00% |
| Front end candies | 0.121 | 41 | 95.12% | 29 | 100.00% |
| Frozen dinners | 0.347 | 32 | 93.75% | 14 | 100.00% |
| Frozen entrees | 0.040 | 177 | 87.57% | 97 | 97.94% |
| Frozen juices | 0.039 | 66 | 89.39% | 37 | 91.89% |
| Grooming products | 0.009 | 30 | 53.33% | 1 | 100.00% |
| Laundry detergents | 0.016 | 18 | 61.11% | 2 | 100.00% |
| Oatmeal | 0.298 | 15 | 80.00% | 0 | – |
| Paper towels | 0.012 | 21 | 66.67% | 10 | 60.00% |
| Refrigerated juices | 0.007 | 57 | 56.14% | 29 | 68.97% |
| Shampoos | 0.075 | 11 | 54.55% | 0 | – |
| Snack crackers | 0.036 | 76 | 88.16% | 37 | 100.00% |
| Soaps | 0.027 | 17 | 64.71% | 0 | – |
| Soft drinks | 0.031 | 285 | 72.63% | 103 | 99.03% |
| Toothbrushes | 0.034 | 22 | 63.64% | 1 | 100.00% |
| Toothpastes | -0.245 | 51 | 72.55% | 10 | 100.00% |
| **Average** | **0.063** | **59** | **75.78%** | **23** | **91.96%** |

Notes: The table reports the estimation results of product-level LPM regressions. The dependent variable in all regressions is a dummy for price small price changes ($\Delta p < 10 \cent$). The main independent variable is the log of the sales volume. For each product category, column 1 presents the average estimated coefficients of the average sales volumes. Column 2 presents the total number of coefficients. Column 3 presents the % of positive coefficients out of all coefficients. Column 4 presents the total number of coefficients that are statistically significant at the 5% level. Column 5 presents the % of coefficients that are positive and statistically significant, at the 5% level.

Table F3. Product-level regressions of the % of small price changes and sales volume by categories, using LPM with extra controls

| Product Category | Average coefficient (1) | No. of coefficients (2) | % positive coefficients (3) | No. of significant coefficients (4) | % positive and significant coefficients (5) |
|---|---|---|---|---|---|
| Analgesics | 0.078 | 24 | 62.50% | 1 | 100.00% |
| Bath Soaps | -0.044 | 1 | 0.00% | 0 | – |
| Bathroom tissues | -0.010 | 23 | 69.57% | 5 | 80.00% |
| Beers | -0.001 | 69 | 57.97% | 8 | 100.00% |
| Bottled juices | 0.019 | 98 | 71.43% | 14 | 92.86% |
| Canned soups | 0.025 | 100 | 84.00% | 13 | 100.00% |
| Canned tuna | 0.004 | 37 | 64.86% | 5 | 40.00% |
| Cereals | 0.052 | 59 | 77.97% | 18 | 100.00% |
| Cheese | 0.008 | 161 | 73.91% | 38 | 92.11% |
| Cigarettes | – | 0 | – | 0 | – |
| Cookies | -0.060 | 109 | 79.82% | 25 | 100.00% |
| Crackers | 0.071 | 50 | 88.00% | 20 | 100.00% |
| Dish detergents | 0.025 | 30 | 83.33% | 9 | 100.00% |
| Fabric softeners | 0.090 | 21 | 100.00% | 9 | 100.00% |
| Front end candies | 0.015 | 41 | 63.41% | 9 | 100.00% |
| Frozen dinners | 0.205 | 32 | 78.13% | 6 | 100.00% |
| Frozen entrees | 0.023 | 177 | 80.23% | 43 | 93.02% |
| Frozen juices | 0.023 | 66 | 74.24% | 22 | 90.91% |
| Grooming products | -0.016 | 30 | 46.67% | 1 | 100.00% |
| Laundry detergents | 0.009 | 18 | 61.11% | 2 | 100.00% |
| Oatmeal | 0.363 | 15 | 66.67% | 0 | – |
| Paper towels | -0.015 | 21 | 33.33% | 6 | 16.67% |
| Refrigerated juices | 0.004 | 57 | 66.67% | 27 | 70.37% |
| Shampoos | 0.074 | 11 | 63.64% | 0 | – |
| Snack crackers | 0.044 | 76 | 85.53% | 33 | 100.00% |
| Soaps | 0.065 | 17 | 64.71% | 3 | 66.67% |
| Soft drinks | 0.185 | 286 | 66.78% | 60 | 96.67% |
| Toothbrushes | 0.012 | 22 | 54.55% | 1 | 100.00% |
| Toothpastes | 0.053 | 50 | 74.00% | 13 | 100.00% |
| **Average** | **0.046** | **59** | **67.61%** | **13** | **89.57%** |

## Appendix G. Robustness: sales volume, revenue, and small price changes

In the paper, we estimate category-level regressions of small price changes where the main independent variables are sales volume and revenue. In this appendix, we conduct two sets of robustness tests. First, we add further controls and estimate the category-level regressions again. Second, because the correlation between sales volume and revenue at the category level is high, the results of category-level regressions could be suspect. We, therefore, pool the data from all categories together and re-estimate the regressions using the pooled data.

In Tables G1 and G2, we present the results of the category-level regression estimations. The regressions we estimate are of the following form:

$$small\ price\ change_{i,s,t} = \alpha + \beta_1 ln\big(average\ sales\ volume_{i,s}\big) + \qquad \text{(G1)}$$

$$\beta_2 ln\big(average\ revenue_{i,s}\big) + \gamma \boldsymbol{X}_{i,s,t} + month_t + year_t + \delta_s + \mu_i + u_{i,s,t}$$

where *small price change* is a dummy that equals 1 if a price change of product *i* in store *s* at time *t* is less or equal to 10¢, and 0 otherwise. The *average sales volume* is the average sales volume of product *i* in store *s* over the sample period. The *average revenue* is the average revenue of product *i* in store *s* over the sample period. $\boldsymbol{X}$ is a matrix of other control variables. *Month* and *year* are fixed effects for the month and the year of the price change. $\delta$ and $\mu$ are fixed effects for stores and products, respectively, and *u* is an i.i.d error term. We estimate separate regressions for each product category, clustering the errors by product. As we do in the paper, we use observations on price changes only if we observe the price in both week *t* and *t*+1 and the post change price remained unchanged for at least 2 weeks.

The coefficient columns in the sales volume and the revenue panels of Table G1 give the coefficients of sales volume and revenue, respectively in a regression that also includes percentage changes in the wholesale price and a dummy for sale- and bounce-back prices as control variables.[3] This does not change the results we report in the paper. 22 of the sales volume coefficients are positive. 14 of the coefficients are statistically significant. 4 of the negative coefficients are statistically significant. Of the revenue coefficients, 21 are negative, and all of them are statistically significant.

---

[3] We do not add the log of the average price because the log of the price plus the log of the sales volume equals the log of the revenue, leading to a perfect multicollinearity.

The coefficient columns in the sales volume and the revenue panels of Table G2, present the coefficients of sales volume and revenue, respectively in a regression in which we also include a dummy for 9-ending prices. We find that 22 of the sales volume coefficients are statistically significant. Of the 22 positive coefficients, 14 are statistically significant. Of the revenue coefficients, 18 of the coefficients are negative, all of which are statistically significant.

Thus, including more controls does not change the conclusions we derive in the paper. The revenue seems to be correlated to small price changes mostly through the sales volume. The effect of the price, holding sales volume constant seems to be mostly negative.

However, the results at the category level are suspect because of the strong correlation between sales volume and revenue. In the paper, we show that the average correlation at the category level is 0.85. The high correlation at the category level is due to the relatively low within-category variation in prices. To attenuate this concern, we pool the data from all categories together. Since the between-categories variation in prices is higher than the within-category variation, we find that in the pooled data, the correlation between sales volume and revenue is 0.70.

Table G3 presents the results of regressions similar to G1, to which we also add fixed effects for the categories. Column 1 gives the results of a regression that includes only the sales volume and the revenue as independent variables. The coefficient of the sales volume, 0.42, is positive and statistically significant, whereas the coefficient of the revenue, −0.40, is negative and statistically significant.

In column 2, we add controls for percentage changes in the wholesale price, and for sale- and bounce-back prices. The coefficient of the sales volume, 0.38, is positive and statistically significant, whereas the coefficient of the revenue, −0.36, is negative and statistically significant.

In column 3, we also add a dummy for 9-ending prices. The coefficient of the sales volume, 0.02, remains positive and statistically significant, whereas the coefficient of the revenue, −0.61, remains negative and statistically significant.

Finally, in column 4, we remove sale prices and focus on regular prices. The coefficient of sales volume, 0.03, is positive and significant. The coefficient of revenue, −0.53, is negative and statistically significant.

Thus, also when we estimate the regressions using the pooled data, we find a positive correlation between sales volume and small price changes. We also find that when we hold the sales volume constant, the correlation between revenue and small price changes is negative.

As an alternative test of the role of revenue, we redefine the average revenue as the product of the average sales volume and the average price. Both are defined the same way as in the paper. We then estimate

$$small\ price\ change_{i,s,t} = \alpha + \beta_1 ln\big(averag\overline{e\ reve}nue_{i,s}\big) + \gamma \boldsymbol{X}_{i,s,t} + month_t +$$
$$year_t + \delta_s + \mu_i + u_{i,s,t} \tag{G2}$$

where $average\ revenue_{i,s}$ is the product of the average sales volume and the average price of product $i$ offered at store $s$, and the other variables are defined as above. As above, we estimate a series of category-level regressions.

Table G4 gives the estimation results. In column 1, the only control variables are the log of the average sales volume, and the dummies for months, years, stores, and products. We find that 28 of the 29 coefficients of the log of the average revenue are positive. All the positive coefficients are statistically significant. The average coefficient is 0.017, suggesting that a 1% increase in the average revenue is associated with a 1.7% increase in the likelihood of a small price change.

In column 2, we add controls for the log of the average price, the log of the absolute change in the wholesale price, and a control for sale- and bounce-back prices (which we identify using the sales filter algorithm of Fox and Syed 2016). We find that all 29 coefficients are positive. 28 of the positive coefficients are statistically significant, and one more is marginally significant. The average coefficient is 0.015, suggesting that a 1% increase in the average revenue is associated with a 1.5% increase in the likelihood of a small price change.

In column 3, we add a control for 9-ending prices. We find that all 29 coefficients are positive. 28 of the positive coefficients are statistically significant, and one more is marginally significant. The average coefficient is 0.015, suggesting that a 1% increase in the average revenue is associated with a 1.5% increase in the likelihood of a small price change.

As a further control for the effects of sales on the estimation results, in column 4 we focus on regular prices by excluding all sale- and bounce-back prices. When we focus on regular prices, we find that all 29 coefficients are positive and statistically significant. The average coefficient is 0.029, suggesting that a 1% increase in the average revenue is associated with a 2.9% increase in the likelihood of a small price change.

Thus, the finding of the positive correlation between revenue and the likelihood of small price changes is robust. However, our previous results suggest that this correlation holds because of the sales volume component rather than the price component of the revenue. Indeed, we also include the average price as a control variable in this regression. If the correlation between the likelihood of a small price change and revenue were to work mainly through the price component of the revenue, then we would expect that the coefficient of the average price in columns 2–4 would be positive and statistically significant, while the average revenue coefficient would be close to 0 and statistically insignificant.

Table G1. Regressions with sales volume and revenue, with extra controls

| Category | Sales Volume | | Revenue | | No. of |
| | Coefficient | Std. | Coefficient | Std. | Observations |
|---|---|---|---|---|---|
| Analgesics | 0.2405*** | 0.0464 | -0.223*** | 0.0467 | 144,461 |
| Bath Soap | -0.3004** | 0.1098 | 0.3289*** | 0.1099 | 15,295 |
| Bathroom Tissues | 0.4332*** | 0.1270 | -0.4187*** | 0.1258 | 149,441 |
| Beer | -0.0613** | 0.0258 | 0.0764*** | 0.0257 | 290,620 |
| Bottled Juice | 0.69*** | 0.0983 | -0.6651*** | 0.0982 | 496,557 |
| Canned Soup | 0.1141 | 0.0688 | -0.1027 | 0.0690 | 495,543 |
| Canned Tuna | 0.4883*** | 0.1066 | -0.4736*** | 0.1071 | 213,043 |
| Cereals | 0.1056** | 0.0400 | -0.0901** | 0.0406 | 357,120 |
| Cheese | 0.3698*** | 0.1062 | -0.3583*** | 0.1067 | 796,150 |
| Cigarettes | -0.3952*** | 0.0589 | 0.405*** | 0.0588 | 36,157 |
| Cookies | -0.0207 | 0.0222 | 0.0431* | 0.0224 | 688,761 |
| Crackers | 0.1497*** | 0.0523 | -0.1203*** | 0.0527 | 245,185 |
| Dish Detergent | 0.4543*** | 0.1332 | -0.4291*** | 0.1329 | 189,633 |
| Fabric Softener | 0.8047*** | 0.1924 | -0.7907*** | 0.1932 | 181,056 |
| Front-End-Candies | 0.3188*** | 0.0481 | -0.3113*** | 0.0490 | 278,853 |
| Frozen Dinners | 0.0474 | 0.0460 | -0.0106 | 0.0456 | 203,191 |
| Frozen Entrees | -0.0026 | 0.0174 | 0.0278 | 0.0170 | 864,832 |
| Frozen Juices | 0.0946 | 0.0527 | -0.075 | 0.0534 | 308,817 |
| Grooming Products | 0.0361 | 0.0335 | -0.0155 | 0.0341 | 269,873 |
| Laundry Detergents | 0.251*** | 0.0537 | -0.2413*** | 0.0532 | 272,765 |
| Oatmeal | -0.0251 | 0.0212 | 0.0415* | 0.0230 | 79,983 |
| Paper Towels | 0.711*** | 0.1851 | -0.6752*** | 0.1856 | 116,204 |
| Refrigerated Juices | 0.0277 | 0.0606 | -0.0063 | 0.0616 | 306,865 |
| Shampoos | 0.0178 | 0.0159 | 0.0017 | 0.0158 | 261,778 |
| Snack Crackers | 0.0477 | 0.0817 | -0.0198 | 0.0826 | 398,665 |
| Soap | 0.4612*** | 0.1564 | -0.4379*** | 0.1567 | 152,379 |
| Soft Drinks | 0.4933*** | 0.0442 | -0.4694*** | 0.0435 | 1,350,618 |
| Toothbrushes | 0.0416 | 0.0401 | -0.0215 | 0.0401 | 125,380 |
| Toothpastes | -0.0764** | 0.0344 | 0.0894*** | 0.0345 | 264,317 |
| **Average** | **0.1902** | **0.0717** | **-0.1704** | **7190.0** | **329,432** |

Notes: The table reports the results of category-level fixed effect regressions of the probability of a small price change. The dependent variable is "small price change," which equals 1 if a price change of product $i$ in store $s$ at time $t$ is less or equal to 10¢, and 0 otherwise. The main independent variables are the log of the average sales volume of product $i$ in store $s$ over the sample period and the log of the average revenue of product $i$ in store $s$ over the sample period. The regressions also include the following independent variables: percentage changes in the wholesale price and a dummy for sale and bounce-back prices, as well as fixed effects for years, months, stores, and products. We estimate separate regressions for each product category, clustering the errors by product. * $p < 10\%$, ** $p < 5\%$, *** $p < 1\%$

Table G2. Regressions with sales volume and revenue, with extra controls, including a control for 9-ending prices

| Category | Sales Volume | | Revenue | | No. of Observations |
|----------|-------------|------|---------|------|---------------------|
| | Coefficient | Std. | Coefficient | Std. | |
| Analgesics | 0.2366*** | 0.0456 | -0.2193*** | 0.0459 | 144,461 |
| Bath Soap | -0.3045*** | 0.1103 | 0.3337*** | 0.1103 | 15,295 |
| Bathroom Tissues | 0.4045*** | 0.1253 | -0.3894*** | 0.1241 | 149,441 |
| Beer | -0.0609** | 0.0257 | 0.0759*** | 0.0256 | 290,620 |
| Bottled Juice | 0.682*** | 0.0931 | -0.6576*** | 0.0930 | 496,557 |
| Canned Soup | 0.0983 | 0.0699 | -0.0842 | 0.0701 | 495,543 |
| Canned Tuna | 0.482*** | 0.1057 | -0.4676*** | 0.1062 | 213,043 |
| Cereals | 0.104*** | 0.0400 | -0.0885** | 0.0406 | 357,120 |
| Cheese | 0.3678*** | 0.1061 | -0.3564*** | 0.1066 | 796,150 |
| Cigarettes | -0.3872*** | 0.0587 | 0.397*** | 0.0586 | 36,157 |
| Cookies | -0.0189 | 0.0218 | 0.0416* | 0.0220 | 688,761 |
| Crackers | 0.1437*** | 0.0520 | -0.1139** | 0.0524 | 245,185 |
| Dish Detergent | 0.4572*** | 0.1321 | -0.4321*** | 0.1318 | 189,633 |
| Fabric Softener | 0.8009*** | 0.1920 | -0.7866*** | 0.1929 | 181,056 |
| Front-End-Candies | 0.3019*** | 0.0483 | -0.2942*** | 0.0491 | 278,853 |
| Frozen Dinners | 0.0337 | 0.0450 | 0.0061 | 0.0444 | 203,191 |
| Frozen Entrees | -0.0045 | 0.0175 | 0.0305* | 0.0171 | 864,832 |
| Frozen Juices | 0.0847 | 0.0533 | -0.0644 | 0.0539 | 308,817 |
| Grooming Products | 0.0299 | 0.0333 | -0.0091 | 0.0339 | 269,873 |
| Laundry Detergents | 0.25*** | 0.0534 | -0.2398*** | 0.0530 | 272,765 |
| Oatmeal | -0.0234 | 0.0207 | 0.04* | 0.0226 | 79,983 |
| Paper Towels | 0.7135*** | 0.1868 | -0.6772*** | 0.1873 | 116,204 |
| Refrigerated Juices | 0.0209 | 0.0595 | 0.0003 | 0.0604 | 306,865 |
| Shampoos | 0.0152 | 0.0159 | 0.0043 | 0.0159 | 261,778 |
| Snack Crackers | 0.0423 | 0.0820 | -0.0142 | 0.0829 | 398,665 |
| Soap | 0.4744*** | 0.1522 | -0.4501*** | 0.1527 | 152,379 |
| Soft Drinks | 0.4319*** | 0.0332 | -0.409*** | 0.0327 | 1,350,618 |
| Toothbrushes | 0.0002 | 0.0398 | 0.0197 | 0.0398 | 125,380 |
| Toothpastes | -0.0752** | 0.0344 | 0.0882** | 0.0345 | 264,317 |
| **Average** | **0.1828** | **0.0708** | **-0.1626** | **0.0710** | **329,432** |

Notes: The table reports the results of category-level fixed effect regressions of the probability of a small price change. The dependent variable is "small price change," which equals 1 if a price change of product $i$ in store $s$ at time $t$ is less or equal to 10¢, and 0 otherwise. The main independent variables are the log of the average sales volume of product $i$ in store $s$ over the sample period and the log of the average revenue of product $i$ in store $s$ over the sample period. The regressions also include the following independent variables: percentage changes in the wholesale price, a dummy for sale and bounce-back prices, and a dummy for 9-ending prices, as well as fixed effects for years, months, stores, and products. We estimate separate regressions for each product category, clustering the errors by product. * $p < 10\%$, ** $p < 5\%$, *** $p < 1\%$

Table G3. Regressions with sales volume and revenue, using a pooled dataset

|  | (1) | (2) | (3) | (4) |
|---|---|---|---|---|
| Log of sales | 0.42*** | 0.38*** | 0.02*** | 0.03*** |
| volume | (0.024) | (0.025) | (0.001) | (0.001) |
| Log of | −0.40*** | −0.36*** | −0.61*** | −0.53*** |
| revenue | (0.024) | (0.025) | (0.016) | (0.029) |
| Observations | 9,553,542 | 9,553,542 | 9,553,542 | 2,328,405 |

Notes: The table reports the results of pooled fixed effect regressions of the probability of a small price change. The dependent variable is "small price change," which equals 1 if a price change of product $i$ in store $s$ at time $t$ is less or equal to 10¢, and 0 otherwise. The main independent variables are the log of the average sales volume and the log of the revenue of product $i$ in store $s$ over the sample period. Column 1 reports the results of the baseline regression that includes only the log of the average sales volume, the log of the average revenue, and the fixed effects for months, years, categories, stores, and products. In column 2, we add the following controls: the log of the average price, the log of the absolute change in the wholesale price, a control for sale- and bounce-back prices (which we identify using a sales filter algorithm) and the competition zone of the store. In column 3, we add a dummy for 9-ending prices as an additional control. In column 4, we focus on regular prices by excluding the sale- and bounce-back prices. We estimate separate regressions for each product category, clustering the errors by product. * $p < 10\%$, ** $p < 5\%$, *** $p < 1\%$

Table G4. Regression with the average revenue constructed as the average sales volume times the average price

| Category | | **(1)** | **(2)** | **(3)** | **(4)** |
|---|---|---|---|---|---|
| Analgesics | Coefficient (Std.) | 0.0147*** (0.0023) | 0.0114*** (0.0021) | 0.0113*** (0.0021) | 0.0193*** (0.0045) |
| | Observations | 144,461 | 144,461 | 144,461 | 44,950 |
| Bath Soap | Coefficient (Std.) | 0.018*** (0.0049) | 0.017*** (0.0047) | 0.0174*** (0.0046) | 0.0579*** (0.0119) |
| | Observations | 15,295 | 15,295 | 15,295 | 3,208 |
| Bathroom Tissues | Coefficient (Std.) | 0.0256*** (0.0057) | 0.0134** (0.0056) | 0.0137*** (0.0056) | 0.0334*** (0.0082) |
| | Observations | 149,441 | 149,441 | 149,441 | 47,041 |
| Beer | Coefficient (Std.) | 0.0085*** (0.0009) | 0.0114*** (0.0008) | 0.0114*** (0.0008) | 0.047*** (0.0045) |
| | Observations | 290,620 | 290,620 | 290,620 | 27,348 |
| Bottled Juice | Coefficient (Std.) | 0.0207*** (0.0049) | 0.0172*** (0.0041) | 0.017*** (0.0042) | 0.0239*** (0.0061) |
| | Observations | 496,557 | 496,557 | 496,557 | 133,714 |
| Canned Soup | Coefficient (Std.) | 0.0117** (0.0048) | 0.01** (0.0044) | 0.0121*** (0.0043) | 0.0132*** (0.0045) |
| | Observations | 495,543 | 495,543 | 495,543 | 176,235 |
| Canned Tuna | Coefficient (Std.) | 0.0153*** (0.0042) | 0.0126*** (0.0039) | 0.0124*** (0.0038) | 0.0197*** (0.0048) |
| | Observations | 213,043 | 213,043 | 213,043 | 64,161 |
| Cereals | Coefficient (Std.) | 0.0162*** (0.0032) | 0.0134*** (0.003) | 0.0133*** (0.003) | 0.0158*** (0.0039) |
| | Observations | 357,120 | 357,120 | 357,120 | 155,367 |
| Cheese | Coefficient (Std.) | 0.0148*** (0.0025) | 0.0084*** (0.0023) | 0.0084*** (0.0023) | 0.0109*** (0.003) |
| | Observations | 796,150 | 796,150 | 796,150 | 224,889 |
| Cigarettes | Coefficient (Std.) | 0.0092** (0.0028) | 0.0095** (0.0028) | 0.0095** (0.0028) | 0.0084** (0.0034) |
| | Observations | 36,157 | 36,157 | 36,157 | 30,262 |
| Cookies | Coefficient (Std.) | 0.0208*** (0.0015) | 0.0178*** (0.0014) | 0.018*** (0.0014) | 0.0368*** (0.0029) |
| | Observations | 688,761 | 688,761 | 688,761 | 132,488 |
| Crackers | Coefficient (Std.) | 0.0291*** (0.0025) | 0.0229*** (0.0022) | 0.0232*** (0.0022) | 0.0366*** (0.0055) |
| | Observations | 245,185 | 245,185 | 245,185 | 50,029 |
| Dish Detergent | Coefficient (Std.) | 0.029*** (0.0036) | 0.0213*** (0.0032) | 0.0212*** (0.0031) | 0.0277*** (0.0037) |
| | Observations | 189,633 | 189,633 | 189,633 | 53,289 |
| Fabric Softener | Coefficient (Std.) | 0.0124*** (0.0037) | 0.0088*** (0.0034) | 0.0089*** (0.0034) | 0.0258*** (0.0044) |
| | Observations | 181,056 | 181,056 | 181,056 | 56,234 |
| Front-End-Candies | Coefficient (Std.) | -0.0016 (0.0033) | 0.0045* (0.0026) | 0.0048*** (0.0026) | 0.0088*** (0.0026) |
| | Observations | 278,853 | 278,853 | 278,853 | 111,635 |
| Frozen Dinners | Coefficient (Std.) | 0.0344*** (0.0028) | 0.0288*** (0.0023) | 0.0308*** (0.0023) | 0.0597*** (0.0053) |
| | Observations | 203,191 | 203,191 | 203,191 | 37,527 |

Table G4. (Cont.)

| Category | | (1) | (2) | (3) | (4) |
|---|---|---|---|---|---|
| Frozen Entrees | Coefficient (Std.) | 0.0187*** (0.0016) | 0.0187*** (0.0015) | 0.0193*** (0.0015) | 0.0361*** (0.0026) |
| | Observations | 864,832 | 864,832 | 864,832 | 213,545 |
| Frozen Juices | Coefficient (Std.) | 0.0203*** (0.0041) | 0.0156*** (0.0037) | 0.0162*** (0.0035) | 0.0269*** (0.0055) |
| | Observations | 308,817 | 308,817 | 308,817 | 87,919 |
| Grooming Products | Coefficient (Std.) | 0.0105*** (0.0015) | 0.0134*** (0.0016) | 0.0135*** (0.0016) | 0.026*** (0.0046) |
| | Observations | 269,873 | 269,873 | 269,873 | 51,819 |
| Laundry Detergents | Coefficient (Std.) | 0.0125*** (0.0024) | 0.008*** (0.0023) | 0.0082*** (0.0023) | 0.0173*** (0.0037) |
| | Observations | 272,765 | 272,765 | 272,765 | 85,184 |
| Oatmeal | Coefficient (Std.) | 0.0238*** (0.0066) | 0.0127** (0.0058) | 0.0129*** (0.0058) | 0.0284*** (0.0082) |
| | Observations | 79,983 | 79,983 | 79,983 | 36,043 |
| Paper Towels | Coefficient (Std.) | 0.025** (0.0081) | 0.0251*** (0.0082) | 0.0254*** (0.0083) | 0.0353*** (0.0081) |
| | Observations | 116,204 | 116,204 | 116,204 | 29,280 |
| Refrigerated Juices | Coefficient (Std.) | 0.0277*** (0.0041) | 0.0179*** (0.0033) | 0.0177*** (0.0033) | 0.0272*** (0.0052) |
| | Observations | 306,865 | 306,865 | 306,865 | 72,031 |
| Shampoos | Coefficient (Std.) | 0.0091*** (0.001) | 0.0119*** (0.001) | 0.0119*** (0.001) | 0.0267*** (0.0031) |
| | Observations | 261,778 | 261,778 | 261,778 | 40,996 |
| Snack Crackers | Coefficient (Std.) | 0.0267*** (0.0028) | 0.0234*** (0.0026) | 0.0236*** (0.0026) | 0.0434*** (0.0045) |
| | Observations | 398,665 | 398,665 | 398,665 | 78,581 |
| Soaps | Coefficient (Std.) | 0.0234*** (0.0041) | 0.0155*** (0.0037) | 0.0162*** (0.0037) | 0.0331*** (0.0058) |
| | Observations | 152,379 | 152,379 | 152,379 | 46,829 |
| Soft Drinks | Coefficient (Std.) | 0.0099*** (0.0021) | 0.0096*** (0.0021) | 0.0099*** (0.0018) | 0.0388*** (0.0028) |
| | Observations | 1,350,618 | 1,350,618 | 1,350,618 | 156,004 |
| Toothbrushes | Coefficient (Std.) | 0.0129*** (0.0018) | 0.0137*** (0.0018) | 0.0137*** (0.0018) | 0.0332*** (0.0047) |
| | Observations | 125,380 | 125,380 | 125,380 | 24,955 |
| Toothpastes | Coefficient (Std.) | 0.0082*** (0.002) | 0.0089*** (0.0016) | 0.0088*** (0.0016) | 0.0274*** (0.0046) |
| | Observations | 264,317 | 264,317 | 264,317 | 56,842 |
| **Average coefficients** | | **0.0175** | **0.0146** | **0.0149** | **0.0291** |

Notes: The table reports the results of category-level fixed effect regressions of the probability of a small price change. The dependent variable is "small price change," which equals 1 if a price change of product $i$ in store $s$ at time $t$ is less or equal to 15¢, and 0 otherwise. The main independent variable is the log of average sales volume of product $i$ in store $s$ over the sample period × the average of the price of product $i$ in store $s$ over the sample period. Column 1 reports the results of the baseline regression that includes only the log of average sales volume and the fixed effects for months, years, stores, and products. In column 2, we add the following controls: the log of the average price, the log of the absolute change in the wholesale price, and a control for sale- and bounce-back prices, which we identify using a sales filter algorithm. In column 3, we add a dummy for 9-ending prices as an additional control. In column 4, we focus on regular prices by excluding the sale- and bounce-back prices. We estimate separate regressions for each product category, clustering the errors by product. * $p < 10\%$, ** $p < 5\%$, *** $p < 1\%$

## Appendix H. Producers' size and the robustness of the correlation between small price changes and sales volumes

Bhattarai and Shoenle (2014) report that large producers, i.e., producers that sell a large number of products, are more likely to have small price changes. Table H1 shows for each of the categories, the % of small price changes by quartiles of producers' size. To find the producers' size, in each category, we find the weekly average number of products per producer. We then average over all weeks to get the average number of products sold by each producer (Bhattarai and Shoenle, 2014).

We find that in our data, there is no clear pattern. Taking the average over all categories, we find that there are 33.23%, 29.48%, 28.64%, and 28.54% small price changes in the first, second, third, and fourth quartiles, respectively. Therefore, in our data, we do not find a correlation between small price changes and producers' size, perhaps because in our data decisions on the timing of price changes are made by the retailer rather than by the producers.

Nevertheless, we divide each category into quartiles by producers' size and estimate:

$$small\ price\ change_{i,s,t} = \alpha + \beta_1 ln\big(average\ sales\ volume_{i,s}\big) + \qquad (H1)$$

$$month_t + year_t + \delta_s + \mu_i + u_{i,s,t}$$

where *small price change* is a dummy that equals 1 if a price change of product *i* in store *s* at time *t* is less or equal to 10¢, and 0 otherwise. As we do in the paper, we use observations on price changes only if we observe the price in both weeks *t* and *t* + 1 and the post change price remained unchanged for at least 2 weeks. The *average sales volume* is the average sales volume of product *i* in store *s* over the sample period. *Month* and *year* are fixed effects for the month and the year of the price change. $\delta$ and $\mu$ are fixed effects for stores and products, respectively. *u* is an i.i.d error term.

In Table H2 we report the estimation results. We find that for the first two quartiles, 28 out the 29 coefficients are positive. In the first quartile, 25 of the positive coefficients are statistically significant, and one more is marginally significant. In the second quartile, 24 of the positive coefficients are statistically significant.

In the third quartile, all 29 of the coefficients are statistically significant. 27 of them are statistically significant, and 2 more are marginally significant. In the fourth quartile, 27 of the coefficients are positive. 24 of the positive coefficients are statistically significant, and one more is marginally significant.

We also find that the sizes of the coefficients are similar across quartiles. The average coefficients are 0.026, 0.026, and 0.029 and 0.027 in the first, second, and third and fourth quartile, respectively.

As a final test, we consider the possibility that by calculating the producers' size at the category level, we might be underestimating the size of producers' that offer products in two or more categories. We, therefore, pool the data from all the product categories together.

Table H3 reports the results of regressions similar to the regressions we report in Table 6 in the paper. I.e.:

$$small\ price\ change_{i,s,t} = \alpha + \beta_1 ln\big(average\ sales\ volume_{i,s}\big) + \quad\quad\text{(H2)}$$

$$\gamma \boldsymbol{X}_{i,s,t} + month_t + year_t + category_i + \delta_s + \mu_i + u_{i,s,t}$$

where *small price change* is a dummy that equals 1 if a price change of product *i* in store *s* at time *t* is less or equal to 10¢, and 0 otherwise. The *average sales volume* is the average sales volume of product *i* in store *s* over the sample period. **X** is a matrix of other control variables. *Month*, *year* and *category* are fixed effects for the month of the price change, the year of the price change, and the product category. $\delta$ and $\mu$ are fixed effects for stores and products, respectively, and *u* is an i.i.d error term.

In column 1, the other extra control variables include the average weekly number of products per producer. The coefficient of the average sales volume is positive and statistically significant ($\beta = 0.027, p < 0.01$). It, therefore, seems that controlling for the size of the producers does not change our main finding: there is a positive correlation between the sales volume and the likelihood of a small price change.

In column 2, we also add a control for the percentage of the products that changed the price in the same week, excluding the current observation. This does not affect the coefficient of the average sales volume.

In column 3, we further add the average size of contemporaneous price changes, excluding the current observation. The coefficient of the average sales volume remains unaffected. In column 4, we add the percentage of the products that are produced by the same producer and that changed price in the same week, excluding the current observation. The coefficient of the average sales volume remains unaffected ($\beta = 0.027, p < 0.01$).

Table H1. Percentage of small price changes by quartiles

| | 1st Quartile | 2nd Quartile | 3rd Quartile | 4th Quartile |
|---|---|---|---|---|
| Analgesics | 15.04% | 12.31% | 12.19% | 11.69% |
| Bath Soaps | 18.46% | 9.98% | 12.38% | 16.81% |
| Bathroom Tissues | 46.13% | 33.03% | 38.34% | 50.17% |
| Beers | 2.47% | 3.95% | 6.76% | 8.48% |
| Bottled Juices | 45.52% | 36.06% | 38.22% | 29.14% |
| Canned Soups | 58.63% | 49.65% | 48.05% | 49.13% |
| Canned Tuna | 61.09% | 54.31% | 55.64% | 55.38% |
| Cereals | 34.68% | 28.89% | 28.57% | 32.99% |
| Cheese | 64.29% | 43.16% | 33.49% | 38.81% |
| Cigarettes | 34.13% | 29.13% | 28.24% | 25.09% |
| Cookies | 27.28% | 29.72% | 23.85% | 29.66% |
| Crackers | 38.88% | 48.10% | 27.38% | 23.88% |
| Dish Detergents | 44.63% | 41.61% | 31.13% | 31.20% |
| Fabric Softener | 48.32% | 28.68% | 27.49% | 28.86% |
| Front-End-Candies | 59.34% | 47.39% | 43.92% | 54.14% |
| Frozen Dinners | 13.01% | 23.86% | 34.66% | 20.28% |
| Frozen Entrees | 20.46% | 15.53% | 15.19% | 16.91% |
| Frozen Juices | 30.44% | 38.18% | 42.51% | 31.82% |
| Grooming Products | 11.00% | 10.97% | 13.81% | 14.41% |
| Laundry Detergents | 27.60% | 16.21% | 16.62% | 16.72% |
| Oatmeal | 37.71% | 46.51% | 38.05% | 40.84% |
| Paper Towels | 55.84% | 52.25% | 52.61% | 57.54% |
| Refrigerated Juices | 32.12% | 35.66% | 32.95% | 28.85% |
| Shampoos | 7.41% | 6.26% | 7.82% | 9.34% |
| Snack Cracker | 38.75% | 27.07% | 20.87% | 23.34% |
| Soaps | 46.30% | 49.18% | 46.42% | 38.02% |
| Soft Drinks | 16.23% | 13.71% | 26.57% | 9.06% |
| Toothbrushes | 13.32% | 8.16% | 11.10% | 12.68% |
| Toothpastes | 14.64% | 15.28% | 15.86% | 22.33% |
| **Average** | **33.23%** | **29.48%** | **28.64%** | **28.54%** |

Notes: The table presents the % of small price changes in each quartile of producers' size by category. To calculate the quartiles of producers' size, we first calculate the size of each producer in each category by finding the number of products sold by each producer in each week and then averaging over all weeks. We then divide producers into quartiles using the average number of products they sell each week.

Table H2. Category-level regressions of small price changes, by quartiles of manufacturers' size

| Category | | **(1)** | **(2)** | **(3)** | **(4)** |
|---|---|---|---|---|---|
| Analgesics | Coefficient (Std.) | 0.0298*** (0.0091) | 0.013** (0.0055) | 0.0235*** (0.0065) | 0.0338*** (0.0061) |
| | Observations | 34,663 | 37,456 | 35,731 | 36,611 |
| Bath Soap | Coefficient (Std.) | 0.0359*** (0.0134) | 0.0336*** (0.0163) | 0.0218* (0.0114) | 0.0479** (0.0233) |
| | Observations | 3,791 | 3,918 | 3,839 | 3,747 |
| Bathroom Tissues | Coefficient (Std.) | 0.0354*** (0.0099) | 0.0904*** (0.0127) | 0.0709*** (0.0165) | 0.0306*** (0.0113) |
| | Observations | 39,311 | 36,016 | 37,423 | 36,691 |
| Beer | Coefficient (Std.) | 0.0075*** (0.0017) | 0.01*** (0.0015) | 0.0153*** (0.0029) | 0.0209*** (0.0031) |
| | Observations | 73,377 | 73,587 | 71,340 | 72,316 |
| Bottled Juice | Coefficient (Std.) | 0.0126 (0.0088) | 0.0303*** (0.0096) | 0.0657*** (0.0092) | 0.0313*** (0.008) |
| | Observations | 130,866 | 123,821 | 118,687 | 123,183 |
| Canned Soup | Coefficient (Std.) | 0.0514*** (0.0087) | -0.0034 (0.0068) | 0.0368*** (0.0065) | 0.0204*** (0.0062) |
| | Observations | 121,911 | 131,751 | 120,162 | 121,719 |
| Canned Tuna | Coefficient (Std.) | 0.0378*** (0.0081) | 0.0071 (0.0081) | 0.0241*** (0.0067) | 0.0312*** (0.0093) |
| | Observations | 55,961 | 58,140 | 50,991 | 47,951 |
| Cereals | Coefficient (Std.) | 0.0202*** (0.0073) | 0.0229*** (0.0063) | 0.0302*** (0.0064) | 0.0202*** (0.0062) |
| | Observations | 94,391 | 85,049 | 87,359 | 90,321 |
| Cheese | Coefficient (Std.) | 0.0264*** (0.0047) | 0.0054 (0.0059) | 0.0229*** (0.0041) | 0.0283*** (0.005) |
| | Observations | 151,338 | 221,454 | 202,915 | 220,443 |
| Cigarettes | Coefficient (Std.) | -0.0025 (0.01) | 0.0158** (0.0064) | 0.0218*** (0.0063) | -0.0065 (0.0075) |
| | Observations | 8,454 | 9,371 | 9,048 | 9,284 |
| Cookies | Coefficient (Std.) | 0.0193*** (0.0023) | 0.0383*** (0.0042) | 0.0264*** (0.0035) | 0.0283*** (0.0043) |
| | Observations | 189,332 | 160,793 | 168,919 | 169,717 |
| Crackers | Coefficient (Std.) | 0.0375*** (0.0048) | 0.0325*** (0.0042) | 0.0345*** (0.0055) | 0.0538*** (0.0095) |
| | Observations | 66,641 | 69,958 | 54,179 | 54,407 |
| Dish Detergent | Coefficient (Std.) | 0.0275*** (0.0093) | 0.0455*** (0.0067) | 0.0394*** (0.007) | 0.0426*** (0.0059) |
| | Observations | 50,346 | 48,380 | 46,539 | 44,368 |
| Fabric Softener | Coefficient (Std.) | 0.0326*** (0.012) | 0.0375*** (0.0073) | 0.029*** (0.0077) | 0.0333*** (0.0096) |
| | Observations | 51,749 | 42,994 | 44,935 | 41,378 |
| Front-End-Candies | Coefficient (Std.) | 0.0058 (0.0054) | 0.0059 (0.0083) | 0.0019 (0.0097) | -0.0165*** (0.0052) |
| | Observations | 70,136 | 64,179 | 66,019 | 78,519 |
| Frozen Dinners | Coefficient (Std.) | 0.0443*** (0.0064) | 0.0472*** (0.0083) | 0.049*** (0.0079) | 0.0294*** (0.0071) |
| | Observations | 49,266 | 50,902 | 50,539 | 52,484 |

Table H2. (Cont.)

| Category | | (1) | (2) | (3) | (4) |
|----------|--|-----|-----|-----|-----|
| Frozen Entrees | Coefficient (Std.) | 0.0271*** (0.0033) | 0.0321*** (0.0041) | 0.0178*** (0.0038) | 0.0191*** (0.0042) |
| | Observations | 231,065 | 205,922 | 218,170 | 209,675 |
| Frozen Juices | Coefficient (Std.) | 0.0168 (0.0116) | 0.0268*** (0.0065) | 0.0245*** (0.0071) | 0.0449*** (0.0072) |
| | Observations | 77,812 | 73,824 | 80,038 | 77,143 |
| Grooming Products | Coefficient (Std.) | 0.0143*** (0.0037) | 0.0225*** (0.0038) | 0.0213*** (0.0045) | 0.0153*** (0.0048) |
| | Observations | 66,700 | 65,306 | 66,455 | 71,412 |
| Laundry Detergents | Coefficient (Std.) | 0.0216*** (0.0065) | 0.0205*** (0.0069) | 0.0297*** (0.0061) | 0.0096 (0.006) |
| | Observations | 76,565 | 67,392 | 65,836 | 62,972 |
| Oatmeal | Coefficient (Std.) | 0.0437** (0.0207) | 0.001 (0.0129) | 0.0338*** (0.0146) | 0.0171 (0.0113) |
| | Observations | 20,184 | 21,624 | 19,863 | 18,312 |
| Paper Towels | Coefficient (Std.) | 0.0323** (0.0145) | 0.0333** (0.0153) | 0.0557** (0.0245) | 0.0761*** (0.0178) |
| | Observations | 26,835 | 31,117 | 29,325 | 28,927 |
| Refrigerated Juices | Coefficient (Std.) | 0.0289*** (0.0094) | 0.0312*** (0.0071) | 0.0256*** (0.008) | 0.0358*** (0.0092) |
| | Observations | 81,959 | 73,635 | 76,575 | 74,696 |
| Shampoos | Coefficient (Std.) | 0.0208*** (0.0032) | 0.0095*** (0.0023) | 0.0082*** (0.003) | 0.0229*** (0.003) |
| | Observations | 64,883 | 63,777 | 66,146 | 66,972 |
| Snack Crackers | Coefficient (Std.) | 0.0375*** (0.0063) | 0.0403*** (0.0066) | 0.0187*** (0.0049) | 0.0264*** (0.005) |
| | Observations | 100,802 | 108,690 | 97,783 | 91,390 |
| Soaps | Coefficient (Std.) | 0.0456*** (0.0078) | 0.0437*** (0.0102) | 0.0277** (0.0118) | 0.0324*** (0.0115) |
| | Observations | 39,742 | 39,140 | 38,084 | 35,413 |
| Soft Drinks | Coefficient (Std.) | 0.0211*** (0.0022) | 0.0274*** (0.0026) | 0.0257*** (0.0036) | 0.0253*** (0.0032) |
| | Observations | 346,043 | 355,732 | 350,552 | 298,291 |
| Toothbrushes | Coefficient (Std.) | 0.0211*** (0.0043) | 0.0146*** (0.0052) | 0.0256*** (0.0058) | 0.0176*** (0.0075) |
| | Observations | 33,948 | 27,258 | 32,415 | 31,759 |
| Toothpastes | Coefficient (Std.) | 0.0149*** (0.0052) | 0.0198*** (0.0045) | 0.0127*** (0.0048) | 0.0089 (0.0063) |
| | Observations | 64,984 | 63,457 | 68,506 | 67,370 |
| **Average coefficients** | | **0.0265** | **0.0260** | **0.0290** | **0.0269** |

<u>Notes</u>: The table reports the results of category-level fixed effect regressions of the probability of a small price change. The dependent variable is "small price change," which equals 1 if a price change of product $i$ in store $s$ at time $t$ is less or equal to 10¢. The main independent variable is the log of the average sales volume of product $i$ in store $s$ over the sample period. The regressions also include fixed effects for months, years, stores, and products. Columns 1, 2, 3, and 4 are for the 1st, 2nd, 3rd, and 4th quartiles of the manufacturers' size, measured using the average number of products they sell each week. We estimate separate regressions for each product category, clustering the errors by product. * $p < 10\%$, ** $p < 5\%$, *** $p < 1\%$

Table H3. Pooled data regressions of small price changes and synchronization

| | (1) | (2) | (3) | (4) |
|---|---|---|---|---|
| Log of sales volume | 0.027*** | 0.027*** | 0.027*** | 0.027*** |
| | (0.001) | (0.001) | (0.001) | (0.001) |
| Observations | 9,553,542 | 9,553,542 | 9,553,536 | 9,392,565 |

Notes: The table reports the results of pooled fixed effect regressions of the probability of a small price change. The dependent variable is "small price change," which equals 1 if a price change of product $i$ in store $s$ at time $t$ is less or equal to 10¢, and 0 otherwise. The main independent variable is the log of the average sales volume product $i$ in store $s$ over the sample period. Column 1 reports the results of the baseline regression that includes only the log of the average sales volume and the fixed effects for months, years, categories, stores, and products. In column 2, we add the average number of products each producer offers each week. In column 3, we add the percentage of the products that changed the price on the same week, excluding the current observation as an additional control. In column 4, we control synchronization by adding the percentage of the products that have been produced by the same producer and that changed price in the same week, excluding the current observation. We estimate a pooled regression, clustering the errors by product. * $p < 10\%$, ** $p < 5\%$, *** $p < 1\%$

**Appendix I. Results of cross-category analysis**

Table I1 shows the number of all price changes and the number of small price changes ($\Delta P \leq 10\cancel{c}$) by product category, the percentage of small price changes out of all price changes, and the average sales volume. The latter is calculated by first finding the average weekly sales volume for each product in each store (*product-store*) in the category, and then averaging over all products.[4]

There is a large cross-category variation in the share of small price changes, ranging from 3.3% in the beer category to 55.2% in the paper towels category. This variation is accompanied by a large variation in the average sales volume. As shown in Figure I2 and table I4, the variation in the category-level average sales volume is mostly driven by the number of choices available to consumers. At the category level, a 1% increase in the number of products is associated with a 0.84% decrease in the expected average sales volumes.

More importantly, however, there is a strong correlation between the average sales volume and the share of small price changes. Figure I1 shows a scatterplot of the category-level average sales volume and the percentage of small price changes, along with a linear regression line (solid line). We find a positive correlation between the two variables. The correlation is even stronger (dashed line) if we exclude paper towels and bathroom tissues, two categories with particularly high values of both average sales volume and percentage of small price changes.

To explore this correlation formally, we run cross-category OLS regressions, where the dependent variable is the category-level percentage of small price changes. See Table I2. In column 1, the independent variable is the average weekly sales volume. The coefficient estimate of 0.95 implies that a one-unit increase in the average weekly sales volume is associated with a 0.95% increase in the percentage of small price changes.

A possible explanation for this correlation could be that categories with low average price levels have higher shares of small price changes and higher sales volume. The regression in column 2 shows that there is indeed a negative correlation between the average price in a category and the percentage of small price changes. However, in column 3, which reports the results of a regression that includes both the average prices and the average sales volume as independent variables, we find that the coefficient of the average sales volume is 0.78 and statistically significant. Thus, we find that sales volume is correlated with small price changes even after controlling for the price level.

---

[4] In calculating the average sales volume, we need to account for missing observations, because a missing observation in week $t$ implies that the product was either out of stock or had 0 sales on that week. Thus, averaging over the available observations can lead to an upward bias for products that are sold in small numbers. Therefore, for each product in each store, we calculate the average by first determining the total number of units sold over all available observations. We then identify the first and last week for which we have observations, and calculate the average for each product-store as $\frac{total\ no.\ of\ units\ sold}{last\ week\ -\ first\ week}$. The resulting figure is smaller than we would obtain if we averaged over all available observations (which would not include observations on weeks with 0 sales).

An alternative explanation could be competition. It could be that products in high sales volume categories face stronger competition, and their producers may want to avoid large price changes that could alienate consumers. On the other hand, it is also possible that competition would have a negative effect on the prevalence of small price changes. Wang and Werning (2022) argue that concentrated markets increase the likelihood of pricing complementarities. This suggests that small price changes might be more likely in markets where producers have greater market power.

In column 4, we look at the correlation between the percentage of small price changes and category-level estimates of own price elasticities which are taken from Hoch et al. (1995). We find that the correlation is negative, but not statistically significant. In column 5, where we report the results of a regression with both the sales volume and the price elasticity as independent variables, we find that the coefficient of the sales volume is 0.80, and statistically significant.

The coefficient of the elasticity is negative and statistically significant. I.e., small price changes are more common in product categories with low rather than high price elasticities (in absolute values), which is consistent with pricing complementarities. However, our results suggest that even after accounting for pricing complementarities, the effect of the sales volumes is positive and statistically significant.

As discussed above, table I1 shows that there is a large variation in the average sales volume across categories. In particular, the average weekly sales volume per store in the categories of bathroom tissues and paper towels, 40.35 and 38.92, respectively, stand out: they both are much larger than the average sales volumes in other categories. In contrast, the weekly average sales volume per store in the categories of bath soaps and shampoos, 0.72 and 0.84, are much smaller than the average in other categories.

To a large extent, these variations in the sales volume can be explained by product variety, which can be measured by looking at the number of Universal Product Codes (UPCs) in each category, which captures the number of options consumers can choose from. Table I3 gives the average sales volumes and the number of UPCs for each category.[5] Figure I2 illustrates that the average sales volume is negatively correlated with the number of UPCs. For example, the product categories with the highest sales volumes, bathroom tissues and paper towels, have a relatively small number of UPCs, 102 and 91, respectively. For comparison, the product categories with the lowest average sales volumes, bath soaps and shampoos, have 495 and 1,905 UPCs, respectively.

The negative correlation between the average sales volume and the number of UPCs is statistically significant. Table I4 reports the results of a category level linear regression of the log of the average sales volume on the log of the number of UPCs. According to the table, the correlation is statistically significant ($\beta = -0.84, p < 0.01$). Thus, a 1% increase in the number of UPCs per category is associated with a decrease of 0.84% in the average weekly sales volume per store.

Thus, the large variation in the sales volume should not be surprising. In categories with many UPCs, it appears that a large number of UPCs sell a small number of units, leading to a low average weekly sales volume per store.

---

[5] It turns out the Dominick's occasionally used different UPCs for the same products, perhaps because a product was re-launched (Mehrhoff, 2018). Whenever possible, we treat re-launches as the same product and, consequently, the number of products that we report might differ from the number reported in previous studies, e.g., Chen et al. (2008). See also the Dominick's data manual (https://www.chicagobooth.edu/-/media/enterprise/centers/kilts/datasets/dominicks-manual-and-codebook_kiltscenter.aspx), p. 9.

Table I1. The proportion of small price changes and the average sales volume by product categories

| Product Category | All price changes (1) | Small price changes (2) | % of small price changes (3) | Average sales volume (4) |
|---|---|---|---|---|
| Analgesics | 144,461 | 16,608 | 11.5% | 1.24 |
| Bath soap | 15,295 | 1,783 | 11.7% | 0.72 |
| Bathroom tissues | 149,441 | 60,263 | 40.3% | 40.35 |
| Beer | 290,620 | 9,526 | 3.3% | 3.58 |
| Bottled juices | 496,557 | 170,762 | 34.4% | 8.27 |
| Canned soups | 495,543 | 281,649 | 56.8% | 12.25 |
| Canned tuna | 213,043 | 111,473 | 52.3% | 9.34 |
| Cereals | 357,120 | 112,298 | 31.4% | 15.02 |
| Cheese | 796,150 | 309,021 | 38.8% | 11.32 |
| Cigarettes | 36,157 | 10,527 | 29.1% | 21.20 |
| Cookies | 688,761 | 161,826 | 23.5% | 4.96 |
| Crackers | 245,185 | 77,658 | 31.7% | 4.89 |
| Dish detergents | 189,633 | 67,109 | 35.4% | 7.38 |
| Fabric softeners | 181,056 | 55,199 | 30.5% | 5.56 |
| Front end candies | 278,853 | 124,432 | 44.6% | 10.70 |
| Frozen dinners | 203,191 | 45,050 | 22.2% | 5.64 |
| Frozen entrees | 864,832 | 127,039 | 14.7% | 6.32 |
| Frozen juices | 308,817 | 106,398 | 34.5% | 16.82 |
| Grooming products | 269,873 | 24,172 | 9.0% | 1.21 |
| Laundry detergents | 272,765 | 51,739 | 19.0% | 6.59 |
| Oatmeal | 79,983 | 34,271 | 42.8% | 7.32 |
| Paper towels | 116,204 | 64,183 | 55.2% | 38.92 |
| Refrigerated juices | 306,865 | 91,124 | 29.7% | 19.80 |
| Shampoos | 261,778 | 14,228 | 5.4% | 0.84 |
| Snack crackers | 398,665 | 93,754 | 23.5% | 6.79 |
| Soaps | 152,379 | 60,635 | 39.8% | 5.02 |
| Soft drinks | 1,350,618 | 206,373 | 15.3% | 13.05 |
| Toothbrushes | 125,380 | 13,306 | 10.6% | 2.09 |
| Toothpastes | 264,317 | 38,894 | 14.7% | 3.31 |
| Total | 9,553,542 | 2,541,300 | 26.6% | 10.02 |

Notes: Column 1 presents the total number of price changes in each category. Column 2 presents the number of small price changes $(\Delta P \leq 10\cancel{c})$. Column 3 presents the % of small price changes out of all price changes. Column 4 presents the average number of units sold per product, per week, per store. The average number of units sold is calculated taking into account that missing observations often imply 0 sales.

Table I2. Cross-category regression of the % of small price changes and sales volume

|  | (1) | (2) | (3) | (4) | (5) |
|---|---|---|---|---|---|
| Average sales volume | 0.95*** (0.262) |  | 0.78** (0.270) |  | 0.80*** (0.268) |
| Average price |  | −4.59** (1.703) | −2.88* (1.621) |  |  |
| Absolute elasticity |  |  |  | −7.78 (6.208) | −11.53** (5.223) |
| Constant | 19.29*** (3.319) | 45.65*** (5.327) | 28.85*** (6.249) | 44.93*** (10.504) | 40.97*** (8.681) |
| $R^2$ | 0.33 | 0.21 | 0.40 | 0.09 | 0.43 |
| Number of categories | 29 | 29 | 29 | 18 | 18 |

<u>Notes</u>: The table presents the results of OLS regressions. The dependent variable is the % of small price changes out of all price changes, in each of the 29 categories. Small price changes are defined as price changes of $\Delta P \leq 10\cent$. The average price is the average price in the product category. The absolute elasticity is the absolute value of the demand price elasticity estimates as reported by Hoch et al. (1995). Columns (4) and (5) contain only 18 observations because Hoch et al. (1995) provide elasticity estimates only for 18 of the 29 product categories. * $p < 10\%$, ** $p < 5\%$, *** $p < 1\%$

Table I3. Category-level average sales volume and the number of UPCs

| Product Category | Average sales volume (1) | Number of UPCs (2) |
|---|---|---|
| Analgesics | 1.24 | 507 |
| Bath soap | 0.72 | 495 |
| Bathroom tissues | 40.35 | 102 |
| Beer | 3.58 | 653 |
| Bottled juices | 8.27 | 445 |
| Canned soups | 12.25 | 413 |
| Canned tuna | 9.34 | 212 |
| Cereals | 15.02 | 399 |
| Cheese | 11.32 | 573 |
| Cigarettes | 21.20 | 78 |
| Cookies | 4.96 | 976 |
| Crackers | 4.89 | 295 |
| Dish detergents | 7.38 | 183 |
| Fabric softeners | 5.56 | 203 |
| Front end candies | 10.70 | 416 |
| Frozen dinners | 5.64 | 254 |
| Frozen entrees | 6.32 | 822 |
| Frozen juices | 16.82 | 161 |
| Grooming products | 1.21 | 962 |
| Laundry detergents | 6.59 | 353 |
| Oatmeal | 7.32 | 93 |
| Paper towels | 38.92 | 91 |
| Refrigerated juices | 19.80 | 227 |
| Shampoos | 0.84 | 1,905 |
| Snack crackers | 6.79 | 382 |
| Soaps | 5.02 | 1,370 |
| Soft drinks | 13.05 | 243 |
| Toothbrushes | 2.09 | 325 |
| Toothpastes | 3.31 | 376 |
| **Average** | **10.02** | **466.00** |

Notes: Column 1 presents the average number of units sold per product, per week, per store. Column 2 presents the number of UPCs in each product category.

Table I4. Average sales volume and the number of UPCs

| | ln (average sales volume) |
|---|---|
| ln (Number of UPCs) | −0.84*** |
| | (0.186) |
| | |
| Constant | 6.77*** |
| | (1.098) |
| | |
| $R^2$ | 0.35 |
| Observations | 29 |

Notes: The table reports the results of a category-level linear regression with robust standard errors. The dependent variable is the log of the average weekly sales volume per store. The independent variable is the log of the number of UPCs in each category. Standard errors are in parentheses. *** $p < 0.01$.

Figure I1. Cross-category correlation between small price changes and sales volume

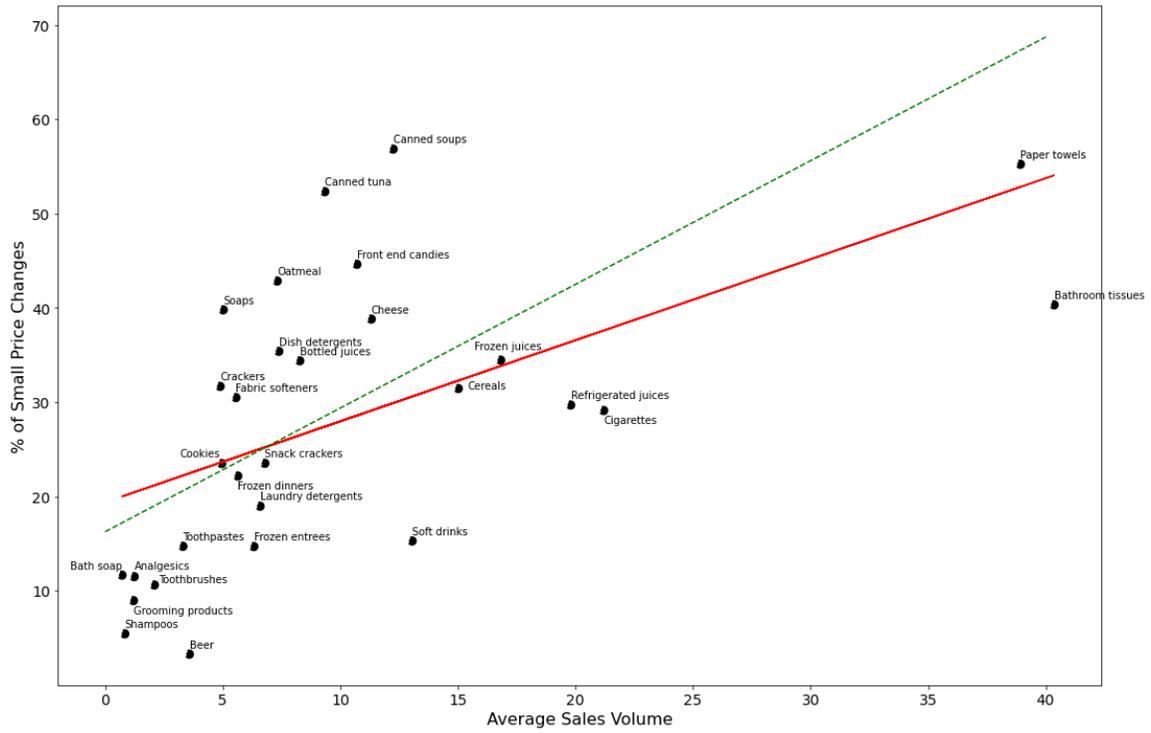

Notes: The red solid line is a linear regression line when all 29 product categories are included. The dotted green line is the linear regression line when two categories, paper towels and bathroom tissues, are excluded.

Figure I2. Average sales volume and the number of UPCs

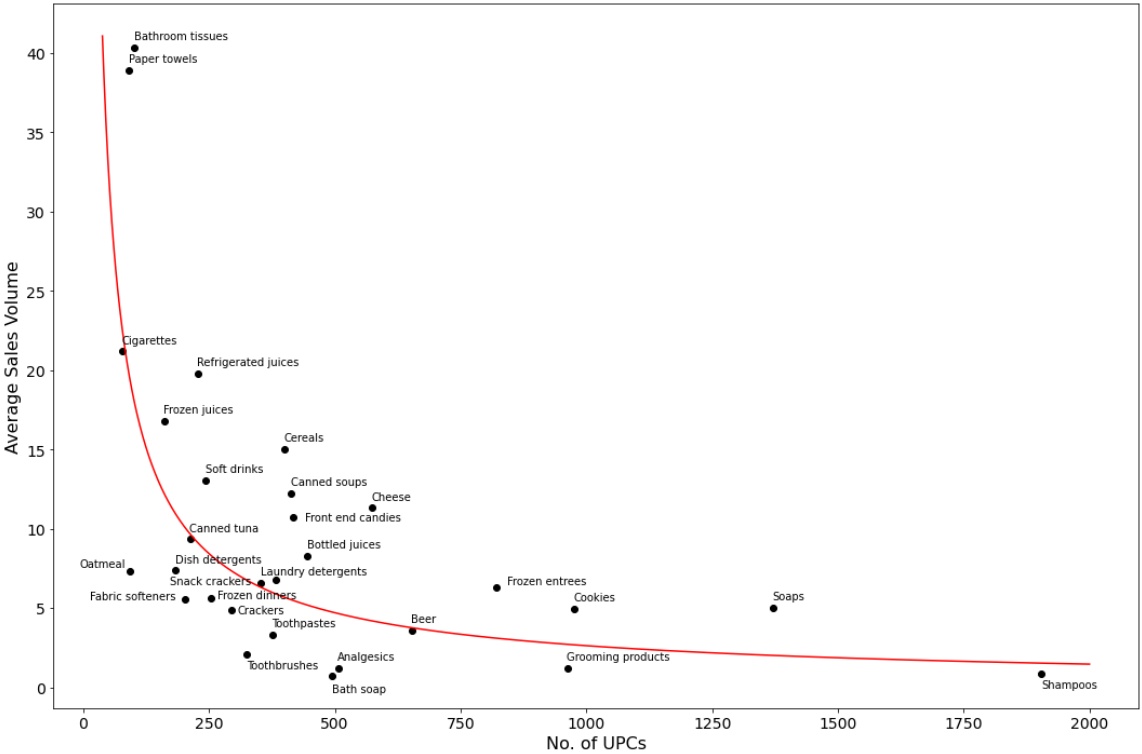

Notes: The *y*-axis gives the average weekly sales volume per store in each of the 29 categories. The *x*-axis gives the number of UPCs per category. The red line gives the prediction line based on a log-log regression specification.

### Appendix J. Sales volume, markup, and small price changes

Our results suggest that small price changes should be relatively common for products with high sales volumes. Yet, in the marketplace there are products with high sales volumes that rarely have small price changes. One example is the iPhone.

A possible explanation is that the likelihood of small price changes is negatively correlated with markups. It is possible that sellers that have high markups are less likely to make small price changes, because the effect of a small price change on the profits of a firm with high markup could be small in percentage terms.

To check this, we take advantage of the fact that Dominick's data contain a proxy for the products' markup (Barsky et al., 2003), and thus estimate a regression that is similar to regression (1) in the paper:

$$small\ price\ change_{i,s,t} = \alpha + \beta_1 ln\big(average\ sales\ volume_{i,s}\big) +$$

$$\beta_2\big(average\ markup_{i,s}\big) + month_t + year_t + \delta_s + \mu_i + u_{i,s,t} \qquad (J1)$$

where *small price change* is a dummy that equals 1 if a price change of product $i$ in store $s$ at time $t$ is less than or equal to 10¢, and 0 otherwise. As we do in the paper, we use observations on price changes only if we observe the price in both weeks $t$ and $t + 1$ and the post change price remained unchanged for at least 2 weeks. The *average sales volume* is the average sales volume of product $i$ in store $s$ over the sample period. The *average markup* is the average markup of product $i$ in store $s$ over the sample period. *Month* and *year* are fixed effects for the month and the year of the price change. $\delta$ and $\mu$ are fixed effects for stores and products, respectively, and $u$ is an i.i.d error term.

The results are summarized in Table J1. Panel A gives information about the coefficients of the sales volumes in each of the 29 product categories. We find that all 29 coefficients are positive. 15 of the coefficients are statistically significant, and 1 more is marginally significant. Panel B gives information about the coefficients of the markup. We find that consistent with the hypothesis that a high markup is associated with a lower frequency of small price changes, the coefficients of the markup are negative in 21 of the 29 product categories, 13 of them statistically significant.

The results, therefore, suggest that there is, indeed, a negative correlation between markups and the frequency of small price changes. Adding the markups to the regression, however, does not affect our main finding. There is a positive correlation between sales volumes and small price changes.

Table J1. Sales volume, markup, and small price changes

| Category | Sales Volume | | Markup | | No. of |
|---|---|---|---|---|---|
| | Coefficient | Std. | Coefficient | Std. | Observations |
| Analgesics | 0.0163*** | 0.0040 | -0.1157*** | 0.0371 | 74,451 |
| Bath Soap | 0.01 | 0.0129 | 0.1211 | 0.1138 | 6,649 |
| Bathroom Tissues | 0.0392*** | 0.0086 | -0.2302*** | 0.0373 | 56,445 |
| Beer | 0.0021*** | 0.0007 | 0.0001 | 0.0011 | 178,518 |
| Bottled Juice | 0.0314*** | 0.0075 | -0.2746*** | 0.0568 | 224,857 |
| Canned Soup | 0.0017 | 0.0090 | 0.0011 | 0.0692 | 233,778 |
| Canned Tuna | 0.0032 | 0.0061 | -0.3597*** | 0.0694 | 112,628 |
| Cereals | 0.0048 | 0.0065 | -0.0681 | 0.0497 | 141,082 |
| Cheese | 0.0109*** | 0.0036 | -0.6253*** | 0.0790 | 357,679 |
| Cigarettes | 0.0037 | 0.0050 | 0.1909*** | 0.0358 | 24,553 |
| Cookies | 0.0086*** | 0.0019 | -0.0247 | 0.0376 | 317,932 |
| Crackers | 0.0012 | 0.0033 | -0.1905*** | 0.0575 | 115,657 |
| Dish Detergent | 0.028*** | 0.0064 | -0.4664*** | 0.1218 | 85,222 |
| Fabric Softener | 0.0085 | 0.0065 | -0.4819*** | 0.1307 | 85,337 |
| Front-End-Candies | 0.0098** | 0.0041 | -0.1295 | 0.0908 | 148,200 |
| Frozen Dinners | 0.0517*** | 0.0069 | -0.1548 | 0.1058 | 52,893 |
| Frozen Entrees | 0.0227*** | 0.0026 | -0.19*** | 0.0476 | 345,223 |
| Frozen Juices | 0.0181** | 0.0071 | -0.3513*** | 0.1189 | 118,582 |
| Grooming Products | 0.0095*** | 0.0033 | 0.0302 | 0.0290 | 101,918 |
| Laundry Detergents | 0.0189*** | 0.0045 | -0.1422*** | 0.0328 | 121,539 |
| Oatmeal | 0.016 | 0.0123 | 0.0513 | 0.0870 | 25,513 |
| Paper Towels | 0.0113 | 0.0141 | -0.7104*** | 0.1077 | 48,198 |
| Refrigerated Juices | 0.0129* | 0.0077 | -0.052 | 0.0665 | 108,964 |
| Shampoos | 0.01*** | 0.0025 | -0.0067 | 0.0121 | 88,163 |
| Snack Crackers | 0.0023 | 0.0029 | -0.0868 | 0.0654 | 176,527 |
| Soap | 0.0258*** | 0.0088 | -0.1312 | 0.0943 | 56,725 |
| Soft Drinks | 0.0306*** | 0.0045 | 0.0029 | 0.0026 | 230,185 |
| Toothbrushes | 0.0137*** | 0.0046 | -0.1526*** | 0.0578 | 52,181 |
| Toothpastes | 0.0015 | 0.0039 | 0.0412 | 0.0296 | 100,831 |
| **Average** | **0.0146** | **0.0059** | **-0.1554** | **0.0636** | **746,413** |

Notes: The table reports the results of category-level fixed effect regressions of the probability of a small price change. The dependent variable is "small price change," which equals 1 if a price change of product $i$ in store $s$ at time $t$ is less or equal to 10¢. The main independent variables are the log of the average sales volume of product $i$ in store $s$ over the sample period, and the average markup of product $i$ in store $s$ over the sample period. The regressions also include fixed effects for months, years, stores, and products. The LHS panel reports the coefficient of the average sales volume. The RHS panel reports the coefficient of the average markups. We estimate separate regressions for each product category, clustering the errors by product. * $p < 10\%$, ** $p < 5\%$, *** $p < 1\%$

**Appendix K. Category level correlation between sales volumes and the size of price changes**

Figure 2 in the paper uses deciles plot to illustrate the correlation between sales volumes and the size of small price changes when we pool data from all categories. Figure K1 illustrates the correlation between sales volumes and the size of small price changes at the category level. The figure depicts, for each category, the scatter plots of the size of price changes, in cents, on the $x$-axis, vs. the average sales volume, on the $y$-axis. The average sales volume is calculated at the store-product level, i.e., for individual goods at each store.

The figure shows that the relationship tends to have a pyramid shape – broad at the bottom, suggesting that relatively large price changes occur at all levels of sales volumes. Small price changes, however, are more likely to occur when the sales volume is high, yielding the pyramid shape.

Figure K1. Sales volume and small price changes

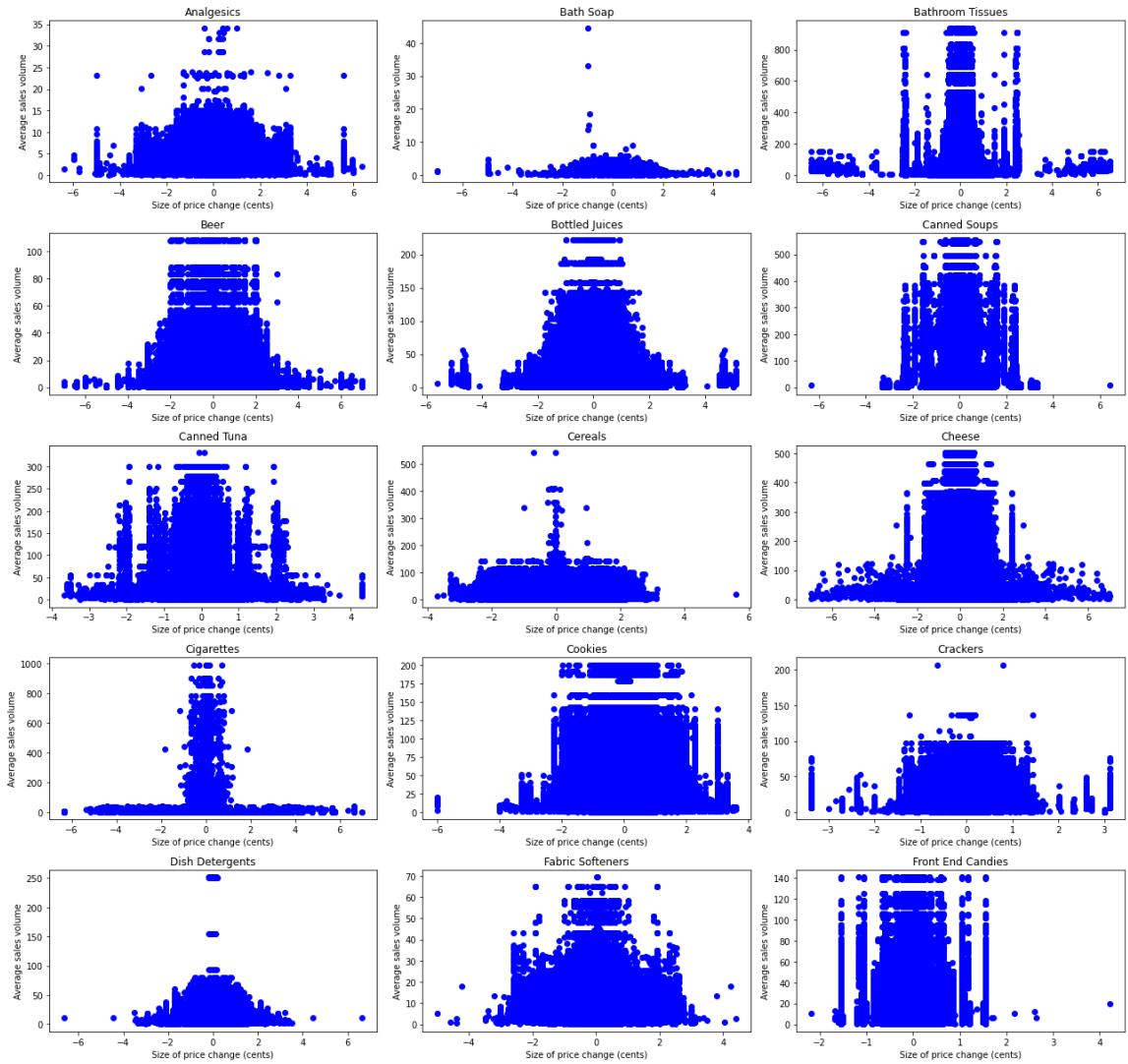

Figure K1. (cont.)

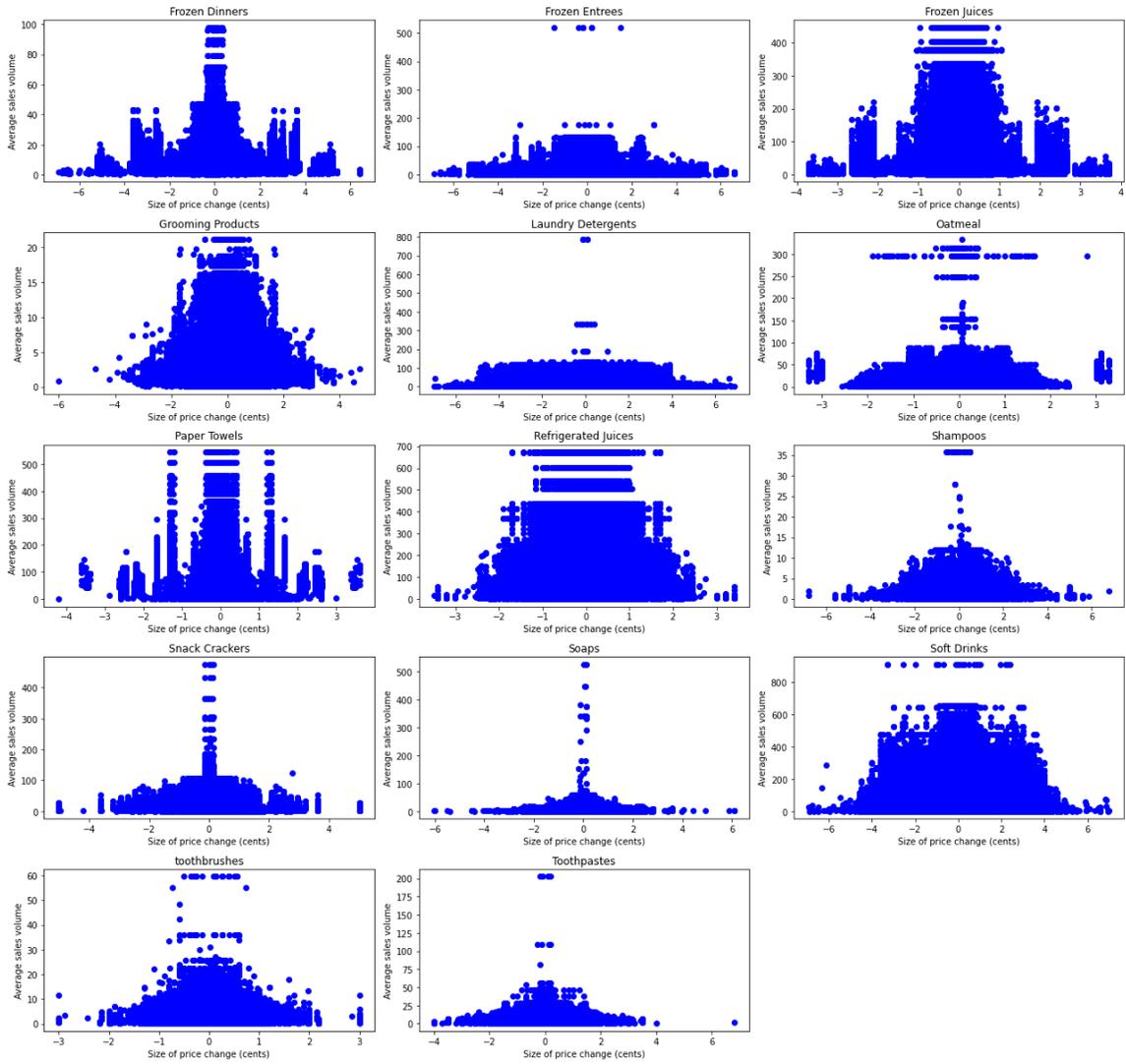

Notes: The figure depicts, for each of the 29 product categories, the correlation between the size of price changes ($x$-axis) and the average sales volume ($y$-axis). The average sales volume is calculated separately for each product in each store.

**Appendix L. Frequency of price changes by size for high, medium, and low sales volume products – in percentage terms**

In the paper, we present Figure 3, which shows that within product categories, price changes in general, and small price changes in particular, are more common among high sales volume products than among middle and low sales volume products. In Figure 3, we measure the size of price changes in cents. However, this has the disadvantage that some price changes that are multiples of 10 cents are much more frequent than other price changes.

In addition, when we measure the size of price changes by cents, we might identify a price change as small because its size is less than 10 cents. In percentage terms, however, this price change might be large. E.g., if a good costs less than 1 dollar.

We therefore generate a figure similar to figure 3, but this time we measure the size of price changes in percentage terms. To draw the figure, we first compute for each product in each store, the average sales volume over the entire sample period. By taking the average over a long period, we obtain an estimate of the expected sales volume that does not depend on transitory shocks or sales. We then group the products into high, medium and low sales volume products. Low sales volume products are products with average sales volume in the lower third of the distribution, high sales volume products have sales volume in the higher third of the distribution, and medium sales volume products have sales volume in between.

Figure L1 shows, for every product category, the frequency of price changes for each size of price change from 1% to 30%. As we do in the paper, we use observations on price changes only if we observe the price in both weeks $t$ and $t + 1$ and the post change price remained unchanged for at least 2 weeks.

The red dashed line depicts the frequency of price changes among high sales volume products, the green dotted line depicts the frequency of price changes among middle sales volume products, while the blue solid line depicts the frequency of price changes among low sales volume products. The shaded area marks the range of small price changes, $\Delta P \leq 5\%$.

We find that in comparison to Figure 3 in the paper, the lines on Figure L1 are smoother, without the peaks at multiples of 10 cents. However, in some categories, there are small peaks, particularly at 20% and 25%, perhaps because sale prices and discounts are often set in percentage terms.

We also find that similar to Figure 3 in the paper, price changes are more common among high sales volume products, and least common among low sales volume products. Focusing on the shaded area, we see that the frequency of small price changes is far greater among the high sales-volume products than among low sales volume products. Indeed, for high sales volume products, in most product categories, the frequency of small price changes exceeds the frequency of large price changes. This is far less common, and less dramatic, among low sales volume products. For the middle sales volume products, the frequency of price changes, and the frequency of small price changes in particular, is in general in between the frequencies of the low and high sales volume products.

Figure L1. Frequency of price changes by size, in % terms, for high, middle, and low sales volume products

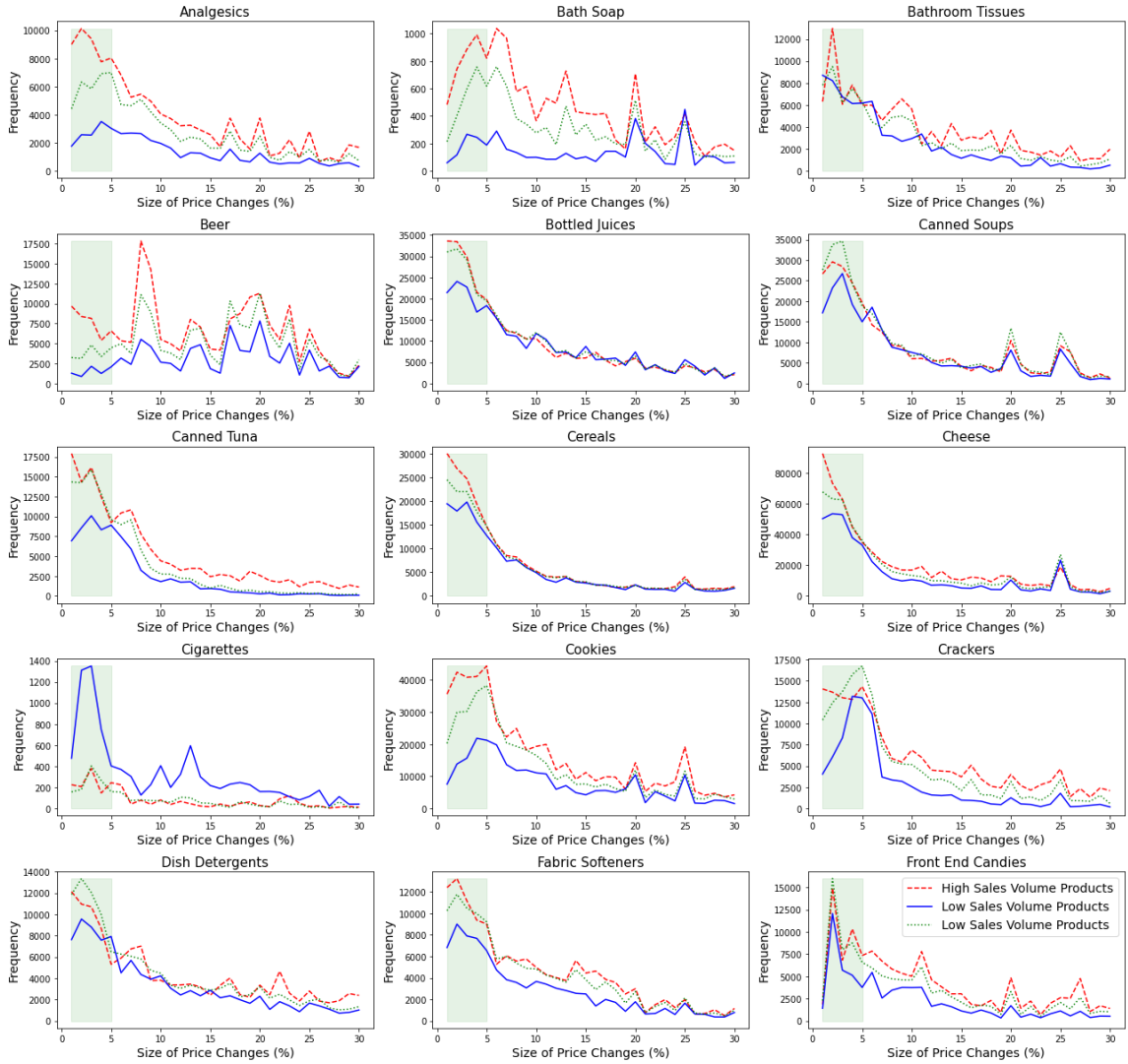

Figure L1 (cont.).

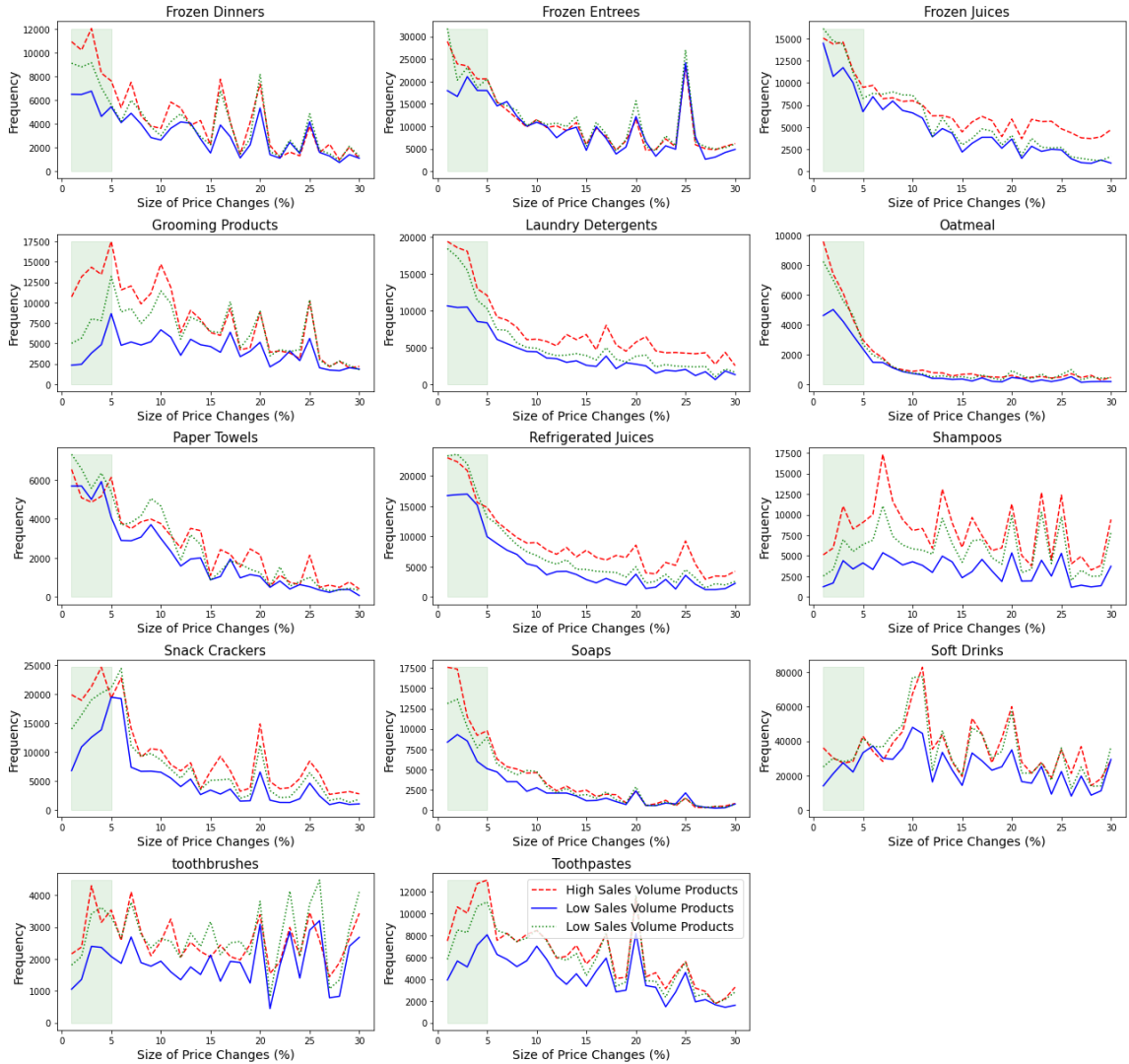

<u>Notes</u>: For each category, the figure shows the frequency of price changes for each size of price change from 1% to 30%, comparing high sales volume products to medium and low sales volume products. To obtain the figures, we compute the average sales volume over the entire sample period for each product, in each store. We then group the products into high, medium, and low sales volume products. High (low) sales volume products are products in the high (low) third of the distribution. Medium sales volume products fall in between. The *y*-axis shows the frequency of price changes. The red dashed line depicts the frequency of price changes for the high sales-volume products, the green dotted line depicts the frequency of price changes for the medium sales-volume products, and the blue solid line depicts the frequency of price changes for the low sales volume products. The shaded area marks the range of small price changes, $\Delta P \leq 5\%$.

**Appendix M. National brand vs. private label products**

It is possible that the correlation between small price changes and sales volumes is an artifact of differences in the patterns of demand between products. In this appendix, therefore, we separate private label and national brand products and analyze them separately, because they tend to have different price levels and different patterns of demand. If our results are an artifact of the pattern of demands, then it is possible that sales volumes would have a different effect on private label products than on national brands.

In the first analysis, we use all price changes, conditional on observing the price one-week before the price change. We then re-estimate the model, using only observations on price changes if the post-change price remained unchanged for at least two weeks. The second analysis is consistent with our analysis in the paper and in the other appendices. However, it has the disadvantage of having too few observations on private label products' price changes, leading to imprecise estimates.

Focusing first on all price changes, Table M1 (M2) presents the results of regressions equivalent to the regressions in Table 3 in the paper. This time, however, we focus on national brand (private label) products. The regressions take the following form:

$$small\ price\ change_{i,s,t} = \alpha + \beta \ln(average\ sales\ volume_{i,s}) + \gamma \mathbf{X}_{i,s,t}$$
$$+ month_t + year_t + \delta_s + \mu_i + u_{i,s,t} \tag{M1}$$

where *small price change* is a dummy that equals 1 if a price change of product *i* in store *s* at time *t* is less or equal to 10¢, and 0 otherwise. The *average sales volume* is the average sales volume of product *i* in store *s* over the sample period. $\mathbf{X}$ is a matrix of other control variables. *Month* and *year* are fixed effects for the month and the year of the price change. $\delta$ and $\mu$ are fixed effects for stores and products, respectively, and *u* is an i.i.d error term. We estimate separate regressions for each product category, clustering the errors by product.

The values in Table M1 are the coefficients of the log of the average sales volume when we focus on the sample of national brand products. In column 1, the only control variables are the log of the average sales volume, and dummies for months, years, stores, and products. Consistent with the results we report in the paper, we find that all the coefficients of the log of the average sales volume are positive. 29 of the 29 coefficients are positive, and 28 of them are statistically significant. The only exception is the coefficient of the highly regulated cigarettes category. The average coefficient is 0.036, suggesting that a 1% increase in the sales volume is associated with a 3.6% increase in the likelihood of a small price change.

In column 2, we add controls for the log of the average price, the log of the absolute change in the wholesale price, and a control for sale- and bounce-back prices, which we identify using the sales filter algorithm of Fox and Syed (2016). All the coefficients are positive and statistically significant: 27 at the 1% level, and two at the 10% level. The average coefficient is 0.029, suggesting that a 1% increase in the sales volume is associated with a 2.9% increase in the likelihood of a small price change.

In column 3, we also add a control for 9-ending prices. All coefficients remain positive and statistically significant: 27 at the 1% level, and two at the 10% level. The average coefficient is 0.025, suggesting that a 1% increase in the sales volume is associated with a 2.5% increase in the likelihood of a small price change.

As a further control, in column 4 we focus on regular prices by excluding all sale and bounce-back prices. When we focus on regular prices, all the coefficients are positive and 28 statistically significant at the 1% level. The average coefficient is 0.045, suggesting that a 1% increase in the sales volume is associated with a 4.5% increase in the likelihood of a small price change.

The values in Table M2 are the coefficients of the log of the average sales volume when we use the sample of private label products. When we focus on private labels, we are left with 24 product categories, because in five categories, which include beers, cigarettes, front-end-candies, frozen dinners and soaps, we have less than 500 observations on private labels.

In column 1, the only control variables are the log of the average sales volume, and dummies for months, years, stores, and products. Consistent with the results we report in the paper, we find that 23 of the 24 estimated coefficients of the log of the average sales volume are positive. 19 of the positive coefficients are statistically significant, and one more is marginally statistically significant. The one negative coefficient (toothbrushes category) is not statistically significant. The average coefficient is 0.039, suggesting that a 1% increase in the sales volume is associated with a 3.9% increase in the likelihood of a small price change.

In column 2, we add controls for the log of the average price, the log of the absolute change in the wholesale price, and a control for sale- and bounce-back prices, which we identify using the sales filter algorithm of Fox and Syed (2016). 23 of the 24 coefficients are positive. 15 of the positive coefficients are statistically significant, and one more is marginally statistically significant. The one negative coefficient (toothbrushes category) is not statistically significant. The average coefficient is 0.030, suggesting that a 1% increase in the sales volume is associated with a 3.0% increase in the likelihood of a small price change.

In column 3, we also add a control for 9-ending prices. Again, 23 of the 24 coefficients are positive. 13 of the positive coefficients are statistically significant, and one more is marginally statistically significant. The one negative coefficient (toothbrushes) is not statistically significant. The average coefficient is 0.024, suggesting that a 1% increase in the sales volume is associated with a 2.4% increase in the likelihood of a small price change.

As a further control for a possible effect of sales on the results, in column 4 we focus on regular prices by excluding all sale- and bounce-back prices. When we focus on regular prices, all the coefficients are positive. 14 of the positive coefficients are statistically significant, and 5 more are marginally statistically significant. The average coefficient is 0.047, suggesting that a 1% increase in the sales volume is associated with a 4.7% increase in the likelihood of a small price change.

We, therefore, find that sales volumes are positively correlated with the likelihood of small price changes among private label products as well as among national brand products.

Focusing on price changes only if the post-change price survived for at least two weeks, Table M3 (M4) presents the results of regressions equivalent to the regressions in Table 3 in the paper. This time, however, we focus on national brand (private label) products.

In column 1 of Table M3, the control variables are the log of the average sales volume, and dummies for months, years, stores, and products. We find that all 29 coefficients of the log of the average sales volume are positive. 17 of the positive coefficients are statistically significant, and 4 more are marginally significant. The average coefficient is 0.016, suggesting that a 1% increase in the sales volume is associated with a 1.6% increase in the likelihood of a small price change.

In column 2, we add controls for the log of the average price, the log of the absolute change in the wholesale price, and a control for sale- and bounce-back prices, which we identify using the sales filter algorithm of Fox and Syed (2016). 26 of the coefficients are positive. 18 of them are statistically significant, and one more is significant at the 10% level. The average coefficient is 0.009, suggesting that a 1% increase in the sales volume is associated with a 0.9% increase in the likelihood of a small price change.

In column 3, we add a control for 9-ending prices. 26 of the coefficients are positive. 13 of them are statistically significant, and 3 more are significant at the 10% level. The average coefficient is 0.009, suggesting that a 1% increase in the sales volume is associated with a 0.9% increase in the likelihood of a small price change.

As a further control, in column 4 we focus on regular prices by excluding all sale and bounce-back prices. 28 of the coefficients are positive. 16 of them are statistically significant, and 3 more are significant at the 10% level. The average coefficient is 0.020, suggesting that a 1% increase in the sales volume is associated with a 2.0% increase in the likelihood of a small price change.

The values in Table M4 are the coefficients of the log of the average sales volume when we use the sample of private-label products. When we focus on private labels, the estimation is imprecise because in many categories we only have a small number of observations. Consequently, we are left with 23 product categories, because in 6 categories, which include bath-soaps, beers, cigarettes, front-end-candies, frozen dinners and soaps, we have less than 500 observations on price changes.

In column 1, the control variables are the log of the average sales volume, and dummies for months, years, stores, and products. We find that 15 of the 23 estimated coefficients of the log of the average sales volume are positive. 5 of the positive coefficients are statistically significant, and 3 more are marginally statistically significant. One of the negative coefficients (cereals) is statistically significant. The average coefficient is 0.013, suggesting that a 1% increase in the sales volume is associated with a 1.3% decrease in the likelihood of a small price change.

In column 2, we add controls for the log of the average price, the log of the absolute change in the wholesale price, and a control for sale- and bounce-back prices, which we identify using the sales filter algorithm of Fox and Syed (2016). 12 of the 23 coefficients are positive. 3 of the positive coefficients are statistically significant, and 4 more are marginally statistically significant. 2 of the negative coefficients (cereals and refrigerated juices) are statistically significant. The average coefficient is 0.007, suggesting that a 1% increase in the sales volume is associated with a 0.7% decrease in the likelihood of a small price change.

In column 3, we add a control for 9-ending prices. 12 of the 23 coefficients are positive. 4 of the positive coefficients are statistically significant, and 1 more is marginally statistically significant. 2 of the negative coefficients (cereals and refrigerated juices) are statistically significant. The average coefficient is 0.007, suggesting that a 1% increase in the sales volume is associated with a 0.7% decrease in the likelihood of a small price change.

As a further control for a possible effect of sales on the estimation results, in column 4 we focus on regular prices by excluding all sale- and bounce-back prices. 12 of the 23 coefficients are positive. 5 of the positive coefficients are statistically significant, and 1 more is marginally statistically significant. 2 of the negative coefficients (cereals and refrigerated juices) are statistically significant. The average coefficient is 0.014, suggesting that a 1% increase in the sales volume is associated with a 1.4% decrease in the likelihood of a small price change.

Table M1. Category-level regressions of small price changes and sales volume, using observations on national brand products

| Category | | **(1)** | **(2)** | **(3)** | **(4)** |
|---|---|---|---|---|---|
| Analgesics | Coefficient (Std.) | 0.0379*** (0.0038) | 0.0305*** (0.0031) | 0.0254*** (0.0029) | 0.0481*** (0.0063) |
| | Observations | 242,823 | 242,823 | 242,823 | 64,271 |
| Bath Soap | Coefficient (Std.) | 0.0455*** (0.0101) | 0.0502*** (0.0102) | 0.0471*** (0.0099) | 0.0892*** (0.0167) |
| | Observations | 30,747 | 30,747 | 30,747 | 6,285 |
| Bathroom Tissues | Coefficient (Std.) | 0.0357*** (0.0057) | 0.0178*** (0.0052) | 0.0155*** (0.0049) | 0.0335*** (0.0071) |
| | Observations | 305,784 | 305,784 | 305,784 | 75,080 |
| Beer | Coefficient (Std.) | 0.023*** (0.0015) | 0.0249*** (0.0012) | 0.0208*** (0.0012) | 0.0687*** (0.005) |
| | Observations | 457,795 | 457,795 | 457,795 | 56,283 |
| Bottled Juice | Coefficient (Std.) | 0.0552*** (0.0049) | 0.037*** (0.0034) | 0.0326*** (0.0035) | 0.0322*** (0.005) |
| | Observations | 838,222 | 838,222 | 838,222 | 212,093 |
| Canned Soup | Coefficient (Std.) | 0.0265*** (0.0044) | 0.0146*** (0.0037) | 0.0154*** (0.0035) | 0.021*** (0.0042) |
| | Observations | 890,105 | 890,105 | 890,105 | 260,495 |
| Canned Tuna | Coefficient (Std.) | 0.0353*** (0.0052) | 0.026*** (0.0044) | 0.0221*** (0.0041) | 0.0323*** (0.0048) |
| | Observations | 355,663 | 355,663 | 355,663 | 110,267 |
| Cereals | Coefficient (Std.) | 0.0235*** (0.0028) | 0.019*** (0.0023) | 0.0184*** (0.0023) | 0.0264*** (0.0038) |
| | Observations | 641,499 | 641,499 | 641,499 | 244,435 |
| Cheese | Coefficient (Std.) | 0.0382*** (0.0032) | 0.0215*** (0.0024) | 0.0184*** (0.0024) | 0.01*** (0.0032) |
| | Observations | 1,382,175 | 1,382,175 | 1,382,175 | 452,595 |
| Cigarettes | Coefficient (Std.) | 0.0152 (0.0093) | 0.0154 (0.0082) | 0.0151* (0.008) | 0.0141 (0.0086) |
| | Observations | 6,982 | 6,982 | 6,982 | 4,120 |
| Cookies | Coefficient (Std.) | 0.0419*** (0.0019) | 0.0369*** (0.0018) | 0.0317*** (0.0017) | 0.0536*** (0.0032) |
| | Observations | 1,172,710 | 1,172,710 | 1,172,710 | 202,932 |
| Crackers | Coefficient (Std.) | 0.0545*** (0.0036) | 0.0432*** (0.0033) | 0.0392*** (0.0031) | 0.0545*** (0.0065) |
| | Observations | 440,282 | 440,282 | 440,282 | 83,083 |
| Dish Detergent | Coefficient (Std.) | 0.0507*** (0.0039) | 0.0359*** (0.0031) | 0.0312*** (0.0031) | 0.0405*** (0.0047) |
| | Observations | 338,430 | 338,430 | 338,430 | 84,418 |
| Fabric Softener | Coefficient (Std.) | 0.0327*** (0.0039) | 0.0215*** (0.0036) | 0.0183*** (0.0037) | 0.0383*** (0.0051) |
| | Observations | 318,661 | 318,661 | 318,661 | 88,496 |
| Front-End-Candies | Coefficient (Std.) | 0.0166*** (0.0039) | 0.0092*** (0.0028) | 0.0082*** (0.0027) | 0.0114*** (0.0032) |
| | Observations | 485,323 | 485,323 | 485,323 | 153,759 |
| Frozen Dinners | Coefficient (Std.) | 0.0534*** (0.0027) | 0.0405*** (0.0024) | 0.0391*** (0.0024) | 0.0902*** (0.0059) |
| | Observations | 502,329 | 502,329 | 502,329 | 72,589 |

Table M1. (Cont.)

| Category | | **(1)** | **(2)** | **(3)** | **(4)** |
|---|---|---|---|---|---|
| Frozen Entrees | Coefficient (Std.) | 0.0352*** (0.0019) | 0.0298*** (0.0017) | 0.029*** (0.0017) | 0.0589*** (0.0032) |
| | Observations | 1,835,884 | 1,835,884 | 1,835,884 | 351,172 |
| Frozen Juices | Coefficient (Std.) | 0.0329*** (0.0037) | 0.0256*** (0.0032) | 0.023*** (0.0031) | 0.0295*** (0.0047) |
| | Observations | 540,070 | 540,070 | 540,070 | 128,103 |
| Grooming Products | Coefficient (Std.) | 0.0421*** (0.0024) | 0.0451*** (0.0022) | 0.0387*** (0.0022) | 0.0651*** (0.0061) |
| | Observations | 639,004 | 639,004 | 639,004 | 94,837 |
| Laundry Detergents | Coefficient (Std.) | 0.0188*** (0.0029) | 0.015*** (0.0025) | 0.012*** (0.0023) | 0.0238*** (0.0049) |
| | Observations | 544,928 | 544,928 | 544,928 | 137,209 |
| Oatmeal | Coefficient (Std.) | 0.0284*** (0.0079) | 0.0177*** (0.0057) | 0.0155*** (0.0056) | 0.0291*** (0.0093) |
| | Observations | 146,887 | 146,887 | 146,887 | 59,961 |
| Paper Towels | Coefficient (Std.) | 0.0381*** (0.0125) | 0.0264* (0.0133) | 0.0252* (0.0136) | 0.0292*** (0.0104) |
| | Observations | 216,280 | 216,280 | 216,280 | 47,035 |
| Refrigerated Juices | Coefficient (Std.) | 0.0312*** (0.0034) | 0.0205*** (0.0028) | 0.018*** (0.0027) | 0.0291*** (0.0044) |
| | Observations | 716,448 | 716,448 | 716,448 | 147,024 |
| Shampoos | Coefficient (Std.) | 0.0328*** (0.0014) | 0.0373*** (0.0014) | 0.0324*** (0.0013) | 0.0675*** (0.0043) |
| | Observations | 701,525 | 701,525 | 701,525 | 85,168 |
| Snack Crackers | Coefficient (Std.) | 0.0431*** (0.0034) | 0.0379*** (0.0031) | 0.0338*** (0.0028) | 0.0658*** (0.0042) |
| | Observations | 751,170 | 751,170 | 751,170 | 133,817 |
| Soaps | Coefficient (Std.) | 0.057*** (0.0053) | 0.0424*** (0.0045) | 0.0347*** (0.0042) | 0.0563*** (0.0058) |
| | Observations | 323,840 | 323,840 | 323,840 | 93,074 |
| Soft Drinks | Coefficient (Std.) | 0.027*** (0.0012) | 0.0255*** (0.0011) | 0.0212*** (0.001) | 0.0608*** (0.0029) |
| | Observations | 3,748,192 | 3,748,192 | 3,748,192 | 301,273 |
| Toothbrushes | Coefficient (Std.) | 0.029*** (0.0032) | 0.0322*** (0.0033) | 0.0269*** (0.0031) | 0.062*** (0.0064) |
| | Observations | 275,080 | 275,080 | 275,080 | 41,256 |
| Toothpastes | Coefficient (Std.) | 0.0286*** (0.0032) | 0.0277*** (0.0027) | 0.0238*** (0.0026) | 0.0581*** (0.0064) |
| | Observations | 570,338 | 570,338 | 570,338 | 86,903 |
| **Average coefficients** | | **0.036** | **0.029** | **0.025** | **0.045** |

Notes: The table reports the results of category-level fixed effect regressions of the probability of a small price change, using observations on national brand products. The dependent variable is "small price change," which equals 1 if a price change of product $i$ in store $s$ at time $t$ is less or equal to 10¢, and 0 otherwise. The main independent variable is the log of the average sales volume of product $i$ in store $s$ over the sample period. Column 1 reports the results of baseline regression that includes only the log of the average sales volume and the fixed effects for months, years, stores, and products. In column 2, we add the following controls: the log of the average price, the log of the absolute change in the wholesale price, a control for sale- and bounce-back prices, which we identify using a sales filter algorithm, and the competition zone of the store. In column 3, we add a dummy for 9-ending prices as an additional control. In column 4, we focus on regular prices by excluding the sale- and bounce-back prices. We estimate separate regressions for each product category, clustering the errors by product. * $p < 10\%$, ** $p < 5\%$, *** $p < 1\%$

Table M2. Category-level regressions of small price changes volume, using observations on private label products

| Category | | **(1)** | **(2)** | **(3)** | **(4)** |
|---|---|---|---|---|---|
| Analgesics | Coefficient (Std.) | 0.0295*** (0.0079) | 0.0137** (0.0049) | 0.0064 (0.0046) | 0.0266* (0.0137) |
| | Observations | 31,617 | 31,617 | 31,617 | 11,090 |
| Bath Soap | Coefficient (Std.) | 0.0096*** (0.0007) | 0.0065 (0.0056) | 0.0072 (0.0064) | |
| | Observations | 4,957 | 4,957 | 4,957 | |
| Bathroom Tissues | Coefficient (Std.) | 0.0553*** (0.0066) | 0.0056 (0.0134) | 0.0031 (0.0114) | 0.0035 (0.0158) |
| | Observations | 18,943 | 18,943 | 18,943 | 6,664 |
| Beer | Coefficient (Std.) | | | | |
| | Observations | | | | |
| Bottled Juice | Coefficient (Std.) | 0.0486*** (0.0053) | 0.0438*** (0.0051) | 0.0355*** (0.0053) | 0.0545*** (0.0099) |
| | Observations | 119,432 | 119,432 | 119,432 | 31,569 |
| Canned Soup | Coefficient (Std.) | 0.0256* (0.0135) | 0.0082 (0.008) | 0.0082 (0.0078) | 0.015 (0.0091) |
| | Observations | 55,997 | 55,997 | 55,997 | 17,732 |
| Canned Tuna | Coefficient (Std.) | 0.0776*** (0.0104) | 0.0449*** (0.0104) | 0.0397*** (0.0113) | 0.0474*** (0.0086) |
| | Observations | 14,748 | 14,748 | 14,748 | 4,410 |
| Cereals | Coefficient (Std.) | 0.0047 (0.0077) | 0.0009 (0.0085) | -0.0109 (0.0102) | 0.0491*** (0.0099) |
| | Observations | 81,201 | 81,201 | 81,201 | 15,169 |
| Cheese | Coefficient (Std.) | 0.0286*** (0.0066) | 0.0146** (0.0059) | 0.0077 (0.0056) | 0.013* (0.0075) |
| | Observations | 414,036 | 414,036 | 414,036 | 64,391 |
| Cigarettes | Coefficient (Std.) | | | | |
| | Observations | | | | |
| Cookies | Coefficient (Std.) | 0.0462*** (0.005) | 0.0374*** (0.0047) | 0.0299*** (0.0044) | 0.0504*** (0.0113) |
| | Observations | 177,701 | 177,701 | 177,701 | 25,616 |
| Crackers | Coefficient (Std.) | 0.0462*** (0.0051) | 0.0423** (0.0075) | 0.034* (0.0086) | 0.054** (0.013) |
| | Observations | 32,249 | 32,249 | 32,249 | 5,487 |
| Dish Detergent | Coefficient (Std.) | 0.0626*** (0.0077) | 0.0575*** (0.007) | 0.056*** (0.0078) | 0.0647*** (0.0099) |
| | Observations | 45,246 | 45,246 | 45,246 | 9,101 |
| Fabric Softener | Coefficient (Std.) | 0.0641*** (0.0095) | 0.0545*** (0.0095) | 0.0465*** (0.01) | 0.0703*** (0.0098) |
| | Observations | 42,022 | 42,022 | 42,022 | 10,860 |
| Front-End-Candies | Coefficient (Std.) | | | | |
| | Observations | | | | |
| Frozen Dinners | Coefficient (Std.) | | | | |
| | Observations | | | | |

Table M2. (Cont.)

| Category | | (1) | (2) | (3) | (4) |
|---|---|---|---|---|---|
| Frozen Entrees | Coefficient (Std.) | 0.0395*** (0.0075) | 0.0445*** (0.0075) | 0.0348*** (0.0064) | 0.0725*** (0.0188) |
| | Observations | 8,736 | 8,736 | 8,736 | 1,560 |
| Frozen Juices | Coefficient (Std.) | 0.0421*** (0.0076) | 0.0365*** (0.0066) | 0.0342*** (0.0065) | 0.0337* (0.0164) |
| | Observations | 118,148 | 118,148 | 118,148 | 21,915 |
| Grooming Products | Coefficient (Std.) | 0.0474*** (0.0107) | 0.0479** (0.0098) | 0.0398*** (0.0073) | 0.081** (0.0311) |
| | Observations | 17,603 | 17,603 | 17,603 | 2,941 |
| Laundry Detergents | Coefficient (Std.) | 0.0686** (0.0299) | 0.0659** (0.029) | 0.0611** (0.0272) | 0.0707*** (0.0248) |
| | Observations | 21,609 | 21,609 | 21,609 | 4,988 |
| Oatmeal | Coefficient (Std.) | 0.0219** (0.0084) | 0.009 (0.0095) | 0.0053 (0.0087) | 0.0635** (0.0259) |
| | Observations | 21,372 | 21,372 | 21,372 | 3,423 |
| Paper Towels | Coefficient (Std.) | 0.0187* (0.0252) | 0.0373** (0.0122) | 0.034** (0.0091) | 0.0504** (0.0104) |
| | Observations | 18,978 | 18,978 | 18,978 | 3,777 |
| Refrigerated Juices | Coefficient (Std.) | 0.0248** (0.011) | 0.0188 (0.0115) | 0.0128 (0.0115) | 0.0417*** (0.0129) |
| | Observations | 83,832 | 83,832 | 83,832 | 14,074 |
| Shampoos | Coefficient (Std.) | 0.0023 (0.0233) | 0.0078 (0.026) | 0.0063 (0.0215) | |
| | Observations | 1,319 | 1,319 | 1,319 | |
| Snack Crackers | Coefficient (Std.) | 0.0525*** (0.0068) | 0.0407*** (0.0057) | 0.0302*** (0.0052) | 0.0688*** (0.0145) |
| | Observations | 49,987 | 49,987 | 49,987 | 9,251 |
| Soaps | Coefficient (Std.) | | | | |
| | Observations | | | | |
| Soft Drinks | Coefficient (Std.) | 0.0568*** (0.0075) | 0.0231*** (0.0046) | 0.0204*** (0.0041) | 0.0324*** (0.0085) |
| | Observations | 511,920 | 511,920 | 511,920 | 38,424 |
| Toothbrushes | Coefficient (Std.) | -0.0029 (0.0125) | -0.0034 (0.013) | -0.0086 (0.0121) | 0.0425 (0.0323) |
| | Observations | 11,756 | 11,756 | 11,756 | 1,876 |
| Toothpastes | Coefficient (Std.) | 0.0735** (0.02) | 0.0533** (0.016) | 0.0388* (0.016) | 0.0235 (0.0394) |
| | Observations | 8,565 | 8,565 | 8,565 | 2,363 |
| **Average coefficients** | | **0.039** | **0.030** | **0.024** | **0.047** |

Notes: The table reports the results of category-level fixed effect regressions of the probability of a small price change, using observations on private label products. We drop categories if we do not have at least 500 observations on private label products. The dependent variable is "small price change," which equals 1 if a price change of product $i$ in store $s$ at time $t$ is less or equal to 10¢, and 0 otherwise. The main independent variable is the log of the average sales volume of product $i$ in store $s$ over the sample period. Column 1 reports the results of baseline regression that includes only the log of the average sales volume and the fixed effects for months, years, stores, and products. In column 2, we add the following controls: the log of the average price, the log of the absolute change in the wholesale price, a control for sale- and bounce-back prices, which we identify using a sales filter algorithm, and the competition zone of the store. In column 3, we add a dummy for 9-ending prices as an additional control. In column 4, we focus on regular prices by excluding the sale- and bounce-back prices. We estimate separate regressions for each product category, clustering the errors by product. * $p < 10\%$, ** $p < 5\%$, *** $p < 1\%$

Table M3. Category-level regressions of small price changes and sales volume, using observations on national brand products, prices that survived for 2 weeks

| Category | | (1) | (2) | (3) | (4) |
|---|---|---|---|---|---|
| Analgesics | Coefficient (Std.) | 0.0166*** (0.0045) | 0.0116*** (0.0043) | 0.0114*** (0.0043) | 0.0077*** (0.0087) |
| | Observations | 67,651 | 67,651 | 67,651 | 20,850 |
| Bath Soap | Coefficient (Std.) | 0.0129 (0.0146) | 0.0063 (0.0148) | 0.0059*** (0.0143) | -0.0203*** (0.0279) |
| | Observations | 5,805 | 5,805 | 5,805 | 1,366 |
| Bathroom Tissues | Coefficient (Std.) | 0.0537*** (0.0097) | 0.0176** (0.0093) | 0.0174*** (0.0092) | 0.0333*** (0.0111) |
| | Observations | 52,445 | 52,445 | 52,445 | 17,009 |
| Beer | Coefficient (Std.) | 0.002*** (0.0006) | 0.0043*** (0.0007) | 0.0043*** (0.0007) | 0.018*** (0.0055) |
| | Observations | 187,691 | 187,691 | 187,691 | 12,080 |
| Bottled Juice | Coefficient (Std.) | 0.0367*** (0.0086) | 0.0176*** (0.0074) | 0.0172*** (0.0075) | 0.0308*** (0.0093) |
| | Observations | 198,936 | 198,936 | 198,936 | 53,339 |
| Canned Soup | Coefficient (Std.) | 0.0005 (0.0099) | -0.006 (0.0096) | -0.0031*** (0.0093) | 0.0137*** (0.008) |
| | Observations | 219,520 | 219,520 | 219,520 | 89,102 |
| Canned Tuna | Coefficient (Std.) | 0.0054 (0.0064) | -0.0036 (0.0057) | -0.0038*** (0.0056) | 0.0089*** (0.0082) |
| | Observations | 108,716 | 108,716 | 108,716 | 30,428 |
| Cereals | Coefficient (Std.) | 0.0111* (0.0064) | 0.0118** (0.006) | 0.0117*** (0.006) | 0.0237*** (0.0072) |
| | Observations | 123,336 | 123,336 | 123,336 | 69,327 |
| Cheese | Coefficient (Std.) | 0.0085* (0.0044) | 0.0041* (0.0036) | 0.0038*** (0.0036) | 0.0143*** (0.0049) |
| | Observations | 291,896 | 291,896 | 291,896 | 80,488 |
| Cigarettes | Coefficient (Std.) | 0.0044 (0.0051) | 0.0021 (0.0049) | 0.0022*** (0.0048) | 0*** (0.0055) |
| | Observations | 24,553 | 24,553 | 24,553 | 20,692 |
| Cookies | Coefficient (Std.) | 0.0069*** (0.0018) | 0.0056*** (0.0019) | 0.0055*** (0.0019) | 0.0048*** (0.0037) |
| | Observations | 296,041 | 296,041 | 296,041 | 61,344 |
| Crackers | Coefficient (Std.) | 0.0006 (0.0034) | 0 (0.0032) | 0.0003*** (0.0032) | 0.0101*** (0.0069) |
| | Observations | 110,219 | 110,219 | 110,219 | 23,427 |
| Dish Detergent | Coefficient (Std.) | 0.0354*** (0.0074) | 0.0297*** (0.0066) | 0.0297*** (0.0065) | 0.029*** (0.0056) |
| | Observations | 72,857 | 72,857 | 72,857 | 24,354 |
| Fabric Softener | Coefficient (Std.) | 0.0177** (0.0074) | 0.0031 (0.0061) | 0.0036*** (0.006) | 0.0278*** (0.0081) |
| | Observations | 75,811 | 75,811 | 75,811 | 24,518 |
| Front-End-Candies | Coefficient (Std.) | 0.01** (0.0041) | -0.003* (0.0033) | -0.0028*** (0.0033) | 0.0017*** (0.003) |
| | Observations | 148,200 | 148,200 | 148,200 | 77,323 |
| Frozen Dinners | Coefficient (Std.) | 0.0512*** (0.0069) | 0.0389*** (0.0062) | 0.0406*** (0.0062) | 0.0758*** (0.0105) |
| | Observations | 52,893 | 52,893 | 52,893 | 12,287 |

## Table M3. (Cont.)

| Category | | **(1)** | **(2)** | **(3)** | **(4)** |
|---|---|---|---|---|---|
| Frozen Entrees | Coefficient (Std.) | 0.0222*** (0.0026) | 0.0163*** (0.0026) | 0.0165*** (0.0026) | 0.0225*** (0.0039) |
| | Observations | 343,898 | 343,898 | 343,898 | 116,594 |
| Frozen Juices | Coefficient (Std.) | 0.0144* (0.0077) | 0.0108** (0.0072) | 0.0123*** (0.007) | 0.0239*** (0.009) |
| | Observations | 102,582 | 102,582 | 102,582 | 34,179 |
| Grooming Products | Coefficient (Std.) | 0.0097*** (0.0034) | 0.0114*** (0.0033) | 0.0116*** (0.0033) | 0.0139*** (0.0115) |
| | Observations | 99,243 | 99,243 | 99,243 | 21,209 |
| Laundry Detergents | Coefficient (Std.) | 0.02*** (0.0044) | 0.0119*** (0.0036) | 0.0121*** (0.0036) | 0.0181*** (0.0058) |
| | Observations | 116,100 | 116,100 | 116,100 | 40,837 |
| Oatmeal | Coefficient (Std.) | 0.0111 (0.0128) | 0.0046 (0.0126) | 0.0025*** (0.0118) | 0.0556*** (0.0114) |
| | Observations | 23,181 | 23,181 | 23,181 | 13,112 |
| Paper Towels | Coefficient (Std.) | 0.028* (0.0163) | 0.0148 (0.0179) | 0.016*** (0.0178) | 0.0202*** (0.0155) |
| | Observations | 46,637 | 46,637 | 46,637 | 8,730 |
| Refrigerated Juices | Coefficient (Std.) | 0.0181** (0.0082) | 0.0113** (0.0084) | 0.0111*** (0.0082) | 0.0222*** (0.0122) |
| | Observations | 99,777 | 99,777 | 99,777 | 21,665 |
| Shampoos | Coefficient (Std.) | 0.0102*** (0.0025) | 0.0107*** (0.0025) | 0.0107*** (0.0025) | 0.0234*** (0.008) |
| | Observations | 87,969 | 87,969 | 87,969 | 16,041 |
| Snack Crackers | Coefficient (Std.) | 0.0025 (0.0028) | 0.0042** (0.0028) | 0.0042*** (0.0028) | 0.021*** (0.005) |
| | Observations | 168,620 | 168,620 | 168,620 | 35,998 |
| Soaps | Coefficient (Std.) | 0.0263*** (0.0088) | 0.0139** (0.008) | 0.0173*** (0.008) | 0.0516*** (0.0117) |
| | Observations | 56,710 | 56,710 | 56,710 | 16,872 |
| Soft Drinks | Coefficient (Std.) | 0.0095*** (0.002) | 0.0079*** (0.0019) | 0.0071*** (0.0019) | 0.0121*** (0.004) |
| | Observations | 183,882 | 183,882 | 183,882 | 39,371 |
| Toothbrushes | Coefficient (Std.) | 0.0135*** (0.0048) | 0.0102*** (0.0049) | 0.0097*** (0.0048) | 0.0182*** (0.0102) |
| | Observations | 49,837 | 49,837 | 49,837 | 12,879 |
| Toothpastes | Coefficient (Std.) | 0.001 (0.004) | 0.0002 (0.0039) | 0.0003*** (0.0039) | 0.0057*** (0.0083) |
| | Observations | 99,045 | 99,045 | 99,045 | 27,348 |
| **Average coefficients** | | **0.0159** | **0.0093** | **0.0095** | **0.0203** |

Notes: The table reports the results of category-level fixed effect regressions of the probability of a small price change, using observations on national brand products. The dependent variable is "small price change," which equals 1 if a price change of product $i$ in store $s$ at time $t$ is less or equal to 10¢, and 0 otherwise. The main independent variable is the log of the average sales volume of product $i$ in store $s$ over the sample period. Column 1 reports the results of baseline regression that includes only the log of the average sales volume and the fixed effects for months, years, stores, and products. In column 2, we add the following controls: the log of the average price, the log of the absolute change in the wholesale price, a control for sale- and bounce-back prices, which we identify using a sales filter algorithm, and the competition zone of the store. In column 3, we add a dummy for 9-ending prices as an additional control. In column 4, we focus on regular prices by excluding the sale- and bounce-back prices. We estimate separate regressions for each product category, clustering the errors by product. * $p < 10\%$, ** $p < 5\%$, *** $p < 1\%$

Table M4. Category-level regressions of small price changes and sales volume, using observations on private label products, prices that survived for 2 weeks

| Category | | (1) | (2) | (3) | (4) |
|---|---|---|---|---|---|
| Analgesics | Coefficient (Std.) | 0.0203*** (0.0111) | 0.019*** (0.0107) | 0.0182*** (0.0112) | 0.0382*** (0.0162) |
| | Observations | 6,800 | 6,800 | 6,800 | 3,879 |
| Bath Soap | Coefficient (Std.) | | | | |
| | Observations | | | | |
| Bathroom Tissues | Coefficient (Std.) | 0.0794*** (0.0263) | -0.0113*** (0.0209) | -0.0113*** (0.0189) | -0.0176*** (0.0067) |
| | Observations | 4,013 | 4,013 | 4,013 | 2,276 |
| Beer | Coefficient (Std.) | | | | |
| | Observations | | | | |
| Bottled Juice | Coefficient (Std.) | 0.0154*** (0.012) | 0.0228*** (0.0117) | 0.0308*** (0.0119) | 0.0986*** (0.0282) |
| | Observations | 25,921 | 25,921 | 25,921 | 6,676 |
| Canned Soup | Coefficient (Std.) | 0.0013*** (0.0169) | -0.0041*** (0.0133) | -0.0002*** (0.0131) | 0.0048*** (0.0178) |
| | Observations | 14,259 | 14,259 | 14,259 | 6,208 |
| Canned Tuna | Coefficient (Std.) | 0.137*** (0.033) | 0.1061*** (0.0256) | 0.1066*** (0.0258) | 0.1109*** (0.0336) |
| | Observations | 3,913 | 3,913 | 3,913 | 1,494 |
| Cereals | Coefficient (Std.) | -0.0686*** (0.0265) | -0.0732*** (0.0244) | -0.0814*** (0.0273) | 0.0235*** (0.0349) |
| | Observations | 17,751 | 17,751 | 17,751 | 3,462 |
| Cheese | Coefficient (Std.) | 0.0115*** (0.0081) | 0.009*** (0.008) | 0.0082*** (0.0079) | -0.0307*** (0.0134) |
| | Observations | 65,783 | 65,783 | 65,783 | 12,270 |
| Cigarettes | Coefficient (Std.) | | | | |
| | Observations | | | | |
| Cookies | Coefficient (Std.) | 0.0387*** (0.0127) | 0.0355*** (0.0128) | 0.0376*** (0.0127) | 0.0503*** (0.0176) |
| | Observations | 21,891 | 21,891 | 21,891 | 4,743 |
| Crackers | Coefficient (Std.) | -0.0006*** (0.0061) | -0.0007*** (0.0056) | -0.0006*** (0.006) | -0.0082*** (0.0284) |
| | Observations | 5,439 | 5,439 | 5,439 | 1,344 |
| Dish Detergent | Coefficient (Std.) | 0.0023*** (0.0175) | 0.0085*** (0.0161) | 0.012*** (0.016) | -0.0118*** (0.0269) |
| | Observations | 12,365 | 12,365 | 12,365 | 2,381 |
| Fabric Softener | Coefficient (Std.) | -0.005*** (0.0071) | -0.0079*** (0.0062) | -0.0075*** (0.0059) | 0.0126*** (0.0249) |
| | Observations | 9,526 | 9,526 | 9,526 | 2,970 |
| Front-End-Candies | Coefficient (Std.) | | | | |
| | Observations | | | | |
| Frozen Dinners | Coefficient (Std.) | | | | |
| | Observations | | | | |

Table M4. (Cont.)

| Category | | **(1)** | **(2)** | **(3)** | **(4)** |
|---|---|---|---|---|---|
| Frozen Entrees | Coefficient (Std.) | 0.0461*** (0.0236) | 0.0467*** (0.024) | 0.0397*** (0.0211) | 0.0328*** (0.0334) |
| | Observations | 1,325 | 1,325 | 1,325 | 450 |
| Frozen Juices | Coefficient (Std.) | 0.0025*** (0.0202) | -0.0032*** (0.0181) | -0.0033*** (0.0176) | 0.0375*** (0.0199) |
| | Observations | 16,000 | 16,000 | 16,000 | 6,338 |
| Grooming Products | Coefficient (Std.) | 0.0262*** (0.0086) | 0.027*** (0.0076) | 0.0276*** (0.0077) | 0.0849*** (0.0298) |
| | Observations | 2,701 | 2,701 | 2,701 | 893 |
| Laundry Detergents | Coefficient (Std.) | 0.0814*** (0.066) | 0.069*** (0.0601) | 0.0674*** (0.0555) | 0.0153*** (0.0245) |
| | Observations | 5,466 | 5,466 | 5,466 | 1,284 |
| Oatmeal | Coefficient (Std.) | 0.0819*** (0.0732) | 0.0648*** (0.0573) | 0.0636*** (0.0595) | -0.0051*** (0.0532) |
| | Observations | 2,342 | 2,342 | 2,342 | 493 |
| Paper Towels | Coefficient (Std.) | 0.1087*** (0.0483) | 0.0922*** (0.0523) | 0.0834*** (0.0476) | 0.069*** (0.0399) |
| | Observations | 1,562 | 1,562 | 1,562 | 513 |
| Refrigerated Juices | Coefficient (Std.) | -0.0393*** (0.0266) | -0.0583*** (0.0232) | -0.06*** (0.0224) | -0.0653*** (0.0435) |
| | Observations | 9,188 | 9,188 | 9,188 | 2,040 |
| Shampoos | Coefficient (Std.) | -0.2464*** (0.1073) | -0.1186*** (0.1087) | -0.1181*** (0.1089) | 0*** (0) |
| | Observations | 224 | 224 | 224 | 58 |
| Snack Crackers | Coefficient (Std.) | -0.0036*** (0.0136) | -0.0088*** (0.013) | -0.01*** (0.013) | -0.0395*** (0.0204) |
| | Observations | 7,907 | 7,907 | 7,907 | 2125 |
| Soaps | Coefficient (Std.) | | | | |
| | Observations | | | | |
| Soft Drinks | Coefficient (Std.) | 0.0765*** (0.0181) | 0.0082*** (0.006) | 0.0081*** (0.0058) | 0.0035*** (0.0064) |
| | Observations | 59,955 | 59,955 | 59,955 | 10,618 |
| Toothbrushes | Coefficient (Std.) | -0.0185*** (0.0133) | -0.0205*** (0.0132) | -0.0215*** (0.0129) | -0.07*** (0.0368) |
| | Observations | 2,348 | 2,348 | 2,348 | 816 |
| Toothpastes | Coefficient (Std.) | -0.0576*** (0.0472) | -0.0386*** (0.0342) | -0.0364*** (0.0335) | -0.0096*** (0.051) |
| | Observations | 1,800 | 1,800 | 1,800 | 691 |
| **Average coefficients** | | **0.0126** | **0.0071** | **0.0066** | **0.0141** |

<u>Notes</u>: The table reports the results of category-level fixed effect regressions of the probability of a small price change, using observations on national brand products. We drop categories if we do not have at least 500 observations on private-label products. The dependent variable is "small price change," which equals 1 if a price change of product $i$ in store $s$ at time $t$ is less or equal to 10¢, and 0 otherwise. The main independent variable is the log of the average sales volume of product $i$ in store $s$ over the sample period. Column 1 reports the results of baseline regression that includes only the log of the average sales volume and the fixed effects for months, years, stores, and products. In column 2, we add the following controls: the log of the average price, the log of the absolute change in the wholesale price, a control for sale- and bounce-back prices, which we identify using a sales filter algorithm, and the competition zone of the store. In column 3, we add a dummy for 9-ending prices as an additional control. In column 4, we focus on regular prices by excluding the sale- and bounce-back prices. We estimate separate regressions for each product category, clustering the errors by product. * $p < 10\%$, ** $p < 5\%$, *** $p < 1\%$

**Appendix N. Sales volumes, small price changes, and holidays**

Levy et al. (2010) argue that menu costs are higher during the holiday season than at other times. As they note, store traffic is higher during holidays than other times and, consequently, tasks such as restocking shelves, running cash registers, cleaning and bagging, etc., become more urgent. Therefore, the opportunity cost of price adjustment increases during holiday periods.

If menu costs are higher, then we should observe fewer small price changes, possibly weakening the correlation between sales volumes and small price changes. To explore this, we focus on the holiday period, which, following Warner and Barsky (1995) and Levy et al. (2010), we define as the period starting the week before Thanksgiving through the week of Christmas, a total of six weeks.

Table N1 presents the results of regressions equivalent to the regressions in Table 3 in the paper. This time, however, we use only observations on the holiday period, defined as above. The regressions take the following form:

$$small\ price\ change_{i,s,t} = \alpha + \beta \ln(average\ sales\ volume_{i,s}) + \gamma \mathbf{X}_{i,s,t} \\ + month_t + year_t + \delta_s + \mu_i + u_{i,s,t} \quad \text{(N1)}$$

where *small price change* is a dummy that equals 1 if a price change of product *i* in store *s* at time *t* is less or equal to 10¢, and 0 otherwise. The *average sales volume* is the average sales volume of product *i* in store *s* over the sample period. **X** is a matrix of other control variables. *Month* and *year* are fixed effects for the month and the year of the price change. $\delta$ and $\mu$ are fixed effects for stores and products, respectively, and *u* is an i.i.d error term. We estimate separate regressions for each product category, clustering the errors by product.

As we do in the paper, we use observations on price changes only if we observe the price in both weeks *t* and *t* + 1 and the post change price remained unchanged for at least 2 weeks. The values in Table N1 are the coefficients of the log of the average sales volume when we use the sample of national brands. In column 1, the control variables are the log of the average sales volume, and the dummies for months, years, stores, and products. We find that 19 out of the 29 coefficients of the log of the average sales volume are positive. Out of the 19 positive coefficients, 4 are statistically significant and 5 are marginally statistically significant. Out of the 10 negative coefficients, 3 are statistically significant and 4 more are marginally significant. The average coefficient is 0.010, suggesting that a 1% increase in the sales volume is associated with a 1.0% increase in the likelihood of a small price change.

In column 2, we add controls for the log of the average price, the log of the absolute change in the wholesale price, and a control for sale- and bounce-back prices, which we identify using the sales filter algorithm of Fox and Syed (2016). 20 of the 29 coefficients are positive. 5 of the positive coefficients are statistically significant and 4 are marginally significant. Out of the 9 negative coefficients, 3 are statistically significant, and 2 more are marginally significant. The average coefficient is 0.005, suggesting that a 1% increase in the sales volume is associated with a 0.5% increase in the likelihood of a small price change.

In column 3, we add a control for 9-ending prices. We find that 19 out of the 29 coefficients of the log of the average sales volume are positive. Out of the 19 positive coefficients, 5 are statistically significant and 3 are marginally statistically significant. Out of the 10 negative coefficients, 3 are statistically significant, and 2 more are marginally significant. The average coefficient is 0.006, suggesting that a 1% increase in the sales volume is associated with a 0.6% increase in the likelihood of a small price change.

As a further control for the possible effects of sales on the results we report, in column 4 we focus on regular prices by excluding all sale- and bounce-back prices. When we focus on regular prices, 18 of the 29 coefficients are positive. 7 of them are statistically significant, and 2 more are marginally significant. Out of the 11 negative coefficients, 3 are statistically significant, and 2 more are marginally significant. The average coefficient is 0.008, suggesting that a 1% increase in the sales volume is associated with a 0.8% increase in the likelihood of a small price change.

We thus find that the positive correlation between sales volumes and small price changes seems to be weaker during the holiday periods, which is consistent with a high cost of price changes during holidays (Levy et al. 2010). However, because the number of observations is relatively small, these results require further research.

Table N1. Category-level regressions of small price changes using observations on products sold in the Thanksgiving-Christmas holiday period

| Category | | (1) | (2) | (3) | (4) |
|---|---|---|---|---|---|
| Analgesics | Coefficient (Std.) | 0.0211** (0.0112) | 0.0156** (0.0105) | 0.0161*** (0.0105) | 0.0352*** (0.0171) |
| | Observations | 8,769 | 8,769 | 8,769 | 2,982 |
| Bath Soap | Coefficient (Std.) | -0.001 (0.0144) | 0.0136* (0.0151) | 0.0106*** (0.0164) | 0.0402 (0.1183) |
| | Observations | 622 | 622 | 622 | 116 |
| Bathroom Tissues | Coefficient (Std.) | 0.0555*** (0.0224) | 0.0007 (0.0185) | 0.0127*** (0.0177) | 0.0078 (0.0259) |
| | Observations | 6,334 | 6,334 | 6,334 | 1,685 |
| Beer | Coefficient (Std.) | -0.0005* (0.0006) | 0.0007* (0.0005) | 0.0007*** (0.0005) | 0.0116* (0.0109) |
| | Observations | 16,773 | 16,773 | 16,773 | 669 |
| Bottled Juice | Coefficient (Std.) | 0.0111* (0.0099) | 0.0042 (0.0096) | 0.0038*** (0.0097) | -0.001 (0.0134) |
| | Observations | 26,881 | 26,881 | 26,881 | 9,610 |
| Canned Soup | Coefficient (Std.) | -0.0123* (0.0104) | -0.0158** (0.0103) | -0.0169*** (0.0098) | 0.0097 (0.0123) |
| | Observations | 28,293 | 28,293 | 28,293 | 8,944 |
| Canned Tuna | Coefficient (Std.) | 0.0006 (0.0103) | -0.0034 (0.0103) | -0.0047*** (0.0103) | 0 (0.0126) |
| | Observations | 12,860 | 12,860 | 12,860 | 3,436 |
| Cereals | Coefficient (Std.) | 0.0204** (0.0148) | 0.0175* (0.0139) | 0.0179*** (0.0139) | 0.0226** (0.013) |
| | Observations | 15,947 | 15,947 | 15,947 | 10,547 |
| Cheese | Coefficient (Std.) | 0.0065* (0.007) | 0.002 (0.0065) | 0.0019*** (0.0065) | 0.0016 (0.0099) |
| | Observations | 42,339 | 42,339 | 42,339 | 10,048 |
| Cigarettes | Coefficient (Std.) | -0.0132* (0.0128) | -0.0101* (0.0097) | -0.0101*** (0.0096) | -0.0096* (0.0095) |
| | Observations | 2,403 | 2,403 | 2,403 | 2,241 |
| Cookies | Coefficient (Std.) | 0.0035 (0.0051) | 0.0029 (0.005) | 0.0022*** (0.0048) | -0.0018 (0.0071) |
| | Observations | 21,508 | 21,508 | 21,508 | 7,429 |
| Crackers | Coefficient (Std.) | 0.0048* (0.0052) | 0.0032 (0.0054) | 0.0041*** (0.0055) | -0.008* (0.0088) |
| | Observations | 7,263 | 7,263 | 7,263 | 1,861 |
| Dish Detergent | Coefficient (Std.) | 0.0376*** (0.009) | 0.0308*** (0.0084) | 0.0325*** (0.0082) | 0.0089** (0.0061) |
| | Observations | 10,342 | 10,342 | 10,342 | 3,385 |
| Fabric Softener | Coefficient (Std.) | 0.008 (0.0182) | -0.0117 (0.0148) | -0.0112*** (0.0149) | -0.0135 (0.0171) |
| | Observations | 8,121 | 8,121 | 8,121 | 3,373 |
| Front-End-Candies | Coefficient (Std.) | -0.0121** (0.0085) | -0.0113*** (0.0045) | -0.0069*** (0.0045) | -0.0031 (0.0038) |
| | Observations | 15,148 | 15,148 | 15,148 | 6,289 |
| Frozen Dinners | Coefficient (Std.) | 0.0132* (0.014) | 0.0056 (0.0142) | 0.0038*** (0.014) | 0.0099** (0.0089) |
| | Observations | 3,534 | 3,534 | 3,534 | 596 |

Table N1. (Cont.)

| Category | | (1) | (2) | (3) | (4) |
|---|---|---|---|---|---|
| Frozen Entrees | Coefficient (Std.) | 0.0069* (0.0068) | -0.0052 (0.0078) | -0.0055*** (0.0076) | -0.0042 (0.0096) |
| | Observations | 21,998 | 21,998 | 21,998 | 9,784 |
| Frozen Juices | Coefficient (Std.) | -0.0021 (0.0134) | 0.0003 (0.015) | -0.0027*** (0.0144) | 0.0017 (0.0231) |
| | Observations | 13,388 | 13,388 | 13,388 | 4,499 |
| Grooming Products | Coefficient (Std.) | 0.0041 (0.008) | 0.0068 (0.0087) | 0.0067*** (0.0085) | 0.0083 (0.0161) |
| | Observations | 9,078 | 9,078 | 9,078 | 2,327 |
| Laundry Detergents | Coefficient (Std.) | 0.0186** (0.0124) | 0.0136* (0.0107) | 0.0137*** (0.0107) | 0.0194** (0.0118) |
| | Observations | 13,130 | 13,130 | 13,130 | 5,979 |
| Oatmeal | Coefficient (Std.) | 0.0219** (0.0237) | 0.0104 (0.0195) | 0.0106*** (0.0191) | 0.0787*** (0.0311) |
| | Observations | 2,854 | 2,854 | 2,854 | 1,082 |
| Paper Towels | Coefficient (Std.) | 0.0636** (0.037) | 0.0578** (0.0359) | 0.0581*** (0.0361) | 0.0614** (0.038) |
| | Observations | 5,633 | 5,633 | 5,633 | 1,284 |
| Refrigerated Juices | Coefficient (Std.) | 0.0026 (0.0143) | 0.0027 (0.014) | 0.0025*** (0.0138) | 0.0259** (0.0184) |
| | Observations | 15,643 | 15,643 | 15,643 | 4,026 |
| Shampoos | Coefficient (Std.) | -0.0042** (0.0029) | 0.0009 (0.0031) | 0.0009*** (0.0031) | 0.0074 (0.011) |
| | Observations | 8,669 | 8,669 | 8,669 | 884 |
| Snack Crackers | Coefficient (Std.) | -0.0053* (0.0046) | -0.0054** (0.005) | -0.0047*** (0.0049) | -0.0022 (0.004) |
| | Observations | 20,929 | 20,929 | 20,929 | 4,015 |
| Soaps | Coefficient (Std.) | 0.0422*** (0.0171) | 0.0371 (0.0171) | 0.042*** (0.0159) | 0.0051 (0.0289) |
| | Observations | 4,592 | 4,592 | 4,592 | 1,032 |
| Soft Drinks | Coefficient (Std.) | 0.0261*** (0.0072) | 0.008** (0.0045) | 0.0106*** (0.005) | -0.0233*** (0.0078) |
| | Observations | 31,935 | 31,935 | 31,935 | 5,062 |
| Toothbrushes | Coefficient (Std.) | -0.0023 (0.0105) | -0.0028*** (0.0101) | -0.0042*** (0.0107) | -0.0199** (0.0142) |
| | Observations | 6,748 | 6,748 | 6,748 | 2,056 |
| Toothpastes | Coefficient (Std.) | -0.0307*** (0.0086) | -0.0227*** (0.0081) | -0.0216*** (0.0079) | -0.0386*** (0.0129) |
| | Observations | 12,819 | 12,819 | 12,819 | 5,182 |
| **Average coefficients** | | **0.0098** | **0.0050** | **0.0056** | **0.0079** |

<u>Notes</u>: The table reports the results of category-level fixed effect regressions of the probability of a small price change, using observations on products sold during the holiday period. We define the holiday period as starting in the week prior to Thanksgiving and continuing through Christmas. The dependent variable is "small price change," which equals 1 if a price change of product $i$ in store $s$ at time $t$ is less or equal to 10¢, and 0 otherwise. The main independent variable is the log of the average sales volume of product $i$ in store $s$ over the sample period. Column 1 reports the results of baseline regression that includes only the log of the average sales volume and the fixed effects for months, years, stores, and products. In column 2, we add the following controls: the log of the average price, the log of the absolute change in the wholesale price, a control for sale- and bounce-back prices, which we identify using a sales filter algorithm and the competition zone of the store. In column 3, we add a dummy for 9-ending prices as an additional control. In column 4, we focus on regular prices by excluding the sale- and bounce-back prices.

We estimate separate regressions for each product category, clustering the errors by product. $* \ p < 10\%$, $** \ p < 5\%$, $*** \ p < 1\%$

**Appendix O. The likelihood of a price change, irrespective of its size**

In the paper, we show that in Barro's (1972) menu cost model, an increase in the sales volume reduces the width of the *S-s* band, leading to (a) more frequent small price changes, and (b) more frequent price changes (of any size). In the paper, we report evidence supporting the first prediction. In this appendix, we show that the data supports the second prediction as well.

As a first test, we look within categories. Table O1 presents the results of regressions equivalent to the regressions in Table 3 in the paper. The regressions take the form:

$$price\ change_{i,s,t} = \alpha + \beta\ ln(\ average\ sales\ volume_{i,s}) + \gamma X_{i,s,t}$$
$$+month_t + year_t + \delta_s + \mu_i + u_{i,s,t} \qquad (O1)$$

where *price change* is a dummy that equals 1 if the price of product *i* in store *s* changed at time *t*, and 0 otherwise. The *average sales volume* is the average sales volume of product *i* in store *s* over the sample period. **X** is a matrix of other control variables. *Month* and *year* are fixed effects for the month and the year of the price change. $\delta$ and $\mu$ are fixed effects for stores and products, respectively, and *u* is an i.i.d error term. We estimate separate regressions for each product category, clustering the errors by product. As we do in the paper, we use observations on price changes only if we observe the price in both weeks *t* and *t* + 1 and the post change price remained unchanged for at least 2 weeks.

The values in the table are the coefficients of the log of the average sales volume. In column 1, the control variables are the log of the average sales volume, and dummies for months, years, stores, and products. We find that 24 of the 29 coefficients of the average sales volume are positive. 20 of the 24 positive coefficients are statistically significant. The 5 negative coefficients are also statistically significant. The average coefficient is 0.004, suggesting that a 1% increase in the average sales volume is associated with a 0.4% increase in the likelihood of a price change.

In column 2, we add controls for the log of the average price, the log of the absolute change in the wholesale price, and a control for sale- and bounce-back prices, which we identify using the sales filter algorithm of Fox and Syed (2016). We find that 24 of the 29 coefficients of the average sales volume are positive. 21 of the 24 positive coefficients are statistically significant. 3 out of the 5 negative coefficients are statistically significant. The average coefficient is 0.005, suggesting that a 1% increase in the average sales volume is associated with a 0.5% increase in the likelihood of a price change.

In column 3, we add a control for 9-ending prices. 24 of the 29 coefficients of the average sales volume are positive. 21 of the 24 positive coefficients are statistically significant. 3 out of the 5 negative coefficients are statistically significant. The average coefficient is 0.005, suggesting that a 1% increase in the average sales volume is associated with a 0.5% increase in the likelihood of a price change.

As a further control for the effects of sales on the results, in column 4 we focus on regular prices by excluding all sale and bounce-back prices. 27 of the 29 coefficients are positive. 25 of the positive coefficients are statistically significant, and 1 more is marginally statistically significant. The average coefficient is 0.002, suggesting that a 1% increase in the sales volume is associated with a 0.2% increase in the likelihood of a price change.

As a second test, we conduct a product-level test, similar to the test that its results are reported in Table 4 in the paper. To conduct the test, we calculate for each product in each of the 29 product categories the average weekly sales volume and the share of small price changes in each store it was offered. Many products in the sample were offered for only short periods of time, or only in a small number of stores. To avoid biases, we drop products for which we do not have information for at least 30 stores.

Using these data, we estimate for each product in each category, an OLS regression with robust standard errors. The dependent variable is the share of price changes out of all observations for the product in each store. The independent variable is the average sales volume of the product in each store. The estimation results are reported in Table O2.

Column 1 presents for each product category, the average of the estimated coefficients. Column 2 presents the total number of coefficients. Column 3 presents the percentage of the positive coefficients. Column 4 presents the number of statistically significant coefficients. Column 5 presents the percentage of positive and significant coefficients out of the total number of statistically significant coefficients.

According to the figures in the table, the average coefficients are positive for all 29

product categories. Further, the number of positive coefficients far exceeds the number of negative coefficients: On average, 83.4% of all the coefficients are positive.

Focusing on statistically significant coefficients, we find a far greater number of positive coefficients than negative coefficients. On average, 90.5% of all statistically significant coefficients are positive. In other words, for the overwhelming majority of the individual products in our sample, we find a positive relationship between sales volume and the likelihood of a price change.

In summary, we find that as predicted by Barro's (1972) model, an increase in the sales volume is associated with an increase in the likelihood of a price change, in addition to an increase in the likelihood of a small price change.

# Table O1. The likelihood of a price change

| Category | | **(1)** | **(2)** | **(3)** | **(4)** |
|---|---|---|---|---|---|
| Analgesics | Coefficient (Std.) | 0.0095*** (0.0006) | 0.0102*** (0.0007) | 0.0103*** (0.0007) | 0.004*** (0.0003) |
| | Observations | 3,060,156 | 3,019,519 | 3,019,519 | 2,615,923 |
| Bath Soap | Coefficient (Std.) | 0.0097*** (0.0008) | 0.0103*** (0.0009) | 0.0102*** (0.0009) | 0.0037*** (0.0004) |
| | Observations | 418,097 | 402,960 | 402,960 | 336,180 |
| Bathroom Tissues | Coefficient (Std.) | -0.0032*** (0.0008) | -0.0026*** (0.0008) | -0.0026*** (0.0008) | 0.0002*** (0.0004) |
| | Observations | 1,159,016 | 1,149,177 | 1,149,177 | 831,301 |
| Beer | Coefficient (Std.) | 0.0307*** (0.0016) | 0.0273*** (0.0014) | 0.0293*** (0.0014) | 0.002*** (0.0003) |
| | Observations | 1,970,266 | 1,940,556 | 1,940,556 | 1,174,512 |
| Bottled Juice | Coefficient (Std.) | -0.0024*** (0.0007) | 0.0005*** (0.0007) | 0.0006*** (0.0007) | 0.0008*** (0.0003) |
| | Observations | 4,325,024 | 4,288,625 | 4,288,625 | 3,205,484 |
| Canned Soup | Coefficient (Std.) | 0.0017*** (0.0006) | 0.0026*** (0.0006) | 0.0025*** (0.0006) | 0.0012*** (0.0003) |
| | Observations | 5,551,684 | 5,518,976 | 5,518,976 | 4,539,808 |
| Canned Tuna | Coefficient (Std.) | 0.0032*** (0.0009) | 0.0052*** (0.001) | 0.0053*** (0.001) | 0.0026*** (0.0004) |
| | Observations | 2,403,558 | 2,383,604 | 2,383,604 | 1,875,309 |
| Cereals | Coefficient (Std.) | -0.001*** (0.0003) | -0.0014*** (0.0003) | -0.0014*** (0.0003) | -0.0003*** (0.0002) |
| | Observations | 4,751,202 | 4,714,708 | 4,714,708 | 4,127,993 |
| Cheese | Coefficient (Std.) | 0.0013*** (0.0005) | 0.0029*** (0.0005) | 0.0032*** (0.0005) | 0.0011*** (0.0003) |
| | Observations | 6,810,625 | 6,763,438 | 6,763,438 | 4,961,570 |
| Cigarettes | Coefficient (Std.) | 0.0069*** (0.0003) | 0.0068*** (0.0003) | 0.0068*** (0.0003) | 0.0061*** (0.0003) |
| | Observations | 1,810,615 | 1,774,701 | 1,774,701 | 1,742,604 |
| Cookies | Coefficient (Std.) | 0.004*** (0.0004) | 0.0055*** (0.0004) | 0.0057*** (0.0004) | 0.0022*** (0.0002) |
| | Observations | 7,635,071 | 7,556,886 | 7,556,886 | 5,821,862 |
| Crackers | Coefficient (Std.) | 0.0103*** (0.0011) | 0.0128*** (0.0012) | 0.013*** (0.0012) | 0.0044*** (0.0004) |
| | Observations | 2,245,703 | 2,224,614 | 2,224,614 | 1,588,598 |
| Dish Detergent | Coefficient (Std.) | 0.0034*** (0.0007) | 0.0042*** (0.0007) | 0.0042*** (0.0007) | 0.0019*** (0.0003) |
| | Observations | 2,183,582 | 2,161,641 | 2,161,641 | 1,744,461 |
| Fabric Softener | Coefficient (Std.) | 0.0021*** (0.0005) | 0.0032*** (0.0007) | 0.0032*** (0.0006) | 0.0018*** (0.0003) |
| | Observations | 2,296,612 | 2,271,465 | 2,271,465 | 1,877,718 |
| Front-End-Candies | Coefficient (Std.) | 0.0022*** (0.0003) | 0.0039*** (0.0004) | 0.0039*** (0.0004) | 0.0033*** (0.0003) |
| | Observations | 4,475,750 | 4,441,325 | 4,441,325 | 3,948,230 |
| Frozen Dinners | Coefficient (Std.) | 0.0005*** (0.0007) | 0.0028*** (0.0007) | 0.0027*** (0.0007) | 0.0027*** (0.0004) |
| | Observations | 1,654,053 | 1,634,182 | 1,634,182 | 1,061,943 |

Table O1. (Cont.)

| Category | | (1) | (2) | (3) | (4) |
|---|---|---|---|---|---|
| Frozen Entrees | Coefficient (Std.) | 0.0025*** (0.0004) | 0.0068*** (0.0005) | 0.0068*** (0.0005) | 0.0044*** (0.0003) |
| | Observations | 7,232,080 | 7,164,744 | 7,164,744 | 5,163,065 |
| Frozen Juices | Coefficient (Std.) | -0.002*** (0.0008) | -0.0004*** (0.0007) | -0.0004*** (0.0007) | 0.001*** (0.0005) |
| | Observations | 2,387,420 | 2,373,678 | 2,373,678 | 1,700,508 |
| Grooming Products | Coefficient (Std.) | 0.011*** (0.0005) | 0.0107*** (0.0005) | 0.0109*** (0.0005) | 0.003*** (0.0002) |
| | Observations | 4,065,694 | 3,980,757 | 3,980,757 | 2,937,437 |
| Laundry Detergents | Coefficient (Std.) | 0.0023*** (0.0005) | 0.0035*** (0.0006) | 0.0036*** (0.0006) | 0.0017*** (0.0003) |
| | Observations | 3,303,174 | 3,258,164 | 3,258,164 | 2,616,474 |
| Oatmeal | Coefficient (Std.) | 0.0008*** (0.0009) | 0.0012*** (0.0012) | 0.0012*** (0.0012) | 0.0013*** (0.0007) |
| | Observations | 981,263 | 973,819 | 973,819 | 839,966 |
| Paper Towels | Coefficient (Std.) | 0.0006*** (0.001) | 0*** (0.001) | 0*** (0.001) | 0*** (0.0005) |
| | Observations | 948,871 | 937,197 | 937,197 | 672,784 |
| Refrigerated Juices | Coefficient (Std.) | -0.0025*** (0.0007) | -0.0019*** (0.0007) | -0.0015*** (0.0007) | 0.0008*** (0.0004) |
| | Observations | 2,182,989 | 2,165,804 | 2,165,804 | 1,363,980 |
| Shampoos | Coefficient (Std.) | 0.0095*** (0.0003) | 0.0092*** (0.0003) | 0.0094*** (0.0003) | 0.0026*** (0.0001) |
| | Observations | 4,676,790 | 4,535,601 | 4,535,601 | 3,330,183 |
| Snack Crackers | Coefficient (Std.) | 0.0046*** (0.0006) | 0.0058*** (0.0007) | 0.006*** (0.0007) | 0.0017*** (0.0003) |
| | Observations | 3,515,192 | 3,484,645 | 3,484,645 | 2,501,842 |
| Soaps | Coefficient (Std.) | 0.0005*** (0.0005) | 0.0009*** (0.0006) | 0.0009*** (0.0006) | 0.0013*** (0.0003) |
| | Observations | 1,835,196 | 1,810,103 | 1,810,103 | 1,464,608 |
| Soft Drinks | Coefficient (Std.) | 0.0019*** (0.0002) | 0.0021*** (0.0002) | 0.0024*** (0.0002) | 0.0013*** (0.0002) |
| | Observations | 10,807,191 | 10,702,594 | 10,702,594 | 5,499,044 |
| Toothbrushes | Coefficient (Std.) | 0.0107*** (0.0006) | 0.011*** (0.0007) | 0.0111*** (0.0007) | 0.0049*** (0.0004) |
| | Observations | 1,854,983 | 1,825,943 | 1,825,943 | 1,354,698 |
| Toothpastes | Coefficient (Std.) | 0.007*** (0.0005) | 0.0072*** (0.0006) | 0.0073*** (0.0006) | 0.0031*** (0.0003) |
| | Observations | 3,003,392 | 2,964,185 | 2,964,185 | 2,234,909 |
| **Average coefficients** | | **0.0043** | **0.0052** | **0.0053** | **0.0022** |

<u>Notes:</u> The table reports the results of category-level fixed effect regressions of the probability of a price change. The dependent variable is "price change," of product $i$ in store $s$ at time $t$ which equals 1 if a price of product $i$ in store $s$ changes at time $t$ and 0 otherwise. The main independent variable is the log of the average sales volume of product $i$ in store $s$ over the sample period. Column 1 reports the results of the baseline regression that includes only the average sales volume and the fixed effects for months, years, stores, and products. In column 2, we add the following controls: the log of the average price, the log of the absolute change in the wholesale price, and a control for sale- and bounce-back prices, which we identify using a sales filter algorithm. In column 3, we add a dummy for 9-ending prices as an additional control. In column 4, we focus on regular prices by excluding the sale- and bounce-back prices. We estimate separate regressions for each product category, clustering the errors by product. * $p < 10\%$, ** $p < 5\%$, *** $p < 1\%$

Table O2. Product-level regressions of the % of small price changes and sales volume by categories, including controls

| Product Category | Average coefficient (1) | No. of coefficients (2) | % positive coefficients (3) | No. of significant coefficients (4) | % positive and significant coefficients (5) |
|---|---|---|---|---|---|
| Analgesics | 0.0037 | 461 | 95.01% | 315 | 99.68% |
| Bath Soaps | 0.0050 | 109 | 90.83% | 59 | 100.00% |
| Bathroom tissues | 0.0021 | 112 | 59.82% | 39 | 53.85% |
| Beers | 0.0079 | 414 | 96.86% | 330 | 100.00% |
| Bottled juices | 0.0061 | 418 | 70.33% | 216 | 79.17% |
| Canned soups | 0.0064 | 368 | 73.37% | 164 | 85.98% |
| Canned tuna | 0.0090 | 219 | 83.56% | 138 | 92.03% |
| Cereals | 0.0012 | 408 | 62.99% | 139 | 67.63% |
| Cheese | 0.0053 | 529 | 72.59% | 286 | 84.27% |
| Cigarettes | 0.0015 | 282 | 92.91% | 163 | 100.00% |
| Cookies | 0.0088 | 877 | 85.52% | 565 | 97.70% |
| Crackers | 0.0115 | 248 | 91.94% | 192 | 96.35% |
| Dish detergents | 0.0056 | 247 | 87.45% | 150 | 94.67% |
| Fabric softeners | 0.0046 | 280 | 84.29% | 143 | 93.01% |
| Front end candies | 0.0148 | 375 | 78.93% | 196 | 94.90% |
| Frozen dinners | 0.0110 | 232 | 96.55% | 175 | 100.00% |
| Frozen entrees | 0.0114 | 750 | 94.80% | 538 | 99.26% |
| Frozen juices | 0.0114 | 155 | 90.32% | 92 | 96.74% |
| Grooming products | 0.0094 | 965 | 91.40% | 626 | 98.40% |
| Laundry detergents | 0.0025 | 514 | 80.16% | 251 | 91.24% |
| Oatmeal | 0.0039 | 85 | 74.12% | 42 | 88.10% |
| Paper towels | 0.0034 | 103 | 59.22% | 41 | 60.98% |
| Refrigerated juices | 0.0030 | 192 | 69.27% | 79 | 64.56% |
| Shampoos | 0.0077 | 1,661 | 87.96% | 834 | 98.92% |
| Snack crackers | 0.0094 | 352 | 89.77% | 235 | 95.74% |
| Soaps | 0.0078 | 270 | 85.56% | 145 | 97.93% |
| Soft drinks | 0.0130 | 1,184 | 89.10% | 778 | 96.27% |
| Toothbrushes | 0.0103 | 333 | 93.39% | 234 | 98.72% |
| Toothpastes | 0.0089 | 467 | 91.86% | 299 | 99.00% |
| **Average** | **0.0071** | **435** | **83.44%** | **257** | **90.52%** |

<u>Notes</u>: Results of product-level regression. The dependent variable in all regressions is the share of price changes for each product at each store. For each product category, column 1 presents the average of the estimated coefficients of the log of the average sales volumes. The regressions also include controls for the median income, the share of ethnic minorities, the unemployment rate, and the pricing zone of the store. Column 2 presents the total number of coefficients. Column 3 presents the % of positive coefficients out of all coefficients. Column 4 presents the total number of coefficients that are statistically significant at the 5% level. Column 5 presents the % of coefficients that are positive and statistically significant, at the 5% level.

## Appendix P. Controlling for peak days

Bonomo et al. (2022) show that the majority of price changes occur during "peak days." Following their definition, for each category in each store we identify peak weeks as the subset of the most active days that jointly account for one-half of all price changes in a store over the entire sample period. We then define a dummy for peak weeks that equals 1 if a week is a peak week and 0 otherwise.

In Tables P1–P4, we present the results of estimating category-level regressions of the following form:

$$small\ price\ change_{i,s,t} = \alpha + \beta_1 ln\big(average\ sales\ volume_{i,s}\big) + \qquad \text{(P1)}$$

$$\beta_2 peak\text{-}days_{s,t} + \gamma \boldsymbol{X}_{i,s,t} + month_t + year_t + \delta_s + \mu_i + u_{i,s,t}$$

where *small price change* is a dummy that equals 1 if a price change of product $i$ in store $s$ at time $t$ is less or equal to 10¢, and 0 otherwise. The *average sales volume* is the average sales volume of product $i$ in store $s$ over the sample period. The variable $peak\text{-}days$ is a dummy that equals 1 if week $t$ in store $s$ is a peak day $\boldsymbol{X}$ is a matrix of other control variables. *Month* and *year* are fixed effects for the month and the year of the price change. $\delta$ and $\mu$ are fixed effects for stores and products, respectively, and $u$ is an i.i.d error term. We estimate separate regressions for each product category, clustering the errors by product. As we do in the paper, we use observations on price changes only if we observe the price in both week $t$ and $t + 1$ and the post-change price remained unchanged for at least 2 weeks.

The coefficient columns in the sales volume and the revenue panels of Table P1 report the coefficients of sales volume and peak-week, respectively in a regression that also includes fixed effects for months, years, stores and products. 27 of the sales volume coefficients are positive. 18 of the coefficients are statistically significant and 2 more are marginally significant. The average coefficient is 0.017, suggesting that a 1% increase in the sales volume is associated with a 1.7% increase in the likelihood of a small price change.

11 of the coefficients of the peak-week dummy are not statistically significant. 13 are negative and statistically significant and only 5 are positive and statistically significant.

The results therefore suggest that the positive correlation between sales volumes and small price changes holds also when we control for peak weeks, but that small price changes are not more common on peak weeks than on other weeks.

Table P2 reports the results when we add controls for the log of the average price, the log of the absolute change in the wholesale price, a control for sale- and bounce-back prices (which we identify using the sales filter algorithm of Fox and Syed 2016), and Dominick's pricing zone. In Table P3, we also add a control for 9-ending prices. In both tables, we find that 26 of the 29 coefficients are positive. 14 of the positive coefficients are statistically significant, and two more are marginally significant. The average coefficient is 0.010, suggesting that a 1% increase in the sales volume is associated with a 1.0% increase in the likelihood of a small price change.

As a further control for the effects of sales on the results, in Table P4 we focus on regular prices by excluding all sale- and bounce-back prices. When we focus on regular prices, we find that 27 of the 29 coefficients are positive. 18 are statistically significant, and 5 more are marginally significant. The average coefficient is 0.021, suggesting that a 1% increase in the sales volume is associated with a 2.1% increase in the likelihood of a small price change.

Thus, adding control for peak weeks does not change our main results regarding the correlation between sales volumes and small price changes.

## Table P1. Controlling for peak days. Baseline regressions

| Category | Sales Volume | | Peak days | | No. of Observations |
|---|---|---|---|---|---|
| | Coefficient | Std. | Coefficient | Std. | |
| Analgesics | 0.0168*** | 0.0040 | 0.003 | 0.0117 | 74,451 |
| Bath Soap | 0.0093 | 0.0128 | 0.0202 | 0.0127 | 6,650 |
| Bathroom Tissues | 0.0576*** | 0.0087 | 0.0036 | 0.0131 | 56,458 |
| Beer | 0.0019*** | 0.0006 | 0.0065*** | 0.0015 | 187,691 |
| Bottled Juice | 0.0366*** | 0.0079 | -0.0508*** | 0.0089 | 224,857 |
| Canned Soup | 0.0033 | 0.0090 | -0.0567*** | 0.0098 | 233,779 |
| Canned Tuna | 0.0094 | 0.0066 | 0.0151 | 0.0108 | 112,629 |
| Cereals | 0.0072 | 0.0066 | -0.0621*** | 0.0141 | 141,087 |
| Cheese | 0.0117*** | 0.0040 | -0.0545*** | 0.0100 | 357,679 |
| Cigarettes | 0.0031 | 0.0049 | 0.0824*** | 0.0050 | 24,553 |
| Cookies | 0.01*** | 0.0019 | -0.0264*** | 0.0079 | 317,932 |
| Crackers | 0.0047 | 0.0033 | -0.0496*** | 0.0128 | 115,658 |
| Dish Detergent | 0.03*** | 0.0071 | -0.0127 | 0.0168 | 85,222 |
| Fabric Softener | 0.0139** | 0.0069 | 0.0274** | 0.0128 | 85,337 |
| Front-End-Candies | 0.0104*** | 0.0040 | -0.0177 | 0.0143 | 148,200 |
| Frozen Dinners | 0.0494*** | 0.0068 | 0.0163* | 0.0084 | 52,893 |
| Frozen Entrees | 0.0258*** | 0.0026 | -0.0829*** | 0.0061 | 345,223 |
| Frozen Juices | 0.0173** | 0.0073 | -0.0562*** | 0.0168 | 118,582 |
| Grooming Products | 0.0091*** | 0.0033 | 0.0093*** | 0.0036 | 101,944 |
| Laundry Detergents | 0.021*** | 0.0046 | 0.0083 | 0.0116 | 121,566 |
| Oatmeal | 0.0189 | 0.0125 | -0.086*** | 0.0217 | 25,523 |
| Paper Towels | 0.0294* | 0.0156 | -0.0887*** | 0.0185 | 48,199 |
| Refrigerated Juices | 0.0142* | 0.0077 | -0.0278*** | 0.0110 | 108,965 |
| Shampoos | 0.0079*** | 0.0025 | 0.0245*** | 0.0033 | 88,193 |
| Snack Crackers | 0.002 | 0.0027 | 0.0018 | 0.0134 | 176,527 |
| Soap | 0.0277*** | 0.0089 | -0.0355*** | 0.0117 | 56,725 |
| Soft Drinks | 0.0307*** | 0.0044 | -0.0431*** | 0.0085 | 243,837 |
| Toothbrushes | 0.0128*** | 0.0046 | 0.0053 | 0.0098 | 52,185 |
| Toothpastes | 0.0009 | 0.0040 | -0.0029 | 0.0093 | 100,845 |
| **Average** | **0.0170** | **0.0061** | **-0.0183** | **0.0109** | **131,496** |

Notes: The table reports the results of category-level fixed effect regressions of the probability of a small price change. The dependent variable is "small price change," which equals 1 if a price change of product $i$ in store $s$ at time $t$ is less or equal to 10¢, and 0 otherwise. The main independent variables are the log of average sales volume of product $i$ in store $s$ over the sample period and a dummy for peak days that equals 1 if it is one of the weeks with the largest number of price changes, so that all the peak weeks account for 50% of all the price changes. The regressions also include fixed effects for stores, products, months, and years. We estimate separate regressions for each product category, clustering the errors by product. * $p < 10\%$, ** $p < 5\%$, *** $p < 1\%$

Table P2. Controlling for peak days with additional controls

| Category | Sales Volume | | Peak days | | No. of Observations |
|---|---|---|---|---|---|
| | Coefficient | Std. | Coefficient | Std. | |
| Analgesics | 0.013*** | 0.0039 | -0.0065 | 0.0131 | 74,451 |
| Bath Soap | 0.0056 | 0.0129 | 0.0097 | 0.0098 | 6,650 |
| Bathroom Tissues | 0.0195** | 0.0083 | -0.0068 | 0.0130 | 56,458 |
| Beer | 0.0042*** | 0.0007 | 0.0049*** | 0.0011 | 187,691 |
| Bottled Juice | 0.02*** | 0.0069 | -0.0343*** | 0.0085 | 224,857 |
| Canned Soup | -0.0035 | 0.0087 | -0.0344*** | 0.0094 | 233,779 |
| Canned Tuna | -0.0012 | 0.0059 | 0.0139 | 0.0110 | 112,629 |
| Cereals | 0.0084 | 0.0065 | -0.1023*** | 0.0146 | 141,087 |
| Cheese | 0.0074** | 0.0034 | -0.0536*** | 0.0097 | 357,679 |
| Cigarettes | 0.0008 | 0.0047 | 0.0818*** | 0.0051 | 24,553 |
| Cookies | 0.009*** | 0.0020 | -0.0326*** | 0.0079 | 317,932 |
| Crackers | 0.0041 | 0.0031 | -0.0476*** | 0.0129 | 115,658 |
| Dish Detergent | 0.0255*** | 0.0063 | -0.034** | 0.0156 | 85,222 |
| Fabric Softener | 0.0018 | 0.0057 | 0.0145 | 0.0110 | 85,337 |
| Front-End-Candies | -0.0036 | 0.0033 | 0.0227* | 0.0121 | 148,200 |
| Frozen Dinners | 0.0331*** | 0.0060 | 0.0496*** | 0.0089 | 52,893 |
| Frozen Entrees | 0.0189*** | 0.0026 | -0.0429*** | 0.0058 | 345,223 |
| Frozen Juices | 0.0121* | 0.0069 | -0.0541*** | 0.0156 | 118,582 |
| Grooming Products | 0.011*** | 0.0033 | 0.0005 | 0.0033 | 101,944 |
| Laundry Detergents | 0.0128*** | 0.0039 | -0.0004 | 0.0105 | 121,566 |
| Oatmeal | 0.0111 | 0.0122 | -0.0598*** | 0.0214 | 25,523 |
| Paper Towels | 0.0171 | 0.0172 | -0.0688*** | 0.0193 | 48,199 |
| Refrigerated Juices | 0.0065 | 0.0079 | -0.0201* | 0.0108 | 108,965 |
| Shampoos | 0.0089*** | 0.0025 | 0.0199*** | 0.0031 | 88,193 |
| Snack Crackers | 0.0035 | 0.0027 | 0.0012 | 0.0135 | 176,527 |
| Soap | 0.0152* | 0.0081 | -0.0331*** | 0.0105 | 56,725 |
| Soft Drinks | 0.0145*** | 0.0019 | -0.0333*** | 0.0075 | 243,837 |
| Toothbrushes | 0.0093** | 0.0047 | 0.0082* | 0.0097 | 52,185 |
| Toothpastes | 0.001 | 0.0038 | -0.0183** | 0.0088 | 100,845 |
| **Average** | **0.0099** | **0.0057** | **-0.0157** | **0.0105** | **131,496** |

<u>Notes</u>: The table reports the results of category-level fixed effect regressions of the probability of a small price change. The dependent variable is "small price change," which equals 1 if a price change of product $i$ in store $s$ at time $t$ is less or equal to 10¢, and 0 otherwise. The main independent variables are the log of average sales volume of product $i$ in store $s$ over the sample period and a dummy for peak days that equals 1 if it is one of the weeks with the largest number of price changes, so that all the peak weeks account for 50% of all the price changes. The regressions also include the following independent variables: the products' average price, percentage changes in the wholesale price and a dummy for sale and bounce-back prices, as well as fixed effects for years, months, stores, and products. We estimate separate regressions for each product category, clustering the errors by product. * $p < 10\%$, ** $p < 5\%$, *** $p < 1\%$

Table P3. Bonomo regressions – including a control for 9-ending prices

| Category | Sales Volume | | Peak days | | No. of Observations |
|---|---|---|---|---|---|
| | Coefficient | Std. | Coefficient | Std. | |
| Analgesics | 0.0128*** | 0.0038 | -0.0076** | 0.0129 | 74,451 |
| Bath Soap | 0.0055 | 0.0125 | 0.0087*** | 0.0100 | 6,650 |
| Bathroom Tissues | 0.0189** | 0.0081 | -0.0208** | 0.0136 | 56,458 |
| Beer | 0.0042*** | 0.0007 | 0.0049*** | 0.0011 | 187,691 |
| Bottled Juice | 0.0199*** | 0.0070 | -0.0346*** | 0.0084 | 224,857 |
| Canned Soup | -0.0008 | 0.0085 | -0.0325*** | 0.0093 | 233,779 |
| Canned Tuna | -0.0015 | 0.0058 | 0.0139** | 0.0108 | 112,629 |
| Cereals | 0.0085 | 0.0065 | -0.1025** | 0.0145 | 141,087 |
| Cheese | 0.0072** | 0.0034 | -0.0557*** | 0.0098 | 357,679 |
| Cigarettes | 0.0009 | 0.0046 | 0.0852*** | 0.0058 | 24,553 |
| Cookies | 0.0091*** | 0.0019 | -0.0347*** | 0.0080 | 317,932 |
| Crackers | 0.0048 | 0.0031 | -0.0528** | 0.0119 | 115,658 |
| Dish Detergent | 0.0258*** | 0.0062 | -0.0356** | 0.0155 | 85,222 |
| Fabric Softener | 0.0023 | 0.0057 | 0.0146** | 0.0110 | 85,337 |
| Front-End-Candies | -0.0034 | 0.0033 | 0.0242** | 0.0121 | 148,200 |
| Frozen Dinners | 0.0342*** | 0.0060 | 0.0555*** | 0.0083 | 52,893 |
| Frozen Entrees | 0.0192*** | 0.0026 | -0.0438*** | 0.0059 | 345,223 |
| Frozen Juices | 0.0131* | 0.0067 | -0.052** | 0.0146 | 118,582 |
| Grooming Products | 0.0112*** | 0.0033 | 0.0011*** | 0.0034 | 101,944 |
| Laundry Detergents | 0.013*** | 0.0038 | 0.0001** | 0.0106 | 121,566 |
| Oatmeal | 0.0091 | 0.0116 | -0.0569** | 0.0218 | 25,523 |
| Paper Towels | 0.0184 | 0.0171 | -0.0751** | 0.0184 | 48,199 |
| Refrigerated Juices | 0.0066 | 0.0078 | -0.024** | 0.0111 | 108,965 |
| Shampoos | 0.0089*** | 0.0025 | 0.0199*** | 0.0031 | 88,193 |
| Snack Crackers | 0.0035 | 0.0027 | 0.0008** | 0.0134 | 176,527 |
| Soap | 0.0184** | 0.0080 | -0.0284*** | 0.0102 | 56,725 |
| Soft Drinks | 0.0137*** | 0.0018 | -0.0331*** | 0.0075 | 243,837 |
| Toothbrushes | 0.0089* | 0.0046 | 0.0056*** | 0.0095 | 52,185 |
| Toothpastes | 0.0012 | 0.0038 | -0.019*** | 0.0088 | 100,845 |
| **Average** | **0.0101** | **0.0056** | **-0.0164** | **0.0104** | **131,496** |

Notes: The table reports the results of category-level fixed effect regressions of the probability of a small price change. The dependent variable is "small price change," which equals 1 if a price change of product $i$ in store $s$ at time $t$ is less or equal to 10¢, and 0 otherwise. The main independent variables are the log of average sales volume of product $i$ in store $s$ over the sample period and a dummy for peak days that equals 1 if it is one of the weeks with the largest number of price changes, so that all the peak weeks account for 50% of all the price changes. The regressions also include the following independent variables: the products' average price, percentage changes in the wholesale price, a dummy for sale and bounce-back prices, a dummy for 9-ending prices that equals 1 if the right-most digit is 9, as well as fixed effects for years, months, stores, and products. We estimate separate regressions for each product category, clustering the errors by product. * $p < 10\%$, ** $p < 5\%$, *** $p < 1\%$

## Table P4. Bonomo regressions – focusing on regular prices

| Category | Sales Volume | | Peak days | | No. of Observations |
|---|---|---|---|---|---|
| | Coefficient | Std. | Coefficient | Std. | |
| Analgesics | 0.014* | 0.0080 | -0.0266 | 0.0189 | 24,729 |
| Bath Soap | -0.0272 | 0.0253 | 0.0082 | 0.0192 | 1,466 |
| Bathroom Tissues | 0.037*** | 0.0094 | -0.0364** | 0.0182 | 19,285 |
| Beer | 0.0182*** | 0.0055 | -0.0087* | 0.0130 | 12,080 |
| Bottled Juice | 0.035*** | 0.0089 | -0.0213 | 0.0164 | 60,015 |
| Canned Soup | 0.0133** | 0.0071 | 0.0095 | 0.0117 | 95,310 |
| Canned Tuna | 0.0132 | 0.0083 | -0.0034 | 0.0202 | 31,922 |
| Cereals | 0.0275*** | 0.0071 | -0.1595*** | 0.0209 | 72,789 |
| Cheese | 0.0139*** | 0.0046 | -0.0516*** | 0.0156 | 92,758 |
| Cigarettes | -0.0006 | 0.0053 | 0.0759*** | 0.0065 | 20,692 |
| Cookies | 0.0088** | 0.0037 | -0.0454*** | 0.0163 | 66,087 |
| Crackers | 0.0174*** | 0.0065 | -0.1153*** | 0.0253 | 24,771 |
| Dish Detergent | 0.0306*** | 0.0061 | -0.1054*** | 0.0305 | 26,735 |
| Fabric Softener | 0.0228*** | 0.0078 | 0.01 | 0.0226 | 27,488 |
| Front-End-Candies | 0.0018 | 0.0030 | -0.0045 | 0.0130 | 77,323 |
| Frozen Dinners | 0.0698*** | 0.0103 | 0.0557*** | 0.0176 | 12,287 |
| Frozen Entrees | 0.0239*** | 0.0039 | -0.0004 | 0.0074 | 117,044 |
| Frozen Juices | 0.0229*** | 0.0086 | -0.0508** | 0.0234 | 40,517 |
| Grooming Products | 0.0158 | 0.0110 | 0.0094 | 0.0107 | 22,102 |
| Laundry Detergents | 0.0175*** | 0.0058 | 0.0001 | 0.0194 | 42,121 |
| Oatmeal | 0.0597*** | 0.0113 | -0.0119 | 0.0269 | 13,605 |
| Paper Towels | 0.0312* | 0.0158 | -0.0668** | 0.0278 | 9,243 |
| Refrigerated Juices | 0.0259** | 0.0123 | -0.0163 | 0.0318 | 23,705 |
| Shampoos | 0.0225*** | 0.0080 | 0.0082 | 0.0096 | 16,099 |
| Snack Crackers | 0.0173*** | 0.0053 | -0.0685** | 0.0282 | 38,123 |
| Soap | 0.0539*** | 0.0119 | -0.0742*** | 0.0210 | 16,882 |
| Soft Drinks | 0.0059* | 0.0033 | 0.0345*** | 0.0119 | 49,989 |
| Toothbrushes | 0.0195* | 0.0102 | -0.0113 | 0.0270 | 13,695 |
| Toothpastes | 0.0058 | 0.0082 | -0.0169 | 0.0195 | 28,039 |
| **Average** | **0.0213** | **0.0084** | **-0.0236** | **0.0190** | **37,824** |

Notes: The table reports the results of category-level fixed effect regressions of the probability of a small price change. The dependent variable is "small price change," which equals 1 if a price change of product $i$ in store $s$ at time $t$ is less or equal to 10¢, and 0 otherwise. The main independent variables are the log of average sales volume of product $i$ in store $s$ over the sample period and a dummy for peak days that equals 1 if it is one of the weeks with the largest number of price changes, so that all the peak weeks account for 50% of all the price changes. The regressions also include the following independent variables: the products' average price, percentage changes in the wholesale price, a dummy for 9-ending prices that equals 1 if the right-most digit is 9, as well as fixed effects for years, months, stores, and products. We exclude observations on sales and bounce back prices. We estimate separate regressions for each product category, clustering the errors by product. * $p < 10\%$, ** $p < 5\%$, *** $p < 1\%$

**Appendix Q. Estimation using only data on products that are sold in single units**

In the paper, we study the correlation between the sales volume and the likelihood of small price changes. There, we define a unit sold the way it is defined by the retailer. I.e., a 6-pack of beer is counted as one unit. However, this might bias the results if products that are sold in packages have different properties than products that are sold in single units. We therefore repeat our estimation, after excluding observations on products that are sold in packages.

We therefore estimate:

$$small\ price\ change_{i,s,t} = \alpha + \beta ln\big(average\ sales\ volume_{i,s}\big) + \gamma X_{i,s,t} +$$
$$month_t + year_t + \delta_s + \mu_i + u_{i,s,t} \tag{Q1}$$

where small price change is a dummy that equals 1 if a price change of product $i$ in store $s$ in week $t$ is less or equal to 10¢, and 0 otherwise. As we do in the paper, we use observations on price changes only if we observe the price in both weeks $t$ and $t+1$ and the post change price remained unchanged for at least 2 weeks.
The average sales volume is the average sales volume of product $i$ in store $s$ over the sample period. By taking the average over a long period, we obtain an estimate of the expected sales volume that does not depend on transitory shocks or sales. **X** is a matrix of other control variables. Month and year are fixed effects for the month (to control for seasonality) and the year of the price change. To control for the differences across stores and products, $\delta$ and $\mu$ are fixed effects for stores and products, respectively, while $u$ is an i.i.d error term.

Table Q1 reports the coefficient estimates of the key variable, average sales volume, for each product category. Column 1 reports the results of baseline regressions that exclude the **X** matrix. I.e, the regressions include only the average sales volume and fixed effects for months, years, stores, and products.

We find that in all 29 product categories, the coefficients are positive and statistically significant. The estimated effect is economically significant. The average coefficient is 0.027, suggesting that an increase of 1% in the sales volume is associated with an increase of 2.7 percentage points in the likelihood that a price change will be small.

In column 2, we add the matrix **X** which includes the following control variables: the log of the average price to control for the price level effect on the size of price changes, the percentage change in the wholesale price, and control for sale- and bounce-back prices, all as defined above. The estimation results are similar to what we report in column 1. The coefficients of the average sales volume are positive and statistically significant in 28 categories, and marginally significant in 1 more. The average coefficient is 0.020. Thus, even after including the controls, we still find that increasing the average sales volume by 1% is associated with an increase of 2.0 percentage points in the likelihood of a small price change.

In column 3, we add a dummy for 9-ending prices as an additional control because when the pre-change price is 9-ending, price changes tend to be larger than when the pre-change price ends in other digits (Levy et al. 2020). Thus, if products with high sales volume tend to have non-9-ending prices, then it might lead to high sales volume products' prices changing by small amounts.

However, adding this dummy does not change the main result appreciably. All 29 coefficients remain positive. 28 are statistically significant, and 1 more is marginally significant. Controlling for 9-ending prices, increasing the average sales volume by 1% is associated with a 2.0 percentage points increase in the likelihood of a small price change, on average.

In column 4, we focus on regular prices by excluding sale- and bounce-back prices. We do this for two reasons. First, sale- and bounce-back prices tend to be large, and therefore, we need to account for them properly. Second, it is often argued that changes in sale prices have a smaller effect on inflation than changes in regular prices (Nakamura and Steinsson 2008, Midrigan 2011, Anderson et al. 2017, Ray et al. 2023).

We find that when we exclude sale prices, all the coefficients remain positive. 28 are statistically significant, and 1 more is marginally significant. The average coefficient is 0.041, implying that for regular prices, an increase of 1% in the average sales volume is associated with an increase of 4.1 percentage points in the likelihood of a small price change.

Table Q1. Category-level regressions of small price changes and sales volume excluding products sold in packages

| Category | | | **(1)** | **(2)** | **(3)** | **(4)** |
|---|---|---|---|---|---|---|
| Analgesics | Coefficient (Std.) | | 0.0262*** (0.0034) | 0.019*** (0.0032) | 0.0113*** (0.0021) | 0.0188*** (0.0031) |
| | Observations | | 144,461 | 144,461 | 144,461 | 144,461 |
| Bath Soap | Coefficient (Std.) | | 0.0293*** (0.008) | 0.0277*** (0.0082) | 0.0174*** (0.0046) | 0.0285*** (0.0081) |
| | Observations | | 15,295 | 15,295 | 15,295 | 15,295 |
| Bathroom Tissues | Coefficient (Std.) | | 0.0328*** (0.0077) | 0.008* (0.0072) | 0.0137*** (0.0056) | 0.0083* (0.0072) |
| | Observations | | 140,505 | 140,505 | 149,441 | 140,505 |
| Beer | Coefficient (Std.) | | 0.013*** (0.0012) | 0.0147*** (0.0012) | 0.0114*** (0.0008) | 0.0147*** (0.0012) |
| | Observations | | 290,591 | 290,591 | 290,620 | 290,591 |
| Bottled Juice | Coefficient (Std.) | | 0.0376*** (0.0051) | 0.0271*** (0.0044) | 0.017*** (0.0042) | 0.026*** (0.0044) |
| | Observations | | 471,256 | 471,256 | 496,557 | 471,256 |
| Canned Soup | Coefficient (Std.) | | 0.0167*** (0.0056) | 0.0077*** (0.0051) | 0.0121*** (0.0043) | 0.0098** (0.0049) |
| | Observations | | 461,989 | 461,989 | 495,543 | 461,989 |
| Canned Tuna | Coefficient (Std.) | | 0.0249*** (0.0055) | 0.0146*** (0.0047) | 0.0124*** (0.0038) | 0.0142*** (0.0046) |
| | Observations | | 206,937 | 206,937 | 213,043 | 206,937 |
| Cereals | Coefficient (Std.) | | 0.021*** (0.0037) | 0.0158** (0.0034) | 0.0133*** (0.003) | 0.0157*** (0.0035) |
| | Observations | | 354,887 | 354,887 | 357,120 | 354,887 |
| Cheese | Coefficient (Std.) | | 0.021*** (0.0029) | 0.012*** (0.0025) | 0.0084*** (0.0023) | 0.0118*** (0.0025) |
| | Observations | | 780,089 | 780,089 | 796,150 | 780,089 |
| Cigarettes | Coefficient (Std.) | | 0.0084** (0.0046) | 0.0073** (0.0045) | 0.0095** (0.0028) | 0.0074** (0.0044) |
| | Observations | | 36,157 | 36,157 | 36,157 | 36,157 |
| Cookies | Coefficient (Std.) | | 0.0276*** (0.0018) | 0.0225*** (0.0017) | 0.018*** (0.0014) | 0.0227*** (0.0017) |
| | Observations | | 668,546 | 668,546 | 688,761 | 668,546 |
| Crackers | Coefficient (Std.) | | 0.0387*** (0.0031) | 0.0301*** (0.0027) | 0.0232*** (0.0022) | 0.0306*** (0.0027) |
| | Observations | | 239,253 | 239,253 | 245,185 | 239,253 |
| Dish Detergent | Coefficient (Std.) | | 0.0394*** (0.0044) | 0.0279*** (0.0036) | 0.0212*** (0.0031) | 0.0278*** (0.0035) |
| | Observations | | 188,737 | 188,737 | 189,633 | 188,737 |
| Fabric Softener | Coefficient (Std.) | | 0.0246*** (0.0048) | 0.0118*** (0.0044) | 0.0089*** (0.0034) | 0.0121*** (0.0043) |
| | Observations | | 178,724 | 178,724 | 181,056 | 178,724 |
| Front-End-Candies | Coefficient (Std.) | | 0.0103*** (0.004) | 0.0161*** (0.0041) | 0.0048*** (0.0026) | 0.0163*** (0.0041) |
| | Observations | | 192,037 | 192,037 | 278,853 | 192,037 |
| Frozen Dinners | Coefficient (Std.) | | 0.0478*** (0.0035) | 0.0385*** (0.0031) | 0.0308*** (0.0023) | 0.0411*** (0.0032) |
| | Observations | | 187,022 | 187,022 | 203,191 | 187,022 |

Table Q1. (Cont.)

| Category | | (1) | (2) | (3) | (4) |
|---|---|---|---|---|---|
| Frozen Entrees | Coefficient (Std.) | 0.0288*** (0.0024) | 0.0281*** (0.002) | 0.0193*** (0.0015) | 0.0289*** (0.002) |
| | Observations | 694,903 | 694,903 | 864,832 | 694,903 |
| Frozen Juices | Coefficient (Std.) | 0.0289*** (0.0049) | 0.0223*** (0.0044) | 0.0162*** (0.0035) | 0.0227*** (0.0043) |
| | Observations | 286,846 | 286,846 | 308,817 | 286,846 |
| Grooming Products | Coefficient (Std.) | 0.0186*** (0.0023) | 0.0208*** (0.0024) | 0.0135*** (0.0016) | 0.021*** (0.0024) |
| | Observations | 269,513 | 269,513 | 269,873 | 269,513 |
| Laundry Detergents | Coefficient (Std.) | 0.0198*** (0.0032) | 0.0094*** (0.0028) | 0.0082*** (0.0023) | 0.0099*** (0.0028) |
| | Observations | 270,780 | 270,780 | 272,765 | 270,780 |
| Oatmeal | Coefficient (Std.) | 0.0283*** (0.0081) | 0.0151** (0.0067) | 0.0129** (0.0058) | 0.0153** (0.0067) |
| | Observations | 79,488 | 79,488 | 79,983 | 79,488 |
| Paper Towels | Coefficient (Std.) | 0.0479*** (0.0114) | 0.026*** (0.0099) | 0.0254*** (0.0083) | 0.0264*** (0.01) |
| | Observations | 111,012 | 111,012 | 116,204 | 111,012 |
| Refrigerated Juices | Coefficient (Std.) | 0.0357*** (0.0047) | 0.0209*** (0.0039) | 0.0177*** (0.0033) | 0.0208*** (0.0039) |
| | Observations | 304,028 | 304,028 | 306,865 | 304,028 |
| Shampoos | Coefficient (Std.) | 0.0164*** (0.0015) | 0.0202*** (0.0016) | 0.0119*** (0.001) | 0.0202*** (0.0016) |
| | Observations | 260,918 | 260,918 | 261,778 | 260,918 |
| Snack Crackers | Coefficient (Std.) | 0.033*** (0.0032) | 0.0284*** (0.003) | 0.0236*** (0.0026) | 0.0285*** (0.003) |
| | Observations | 390,331 | 390,331 | 398,665 | 390,331 |
| Soaps | Coefficient (Std.) | 0.0373*** (0.0055) | 0.0224*** (0.005) | 0.0162*** (0.0037) | 0.0234*** (0.0049) |
| | Observations | 151,326 | 151,326 | 152,379 | 151,326 |
| Soft Drinks | Coefficient (Std.) | 0.026*** (0.0015) | 0.0243*** (0.0014) | 0.0099*** (0.0018) | 0.0238*** (0.0013) |
| | Observations | 1,247,126 | 1,247,126 | 1,350,618 | 1,247,126 |
| Toothbrushes | Coefficient (Std.) | 0.0212*** (0.0029) | 0.0204*** (0.0029) | 0.0137*** (0.0018) | 0.0198*** (0.0029) |
| | Observations | 121,951 | 121,951 | 125,380 | 121,951 |
| Toothpastes | Coefficient (Std.) | 0.0124*** (0.0026) | 0.0113*** (0.0022) | 0.0088*** (0.0016) | 0.0113*** (0.0022) |
| | Observations | 263,971 | 263,971 | 264,317 | 263,971 |
| **Average coefficients** | | **0.0267** | **0.0197** | **0.0199** | **0.0415** |

Notes: The table reports the results of category-level fixed effect regressions of the probability of a small price change. The dependent variable is "small price change," which equals 1 if a price change of product $i$ in store $s$ at time $t$ is less or equal to 10¢, and 0 otherwise. The main independent variable is the log of average sales volume of product $i$ in store $s$ over the sample period. Column 1 reports the results of the baseline regression that includes only the log of average sales volume and the fixed effects for months, years, stores, and products. In column 2, we add the following controls: the log of the average price, the log of the absolute change in the wholesale price, and a control for sale- and bounce-back prices, which we identify using a sales filter algorithm. In column 3, we add a dummy for 9-ending prices as an additional control. In column 4, we focus on regular prices by excluding the sale- and bounce-back prices. We estimate separate regressions for each product category, clustering the errors by product. * $p < 10\%$, ** $p < 5\%$, *** $p < 1\%$

### Appendix R. Storable vs. non-storable products

It's possible that retailers have different strategies for storable vs. non-storable products. To test whether this has an effect on the correlation between small price changes and sales volumes, we estimate the following regression, using pooled data from all product categories:

$$small\ price\ change_{i,s,t} = \alpha + \beta_1 ln(average\ sales\ volume_{i,s}) +$$
$$\beta_2 ln(average\ sales\ volume_{i,s}) \times non\text{-}storable_i + \beta_3 non\text{-}storable_i + \gamma X_{i,s,t} +$$
$$month_t + year_t + \kappa_i + \delta_s + \mu_i + u_{i,s,t} \qquad \text{(R1)}$$

where small price change is a dummy that equals 1 if a price change of product $i$ in store $s$ in week $t$ is less or equal to 10¢, and 0 otherwise. As we do in the paper, we use observations on price changes only if we observe the price in both weeks $t$ and $t + 1$ and the post change price remained unchanged for at least 2 weeks. The average sales volume is the average sales volume of product $i$ in store $s$ over the sample period.[6] By taking the average over a long period, we obtain an estimate of the expected sales volume that does not depend on transitory shocks or sales. $\mathbf{X}$ is a matrix of other control variables. Month and year are fixed effects for the month (to control for seasonality) and the year of the price change. To control for the differences across stores and products, $\kappa$, $\delta$ and $\mu$ are fixed effects for categories, stores and products, respectively, while $u$ is an i.i.d error term. $Non\text{-}storable$ is a dummy for products that have a high cost of storage. It equals 1 if a product belongs to either of the cheese, frozen dinners, frozen entrees, frozen juices, or refrigerated juices categories.

Table R1 reports the coefficient estimates of the key variables, average sales volume, and the interaction between the average sales volume and the dummy for non-storable products. Column 1 reports the results of baseline regressions that exclude the matrix $\mathbf{X}$.

---

[6] In calculating the average sales volume, we need to account for missing observations, because a missing observation in week $t$ implies that the product was either out of stock or had 0 sales on that week. Thus, averaging over the available observations can lead to an upward bias for products that are sold in small numbers. Therefore, for each product in each store, we calculate the average by first determining the total number of units sold over all available observations. We then identify the first and last week for which we have observations, and calculate the average for each product-store as $\frac{total\ no.\ of\ units\ sold}{last\ week\ -\ first\ week}$. The resulting figure is smaller than we would obtain if we averaged over all available observations (which would not include obsservations on weeks with 0 sales).

I.e, the regressions include only the average sales volume, the interaction between the average sales volume and the dummy for non-storable products, the dummy for non-storable products, and fixed effects for months, years, stores, categories, and products.

We find that the coefficient of the sales volume is positive and statistically significant. Its value is similar to the value we report in the paper, 0.025. The value of the coefficient of the interaction between the sales volume and the dummy for non-storable products is small, 0.004, positive and statistically significant. Thus, the results suggests that the correlation between sales volumes and small price changes might be slightly stronger for products that are harder to store than for other products.

In column 2, we add the matrix **X** which includes the following control variables: the log of the average price to control for the price level effect on the size of price changes, the percentage change in the wholesale price, and control for sale- and bounce-back prices, all as defined above. We find that the coefficient of the interaction term is now negative, but it is not statistically significant.

In column 3, we add a dummy for 9-ending prices as an additional control because when the pre-change price is 9-ending, price changes tend to be larger than when the pre-change price ends in other digits (Levy et al. 2020). Thus, if products with high sales volume tend to have non-9-ending prices, then it might lead to high sales volume products' prices changing by small amounts. According to our estimates, the coefficient of the interaction term remains negative, and it is not statistically significant.

In column 4, we focus on regular prices by excluding sale- and bounce-back prices. We do this for two reasons. First, sale- and bounce-back prices tend to be large, and therefore, we need to account for them properly. Second, it is often argued that changes in sale prices have a smaller effect on inflation than changes in regular prices (Nakamura and Steinsson 2008, Midrigan 2011, Anderson et al. 2017, Ray et al. 2023).

We find that when we exclude sale prices, the results remain similar to our findings in columns 2 and 3. The coefficient of the interaction term remains small, negative and statistically insignificant. We therefore conclude that the correlation between the likelihood of a small price change and the sales volumes is similar across storable and less storable products.

Table R1. Pooled regressions of small price changes and sales volume, with controls for non-durable products

|  | (1) | (2) | (3) | (4) |
|---|---|---|---|---|
| Average sales volume | 0.025*** | 0.018*** | 0.018*** | 0.018*** |
|  | (0.001) | (0.001) | (0.001) | (0.001) |
| Average sales volume × non-storable | 0.004** | −0.003 | −0.003 | −0.003 |
|  | (0.002) | (0.002) | (0.002) | (0.002) |
| Observations | 9,553,542 | 9,553,542 | 9,553,542 | 2,328,405 |

Notes: The dependent variable is "small price change," which equals 1 if a price change of product $i$ in store $s$ at time $t$ is less or equal to 10¢, and 0 otherwise. The main independent variable is the log of the average sales volume of product $i$ in store $s$ over the sample period. Non-storable is a dummy for products that are costly to store. Column 1 reports the results of baseline regression that includes only the average sales volume and the fixed effects for months, years, stores, and products. In column 2, we add the following controls: the log of the average price, the log of the absolute change in the wholesale price, and control for sale- and bounce-back prices, which we identify using a sales filter algorithm. In column 3, we add a dummy for 9-ending prices as an additional control. In column 4, we focus on regular prices by excluding the sale- and bounce-back prices. All regressions also include a dummy for non-storable products, and fixed effects for categories, stores, products, years, and months. We estimate separate regressions for each product category, clustering the errors by product.  * $p < 10\%$, ** $p < 5\%$, *** $p < 1\%$

**Appendix S. The correlation between the sales volume and the likelihood of price increases vs. decreases**

Our model implies that the correlation between the sales volume and the likelihood of a small price change is symmetric. Products with high sales volumes should be more likely to both increase and decrease than products with lower sales volumes. However, empirical evidence suggests that this might not be the case (Peltzman, 2000). For example, if shoppers are not attentive to small price changes (Chen et al., 2008, Chakraborty et al., 2015), then retailers may gain from small price increases and lose from small price decreases.

Therefore, in Tables S1–S4, we present the results of the category-level regression estimations. The regressions we estimate are of the following form:

$$small\ price\ change_{i,s,t} = \alpha + \beta_1 ln\big(average\ sales\ volume_{i,s}\big) + \qquad \text{(S1)}$$

$$\beta_2 ln\big(average\ revenue_{i,s}\big) + \gamma \boldsymbol{X}_{i,s,t} + month_t + year_t + \delta_s + \mu_i + u_{i,s,t}$$

where *small price increase (decrease)* is a dummy that equals 1 if a price change of product *i* in store *s* at time *t* is less or equal to 10¢, and 0 otherwise. The *average sales volume* is the average sales volume of product *i* in store *s* over the sample period. The *average revenue* is the average revenue of product *i* in store *s* over the sample period. $\boldsymbol{X}$ is a matrix of other control variables. *Month* and *year* are fixed effects for the month and the year of the price change. $\delta$ and $\mu$ are fixed effects for stores and products, respectively, and *u* is an i.i.d error term. We estimate a separate regression for each product category, clustering the errors by product. As we do in the paper, we use observations on price changes only if we observe the price in both weeks *t* and *t*+1 and the post-change price remained unchanged for at least 2 weeks.

Table S1 reports the results of baseline regressions that exclude the matrix $\boldsymbol{X}$. I.e, the regressions include only the average sales volume and fixed effects for months, years, stores, and products.

For price increases, we find that in all 29 product categories, the coefficients are positive and statistically significant. For price decreases, 27 of the coefficients are positive, and 20 of them are statistically significant. Two more are marginally significant.

It therefore seems that the correlation between price increases and the likelihood of small price changes is stronger than the correlation between price decreases and the likelihood of small price changes. This is also corroborated by the size of the coefficients. In 21 categories, the coefficients of price increases are larger than the coefficients of price decreases, yielding an average coefficient of 0.0195 for price increases and 0.0146 for price decreases.

In Table S2, we add the **X** matrix which includes the following control variables: the log of the average price to control for the price level effect on the size of price changes, the percentage change in the wholesale price, and control for sale- and bounce-back prices, all as defined above. The results are similar to what we report above. When we focus on price increases, we find that all the coefficients are positive, and that 28 of them are statistically significant. When we focus on price decreases, we find that 27 of the coefficients are positive, 19 of them are statistically significant, and 2 more are marginally significant. Again, the average coefficient of price increases, 0.0162, is larger than the average coefficients of price decreases, 0.0127.

In Table S3, we add a dummy for 9-ending prices as an additional control because when the pre-change price is 9-ending, price changes tend to be larger than when the pre-change price ends in other digits (Levy et al. 2020). Thus, if products with high sales volume tend to have non-9-ending prices, then it might lead to high sales volume products' prices changing by small amounts. The results remain almost unchanged relative to the figures presented in Table S2.

In Table S4, we focus on regular prices by excluding sale- and bounce-back prices. We do this for two reasons. First, sale- and bounce-back prices tend to be large, and therefore, we need to account for them properly. Second, it is often argued that changes in sale prices have a smaller effect on inflation than changes in regular prices (Nakamura and Steinsson 2008, Midrigan 2011, Anderson et al. 2017, Ray et al. 2023).

We find that when we exclude sale prices, 29 of the coefficients of the price increase regressions are positive, and all of them are statistically significant. In the regressions of price decreases, 26 of the coefficients are positive and 20 of them are statistically significant. 4 more are marginally significant. The average coefficient of the price increase regressions is 0.0303, again higher than the average coefficient of the price decrease regressions, 0.0213.

We conclude that the correlation is stronger for price increases than for price decreases. Therefore, although our model suggests a symmetric correlation, it seems that there are other forces at play as well. One possibility is consumer inattention, which makes small price increases more profitable than small price decreases, as in Chen et al. (2008).

Table S1. Category-level regressions of small price changes and sales volume, price increases vs. price decreases

| Category | Price Increase | | | Price Decrease | | |
|---|---|---|---|---|---|---|
| | Coefficient | Std. | Obs. | Coefficient | Std. | Obs. |
| Analgesics | 0.0107*** | 0.0378 | 93,254 | 0.021*** | 0.0034 | 51,207 |
| Bath Soap | 0.0211*** | 0.0870 | 9,877 | 0.0047 | 0.0058 | 5,418 |
| Bathroom Tissues | 0.0284*** | 0.0498 | 96,660 | 0.0156* | 0.0087 | 52,781 |
| Beer | 0.013*** | 0.0273 | 128,309 | 0.0049*** | 0.0007 | 162,311 |
| Bottled Juice | 0.023*** | 0.0800 | 298,844 | 0.0173*** | 0.0061 | 197,713 |
| Canned Soup | 0.019*** | 0.0390 | 334,515 | 0.0015 | 0.0066 | 161,028 |
| Canned Tuna | 0.0164*** | 0.0569 | 110,869 | 0.0139*** | 0.0044 | 102,174 |
| Cereals | 0.02*** | 0.0326 | 262,840 | 0.0077 | 0.0049 | 94,280 |
| Cheese | 0.0176*** | 0.0842 | 506,336 | 0.0132*** | 0.0032 | 289,814 |
| Cigarettes | 0.0128*** | 0.0386 | 27,370 | -0.0024 | 0.0035 | 8,787 |
| Cookies | 0.0234*** | 0.0166 | 440,768 | 0.0178*** | 0.0019 | 247,993 |
| Crackers | 0.032*** | 0.0804 | 152,814 | 0.0267*** | 0.0035 | 92,371 |
| Dish Detergent | 0.0304*** | 0.0928 | 120,854 | 0.0307*** | 0.0047 | 68,779 |
| Fabric Softener | 0.0123*** | 0.1206 | 110,126 | 0.0146*** | 0.005 | 70,930 |
| Front-End-Candies | 0.0073** | 0.0199 | 168,056 | -0.0145*** | 0.0036 | 110,797 |
| Frozen Dinners | 0.0262*** | 0.0252 | 142,131 | 0.0523*** | 0.0044 | 61,060 |
| Frozen Entrees | 0.0178*** | 0.0097 | 593,786 | 0.0242*** | 0.0023 | 271,046 |
| Frozen Juices | 0.0196*** | 0.0361 | 201,311 | 0.0256*** | 0.0059 | 107,506 |
| Grooming Products | 0.0084*** | 0.0254 | 177,107 | 0.0117*** | 0.0022 | 92,766 |
| Laundry Detergents | 0.0102*** | 0.0369 | 166,698 | 0.0164*** | 0.0031 | 106,067 |
| Oatmeal | 0.0246*** | 0.0159 | 55,650 | 0.0185** | 0.0086 | 24,333 |
| Paper Towels | 0.0362*** | 0.1112 | 71,451 | 0.0106 | 0.0095 | 44,753 |
| Refrigerated Juices | 0.0348*** | 0.0371 | 195,097 | 0.0161*** | 0.0047 | 111,768 |
| Shampoos | 0.0096*** | 0.0136 | 174,176 | 0.0068*** | 0.0015 | 87,602 |
| Snack Crackers | 0.0292*** | 0.0661 | 253,228 | 0.0252*** | 0.0037 | 145,437 |
| Soap | 0.0236*** | 0.0198 | 94,977 | 0.0218*** | 0.0061 | 57,402 |
| Soft Drinks | 0.0139*** | 0.1117 | 1,037,125 | 0.0048* | 0.0028 | 313,493 |
| Toothbrushes | 0.0144*** | 0.0322 | 83,428 | 0.0105*** | 0.0028 | 41,952 |
| Toothpastes | 0.0091*** | 0.0354 | 189,477 | 0.0069 | 0.0042 | 74,840 |
| **Average** | **0.0195** | **0.0035** | **217,143** | **0.0146** | **0.0044** | **112,290** |

<u>Notes</u>: The table reports the results of category-level fixed effect regressions of the probability of a small price change. We estimate separate regressions for price increases and for price decreases. The dependent variable is "small price change," which equals 1 if a price change of product $i$ in store $s$ at time $t$ is less or equal to 10¢, and 0 otherwise. The main independent variables are the log of average sales volume of product $i$ in store $s$ over the sample period and the log of the average revenue of product $i$ in store $s$ over the sample period. The regressions also includes fixed effects for years, months, stores, and products. We estimate separate regressions for each product category, clustering the errors by product. * $p < 10\%$, ** $p < 5\%$, *** $p < 1\%$

Table S2. Category-level regressions of small price changes and sales volume, price increases vs. price decreases, with extra controls

| Category | Price Increase | | | Price Decrease | | |
|---|---|---|---|---|---|---|
| | Coefficient | Std. | Obs. | Coefficient | Std. | Obs. |
| Analgesics | 0.0097*** | 0.0024 | 93,254 | 0.0159*** | 0.0031 | 51,207 |
| Bath Soap | 0.0201*** | 0.0064 | 9,877 | 0.0042 | 0.006 | 5,418 |
| Bathroom Tissues | 0.0086 | 0.0058 | 96,660 | 0.0124 | 0.0079 | 52,781 |
| Beer | 0.0167*** | 0.0014 | 128,309 | 0.0077*** | 0.0007 | 162,311 |
| Bottled Juice | 0.0175*** | 0.0037 | 298,844 | 0.0172*** | 0.0055 | 197,713 |
| Canned Soup | 0.016*** | 0.004 | 334,515 | 0.0023 | 0.0063 | 161,028 |
| Canned Tuna | 0.0125*** | 0.0043 | 110,869 | 0.0125*** | 0.0042 | 102,174 |
| Cereals | 0.0164*** | 0.0028 | 262,840 | 0.0053 | 0.0048 | 94,280 |
| Cheese | 0.0102*** | 0.0022 | 506,336 | 0.0082*** | 0.003 | 289,814 |
| Cigarettes | 0.0131*** | 0.0028 | 27,370 | -0.0013 | 0.0037 | 8,787 |
| Cookies | 0.0199*** | 0.0016 | 440,768 | 0.0178*** | 0.0019 | 247,993 |
| Crackers | 0.0239*** | 0.0025 | 152,814 | 0.0237*** | 0.0034 | 92,371 |
| Dish Detergent | 0.0225*** | 0.0033 | 120,854 | 0.0245*** | 0.0044 | 68,779 |
| Fabric Softener | 0.0094*** | 0.0034 | 110,126 | 0.0096** | 0.0047 | 70,930 |
| Front-End-Candies | 0.0124*** | 0.0032 | 168,056 | -0.0082*** | 0.0027 | 110,797 |
| Frozen Dinners | 0.0231*** | 0.0024 | 142,131 | 0.0456*** | 0.004 | 61,060 |
| Frozen Entrees | 0.0193*** | 0.0015 | 593,786 | 0.023*** | 0.0022 | 271,046 |
| Frozen Juices | 0.0142*** | 0.0036 | 201,311 | 0.0232*** | 0.0053 | 107,506 |
| Grooming Products | 0.0117*** | 0.0017 | 177,107 | 0.0149*** | 0.0022 | 92,766 |
| Laundry Detergents | 0.0068*** | 0.0023 | 166,698 | 0.0102*** | 0.0031 | 106,067 |
| Oatmeal | 0.0193*** | 0.0062 | 55,650 | 0.007 | 0.0088 | 24,333 |
| Paper Towels | 0.0332*** | 0.0083 | 71,451 | 0.0116 | 0.0102 | 44,753 |
| Refrigerated Juices | 0.0221*** | 0.0034 | 195,097 | 0.0105** | 0.0043 | 111,768 |
| Shampoos | 0.0128*** | 0.0011 | 174,176 | 0.0086*** | 0.0015 | 87,602 |
| Snack Crackers | 0.0244*** | 0.0025 | 253,228 | 0.0247*** | 0.0036 | 145,437 |
| Soap | 0.016*** | 0.0038 | 94,977 | 0.0146*** | 0.0057 | 57,402 |
| Soft Drinks | 0.0126*** | 0.0017 | 1,037,125 | 0.0052* | 0.0027 | 313,493 |
| Toothbrushes | 0.0154*** | 0.0021 | 83,428 | 0.0119*** | 0.0028 | 41,952 |
| Toothpastes | 0.0107*** | 0.0019 | 189,477 | 0.0058* | 0.003 | 74,840 |
| **Average** | **0.0162** | **0.0032** | **217,143** | **0.0127** | **0.0042** | **112,290** |

The table reports the results of category-level fixed effect regressions of the probability of a small price change. We estimate separate regressions for price increases and for price decreases. The dependent variable is "small price change," which equals 1 if a price change of product $i$ in store $s$ at time $t$ is less or equal to 10¢, and 0 otherwise. The main independent variables are the log of average sales volume of product $i$ in store $s$ over the sample period and the log of the average revenue of product $i$ in store $s$ over the sample period. The regressions also include the following independent variables: percentage changes in the wholesale price, a dummy for sale and bounce-back prices, as well as fixed effects for years, months, stores, and products. We estimate separate regressions for each product category, clustering the errors by product. * $p < 10\%$, ** $p < 5\%$, *** $p < 1\%$

Table S3. Category-level regressions of small price changes and sales volume, price increases vs. price decreases, with extra controls and a dummy for 9-ending prices

| Category | Price Increase | | | Price Decrease | | |
|---|---|---|---|---|---|---|
| | Coefficient | Std. | Obs. | Coefficient | Std. | Obs. |
| Analgesics | 0.0097*** | 0.0024 | 93,254 | 0.0155*** | 0.0031 | 51,207 |
| Bath Soap | 0.0221*** | 0.0064 | 9,877 | 0.0044 | 0.0061 | 5,418 |
| Bathroom Tissues | 0.0091 | 0.0058 | 96,660 | 0.0119 | 0.0079 | 52,781 |
| Beer | 0.0167*** | 0.0014 | 128,309 | 0.0077*** | 0.0008 | 162,311 |
| Bottled Juice | 0.0175*** | 0.0037 | 298,844 | 0.0155*** | 0.0056 | 197,713 |
| Canned Soup | 0.0182*** | 0.0039 | 334,515 | 0.0039 | 0.0062 | 161,028 |
| Canned Tuna | 0.0123*** | 0.0043 | 110,869 | 0.0125*** | 0.0042 | 102,174 |
| Cereals | 0.0164*** | 0.0028 | 262,840 | 0.0049 | 0.0048 | 94,280 |
| Cheese | 0.01*** | 0.0022 | 506,336 | 0.0074*** | 0.003 | 289,814 |
| Cigarettes | 0.0129*** | 0.0028 | 27,370 | -0.0011 | 0.0038 | 8,787 |
| Cookies | 0.02*** | 0.0016 | 440,768 | 0.0174*** | 0.0019 | 247,993 |
| Crackers | 0.0241*** | 0.0025 | 152,814 | 0.0232*** | 0.0034 | 92,371 |
| Dish Detergent | 0.0227*** | 0.0033 | 120,854 | 0.0236*** | 0.0042 | 68,779 |
| Fabric Softener | 0.0094*** | 0.0034 | 110,126 | 0.0098** | 0.0047 | 70,930 |
| Front-End-Candies | 0.0143*** | 0.0032 | 168,056 | -0.0076*** | 0.0026 | 110,797 |
| Frozen Dinners | 0.0244*** | 0.0023 | 142,131 | 0.0453*** | 0.004 | 61,060 |
| Frozen Entrees | 0.0193*** | 0.0015 | 593,786 | 0.0235*** | 0.0022 | 271,046 |
| Frozen Juices | 0.015*** | 0.0036 | 201,311 | 0.0226*** | 0.0052 | 107,506 |
| Grooming Products | 0.0118*** | 0.0017 | 177,107 | 0.0144*** | 0.0022 | 92,766 |
| Laundry Detergents | 0.0071*** | 0.0023 | 166,698 | 0.0102*** | 0.0031 | 106,067 |
| Oatmeal | 0.0195*** | 0.0063 | 55,650 | 0.0052 | 0.0089 | 24,333 |
| Paper Towels | 0.0336*** | 0.0086 | 71,451 | 0.0114 | 0.0099 | 44,753 |
| Refrigerated Juices | 0.0221*** | 0.0034 | 195,097 | 0.0094** | 0.0044 | 111,768 |
| Shampoos | 0.0128*** | 0.0011 | 174,176 | 0.0087*** | 0.0015 | 87,602 |
| Snack Crackers | 0.0244*** | 0.0026 | 253,228 | 0.0246*** | 0.0036 | 145,437 |
| Soap | 0.016*** | 0.0038 | 94,977 | 0.0172*** | 0.0055 | 57,402 |
| Soft Drinks | 0.0125*** | 0.0016 | 1,037,125 | 0.0047* | 0.0025 | 313,493 |
| Toothbrushes | 0.0154*** | 0.0021 | 83,428 | 0.0104*** | 0.0028 | 41,952 |
| Toothpastes | 0.0107*** | 0.0019 | 189,477 | 0.0057* | 0.003 | 74,840 |
| **Average** | **0.0166** | **0.0032** | **217,143** | **0.0125** | **0.0042** | **112,290** |

The table reports the results of category-level fixed effect regressions of the probability of a small price change. We estimate separate regressions for price increases and for price decreases. The dependent variable is "small price change," which equals 1 if a price change of product $i$ in store $s$ at time $t$ is less or equal to 10¢, and 0 otherwise. The main independent variables are the log of average sales volume of product $i$ in store $s$ over the sample period and the log of the average revenue of product $i$ in store $s$ over the sample period. The regressions also include the following independent variables: percentage changes in the wholesale price, a dummy for sale and bounce-back prices, and a dummy for 9-ending prices, as well as fixed effects for years, months, stores, and products. We estimate separate regressions for each product category, clustering the errors by product. * $p$ < 10%, ** $p$ < 5%, *** $p$ < 1%

Table S4. Category-level regressions of small price changes and sales volume, price increases vs. price decreases, focusing on regular prices

| Category | Price Increase | | | Price Decrease | | |
|---|---|---|---|---|---|---|
| | Coefficient | Std. | Obs. | Coefficient | Std. | Obs. |
| Analgesics | 0.0157*** | 0.0047 | 33,833 | 0.0346*** | 0.0087 | 11,117 |
| Bath Soap | 0.052*** | 0.0143 | 2,610 | 0.0209 | 0.0263 | 598 |
| Bathroom Tissues | 0.0303*** | 0.0105 | 27,822 | 0.0188* | 0.0096 | 19,219 |
| Beer | 0.0523*** | 0.0049 | 16,369 | 0.0311*** | 0.0059 | 10,979 |
| Bottled Juice | 0.0202*** | 0.006 | 84,037 | 0.0219** | 0.0087 | 49,677 |
| Canned Soup | 0.0188*** | 0.0044 | 121,223 | -0.0004 | 0.007 | 55,012 |
| Canned Tuna | 0.018*** | 0.0059 | 35,488 | 0.0218*** | 0.006 | 28,673 |
| Cereals | 0.0174*** | 0.0043 | 112,141 | 0.0159*** | 0.0062 | 43,226 |
| Cheese | 0.0163*** | 0.0031 | 145,646 | 0.0022 | 0.0048 | 79,243 |
| Cigarettes | 0.0123*** | 0.0031 | 24,297 | -0.0053 | 0.0042 | 5,965 |
| Cookies | 0.0371*** | 0.0033 | 97,877 | 0.0227*** | 0.0036 | 34,611 |
| Crackers | 0.0381*** | 0.0056 | 35,793 | 0.0266*** | 0.0081 | 14,236 |
| Dish Detergent | 0.0285*** | 0.0047 | 29,978 | 0.0256*** | 0.0056 | 23,311 |
| Fabric Softener | 0.016*** | 0.0057 | 31,744 | 0.035*** | 0.0064 | 24,490 |
| Front-End-Candies | 0.0126*** | 0.0028 | 65,667 | -0.0001 | 0.0027 | 45,968 |
| Frozen Dinners | 0.0609*** | 0.0067 | 19,262 | 0.06*** | 0.0063 | 18,265 |
| Frozen Entrees | 0.0411*** | 0.0036 | 117,948 | 0.0266*** | 0.0034 | 95,597 |
| Frozen Juices | 0.025*** | 0.0061 | 50,141 | 0.027*** | 0.0068 | 37,778 |
| Grooming Products | 0.0268*** | 0.0043 | 37,589 | 0.0169* | 0.0094 | 14,230 |
| Laundry Detergents | 0.0128*** | 0.0044 | 47,061 | 0.0193*** | 0.0048 | 38,123 |
| Oatmeal | 0.0297*** | 0.0107 | 22,934 | 0.0258*** | 0.0116 | 13,109 |
| Paper Towels | 0.0368*** | 0.0096 | 16,360 | 0.0219* | 0.0112 | 12,920 |
| Refrigerated Juices | 0.0329*** | 0.0056 | 44,566 | 0.0121* | 0.0067 | 27,465 |
| Shampoos | 0.0272*** | 0.0035 | 29,135 | 0.0171*** | 0.0065 | 11,861 |
| Snack Crackers | 0.0445*** | 0.0046 | 55,142 | 0.0231*** | 0.0076 | 23,439 |
| Soap | 0.038*** | 0.0064 | 28,658 | 0.0174** | 0.0084 | 18,171 |
| Soft Drinks | 0.0539*** | 0.0033 | 86,187 | 0.0112*** | 0.0035 | 69,817 |
| Toothbrushes | 0.0351*** | 0.0056 | 18,109 | 0.0349*** | 0.0082 | 6,846 |
| Toothpastes | 0.0277*** | 0.0045 | 41,918 | 0.0341*** | 0.0083 | 14,924 |
| **Average** | **0.0303** | **0.0056** | **51,018** | **0.0213** | **0.0075** | **29,271** |

The table reports the results of category-level fixed effect regressions of the probability of a small price change. We estimate separate regressions for price increases and for price decreases. The dependent variable is "small price change," which equals 1 if a price change of product $i$ in store $s$ at time $t$ is less or equal to 10¢, and 0 otherwise. The main independent variables are the log of average sales volume of product $i$ in store $s$ over the sample period and the log of the average revenue of product $i$ in store $s$ over the sample period. The regressions also include the following independent variables: percentage changes in the wholesale price, a dummy for sale and bounce-back prices, and a dummy for 9-ending prices, as well as fixed effects for years, months, stores, and products. We estimate separate regressions for each product category, clustering the errors by product. * $p < 10\%$, ** $p < 5\%$, *** $p < 1\%$